\documentclass[11pt,a4paper]{article} 
\usepackage{amsmath,amsfonts,latexsym,amssymb,hhline,stmaryrd,color,verbatim,graphicx,epstopdf,slashed,multirow}
\usepackage{jheppub}
\pdfoutput=1

\newcommand{\FR}{{\sc FeynRules}}
\newcommand{\MG}{{\sc MG5\_aMC}}
\newcommand{\MA}{{\sc MadAnalysis~5}}
\newcommand{\MD}{{\sc MadDM}}
\newcommand{\MO}{{\sc MicrOMEGAs}}
\newcommand{\MN}{{\sc MultiNest}}
\newcommand{\PY}{{\sc Pythia 6}}
\newcommand{\DEL}{{\sc Delphes 3}}
\newcommand{\FJ}{{\sc FastJet}}
\newcommand{\MET}{{\slashed{E}_T}}
\newcommand{\GeV}{{\rm\ GeV}}

\newcommand{\TeV}{{\rm\ TeV}}
\newcommand{\fb}{{\rm\ fb}}
\newcommand{\ie} {{\it i.e.}}
\newcommand{\eg} {{\it e.g.}}
\newcommand{\etc} {{\it etc.}}

\newcommand{\be}{\begin{equation}} \newcommand{\ee}{\end{equation}}
\newcommand{\ba}{\begin{eqnarray}} \newcommand{\ea}{\end{eqnarray}}

\newcommand{\dm}{X}
\newcommand{\dmbar}{\bar{X}}
\newcommand{\y}{Y_0}
\newcommand{\mdm}{m_X} 
\newcommand{\my}{m_Y} 
\newcommand{\gx}{g_X}
\newcommand{\wy}{\Gamma_Y} 

\newcommand{\gsm}{g_{t}} 
\newcommand{\gglu}{g_{g} }
\newcommand{\ggam}{g_{\gamma} }

\definecolor{jcyan}{RGB}{40,140,160}

\preprint{CP3-16-26, LPSC16099, TTK-16-13} 

\title{A comprehensive approach to dark matter studies: exploration of simplified top-philic models}

\author[1]{Chiara Arina}
\author[1]{\!\!, Mihailo Backovi\'c}
\author[2]{\!\!, Eric Conte}
\author[3,4]{\!\!, Benjamin Fuks}
\author[5,6]{\!\!, Jun Guo}
\author[7]{\!\!, Jan~Heisig}
\author[1]{\!\!, Beno\^{i}t Hespel}
\author[7]{\!\!, Michael Kr\"{a}mer}
\author[1]{\!\!, Fabio Maltoni}
\author[1]{\!\!, Antony Martini}
\author[8,9]{\!\!, Kentarou Mawatari}
\author[10]{\!\!, Mathieu Pellen}
\author[1]{\! and Eleni Vryonidou}

\affiliation[1]{Centre for Cosmology, Particle Physics and Phenomenology (CP3), Universit\'e catholique de Louvain, B-1348 Louvain-la-Neuve, Belgium}
\affiliation[2]{Groupe de Recherche de Physique des Hautes \'Energies (GRPHE), Universit\'e de Haute-Alsace, IUT Colmar, F-68008 Colmar Cedex, France}
\affiliation[3]{Sorbonne Universit\'es, UPMC Univ.~Paris 06, UMR 7589, LPTHE, F-75005, Paris, France}
\affiliation[4]{CNRS, UMR 7589, LPTHE, F-75005, Paris, France}
\affiliation[5]{State Key Laboratory of Theoretical Physics, Institute of Theoretical Physics, Chinese Academy of Sciences, Beijing 100190, P.~R.~China}
\affiliation[6]{Institut Pluridisciplinaire Hubert Curien/D\'epartement Recherches Subatomiques, Universit\'e de Strasbourg/CNRS-IN2P3, F-67037 Strasbourg, France}
\affiliation[7]{Institute for Theoretical Particle Physics and Cosmology, RWTH Aachen University, D-52056 Aachen, Germany}
\affiliation[8]{Laboratoire de Physique Subatomique et de Cosmologie, Universit\'e Grenoble-Alpes,\\ CNRS/IN2P3, 53 Avenue des Martyrs, F-38026 Grenoble, France}
\affiliation[9]{Theoretische  Natuurkunde  and  IIHE/ELEM,  Vrije  Universiteit Brussel, and International Solvay Institutes,
Pleinlaan 2, B-1050  Brussels, Belgium}
\affiliation[10]{Universit\"at W\"urzburg, Institut f\"ur Theoretische Physik und Astrophysik, Emil-Hilb-Weg 22, 97074 W\"urzburg, Germany}

\abstract{
Studies of dark matter lie at  the interface of collider physics, astrophysics and cosmology. Constraining models featuring dark matter candidates entails the  capability to provide accurate predictions for large sets of observables and  compare them to a wide  spectrum of data. We present a framework which, starting from a model lagrangian, allows one to consistently and systematically make predictions, as well as to confront those predictions with a multitude of experimental results.
As an application, we consider a class of simplified dark matter models where a scalar mediator couples only to the top quark and a fermionic dark sector ($i.e.$ the simplified top-philic dark matter model). We study in detail the complementarity of relic density, direct/indirect detection and collider searches in constraining the multi-dimensional model parameter space, and efficiently identify regions where individual approaches to dark matter detection provide the most stringent bounds. In the context of collider studies of dark matter,  we point out the complementarity of LHC searches in probing different regions of the model parameter space with final states involving top quarks, photons, jets and/or missing energy.  Our study of dark matter production at the LHC goes beyond the tree-level approximation and we show examples of how higher-order corrections to dark matter production processes can affect the interpretation of the experimental results.}

\begin{document}
\vspace*{-1.5cm}
\maketitle


\section{Introduction}
\label{sec:intro}
Evidence for the existence of dark matter (DM), although indirect, is quite convincing~\cite{Bertone:2004pz,Bertone:1235368,Drees:2012ji}. 
Measurements
of the cosmic microwave background and baryonic acoustic oscillations predict a
dominant dark matter component in the matter budget of the Universe (in the
framework of standard cosmology). In addition, detection of gravitational anomalies,
such as the flattening of galaxy rotation curves and
the presence of gravitational lensing in the absence of visible matter
(\eg~the bullet cluster~\cite{Clowe:2006eq}), strongly favours
gravitational interactions of dark matter as plausible explanations.

The many hints for dark matter sparked a huge endeavour to detect it and measure
its properties, leading to a number of experiments and searches which exploit very different ideas and approaches to dark matter detection.
The experiments can be broadly grouped into three categories:

\begin{itemize} 
\item  A wide range of underground nuclear
recoil experiments aimed at detecting galactic dark matter scattering off atomic
nuclei; 
\item Searches for dark matter
annihilation in the galaxy or nearby dense sources via measurements of, for
instance, gamma-rays; 
\item Collider searches in channels with large missing transverse energy ($\MET$). 
\end{itemize}

 However, despite an enormous
experimental effort, the detection of the dark matter particles remains elusive. In fact,
 there is no clear indication that dark matter interacts with ordinary matter
via forces other than gravity, and 
current experimental results are not able to
put stringent bounds on the dark matter properties and couplings in a model-independent
way.

As so little is known about the true nature of dark matter, it is a useful strategy to try and constrain viable dark matter scenarios in the most model-independent way (\ie~via
simplified models), confronting them with results from collider experiments,
direct dark matter searches, astrophysical observations and cosmology. 
If or when a signal is observed, the aforementioned approach will help us to
determine more accurately both the particle properties (mass, couplings, \etc)
and astroparticle properties (halo properties, thermal relic density, \etc) of dark matter. Conversely, if searches result only in limits on dark matter parameters,
combining constraints from different approaches aids us in excluding
specific scenarios and hence narrow down the scope of viable dark matter
theories.

Recent collider searches have focused mostly on studies of dark matter in the
simplified model framework, where a single dark matter candidate of arbitrary
spin couples to visible matter (\eg\ quarks) via an $s$-channel or a $t$-channel mediator, whose quantum numbers are fixed by assumed local and global symmetries~\cite{Abercrombie:2015wmb}. The minimal
implementations of simplified dark matter models involve four basic
parameters: the mass $\mdm$ of the dark matter particle, the mass $\my$ of the
mediator, the coupling constant $\gx$ of the dark matter to the mediator and the
universal coupling $g_{\rm SM}$ of the mediator to the visible sector (the width of the mediator is a derived quantity). Fast and
efficient studies of the full simplified model parameter space require parameter
scanning technology beyond simple sequential grids, due to the relatively high
dimensionality of the parameter space. Past studies of simplified dark matter
models have hence been limited to explorations of the parameter space in
two-dimensional projections while keeping the remaining parameters fixed (see
\eg~the works of refs.~\cite{Buckley:2014fba, Haisch:2015ioa, Heisig:2015ira, Bell:2015rdw, %
Brennan:2016xjh, Pree:2016hwc, Jackson:2013pjq, Abdallah:2015ter, %
Godbole:2015gma, Xiang:2015lfa,DiFranzo:2013vra} and the references therein).

In this paper we illustrate how comprehensive studies of simplified dark matter models can be performed,
exploring their full four-dimensional parameter space while taking into account
constraints from collider physics, astroparticle physics and cosmology. For
concreteness, we focus on a class of simplified models where the dark matter
dominantly couples via a scalar mediator to top quarks (\ie~`top-philic dark matter' scenarios). Yet, the methodology we employ is general and can be applied to other scenarios as well. We  provide detailed examinations of the two-dimensional
projections of the full  parameter space, and we demonstrate
that striking features in the structure of the viable parameter space emerge
through the combination of all current constraints. We also stress that in addition to collider searches for dark matter in channels with large missing energy, in this study we also consider resonance searches in channels with fully reconstructed final states, which can be useful to constrain the properties of the mediators.

We perform the study of simplified top-philic dark matter models
by using a combination of simulation tools, including the {\sc MadGraph5\_aMC@NLO} (\MG\ shorthand) event generator~\cite{Alwall:2014hca}, the \FR\
package~\cite{Alloul:2013bka,Degrande:2014vpa}, 
the \MA\ platform~\cite{Conte:2012fm,%
Conte:2014zja,Dumont:2014tja}, the \DEL~detector simulator~\cite{deFavereau:2013fsa}  and the \MD\ program~\cite{Backovic:2013dpa, Backovic:2015cra},
together with an efficient parameter sampling technology based on the \MN\
algorithm~\cite{Feroz:2007kg,Feroz:2008xx}. We
explore the full four-dimensional parameter space of
the model in the light of existing collider and astroparticle constraints. Our
analysis thus also represents a proof of concept for a unified numerical
framework for comprehensive dark matter studies at the interface of collider
physics, astrophysics and cosmology. This has direct implications for dark matter
searches at colliders, as comprehensive phenomenological studies of dark matter models
can be used to drive the experimental
efforts towards the regions of the parameter space that are not already ruled out by
astrophysical and cosmological constraints. 
In addition, we have also implemented
previously unavailable experimental analyses into the \MA\ platform, providing
an added benefit of our work for future collider studies which go beyond
searches for dark matter.

The article is organised as follows. Section~\ref{sec:model} describes the details
of the simplified top-philic dark matter model under consideration and discusses
the constraints on the model parameter space that are
implemented in our analysis setup. All cosmology and astrophysics constraints
are discussed in section~\ref{sec:cosmo}. More precisely, the relic density
constraints are illustrated
in section~\ref{sec:oh2}. We discuss the direct detection constraints in section~\ref{sec:dd},
while constraints from gamma-ray flux measurements are detailed in
section~\ref{sec:dm_id}. Collider constraints are investigated in
section~\ref{sec:collider}. We study constraints from searches with and without missing transverse energy in section~\ref{sec:met} and \ref{sec:nonmet} respectively. Section~\ref{sec:combined} is then dedicated to a detailed discussion of the
overall combined information coming from all the considered data. Before concluding in
section~\ref{sec:concl}, we
briefly elaborate in section~\ref{sec:diphoton} on whether the potential diphoton
excess observed by the ATLAS and CMS collaborations at a diphoton invariant mass of
$m_{\gamma \gamma} \simeq 750$~GeV~\cite{atlasDigamma,CMS:2015dxe} could be interpreted within the
considered class of simplified top-philic dark matter models. We provide more information on the
mediator width in appendix~\ref{sec:ywidth}. As a validation of our calculations, we perform a
detailed comparison between \MD\ and \MO\ in appendix~\ref{sec:cc}, give details
on the annihilation cross section of dark matter in the top-philic model in
appendix~\ref{sec:dmann} and present the validation of the CMS
 $t\bar{t}+\MET$ and monojet implementation in the \MA\ framework in
appendix~\ref{sec:recast}.

\section{Simplified top-philic dark matter model and its numerical implementation}\label{sec:model}
The simplified top-philic dark matter model that we consider is constructed by
supplementing the Standard Model (SM) with a Dirac-type fermionic dark matter candidate
$\dm$ and a scalar mediator $\y$. The interactions of the two particles are
described by the Lagrangian
\begin{equation}
	{\cal L}^{\y}_{t,\dm} = -\Big(\gsm\, \frac{y_t}{\sqrt{2}}\,  \bar{t}t +  \gx\, \dmbar \dm \Big)\y\,, \label{eq:lagrangian}
\end{equation}
where the new physics interaction strengths are denoted by $\gsm$ and $\gx$ for
the mediator couplings to the Standard Model sector and to dark matter
respectively. We have assumed an ultraviolet-complete description of the scalar
theory where the mediator couples to quarks with a strength proportional to the
Standard Model Yukawa couplings, so that we neglect all light quark flavour couplings and only
include the coupling of the mediator to the top quark,
$y_t = \sqrt{2}m_t/v$ where $v=246$~GeV is
the Higgs vacuum expectation value and $m_t$ is the top quark mass.
Note that the model in eq.~\eqref{eq:lagrangian} is neither complete, nor stable under radiative corrections. Couplings to the top quark induce a mixing with the standard model Higgs, which we set to zero by construction. In addition, loop corrections will also generate finite couplings to pairs of $W$ and $Z$ bosons, which we will omit in the following.

The model contains four free parameters (two couplings and two masses),
\begin{equation}
 \{\gsm,\,\gx,\,\mdm,\,\my\}\,,
\end{equation}
%
while the width $\wy$ is fixed by the remaining model parameters. 
In addition to the Lagrangian of
eq.~\eqref{eq:lagrangian}, we could also have considered mediator couplings to
leptons. They however cannot be well constrained by LHC searches and dark matter
direct detection data, and we have excluded them from our model description.
We will nonetheless comment on their relevance for relic density predictions and
dark matter indirect detection signals in sections~\ref{sec:oh2}
and~\ref{sec:dm_id}.

The Lagrangian of eq.~\eqref{eq:lagrangian} induces dimension-five couplings of
the mediator to gluons and photons via loop diagrams of top quarks.  The loop-induced
operators can be relevant in the context of both astrophysical and collider
searches for dark matter. The couplings of the mediator to gluons and photons
are given, at the leading order (LO), by the effective operators
\begin{equation}
	{\cal L}^{\y}_{g} =-\frac{1}{4} \frac{\gglu(Q^2)}{v}\, G^a_{\mu\nu} G^{a,\mu\nu} \y
        \qquad\text{and}\qquad
	{\cal L}^{\y}_{\gamma} =-\frac{1}{4} \frac{\ggam(Q^2)}{v}\, F_{\mu\nu} F^{\mu\nu} \y\,,
\end{equation}
with the effective couplings being
\begin{equation}
	\gglu(Q^2) = \gsm \frac{\alpha_{s}}{3\pi}\,\frac{3}{2}  F_S \Big( \frac{4 m^2_t}{Q^2}\Big)
        \qquad\text{and}\qquad
	\ggam(Q^2) = \gsm \frac{8\alpha_{e}}{9\pi}\,\frac{3}{2} F_S \Big( \frac{4 m^2_t}{Q^2}\Big)\,. \label{eq:couplings}
\end{equation}
In the above expressions, $Q^2$ denotes the virtuality of the $s$-channel resonance, while $F_S$ is  the one-loop form factor
\begin{equation}\label{eq:fsg}
F_S(x) = x\Big[1 + (1-x) \, {\rm \arctan}^2 \Big( \frac{1}{\sqrt{x-1}} \Big)\Big]\,,
\end{equation}
with $F_S(x)\to2/3$ for $x\gg1$.
Eq.~\eqref{eq:couplings} contrasts with the Standard Model Higgs case where the effective
Higgs-photon coupling receives contributions from vector-boson loop-diagrams
that are absent in our simplified dark matter model setup. As a result,
the gluon and photon effective couplings to $\y$ are characterised by a larger hierarchy compared to their Higgs counterparts. 

The tree-level partial decay widths of the scalar mediator are given
by
\begin{align}
 \Gamma(\y \to t \bar{t}) &= \gsm^2 \frac{3 y_t^2 \my}{16\pi} \beta_t^3\, \Theta(\my- 2 m_t)\,,  \label{eq:y0widthtt}\\
 \Gamma(\y \to \dm \dmbar) &= \gx^2 \frac{\my}{8 \pi} \beta_X^3\, \Theta(\my - 2 \mdm)\,, \label{eq:y0widthdm}
\end{align}
where $\beta_{t,\dm}=\sqrt{1-4m_{t,\dm}^2/\my^2}$ and $\Theta(x)$ is the
Heaviside step function, and we ignored the top quark width in the expression for $ \Gamma(\y \to t \bar{t})$.
The loop-induced $\y$ partial widths are
\begin{align}
 \Gamma(\y  \to g g) &= \gsm^2 \frac{\alpha_s^2 \my^3}{72 \pi^3 v^2} \Big| \frac{3}{2} F_S \Big( \frac{4 m_t^2}{\my^2} \Big) \Big|^2\,,  \label{eq:y0widthgg}\\
 \Gamma(\y  \to \gamma \gamma ) &= \gsm^2 \frac{\alpha_e^2 \my^3}{81 \pi^3 v^2} \Big| \frac{3}{2} F_S \Big( \frac{4 m_t^2}{\my^2} \Big) \Big|^2\,.  \label{eq:y0widthph}
\end{align}
The $\y$ partial width to photons 
is by construction always smaller than the partial decay width into a pair of gluons by
virtue of $\alpha_s^2/\alpha_e^2 \sim 100$.
In addition to a coupling suppression, other decay processes such as the loop-induced
$\y$ decays into $Z\gamma$, $ZZ$ and $hh$ final states receive a kinematic suppression. Couplings of $\y$ to $ZZ$ and $hh$ could also appear at tree level in our model, but in the spirit of simplified models,  we define them to be vanishing. 
In the following we hence safely approximate the total
decay width for the mediator to be the sum of eqs.~\eqref{eq:y0widthtt},
\eqref{eq:y0widthdm} and~\eqref{eq:y0widthgg}.

\begin{figure}[h!]
 \includegraphics[width=0.32\textwidth]{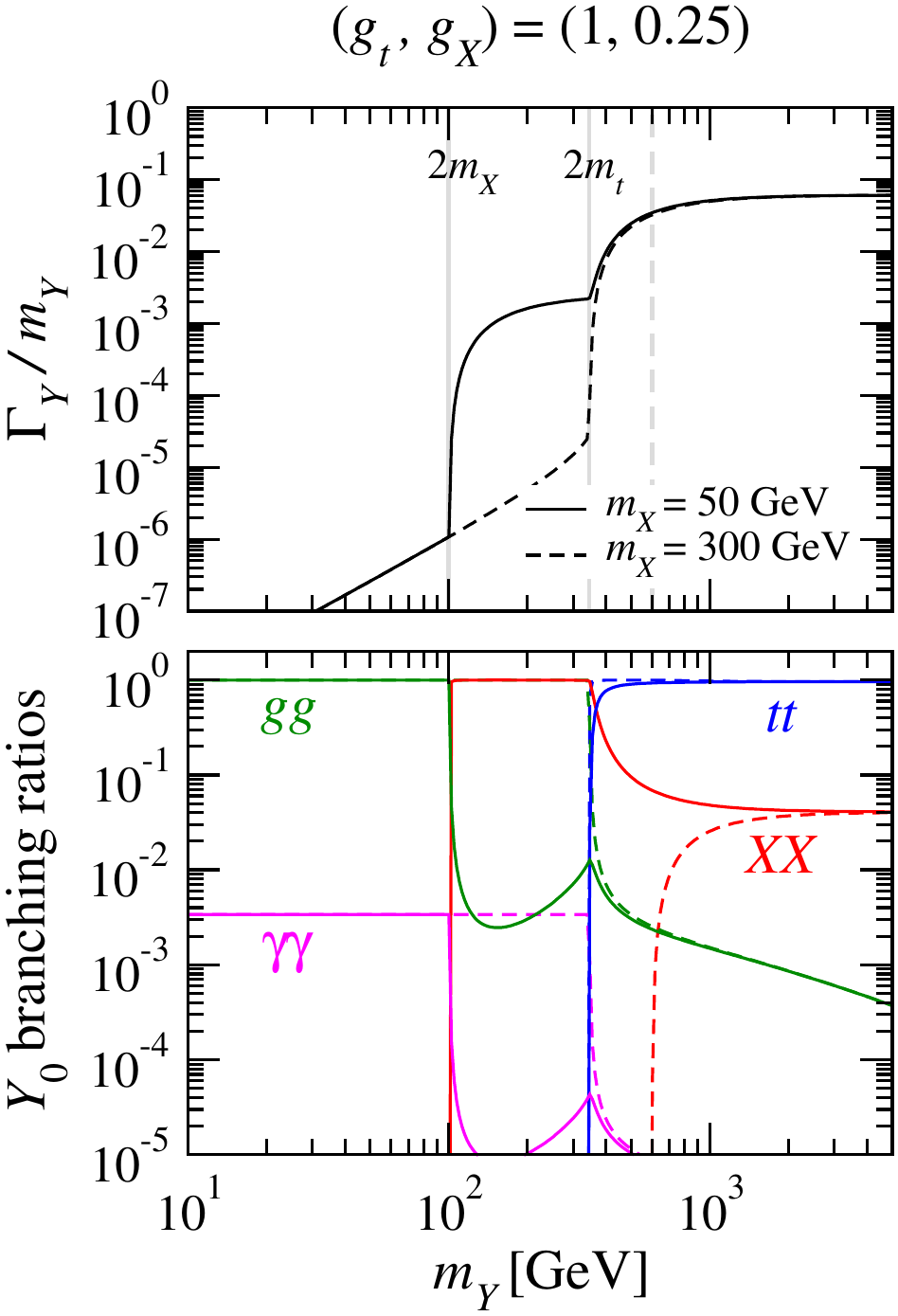}\quad
 \includegraphics[width=0.32\textwidth]{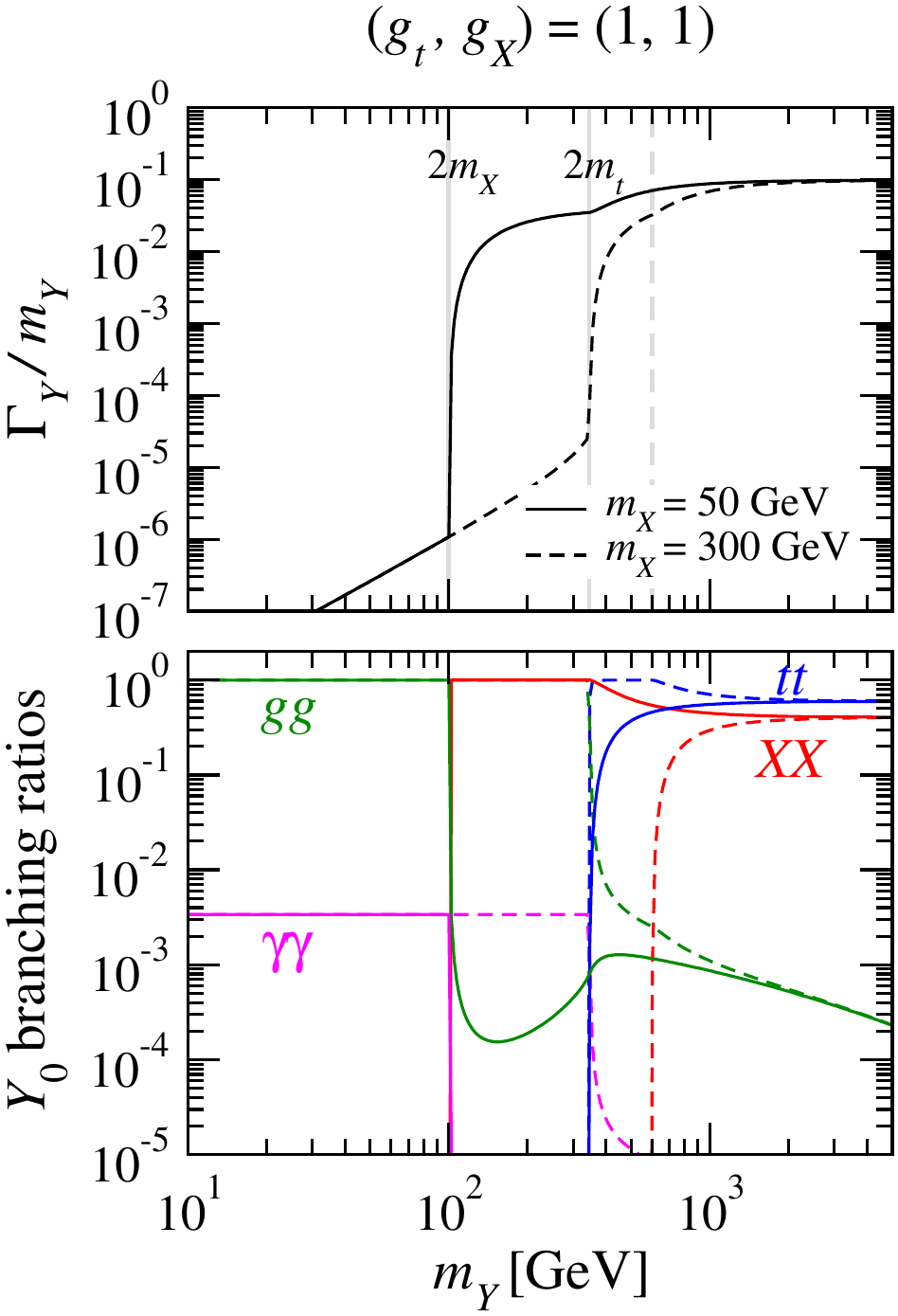}\quad
 \includegraphics[width=0.32\textwidth]{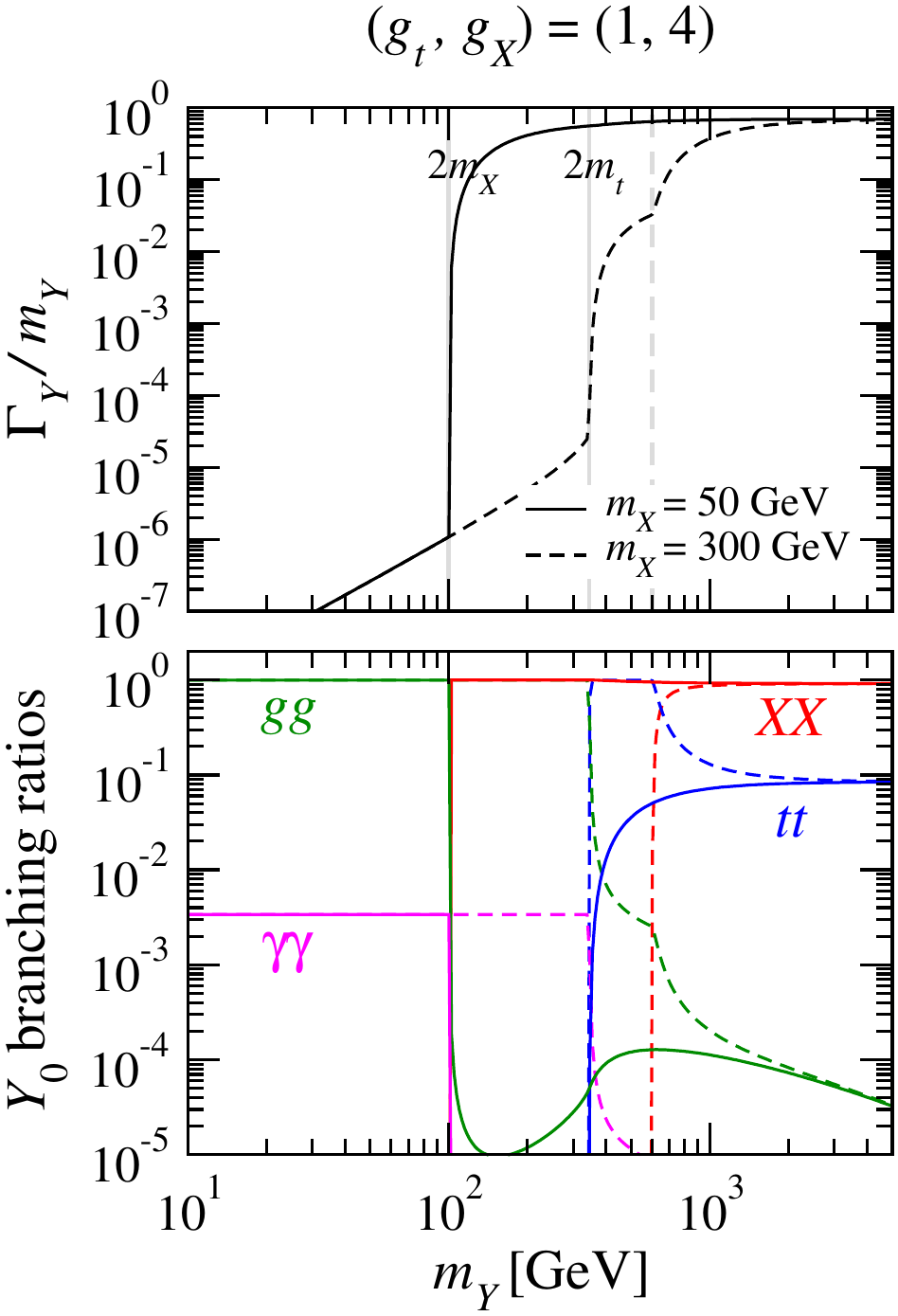}
\caption{Ratio of the mediator width to its mass $\wy/\my$ (upper
panels) and mediator branching ratios (lower panels) as a function of the mediator mass for different coupling
choices and a dark matter mass fixed to $\mdm=50$~GeV (solid lines) and 300~GeV
(dashed lines).}
\label{fig:width}
\end{figure}

The total decay width and the branching ratios of the mediator into $t\bar{t},
\dm \dmbar, gg$ and $\gamma\gamma$ final states are displayed in
figure~\ref{fig:width} for different choices of new physics couplings and
masses. Light mediators with masses below the top-quark pair or the dark matter pair decay
thresholds are narrow states, while above these thresholds, large $\wy/\my$
values are possible in particular for large couplings. For mediators with $\my \lesssim m_t, \mdm$, the dominant decay channel is into a pair of gluons.  In contrast, heavy mediators with mass $\my > m_t, \mdm$
decay predominantly into pairs of top quarks and/or dark matter particles, where the
exact details of the partial width values strongly depend on the masses and couplings. The
branching ratio of $\y$ to photons is always
suppressed, as argued above. We  present in appendix~\ref{sec:ywidth}
the dependence of the $\Gamma_Y/m_Y$ ratio on the $g_t$ and $g_X$ couplings for
different mass choices and on the $m_Y$ and $m_X$ masses for different coupling choices.

\begin{table}
\center
\begin{tabular}{l|p{1.30cm}clp{3.3cm}}
\hline\\[-4mm]
  & &  \raisebox{-.4\height}{\includegraphics[height=1cm]{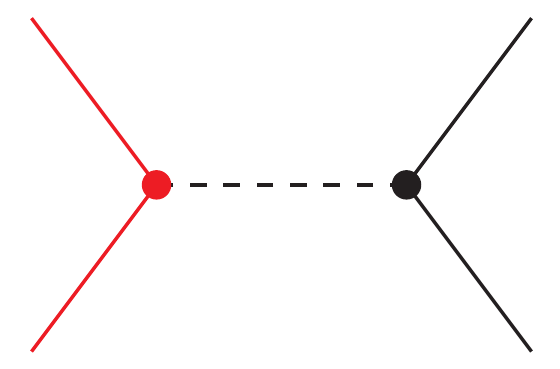}} & $\mdm>m_t$ &  \\
  Cosmology &relic$\qquad~$ indirect& \raisebox{-.4\height}{\includegraphics[height=1cm]{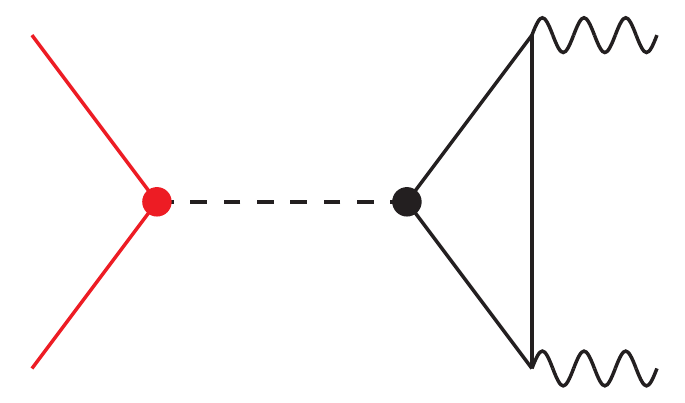}} & $\mdm<m_t$ & Planck, FermiLAT\\
  Astrophysics && \raisebox{-.4\height}{\includegraphics[height=1cm]{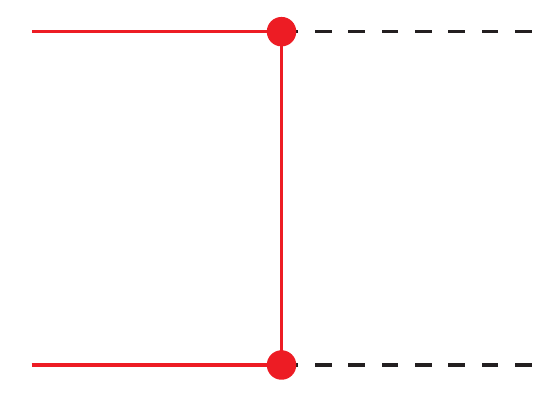} } & $\mdm>\my$ & \\[5mm]
\cline{2-5}\\[-2mm]
 & direct & \raisebox{-.4\height}{\includegraphics[height=1cm]{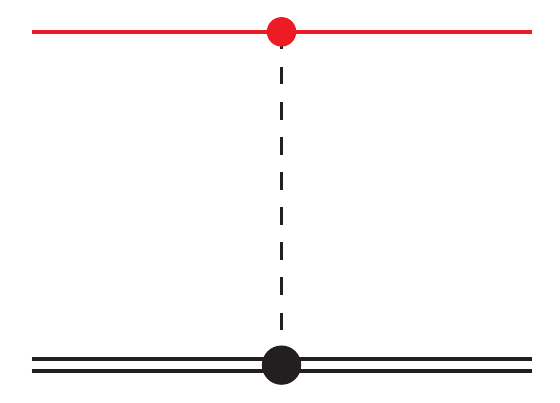}} & $\mdm>1$~GeV & LUX, CDMSLite \\[5mm]
\hline \\[-2mm]
 \multirow{5}{*}{Colliders}
 & \multirow{2}{*}{$\MET$}
  &  \raisebox{-.4\height}{\includegraphics[height=1cm]{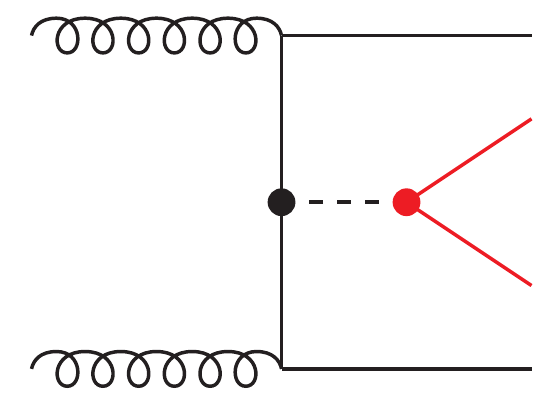}} & $\my>2\mdm$ & $+ t\bar t$ \\ 
  && \raisebox{-.4\height}{\includegraphics[height=1cm]{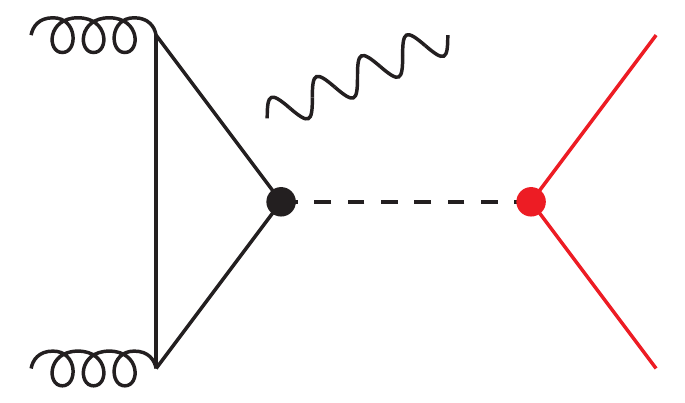}} & $\my>2\mdm$ & $+ j$,
       $+ Z$, $+ h$ \\[2mm]
\cline{2-5}\\[-2mm]
 & \multirow{3}{*}{no $\MET$}
  &   \raisebox{-.4\height}{\includegraphics[height=1cm]{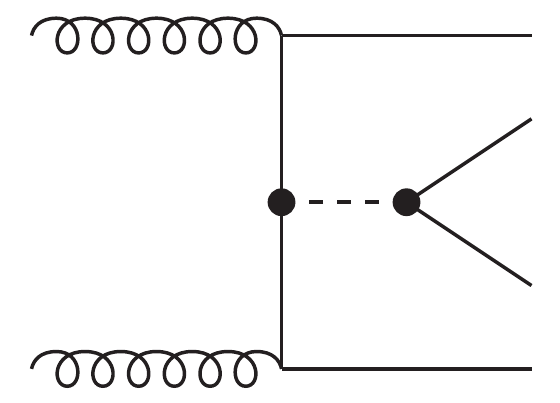}}
  & $\my>2m_t$ & 4$t$ \\
  &&  \raisebox{-.4\height}{\includegraphics[height=1cm]{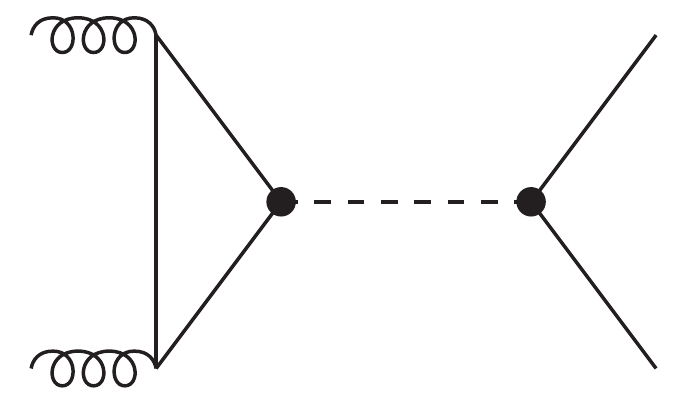}}
  & $\my>2m_t$ & $t\bar t$ \\
  &&  \raisebox{-.4\height}{\includegraphics[height=1cm]{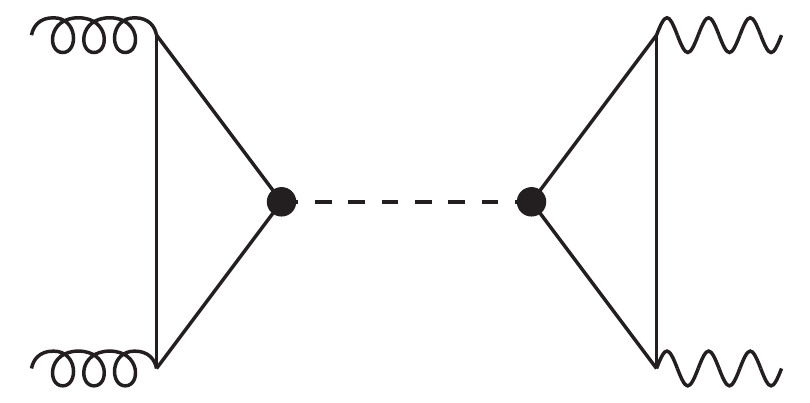}}
  & $\my<2m_X, 2m_t$ & $jj$, $\gamma\gamma$ \\[4mm]
\hline
\end{tabular}
\caption{Signatures of our simplified top-philic dark matter model. 
 }
\label{tab:sig}
\end{table}

Our top-philic dark matter model can be probed in different ways
including astrophysical and collider searches, as listed in table~\ref{tab:sig}.
The relative importance of the various searches depends on the hierarchy of the
dark matter, mediator and top-quark masses, as well as on the hierarchy between the
couplings. Starting with the dark matter relic density, the annihilation cross
section is dominated by subprocesses with
top-quark final states for $\mdm > m_t$, and by annihilation into gluons and to
a lesser extent photons for light dark matter particles with $\mdm < m_t$. If
the mediator is lighter than the dark matter state, an additional annihilation
channel into a pair of mediators can open up. The annihilation mechanisms into
top-quarks, gluons/photons and mediators moreover provide an opportunity to
indirectly search for dark matter,
\eg~in gamma-ray data. The interactions of the dark matter
particles with nuclei, relevant for direct detection experiments, proceed via
mediator exchanges. The mediator-nucleon coupling is in turn dominated by the
scattering off gluons through top-quark loops.

Dark matter production at the LHC proceeds either through the production of the
mediator in association with top quarks, or from gluon-fusion through top-quark
loops. 
Searches at the LHC can be classified into two categories regarding the nature
of the final states that can contain missing transverse energy $\MET$ or not. 
Searches involving missing
energy may include final state systems containing a top-quark pair and probe in
this way the associated production of a top-antitop-mediator system where the
mediator subsequently decays into a pair of dark matter particles.
Alternatively, the mediator can be produced via gluon fusion
through top-quark loops, where the probe of the associated events consists of tagging an extra
radiated object. This yields the well-known monojet, mono-$Z$ and mono-Higgs
signatures. We do not consider the monophoton channel, as photon emission is forbidden at LO in our simplified
model by means of charge conjugation invariance.
The second search category is related to final states without any missing energy, \ie~when the
mediator decays back into Standard Model particles. This
includes decays into top-quarks, leading to final states comprised of four top quarks,
into a top-quark pair, as well as into a dijet or a diphoton system
via a loop-induced decay.
This is, however, relevant only for on-shell (or close to on-shell) mediator production.

We proceed with a description of the numerical setup for our calculations. In the following sections, we explore the full four-dimensional model parameter
space and present results in terms of two-dimensional projections. We perform
the four-dimensional sampling using the \MN\ algorithm~\cite{Feroz:2007kg, Feroz:2008xx}, where we assume Jeffeys' prior on all the free parameters in order not to favour a particular mass or coupling scale. The choice of prior ranges for
the parameters is summarised in table~\ref{tab:param}, in which we have chosen
to limit the coupling values to a maximum of $\pi$ to ensure perturbativity.  We implement the relic density constraints into {\sc MultiNest} using 
 a Gaussian likelihood profile, while for the direct detection limits we assume a step
 likelihood function smoothed with half a Gaussian.  In addition, the sampling imposes that the model is consistent with values of $\wy$ such that the mediator $\y$ decays promptly within the LHC detectors. Table~\ref{tab:co}
summarises the constraints that we have imposed on the model parameter space.

\begin{table}
\begin{center} \renewcommand\arraystretch{1.2}
\begin{tabular}{c c}
\hline
{\sc MultiNest} parameter & Prior \\
\hline \hline
$\log(\mdm/\GeV)$ & $0 \to 3$ \\
$\log(\my/ \GeV)$ & $0 \to 3.7$ \\
$\log(\gx)$ & $-4 \to \log(\pi)$ \\
$\log(\gsm)$ & $-4 \to \log(\pi)$ \\
\hline
\end{tabular}
\end{center} \renewcommand\arraystretch{1.0}
\caption{{\sc MultiNest} parameters and prior ranges for the four free parameters. All priors are uniform over the indicated range.\label{tab:param}}
\end{table}

Throughout our study, we assume that $\dm$ is the dominant dark matter component, namely
that it fully accommodates a relic density $\Omega_{\rm DM} h^2$ as measured by
the Planck satellite~\cite{Planck:2015xua}. Concerning the direct detection of
dark matter, we consider the currently most stringent bounds on the
spin-independent (SI) nucleon-DM cross section as measured by LUX for
dark matter with \mbox{$\mdm>8$~GeV}~\cite{Akerib:2013tjd} and by CDMSLite for \mbox{1~GeV$<\mdm<8$~GeV}~\cite{Agnese:2015nto}. In
section~\ref{sec:dm_id}, we focus on indirect detection constraints that are
imposed on the basis of the gamma-ray measurements achieved by the Fermi-LAT
telescope~\cite{Ackermann:2015zua,Ackermann:2015lka}. Those bounds are however
not applied at the level of the likelihood function encoded in our \MN\ routine, and we have chosen instead to reprocess the scan results for those
parameter points that are consistent with both the relic density and direct
detection considerations. For the purpose of the relic density and direct
detection cross section calculations, we utilise both the
\MD~\cite{Backovic:2013dpa, Backovic:2015cra} and \MO~\cite{Belanger:2014vza}
numerical packages, although we only present the results obtained with \MD. The
consistency checks that we have performed with both codes are detailed in
appendix~\ref{sec:cc}.

\begin{table}
\footnotesize
\begin{center} \renewcommand\arraystretch{1.2}
\begin{tabular}{ c   c  c  l }
\hline
 &  Observable & Value/Constraint  & Comment\\
  \hline\hline
{\it \underline{Measurement}} &  $\Omega_{\rm DM} h^2$  & $ 0.1198 \pm 0.0015 $  & Planck 2015~\cite{Planck:2015xua}  \\
\hline
{\it \underline{Limits}} &   $\wy/\my$  & $ < 0.2 $  &  Narrow width approximation\\ 
&    $\wy$  & $ >   10^{-11}$ GeV & Ensures prompt decay  at colliders  \\ 
 &   $ \sigma_{n}^{\rm SI}  $  & $ < \sigma^{\rm SI}_{\rm LUX}$ (90\% CL)  & LUX bound~\cite{Akerib:2013tjd} ($m_X > 8$ GeV)\\
 &   $ \sigma_{n}^{\rm SI}  $  & $ < \sigma^{\rm SI}_{\rm CDMS}$ (95\% CL)  & CDMSlite bound~\cite{Agnese:2015nto} ($1 \, {\rm GeV} < m_X < 8\,  {\rm GeV}$) \\
 \hline
\end{tabular}
\end{center}\renewcommand\arraystretch{1.0}
 \caption{Summary of the observables and constraints used in this analysis and
 encoded into our \MN\ routine. The relic density constraints assume
 a Gaussian likelihood function, while the direct detection limits use step
 likelihood functions smoothed with half a Gaussian. \label{tab:co}}
\end{table}

We derive collider constraints on the simplified top-philic dark matter model
using the \MG~\cite{Alwall:2014hca} framework and the recast functionalities of
\MA~\cite{Conte:2012fm,Conte:2014zja,Dumont:2014tja} (where appropriate). We apply the LHC
constraints on the top-philic dark matter model with two different procedures.
On one side, similarly to what has been performed for the indirect detection
bounds, we reprocess the scenarios that accomodate the observed relic density
and that are compatible with LUX and CDMSLite data. However, we also study the
collider bounds on the parameter space independently of any astrophysics and
cosmology consideration and by relaxing the narrow width requirement
(allowing $\wy/\my$ to be of ${\cal O}(1)$) as well. In order to increase
the sensitivity of the LHC searches, we allow for wider coupling ranges of
$10^{-2}<\gx<2 \pi$ and $10^{-2} < \gsm < 2 \pi$. The collider study without any cosmological
and astrophysical constraint therefore includes the cases where the dark matter is not a
standard thermal relic (\ie \, its relic density is a result of a non-thermal
mechanism or a non-standard evolution of the Universe). Details are provided in
section~\ref{sec:collider} and appendix~\ref{sec:recast} for what concerns the
validation of the CMS analyses that we have implemented in \MA\ for this work.

In conclusion to this section, we point out that even though our current work focuses on a dark matter candidate which is a Dirac fermion,  a more general implementation of simplified dark matter models in \FR~\cite{Alloul:2013bka,Degrande:2014vpa} can also account for pseudoscalar mediators as well as for $CP$-mixed states and for dark matter particles which are real or complex scalars \cite{Backovic:2015soa, Mattelaer:2015haa, Neubert:2015fka}. The corresponding model files
have been used in this work and can be downloaded from the \FR\ model repository~\cite{FR-DMsimp:Online}
that also includes a model where the mediator is a spin-1 state that
couples to either a fermionic or a scalar dark matter
candidate~\cite{Backovic:2015soa}. All the models allow for the
automated calculation of next-to-leading-order (NLO) effects
and loop-induced leading-order (LO) processes in QCD in the
context of LHC predictions.

\section{Cosmological and astrophysical constraints}
\label{sec:cosmo}
We begin our analysis of the simplified top-philic dark matter model with a detailed discussion of the cosmological and astrophysical constraints. 
\subsection{Constraints from dark matter relic density}\label{sec:oh2}
Dark matter annihilation in the early Universe is determined, in the simplified top-philic dark matter
model, by a combination of three processes,
\begin{equation*}
	\dm \dmbar \rightarrow t \bar{t} \quad ({\rm I})\, , \qquad
	\dm \dmbar \rightarrow g g \quad  ({\rm II})\,,\qquad\text{and}\qquad
	\dm \dmbar \rightarrow \y \y \quad ({\rm III})\,,
\end{equation*}
where we have omitted the annihilation into photons as it is always
suppressed compared to the annihilation into gluons. The analytic expressions
for the thermally averaged annihilation cross section in the non-resonant region
$\langle \sigma v_{\rm rel}  \rangle$ corresponding to each of the processes listed above
are provided in appendix~\ref{sec:dmann}. The first two processes proceed via
an $s$-channel $\y$ exchange (first two rows of table~\ref{tab:sig}), while the
third process consists of a $t$-channel $\dm$ exchange (third row of
table~\ref{tab:sig}). The resonance structure of the $s$-channel processes
implies that the width of $\y$ potentially plays an important role in the
determination of the relic density assuming a dominant annihilation via the
processes (I) and (II), while the effects of the $\y$ width are mostly
negligible if the annihilation dominantly proceeds via the $t$-channel $\dm$
exchange process (III).

According to the hierarchy between the dark matter mass $\mdm$, the mediator
mass $\my$ and the top quark mass $m_t$, different situations can occur.
Qualitatively, one expects that:
\begin{itemize}
\item for $\my \gtrsim \mdm \gtrsim m_t$: process (I) is dominant as the
tree-level annihilation into a pair of top quarks is kinematically allowed, the
annihilation into gluons being loop suppressed, and the one into a pair of
mediators kinematically suppressed;
\item for $ \mdm \lesssim m_t,\, \my$: dark matter annihilates into a pair of
gluons as in process (II),  since it is the only kinematically allowed channel;
\item for $m_t \gtrsim \mdm \gtrsim \my$: relic density is determined by
process (III) since annihilation into top quarks is kinematically forbidden and
the one into gluons occurs away from the resonant pole of $\my$;
\item for $ \mdm > m_t,\my$ and $\my < 2m_t$: similarly to the case above, the
dominant annihilation mechanism is process (III), as annihilation into top
quarks occurs far from the resonant pole and is suppressed kinematically;
\item for $ \mdm > m_t,\my$ and $\my > 2m_t$: processes (I) and (III) are
competitive and the dominant process among the two is determined by the
hierarchy between the $\gsm$ and $\gx$ couplings.
\end{itemize}

Requiring our simplified top-philic dark matter model to result in a dark matter
relic density consistent with the most recent Planck
measurements~\cite{Planck:2015xua} implies strong constraints on the viable
regions of the parameter space. As an illustration, we consider the region of
the parameter space in which $m_t \gtrsim \mdm \gtrsim \my$, where we expect
the dominant annihilation mechanism of dark matter to be process (III) and to
give rise to a pair of mediators. In this region, the thermally averaged
annihilation cross section approximately reads
\begin{equation}
\langle \sigma v_{\rm rel} \rangle_{\rm ann} \sim \frac{\gx^4} { \mdm^2 } \sim 10^{-9} \GeV^{-2},
\end{equation}
so that it is clear that imposing that the relic density predictions agree with
Planck data leads to a stringent constraint on the ratio $\gx^2/\mdm$. The
argument is more involved in parameter space regions where the total mediator width $\wy$
plays a role, as the relevant quantity involved in the relic density calculation
is in general not $\langle \sigma v_{\rm rel}  \rangle_{\rm ann} $ but
$\int {\rm d}x \langle \sigma v_{\rm rel}  \rangle_{\rm ann}(x)$ where $x \equiv \mdm/T$
and $\langle \sigma v_{\rm rel}  \rangle_{\rm ann}$ is a non trivial function
of $x$. This is especially true, for instance, for the Breit-Wigner-type
amplitudes that appear in processes (I) and (II).

\begin{figure}
\includegraphics[width=3in]{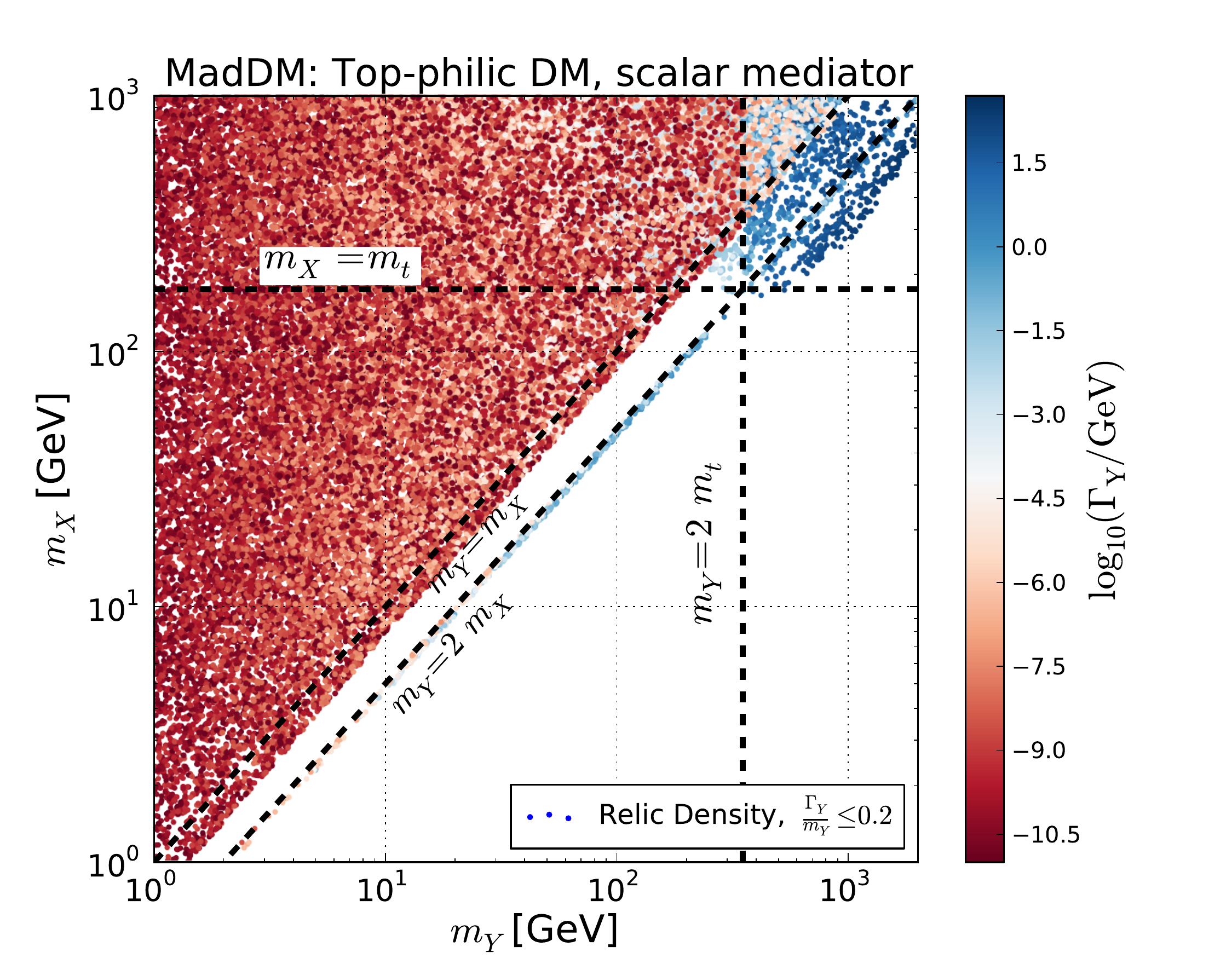}
\includegraphics[width=3in]{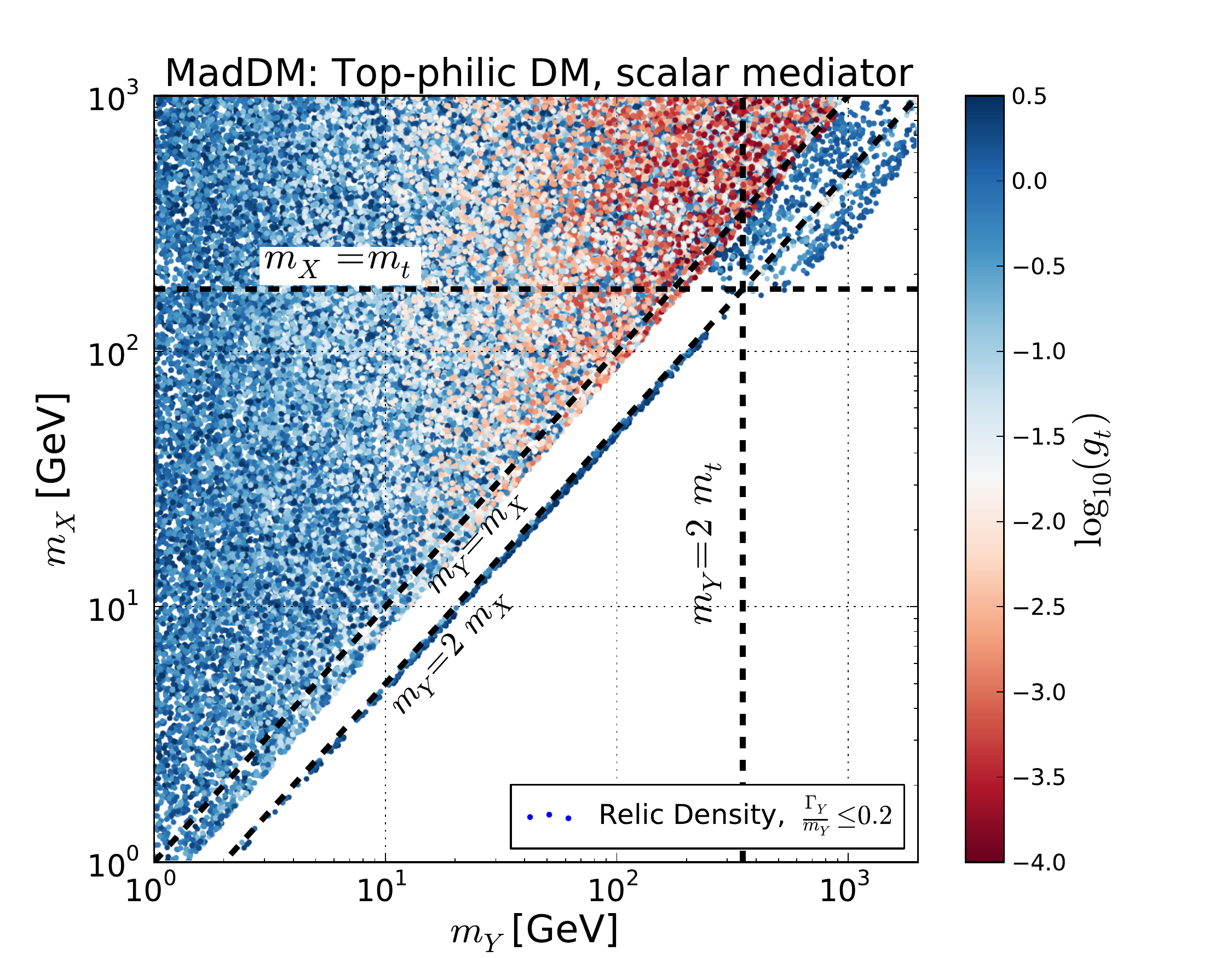}
\includegraphics[width=3in]{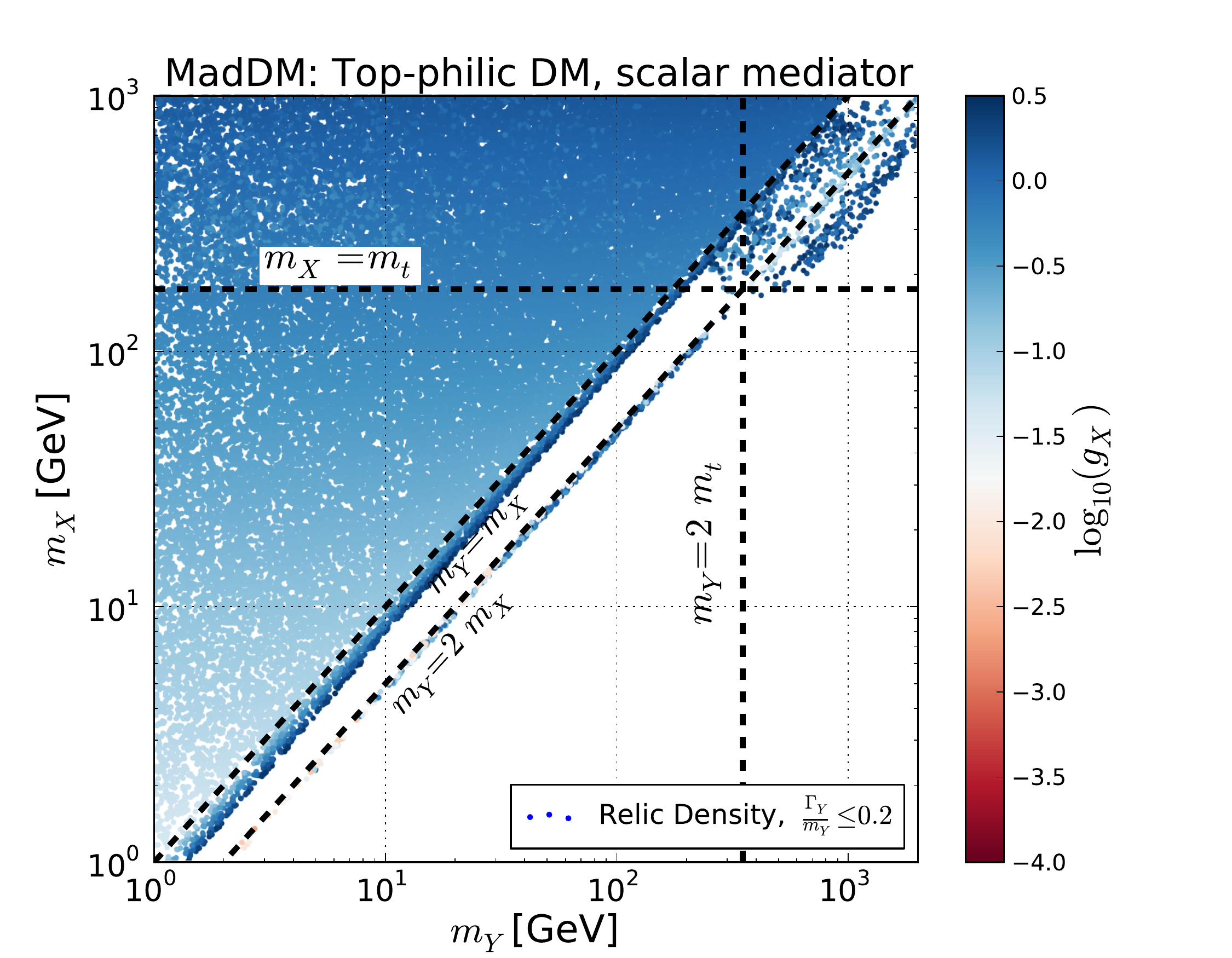}
\includegraphics[width=3in]{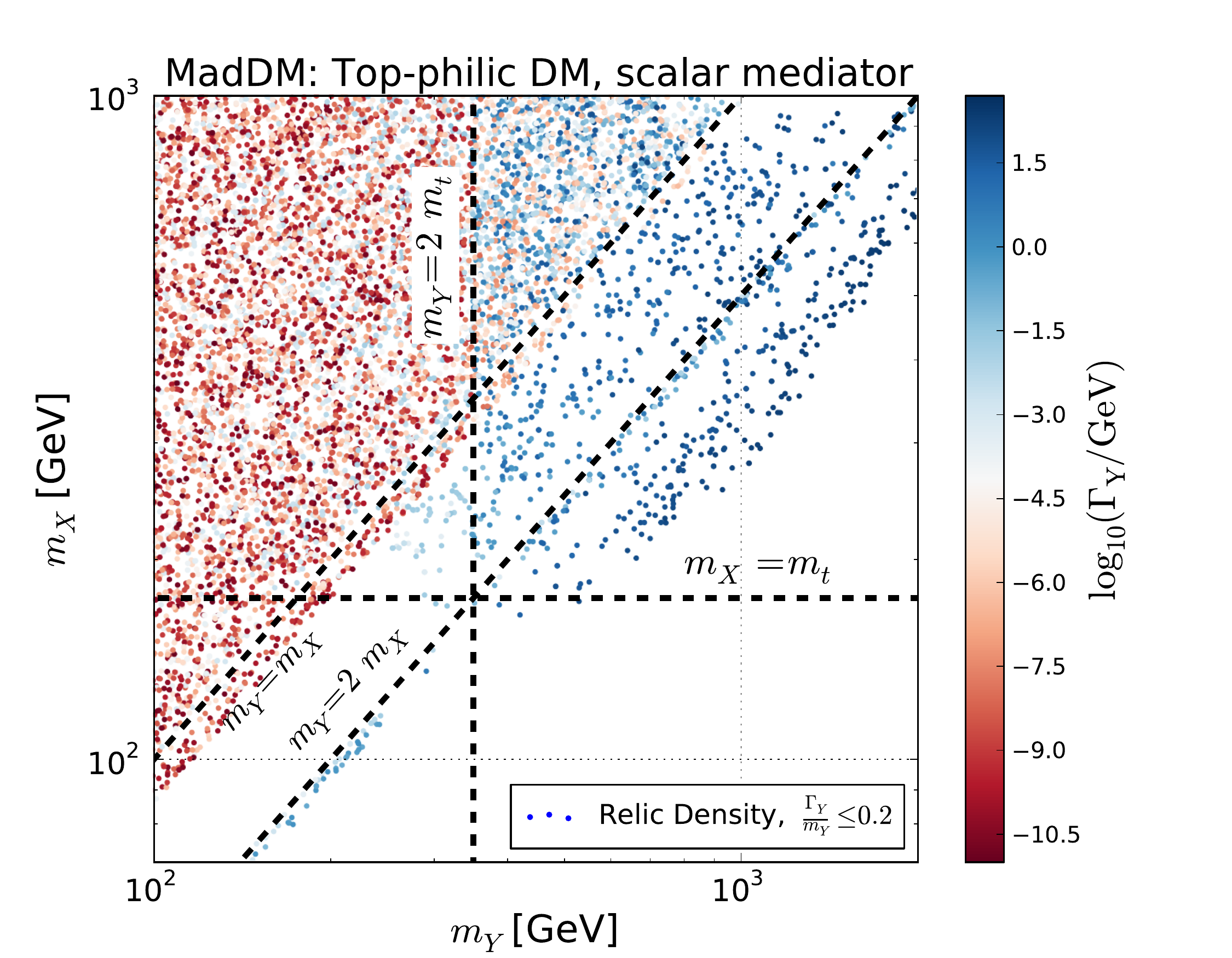}
\caption{Results of our four-dimensional parameter scan using \MD\ projected
  onto the $(\my,\mdm)$ plane. The first three panels show the projections with
  the colourmap representing the values of $\wy$, $\gx$ and $\gsm$ respectively.
  The right-most panel shows the zoomed-in upper right region of the left-most
  panel. All represented points feature a relic density in agreement with Planck
  data and $\wy/\my \leq 0.2$.} \label{fig:scan1maddm}
\end{figure}

In order to provide a more detailed quantitative analysis, we have performed a
four-dimensional scan the top-philic dark matter model parameter space and
examined the effects of imposing relic density constraints on the allowed/ruled
out parameter sets. Figure~\ref{fig:scan1maddm} reveals the rich structure of
the four-dimensional parameter space allowed by relic density measurements. The
bulk of the allowed parameter points lies in the region where $\mdm > \my$, and
the annihilation cross section is dominantly driven by process (III). This region
of the parameter space has the particularity of not being reachable by
traditional monojet, monophoton, mono-$Z$ and mono-Higgs searches at colliders.
The decay of the mediator into a pair of dark matter particles is indeed not
kinematically allowed, so that any new physics signal will not contain a large
amount of missing energy. The model can however be probed at colliders via
dijet, diphoton, $t\bar{t}$ (plus jets) and four-top analyses. We elaborate on this point more in
section~\ref{sec:nonmet}. The characteristic mediator width $\wy$ in this
region tends to be extremely small, with values of at most $10^{-4}$~GeV as
shown in the top left panels of figure~\ref{fig:scan1maddm}. This is expected as
the width is mostly controlled by the decays into gluons, and into top quarks in the
regions where this decay is kinematically allowed, the decay into a pair of dark
matter particles being forbidden.

In the region where $\mdm \gtrsim m_t$ and $\my \gtrsim 2m_t$, the mediator decay
into a $t\bar{t}$ final state is kinematically allowed and the dark matter
annihilation cross section is driven by the $\dm \bar\dm \to \y \to t\bar{t}$
process. The only other parameter space region that is not ruled out by the
relic density data is centered around the resonance region where
$\my\sim 2\mdm$. The extension of the region away from the resonance pole is due
to the $\y$ width that can reach $ {\cal O}(10)$~GeV. The resonant region extends to
lower $\mdm$ and $\my$ values, and is the only allowed region when both $\mdm$
and $\my$ are smaller than $m_t$ (but with $\y$ decays into a pair of dark
matter particles being allowed). This has interesting implications for LHC
searches as the low dark matter/mediator mass region is the one where colliders have the
best sensitivity, in particular through monojet searches (see for instance
section~\ref{sec:monojres}).

Relic density constraints favour $\gx$ couplings of $ {\cal O}(1)$ in most of the
scanned parameter range as evident in the second and third panels of
figure~\ref{fig:scan1maddm}, regardless of the actual value of the $\gsm$
coupling which is irrelevant in the $\mdm \gtrsim \my$ region (upper right panel
of figure~\ref{fig:scan1maddm}) as it does not enter the calculation of the
relic density.

The structure of the ruled out parameter space regions shows several other interesting qualities. The most striking feature is that almost the
entire region where $\my \gtrsim 2 \mdm$ does not lead to predictions of a dark
matter relic density in agreement with the observations. There are also no allowed points for
$\mdm \lesssim m_t, $ except very close to the resonance line. This region is
 characterised by a dominant mediator decay into gluons, which results in
typical $\wy/\my \ll 1$, a small total dark matter annihilation cross section,
and hence an overproduced dark matter. The upper limit imposed on the size of
the couplings (see table~\ref{tab:param}) is largely responsible for the absence
of allowed points in the region. For instance, taking any $\mdm$ value so that
the predicted relic density agrees with the observed value,  an increase in $\my$ will result
in a decrease of the annihilation cross section, in turn leading to a higher
relic density. The only way (away from the resonance) to restore the correct
relic density is then to increase the size of $\gx$ and/or $\gsm$. However, our
results show that even for couplings of $ {\cal O}(1)$, the cross section in this region
is too small not to overproduce  dark matter.

The region of parameter space between $\my \sim \mdm $ and $\my \sim 2 \mdm$ is
consistent with the above-mentioned argument. This strip of the ruled out
parameter space can be seen as a part of the larger ruled out region for which
$\my \gtrsim \mdm$. Tuning $\mdm$ to be close to $\my/2$ and assuming a
relatively small $\wy$ value is the only way to enhance the dark
matter annihilation cross section and not overproduce dark matter.

\begin{figure}
\includegraphics[width=3in]{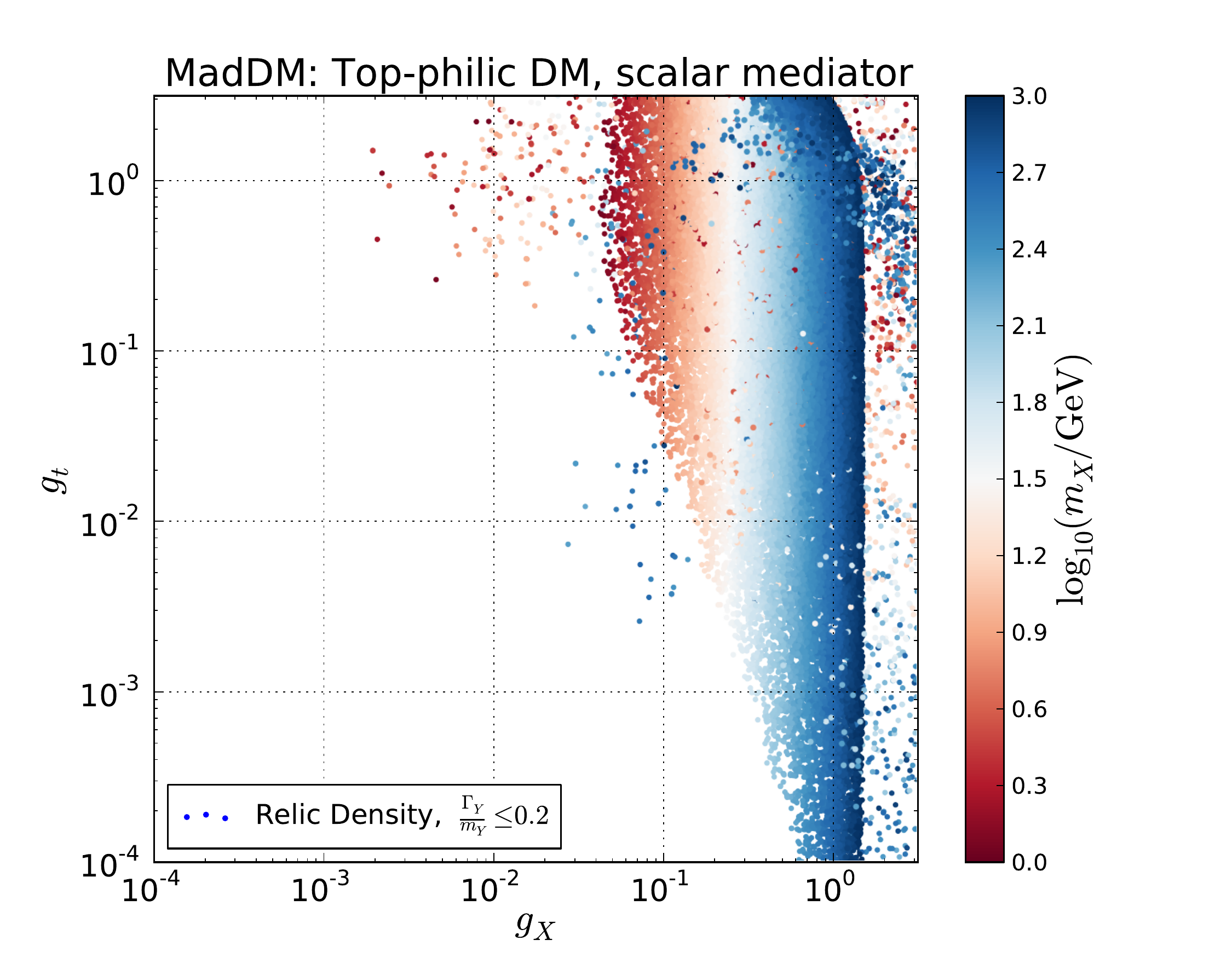}
\includegraphics[width=3in]{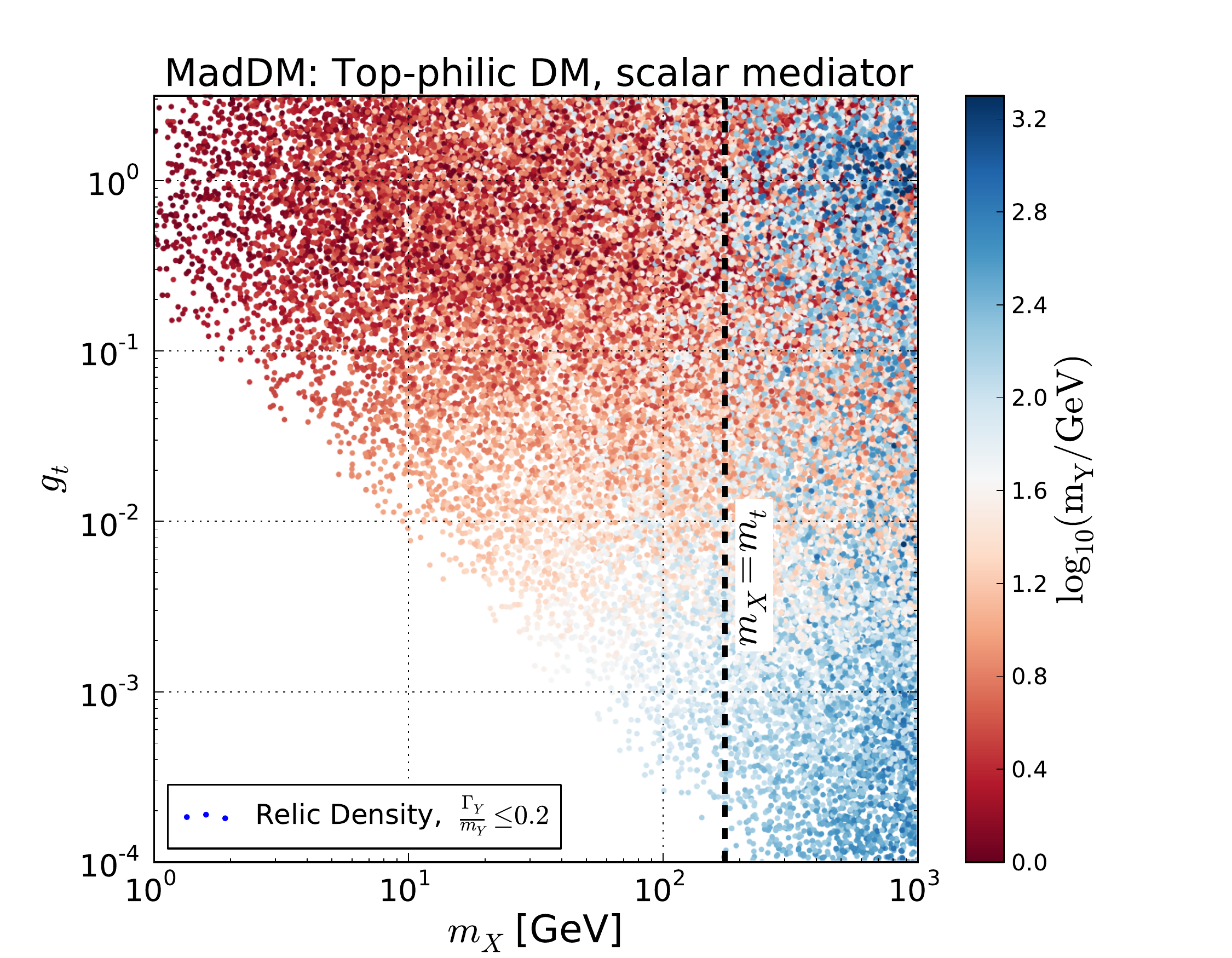}
\caption{Results of our four-dimensional parameter scan using \MD, projected
  onto the $(\gx, \gsm)$ plane (left) and $(\mdm, \gsm)$ plane (right). All
  represented points feature a relic density in agreement with Planck data and a
  narrow width mediator.} \label{fig:scan2}
\end{figure}

In addition to projections of the allowed parameter space onto the $(\my,\mdm)$
plane, we have also studied several other projections. 
Figure~\ref{fig:scan2} shows the projections of our results  onto the $(\gx,\gsm)$
plane (left) and $(\mdm, \gsm)$ plane (right), where we show $\mdm$ and $\my$ as a
colourmap in the first and second panel respectively. Regardless of the value of
$\mdm$ and $\my$ in the considered scan range, there are no solutions for
$\gx$ and $\gsm$ which satisfy the relic density constraint in the region where
$\gsm \lesssim 10^{-2}$ and $\gx \lesssim 10^{-1}$. This finding is consistent
with the left lower panels of figure~\ref{fig:scan1maddm} where we have found
that a correct relic density favours $\gx$ couplings of $ {\cal O}(1)$. Furthermore, we
can observe that only couplings of $ {\cal O}(1)$ result in $\wy \gtrsim  {\cal O}(1) \GeV$,
while in the majority of the allowed $(\gx, \gsm)$ parameter space regions
$\wy/\my \ll 1$. 

We find no striking features in the $(\mdm, \gsm)$
projection of the scanned parameter space. The unpopulated regions in the lower left
corners are artifacts of the lower limit on the coupling size of $10^{-4}$.

As a validation, we have cross checked our calculations with the \MO\ code. The results obtained with \MD\ and \MO\ agree in most of the parameter space, except in
the region where $\gsm$ and $\mdm$ are small. Some numerical discrepancies are
expected to occur in this region, as shown in appendix~\ref{sec:cc} and by comparing
figures~\ref{fig:scan1maddm} and~\ref{fig:scan1momegas}.

As a last remark, allowing the scalar mediator to couple to all quarks and
leptons would only have a minor impact on our results. The region dominated by
the process (III) will indeed stay unchanged, since it is insensitive to the
coupling between $\y$ and the Standard Model fermions. As far as it concerns
dark matter annihilation via an $s$-channel $\y$ exchange, one would have to sum
up over all the possible final states kinematically open. This would increase
the total annihilation cross-section and decrease $\Omega_{\rm DM} h^2$,
implying that the constraint of having $\Omega_{\rm DM} h^2 \sim 0.12$ leads to
a rescaling of all fermionic couplings towards smaller values with respect to
the  $\gsm$ values shown in this work. The major difference would reside in a
potentially larger decay width for $\y$ and hence wider ``bands'' of the resonance 
regions of the allowed parameter space.

\subsection{Constraints from direct detection}\label{sec:dd}
Simplified models of dark matter which feature couplings to quarks and gluons
can also be bounded by results from underground direct dark matter detection
experiments. In top-philic dark matter scenarios, dark matter scatters off
nucleons via the $t$-channel exchange of $\y$, where the scattering off gluons
via triangle top loops accounts for the dominant contribution to the DM-nucleon
scattering rate.

The spin independent (SI) dark matter-nucleon cross section is given by
\begin{equation}\label{eq:dd}
	\sigma_{SI} ^{n}= \frac{4}{\pi} \left(\frac{\mdm m_{n}}{\mdm + m_{n}} \right)^2 \left[ \frac{2}{27}\frac{m_{n}}{m_t}\frac{\gx \gsm }{\my^2 } f_{G} \right]^2 \,,
\end{equation}
where $f_G \equiv 1-\sum_{q \leq 3} f_q =0.921$~\cite{Beringer:1900zz}%
\footnote{The gluon form factor suffers from relatively large uncertainties on
the strange quark content of the nucleons, which we here omit.} is the gluon
form factor and the sum runs over the light quarks $q=u,d,s$, where
$m_{n} \approx 0.938\GeV$ is the nucleon mass and $m_t = 173 \GeV$ is the top
quark mass. The expression in eq.~\eqref{eq:dd} does not depend on $\wy$,
simplifying the constraints which can possibly be derived from direct detection.
For instance, considering a scenario in which generic $\mdm$ and $\my$ masses
are fixed and where the dominant annihilation process is process (I), direct
detection directly constraints the product $\gx \gsm$. Extracting the constraint
on this quantity in a generic fashion is much more complicated in the case of
dark matter annihilation in the early Universe and at colliders, as the processes
involved in dark matter relic density and dark matter production calculations
intrinsically depend on a quantity which is proportional to $\gx\gsm/\wy$.

The running of the $\gx$ and $\gsm$ couplings could have an effect on the value
of the spin independent DM-nucleon scattering cross
section~\cite{D'Eramo:2014aba,Vecchi:2013iza}. However, a proper inclusion of
the running couplings would require a careful treatment of the renormalisation
group evolution via multiple energy scales which is beyond the scope of our
current effort. Instead, we restrict here our calculations to constant $\gx$ and
$\gsm$ values. The effect of the running couplings would then be equivalent to a rescaling of $\gx$ and $\gsm$ to different values.

\begin{figure}
\begin{center}
\includegraphics[width=2.9in]{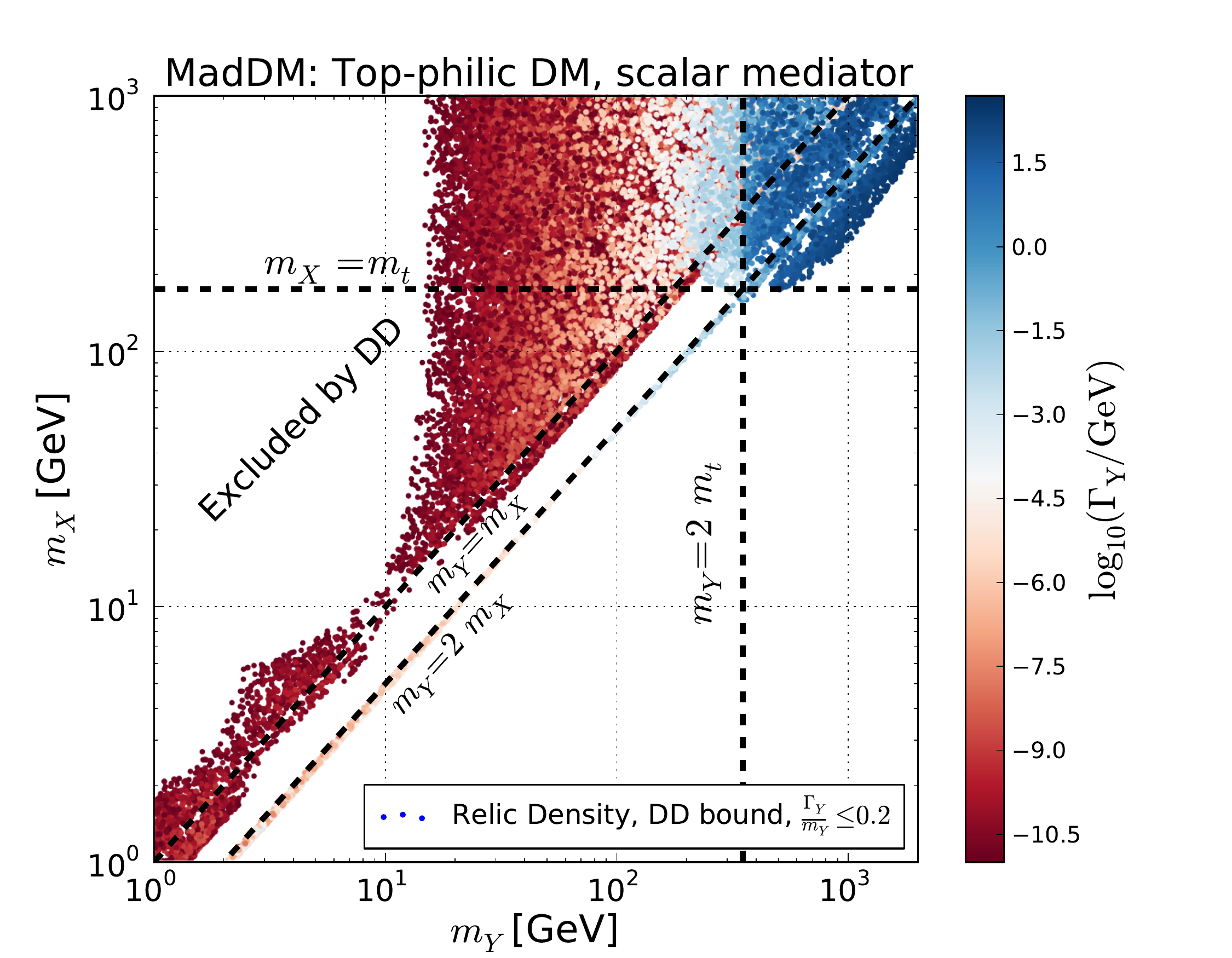}
\includegraphics[width=2.9in]{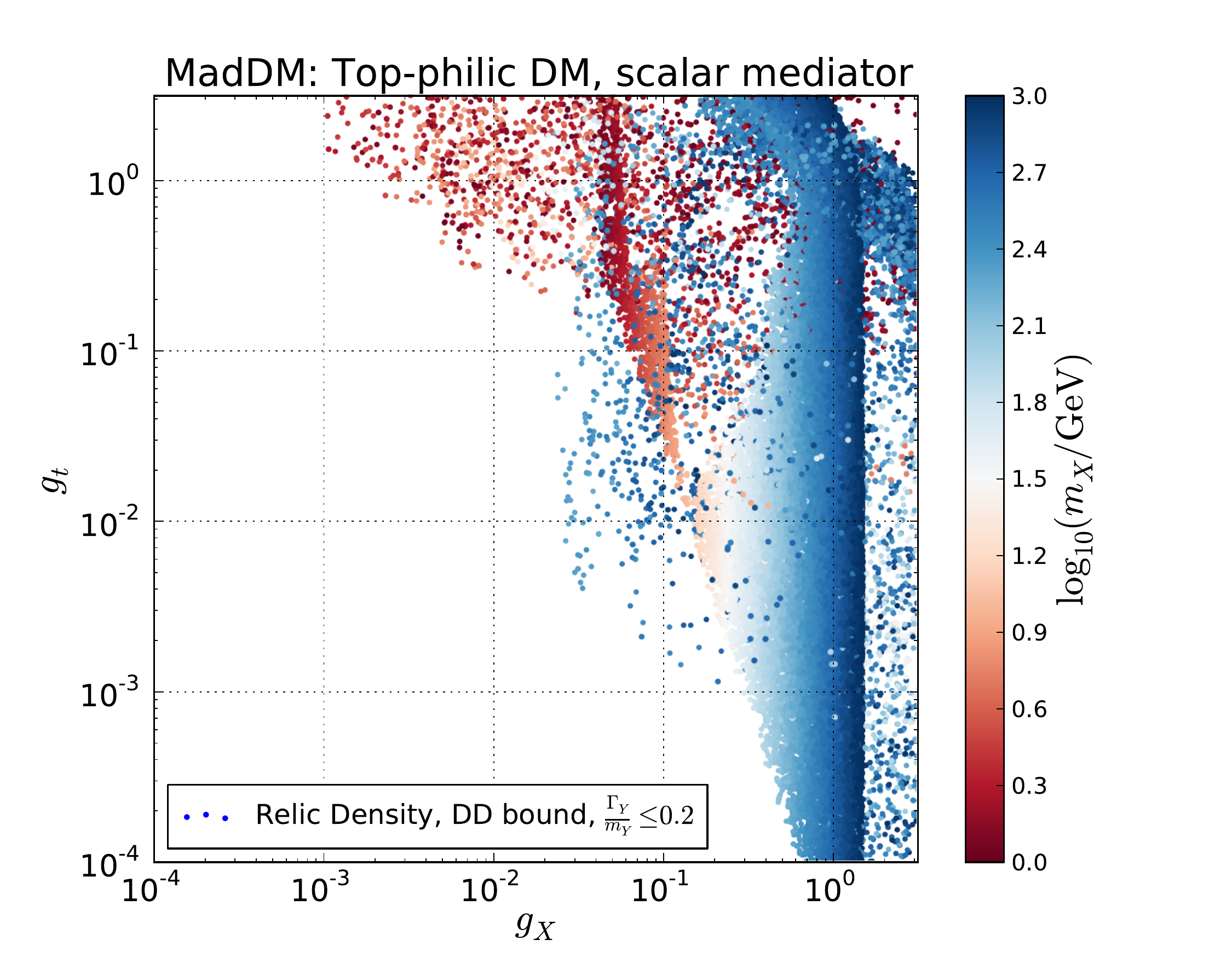}
\includegraphics[width=2.9in]{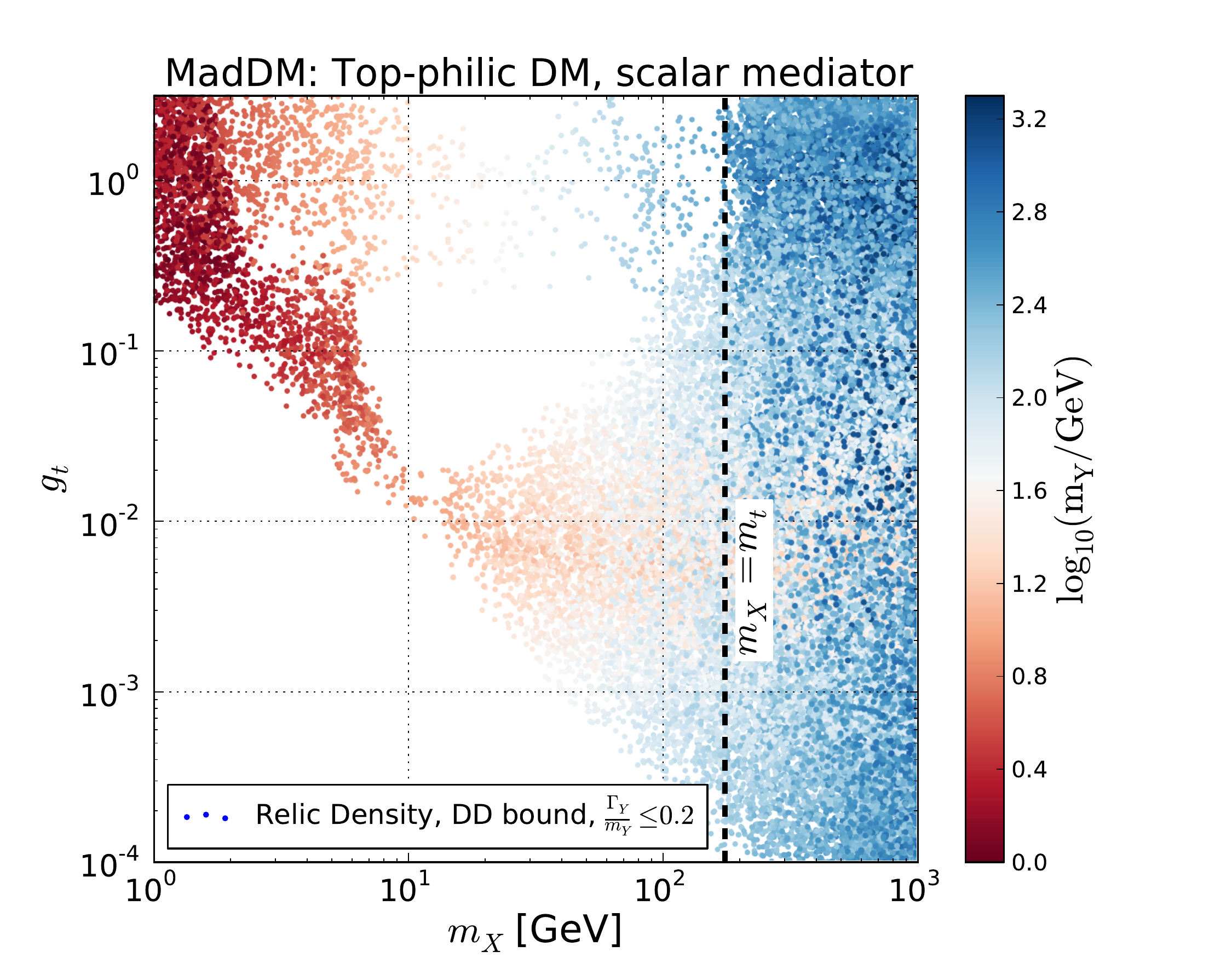}
\end{center}
\caption{Results of our four-dimensional parameter scan using \MD. The top left
panel shows the projection of the scan into the $(\my, \mdm)$ plane with a
colourmap representing the values of $\wy$. The top right panel shows the
projection of the allowed points into the $(\gx,\gsm)$ plane with a colourmap
given by $\mdm$. Finally the lower panel shows a projection onto the
$(\mdm,\gsm)$ plane with a colourmap denoting the values of $\my$. All
represented points feature a relic density in agreement with Planck data, a
narrow width mediator and accommodate the direct detection constraints.}
\label{fig:scan3}
\end{figure}

Next, we have repeated the four-dimensional parameter scan from
section~\ref{sec:oh2} including into the {\sc MultiNest}  likelihood function also bounds
stemming from direct detection. Figure~\ref{fig:scan3} shows the results of the
scan, projected onto three different planes, where we removed the points excluded by the 95\% confidence limit (CL) bound from LUX and CDMSLite. Direct detection rules out a major
portion of $(\my,\mdm)$ space allowed by the relic density constraints
(regardless of the coupling value) in the region where $\mdm \gtrsim \my$, where
collider bounds are irrelevant. Figure~\ref{fig:scan3}
hence serves as a good example for the complementarity among direct detection, relic
density and collider bounds. In the $(\gx, \gsm)$ plane, direct
detection does not rule out a well defined-region (top-right panel of
figure~\ref{fig:scan3}), indicating that for any pair of couplings $(\gx,\gsm)$
in the range of $[10^{-4}, \pi]$ which are allowed by the relic density
constraint, it is always possible to find a pair of $(\my,\mdm)$ values which
are not ruled out by direct detection data. In the $(\mdm, \gsm)$ projection, we
finally observe that direct detection rules out a well-defined portion of the
parameter space. Furthermore, the constraint also rules out small width points
for $\gx \gtrsim 0.1$ and $\mdm \gtrsim m_t$. Direct detection bounds are indeed
more sensitive to dark matter masses in the ballpark of 10 to 200 GeV and
quickly deteriorate at larger dark matter masses, since the event rate in the
detector scales as $1/\mdm^2$. We also see that the direct detection exclusion limit is able
to rule out a large portion of the parameter space where $\y$ is light, below
30~GeV, while the sensitivity is quickly lost for heavier masses of the scalar
mediator. This can be understood by the $1/\my^2$ dependence of the SI elastic
cross section of eq.~\eqref{eq:dd}. Both mass dependences are illustrated
by the lower panel of figure~\ref{fig:scan3}.

\subsection{Constraints from indirect detection}\label{sec:dm_id}
Top-philic dark matter annihilation in the present Universe could result in
fluxes of cosmic rays and prompt gamma-rays, which can also be used to infer
useful limits on the model parameter space. The annihilation of a $\dm \dmbar$
pair in the galactic halo (or in dense environments of galactic centers) and
the subsequent production of a secondary gamma ray flux is dictated by the same
processes (I), (II) and (III) that set the relic abundance. These processes
give rise to a continuum of secondary photons due to the decay and subsequent
QED showering of the pair-produced top quarks, gluons and/or mediators. 
As already mentioned in section~\ref{sec:model}, a direct coupling of the mediator to a
pair of prompt photons is induced at higher order in perturbation theory via a
loop of top quarks. Hence, analogously to process (II), the process
$\dm \dmbar \to \gamma\gamma$ exists and yields the production of two
monochromatic photons that could be detected in searches for lines in the
gamma-ray spectrum.\footnote{Dark matter annihilation into two prompt photons is
always suppressed by a factor $8\alpha_e^2/9 \alpha_s^{2}$ with respect to
annihilation into a pair of jets in the considered class of scenarios.} Finally,
photons arising from process (III) and the subsequent decay of the mediator into
two photons do not provide a signal line as the mediators are in general not
produced at rest in the annihilation process.

Similarly to the relic density case, measurements of the gamma-ray fluxes can
potentially constrain the coupling $\gx$ for the $t$-channel process (III) or
the product of couplings $\gx \gsm$ in the case of an $s$-channel annihilation
via the processes (I) and (II). However, it is important to highlight the
differences between factors which are constrained by the dark matter relic
density and by its indirect detection. The relic density is an integrated result
over the thermal history of the Universe. Hence, the width of the resonance is
important, even if $|\my - 2\mdm|\gg\wy$ (except in the case where $\my\ll2\mdm$).
Conversely, the characteristic velocity of the dark matter particles today is of
the order of $v \sim 10^{-3}$, implying highly non-relativistic dark matter
annihilation. The width of the mediator in an $s$-channel dark matter
annihilation process is hence relevant for indirect detection only in the case
of $\lvert \my - 2\mdm \rvert \lesssim \wy$.

\begin{figure}
\includegraphics[width=3in]{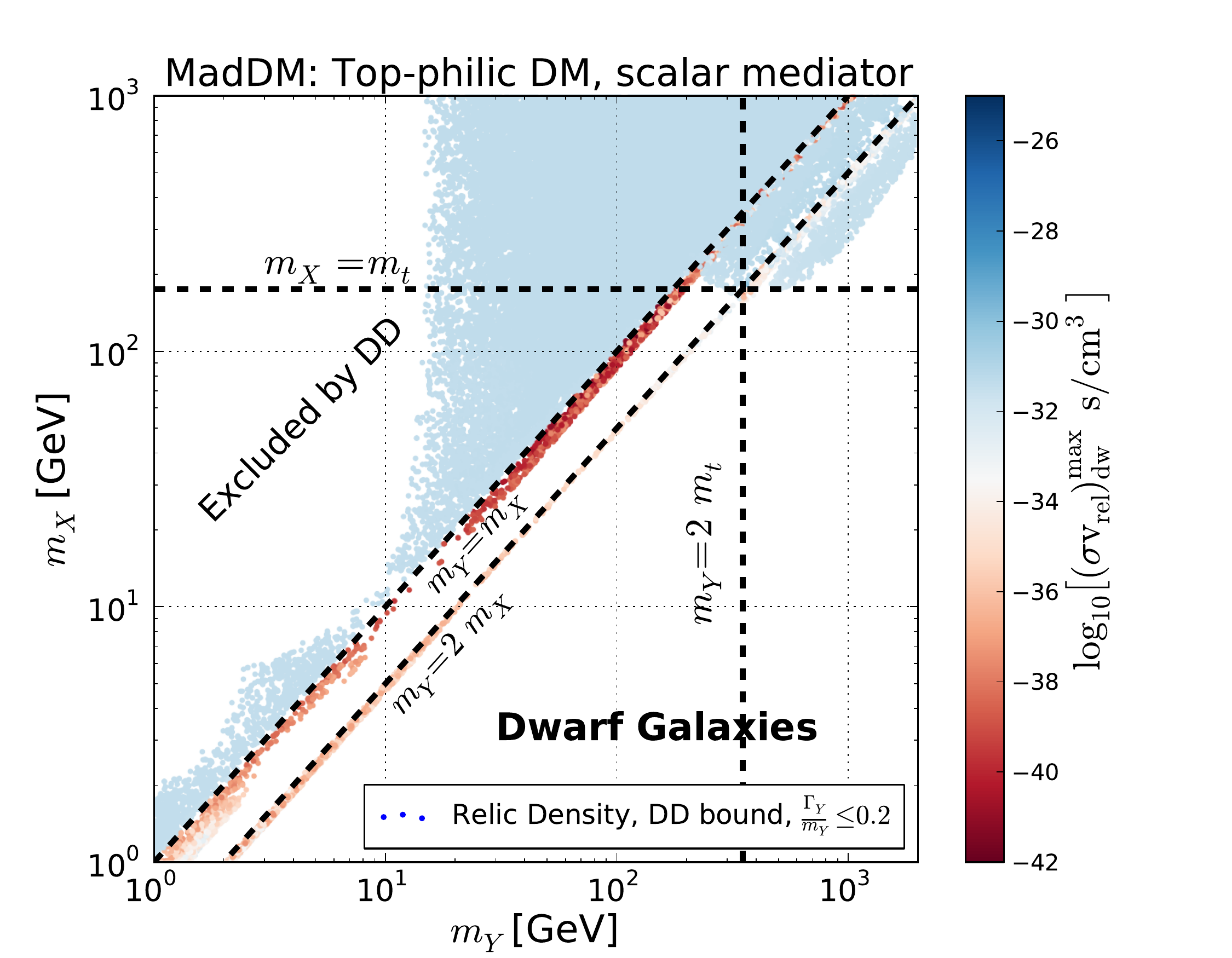}
\includegraphics[width=3in]{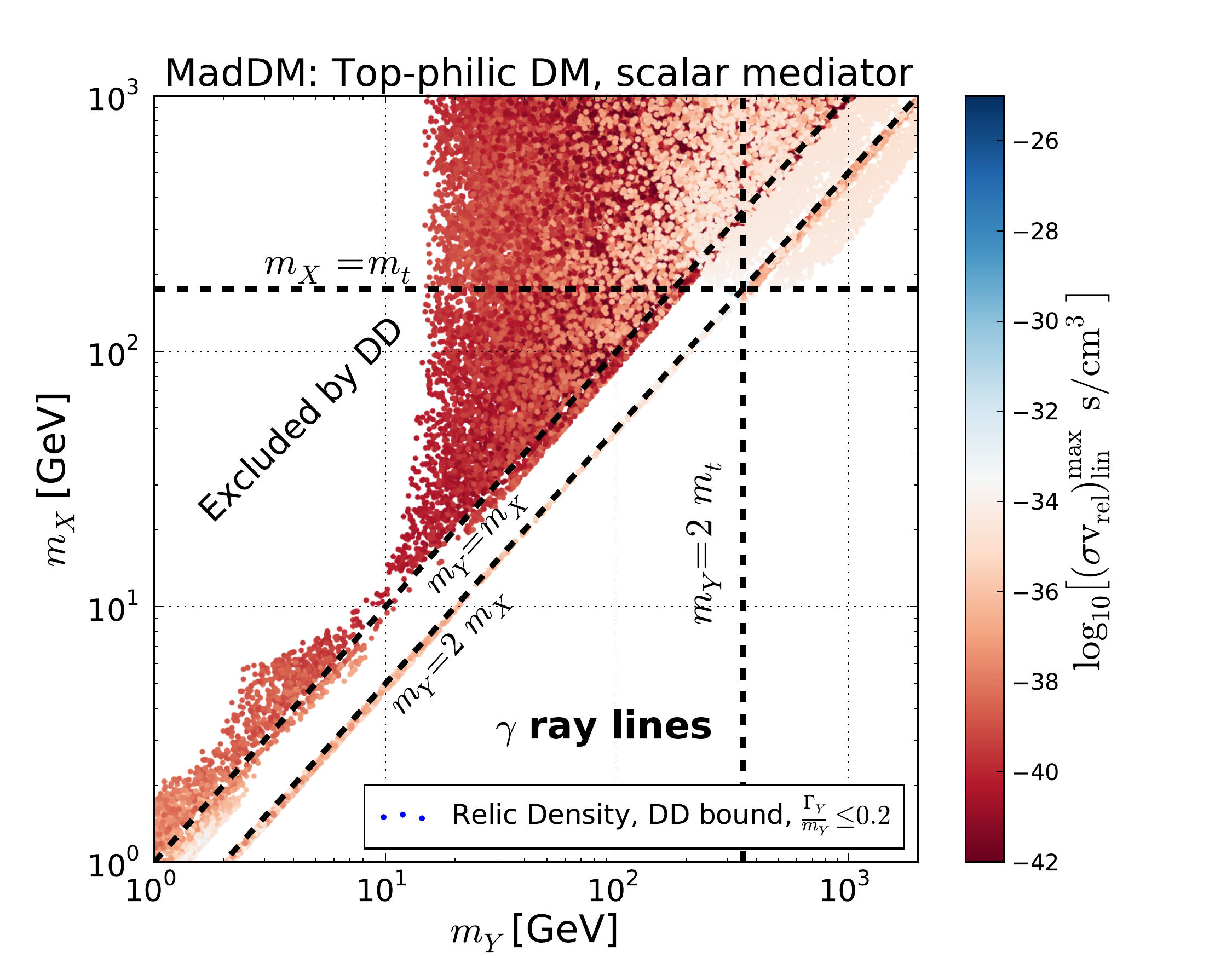}
\caption{Dark matter annihilation cross section at present time that is relevant
 for gamma-ray limits extracted from dwarf spheroidal galaxies measurements
 (left) and gamma-ray line searches (right). We show a maximal estimate of
 $(\sigma v_\text{rel})_\text{tot}$ and $(\sigma v_\text{rel})_{\gamma\gamma}$
 obtained by choosing $v_\text{rel}=2v_\infty$, where $v_\infty$ is the escape
 velocity for dwarf spheroidal galaxies and the galactic center, respectively.
 All represented points are compatible with the relic density, a narrow width
 mediator and the direct detection requirements.} \label{fig:id}
\end{figure}

Searches for gamma-ray signals of dark matter annihilation weakly constrain our
simplified top-philic dark matter model. We have investigated results from
gamma-ray line searches in the inner galactic region~\cite{Ackermann:2015lka},
as well as continuum gamma-ray measurements from dwarf spheroidal
galaxies~\cite{Ackermann:2015zua} and found no meaningful exclusion of the
parameter space once the relic density and direct detection constraints are
imposed. The lack of additional useful bounds is expected, as the
annihilation of dark matter in the present Universe is $p$-wave suppressed, \ie\
$\sigma v_\text{rel} \propto v_\text{rel}^2$ for all three annihilation channels
(see appendix~\ref{sec:dmann} for more detail). This contrasts with scenarios in
which the mediator is a pseudoscalar state that implies that the $p$-wave
suppression at low dark matter velocity is only present for process (III), so
that the gamma-ray constraints should be significantly stronger.

The gamma-ray line searches constrain the velocity-averaged cross section for
the direct dark matter annihilation into two photons. Due to its $p$-wave
suppression, this quantity is very sensitive to the choice of the velocity
distribution of the dark matter in the galaxy which is subject to large
uncertainties (see \eg~ref.~\cite{Mao:2012hf}). We adopt a conservative
viewpoint here, evaluating the annihilation cross section at the highest
possible velocity $v_\text{rel}=2v_\infty$ with $v_\infty$ being the escape
dark matter velocity for our galaxy which we take to be
$v_\infty~=~550$~km/s~\cite{Lisanti:2010qx}. The left panel of
figure~\ref{fig:id} shows the respective result for
$(\sigma v_\text{rel} )_{\gamma\gamma}$. The limits from gamma-ray line
searches lie between $2\times 10^{-32}\,\text{cm}^3\text{s}^{-1}$
(for dark matter masses around 1~GeV) and
$4\times 10^{-28}\,\text{cm}^3\text{s}^{-1}$ (for dark matter masses around
500~GeV).

Searches for gamma-ray signals in dwarf spheroidal galaxies constrain the total
the annihilation cross section at (two times) the escape velocity, the escape
velocity of the considered dwarf spheroidal galaxies being typically much
smaller and of the order of 10~km/s~\cite{Walker:2009zp},
which leads to a heavy suppression of the dark matter annihilation cross
section. The right panel of figure~\ref{fig:id} shows the annihilation cross
section evaluated for $v_\infty=50\,$km$/$s. The cross sections are much smaller
than the constraints which are around
$10^{-26}\,\text{cm}^3\text{s}^{-1}$, the exact details depending on the dark
matter mass and the relevant annihilation processes.

In cases where we would have allowed for leptonic couplings of the scalar
mediator $\y$, our general conclusion about the poor ability of indirect dark
matter searches to constrain the model parameter space remains unchanged. Dark
matter annihilation into leptonic final states could give rise to additional
continuum gamma-ray or positron fluxes, but the overall normalisation of
$\langle \sigma v_\text{rel} \rangle$ would not change significantly and remain
four to five orders of magnitude below the current bounds. Even under the most
aggressive assumptions, all obtained bounds would still be far from being able
to constrain a top-philic dark matter model with scalar mediators.

\section{Collider constraints}\label{sec:collider}

\begin{figure}[h!]
\center
 \includegraphics[height=0.45\textwidth]{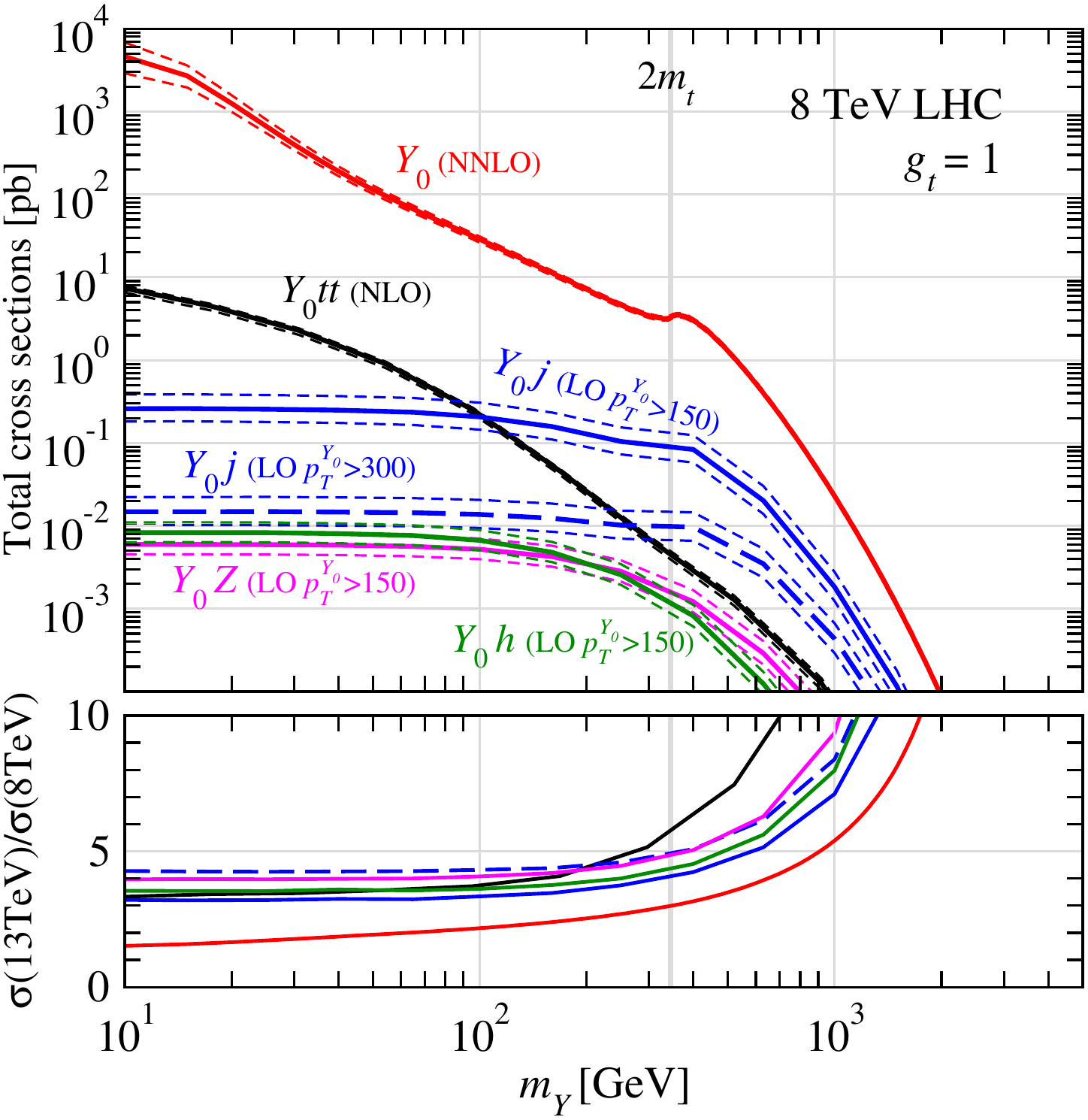}\qquad
 \includegraphics[height=0.45\textwidth]{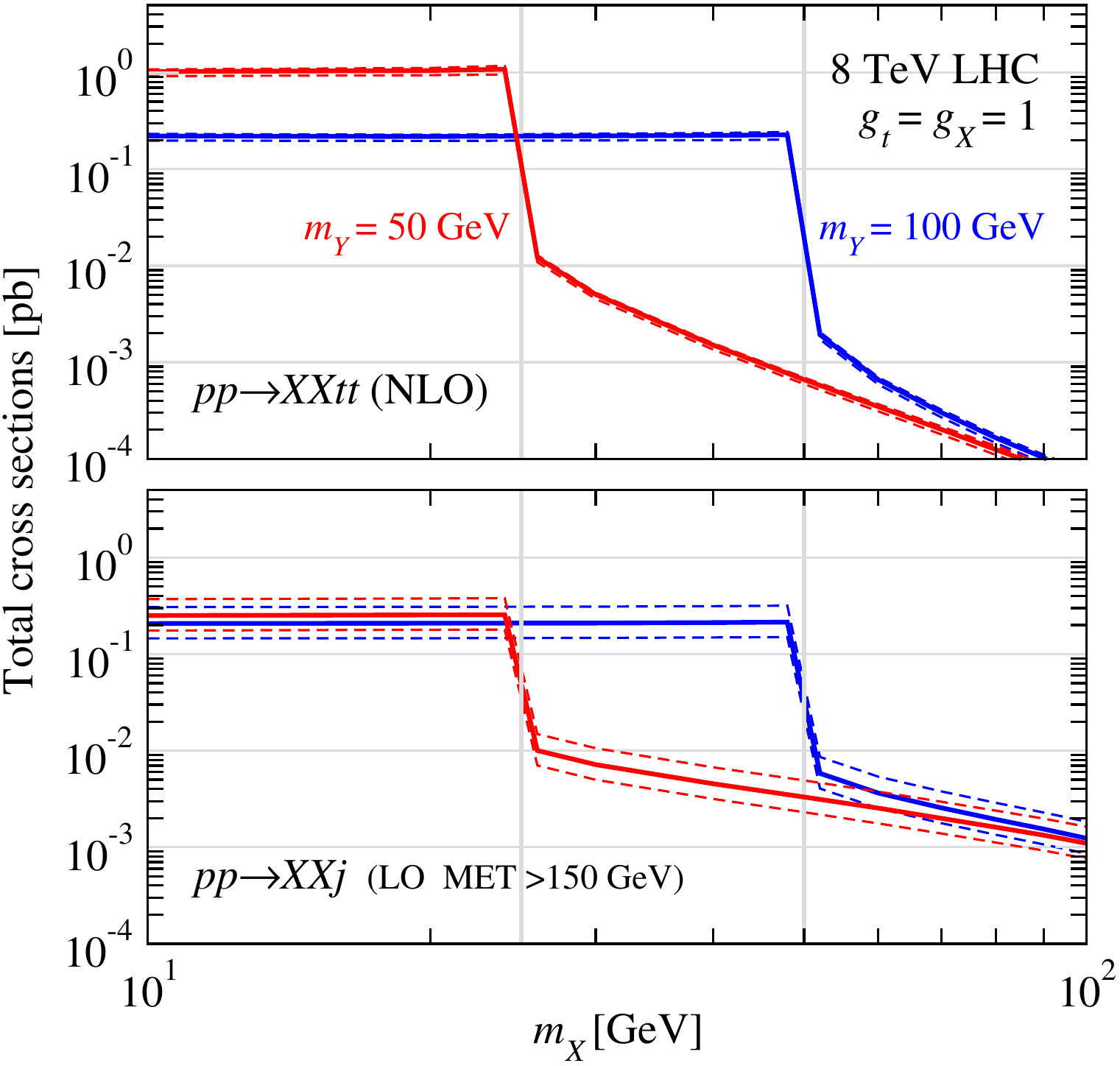}
  \caption{Left: Total cross sections (with scale uncertainties) for various
    mediator production channels (with \mbox{$g_t=1$}) at a centre-of-mass
    energy of $\sqrt{s}=8$~TeV as a function of the mediator mass. The
    NNLO cross section for single mediator production $\sigma(Y_0)$ is
    taken from the Higgs Cross Section Working Group
    report, the $Y_0t\bar t$ one is computed at NLO
    accuracy and all other loop-induced processes are evaluated at LO
    accuracy. 
	The monojet ($Y_0j$),
    mono-$Z$ ($Y_0Z$) and mono-Higgs ($Y_0h$) cross sections include a
    transverse momentum cut on the mediator as indicated in the figure.
    In the lower panel, we show the ratios of the cross sections evaluated at a
    centre-of-mass energy of $\sqrt{s}=13$~TeV over those at 8~TeV.
    Right: Cross sections for $t\bar t+\MET$ and monojet (with $\MET>150$~GeV)
    production for a mediator mass of $m_Y=50$ and 100~GeV and at a
    centre-of-mass energy of $\sqrt{s}=8$~TeV given as a function of the dark
    matter mass.
	}
\label{fig:xsec}
\end{figure}

As discussed in section~\ref{sec:model}, simplified top-philic dark matter
scenarios can be probed at colliders through the production of the mediator
either in association with a top-quark pair or through a top-quark loop.
Depending on the mass and coupling hierarchy, the mediator decays either into
a pair of dark matter particles, which results in signatures including missing
transverse energy ($\MET$), or into Standard Model final states. The size
of the cross sections associated with these two classes of mediator production
mechanisms is depicted in figure~\ref{fig:xsec} where we present their
dependence on the mediator and dark matter masses $\my$ and $\mdm$. For the case
where the mediator is singly produced, we use the Higgs cross section values that are
reported
in the Higgs Cross Section Working Group documentation~\cite{Heinemeyer:2013tqa}
and that are evaluated at the next-to-next-to-leading order (NNLO) accuracy in
QCD. For all the other cases, the hard-scattering cross section is convoluted
with the NNPDF~2.3~\cite{Ball:2012cx} set of parton distribution functions
(PDF) within \MG, the PDFs being accessed via the LHAPDF
library~\cite{Whalley:2005nh,Buckley:2014ana}. We employ a five-flavour-number
scheme, and leading-order (LO) and next-to-leading-order (NLO) PDFs are used
where relevant.
The renormalisation and factorisation scales are set to half the sum of the
transverse mass of all the final-state particles both for LO and NLO calculations,
and the scale uncertainty is
estimated by varying the two scales independently by a factor of two up and down.
Additional details on the calculation of the $Y_0t\bar t$ cross section are
provided in ref.~\cite{Backovic:2015soa} while loop-induced processes are
extensively documented in ref.~\cite{Mattelaer:2015haa}.

All the cross sections shown in figure~\ref{fig:xsec} are proportional to
$g_t^2$ and we therefore arbitrarily choose $g_t=1$ as a benchmark. In this
case, sizeable cross sections of $10^1-10^3$~pb are expected for the production
 of light mediators with $m_Y \lesssim 100$~GeV at a
centre-of-mass energy of 8~TeV (left panel), the
dominant mechanism being the loop-induced $gg\to Y_0$ production mode. Requiring
an extra hard jet in the final state reduces the cross section by a factor
which depends on the missing energy (or the jet transverse momentum $p_T$)
selection, and the production rates are not sensitive to the mediator mass as
soon as the latter is smaller than the $\MET$ selection threshold. The cross
sections for producing the mediator in association with a Standard Model Higgs
or $Z$ boson are further suppressed. In contrast, the cross section related to
the production of the mediator in association with a top-quark pair
is significant for light mediators, but falls off
quickly with the increase in the mediator mass due to phase-space suppression.
As a result, a change in the collider energy from 8 to 13~TeV is important
for heavy mediators and the cross section can be enhanced by about an
order of magnitude. In the right panel of figure~\ref{fig:xsec}, we further show
first that the cross sections are constant when the dark matter particle pair is
produced through the decay of an on-shell mediator, and next that they are
considerably suppressed when the mediator is off-shell, especially for the $t\bar t X\bar X$ channel.

As already mentioned, the collider searches which provide the most relevant
constraints on simplified top-philic dark matter models are based on the
production channels shown in figure~\ref{fig:xsec} and can in general be divided
into two categories. The first category involves signals with missing transverse
energy originating from the production of dark matter particles that do not
leave any trace in the detectors and that are accompanied by
one or more Standard Model states. The most relevant searches of this type
are the production of dark matter in association with a top-quark
pair and the loop-induced production of dark matter in association with a jet, a
$Z$ boson or a Higgs boson. This is discussed in section~\ref{sec:met}.
The second category of searches relies on $\y$ resonant
contributions to Standard Model processes.
In our scenario, dijet, diphoton, top-pair and four-top
searches are expected to set constraints on the model parameter space. This is
discussed in section~\ref{sec:nonmet}. As shown below, missing-energy-based
searches and resonance searches are complementary and necessary for the best
exploration of the model parameter space at colliders.

In the rest of this section, we study collider constraints independently from
the cosmological and astrophysical ones, and we dedicate
section~\ref{sec:combined}
to their combination. We moreover allow the mediator couplings to be as large as
$2\pi$ 
and do not impose any constraint on the mediator width over mass ratio.
We summarise the relevant 8~TeV LHC constraints used in this study in
table~\ref{tab:colliderlimits} and give details on the $t\bar{t}+\MET$ and
monojet searches that have been recast in the \MA~framework in
appendix~\ref{sec:recast}.

 \begin{table}
 \center
 \begin{small}
 \begin{tabular}{l|lll}
 \hline
  Final state & Imposed constraint & Reference & Comments \\
 \hline\hline
  $\MET+t\bar t$
   & See appendix~\ref{sec:ttmet} & CMS~\cite{CMS:2014pvf}
   & Semileptonic top-antitop decay\\
  $\MET+j$
   & See appendix~\ref{sec:monoj} & CMS~\cite{Khachatryan:2014rra} & \\
  $\MET+Z$ 
   & $\sigma(\MET>150~{\rm GeV})<0.85$\,fb & CMS~\cite{Khachatryan:2015bbl}
   & Leptonic $Z$-boson decay\\  
  $\MET+h$ 
   &  $\sigma(\MET>150~{\rm GeV})<3.6$\,fb & ATLAS~\cite{Aad:2015dva}
   & $h\to b\bar b$ decay\\
 \hline
  $jj$
   & $\sigma(m_Y=500~{\rm GeV})<10$\,pb & CMS~\cite{CMS:2015neg}
   & Only when $m_Y>500$\,GeV\\
  $\gamma\gamma$
   & $\sigma(m_Y=150~{\rm GeV})<30$\,fb & CMS~\cite{Khachatryan:2015qba}
   & Only when $m_Y>150$\,GeV\\
  $t\bar t$ 
   & $\sigma(m_Y=400~{\rm GeV})<3$\,pb & ATLAS~\cite{Aad:2015fna}
   & Only when $m_Y>400$\,GeV \\
  $t\bar t t\bar t$ 
   & $\sigma<32$\,fb & CMS~\cite{Khachatryan:2014sca}
   & Upper limit on the SM cross section \\
 \hline
 \end{tabular}
 \caption{Summary of the 8\,TeV LHC constraints used in this paper.}
 \label{tab:colliderlimits}
 \end{small}
 \end{table}

\subsection{Constraints from searches with missing transverse energy}
\label{sec:met}

\subsubsection{The $t\bar t+\MET$ final state}
\label{sec:ttmetres}

Dark matter production in association with a top-quark pair ($t\bar{t}+\MET$)
has been explored by both the ATLAS~\cite{Aad:2014vea} and
CMS~\cite{Khachatryan:2015nua} collaborations within the 8~TeV LHC dataset, and
limits have been derived in particular in
the effective field theory approach~\cite{Cheung:2010zf,Lin:2013sca}. Such analyses could however be used to
derive constraints in other theoretical contexts, and we choose to recast the CMS
search to constrain the parameters of the simplified top-philic dark matter
model under consideration. In this work, we simulate $t\bar{t}X\bar{X}$ events at the NLO
accuracy in QCD by making use of \MG. The first study of the genuine NLO effects
on the production of a system composed of a pair of top quarks and a pair of
dark matter particles has been presented in ref.~\cite{Backovic:2015soa} in
which NLO $K$-factors have been investigated both at the total cross-section and
differential distribution level for a series of representative benchmark
scenarios. Here, we explore the impact of the NLO corrections on the
exclusion limits originating from the $t\bar{t}+\MET$ channel.

\begin{figure}
\center \includegraphics[width=0.48\textwidth]{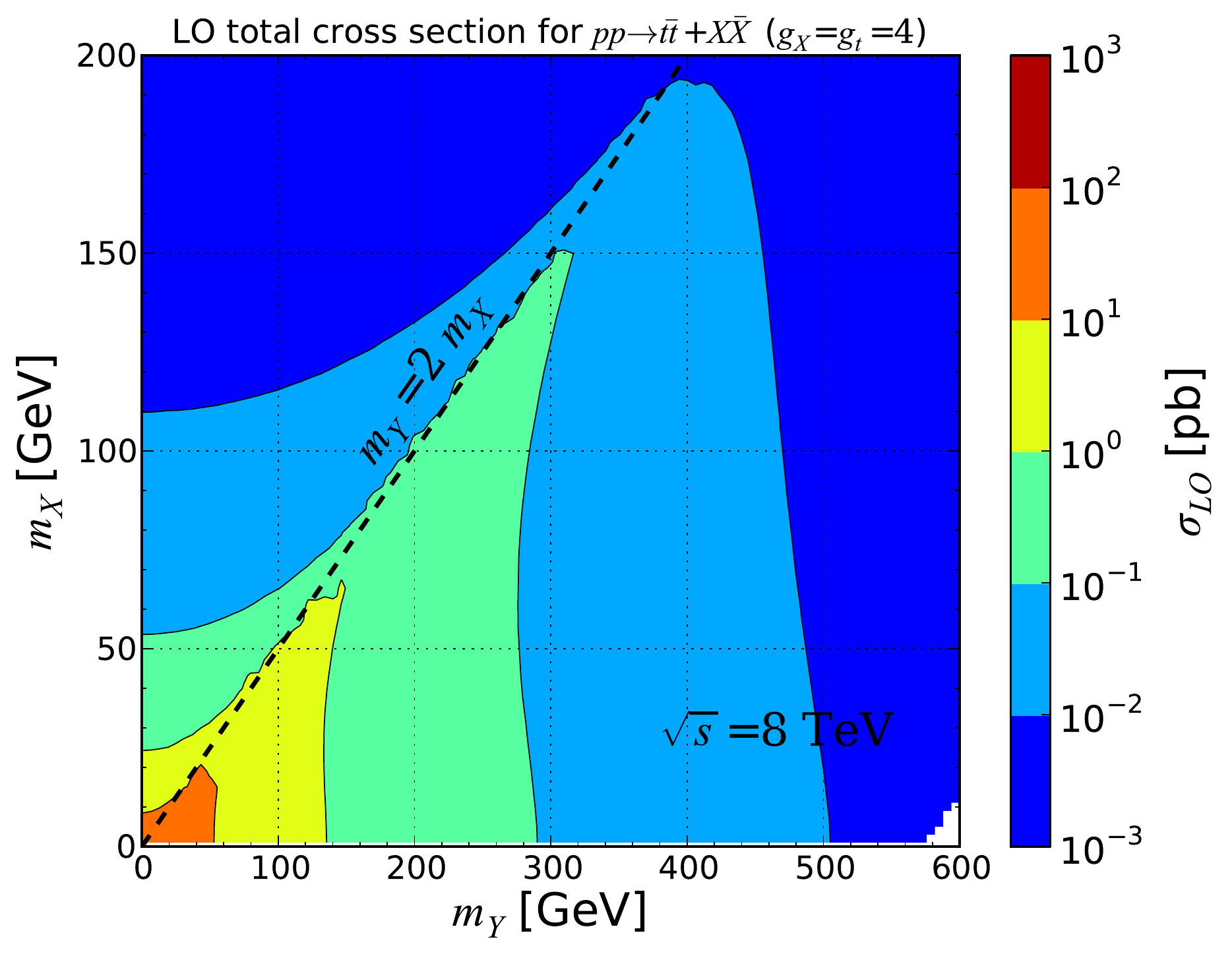}\quad 
\includegraphics[width=0.48\textwidth]{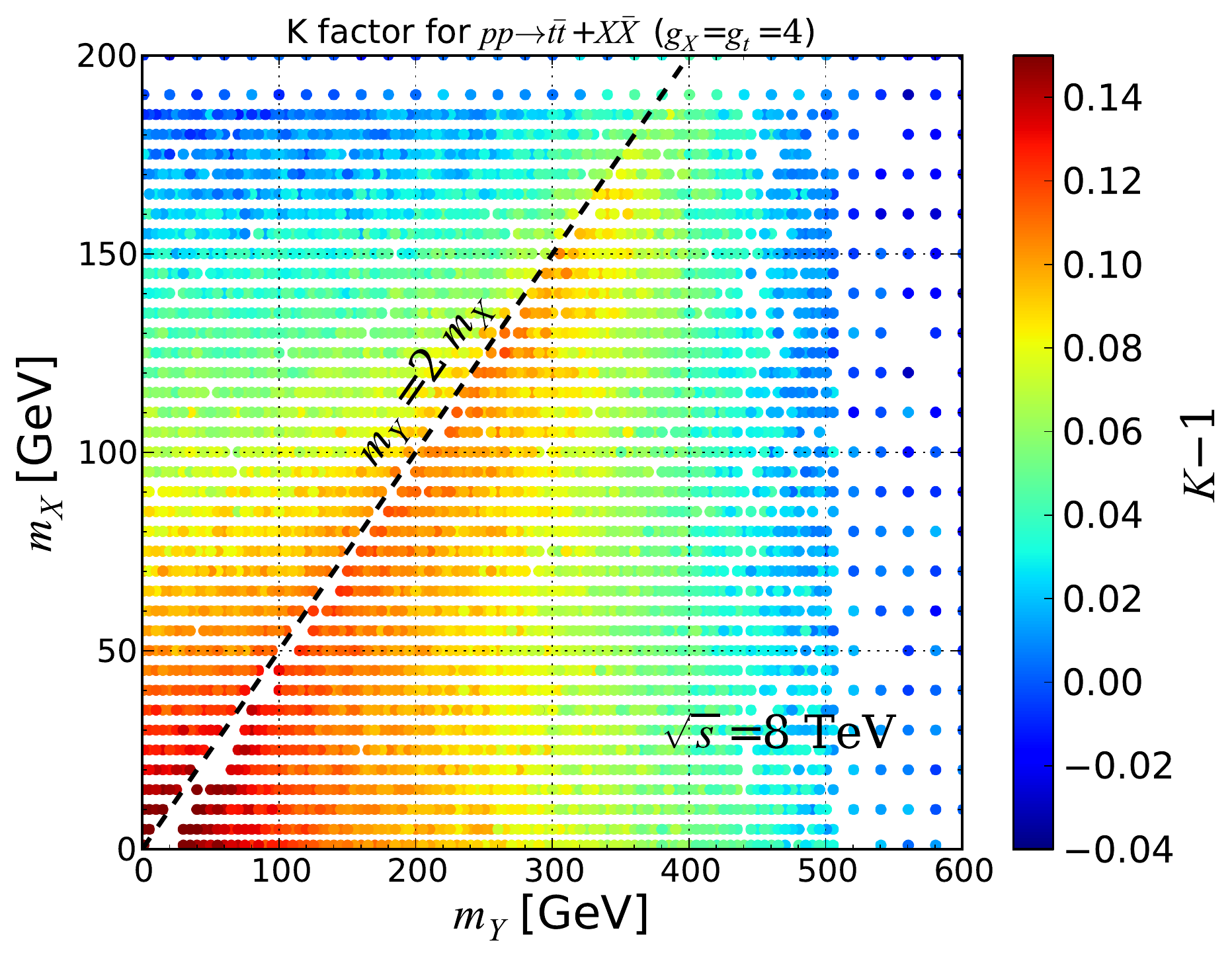} 
 \caption{LO cross sections (left) and corresponding $K$-factors (right) for $pp\to t\bar{t}X\bar{X}$ at $\sqrt{s}=8$\,TeV as a function of the mediator and dark matter masses.
 The top and dark matter couplings to the mediator are set to 4.
 } 
\label{fig:ttmet_sigma}
\end{figure}

In order to examine the reach of the CMS search, we start by performing a
two-dimensional scan of the mediator and dark matter masses with fixed mediator couplings, similar to figure~7 in ref.~\cite{Haisch:2015ioa}.
The same scan is performed at both LO and NLO accuracy concerning
the simulation of the hard scattering process, which allows us to determine the impact
of the QCD corrections on the exclusion bounds. Before presenting the results
for the excluded regions and to facilitate the discussion, we show the
dependence of the LO cross section for $g_t=g_X=4$ on the new physics masses and
the corresponding $K$-factors in figure~\ref{fig:ttmet_sigma}. The cross section
is the largest in the low mass regions where the mediator can resonantly decay
to a pair of dark matter particles,  and falls steeply in the off-shell regions.
In particular, the region where
$2\mdm<\my<2m_t$ is characterised by mediator decays either into a pair
of dark matter particles or into a pair of gluons. These two decay rates are
related by (see section~\ref{sec:model})
\be
  \frac{\Gamma(Y_0\to gg)}{\Gamma(Y_0\to X\bar X)} = 
    \frac{g_t^2}{g_X^2}\frac{\alpha_s^2}{9\pi^2\beta_X^3}\frac{m_Y^2}{v^2}
     \sim \bigg(\frac{g_t}{g_X}\bigg)^2\times 10^{-5} \frac{\my}{\rm GeV},
\ee
which suggests that the decay rate into a pair of dark matter particles is
always significantly higher, except in the case of a large hierarchy between the
couplings ($\gsm /\gx \gtrsim 100$). For $\my>2m_t$, the $Y_0\to t\bar t$ decay
mode is open, and $t\bar t+\MET$ production turns out to be suppressed by the
visible decay channels of the mediator, unless $g_X>g_t$. Such a feature has
already been illustrated
in figure~\ref{fig:width}. The NLO $K$-factors related to $t\bar{t}+\MET$
production (right panel of
figure~\ref{fig:ttmet_sigma}) are found to vary from 0.96 to 1.15 in the range
of masses examined here, the QCD corrections being more important in the low
mass region.

\begin{figure}
  \center
   \includegraphics[width=0.45\textwidth]{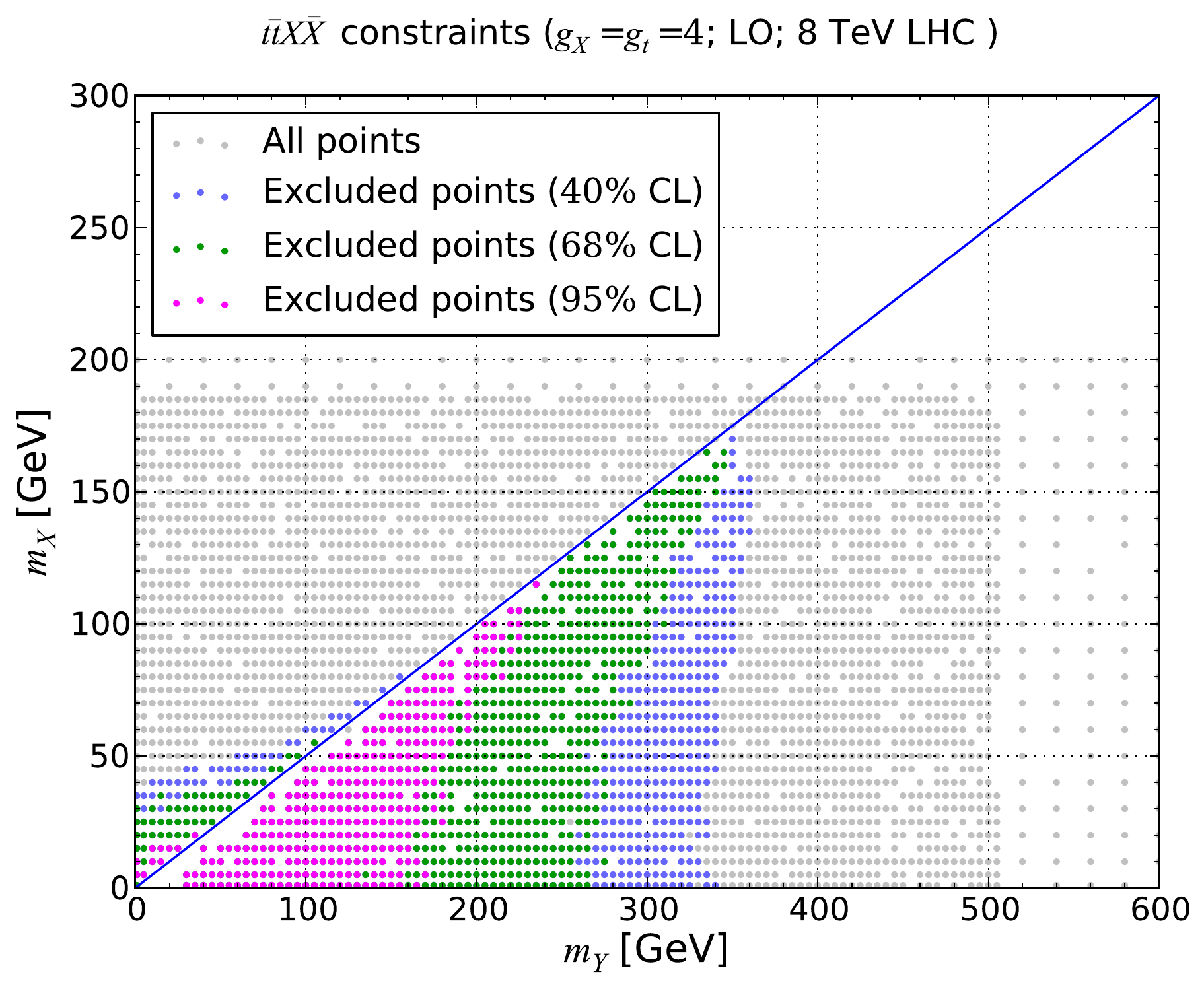}
   \quad
   \includegraphics[width=0.45\textwidth]{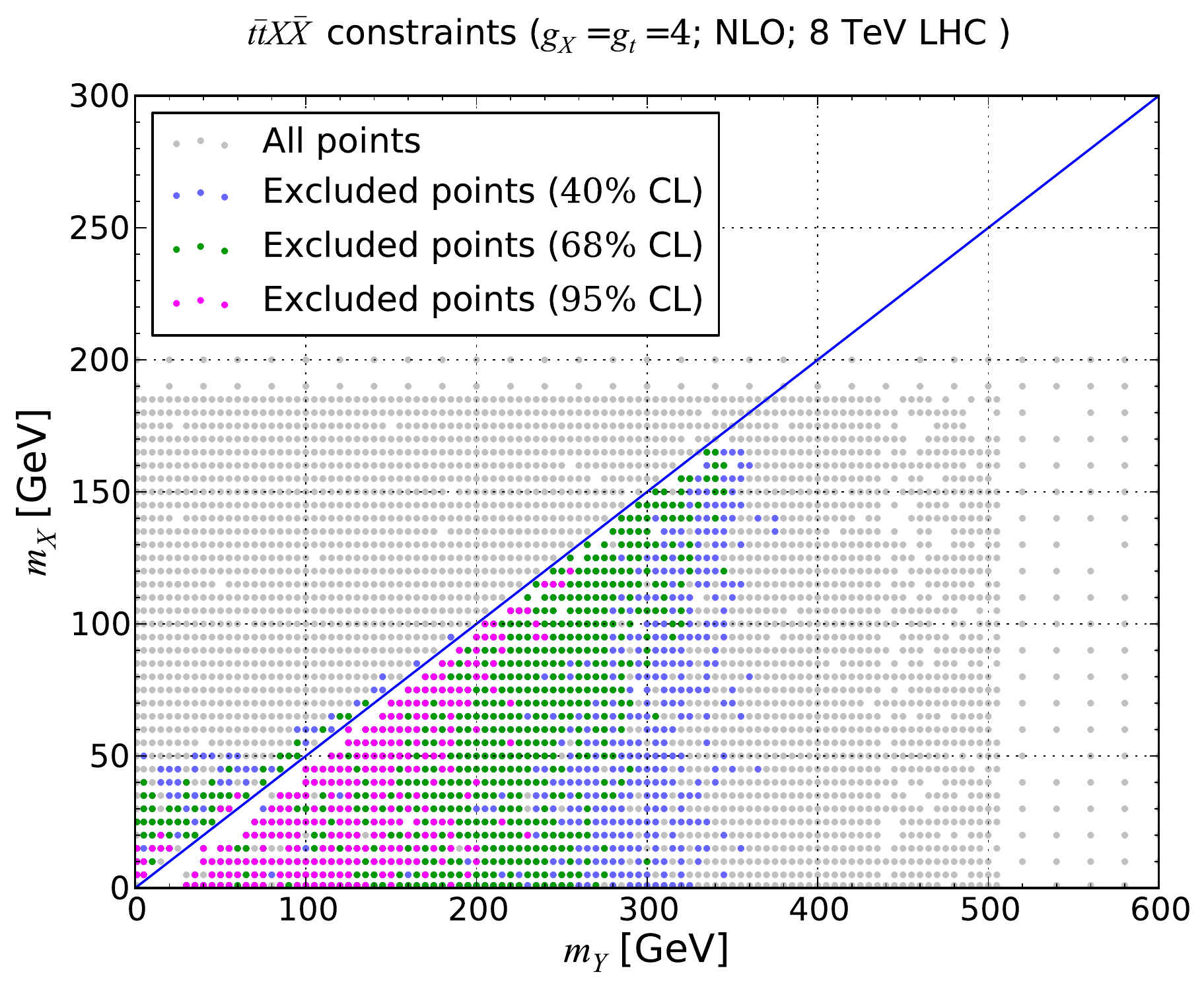}
 \caption{Constraints on simplified top-philic dark matter scenarios from the
  CMS 8TeV $t\bar{t}+\MET$ analysis~\cite{Khachatryan:2015nua}. The top and dark
  matter couplings to the mediator are set to 4 while the mediator and dark
  matter masses are allowed to vary freely. LO and NLO exclusions are
  respectively shown in the left and right panels of the figure.
 }
\label{fig:ttmet_g4}
\end{figure}

The results for the exclusion regions are shown in figure~\ref{fig:ttmet_g4}
when LO (left panel) and NLO (right panel) simulations are used; see more details on the recasting procedure in appendix~\ref{sec:ttmet}.
Setups excluded
at the 40\%, 68\% and 95\% confidence level (CL) are marked separately in the
figures. As expected from the total cross section results, all excluded points
(at the 95\% CL) lie in the triangular low-mass region where the mediator
resonantly decays into a dark matter particle pair. The exclusion region reaches
mediator masses of about 200--250~GeV if close to threshold ($\my\sim2\mdm$). This region
is in fact not exactly triangular as for a given mediator mass, not all dark
matter masses below $\my/2$ are excluded. This is related to the parametric
choice of
$g_t=g_X=4$ for which the mediator width can become large. In this case, the
narrow width approximation is not valid and the $t\bar{t}+\MET$ cross section
acquires a dependence on the dark matter mass even in the resonant region.

Comparing the LO and NLO results, we observe that in
the low mass resonant region where the $K$-factor is small and of about 1.10,
the exclusion contours are mildly modified and this small 10\% shift in the
cross section does not lead to any significant change. For larger mediator masses,
the $K$-factors are  $\sim 1$ and therefore do not imply a modification of the exclusion regions, if the central prediction at the 
default choice of scale is considered. However, the inclusion of NLO corrections significantly 
reduces the theoretical error and thus leads to sharper exclusion bounds as discussed below.

\begin{figure}
\center
\includegraphics[width=0.6\textwidth]{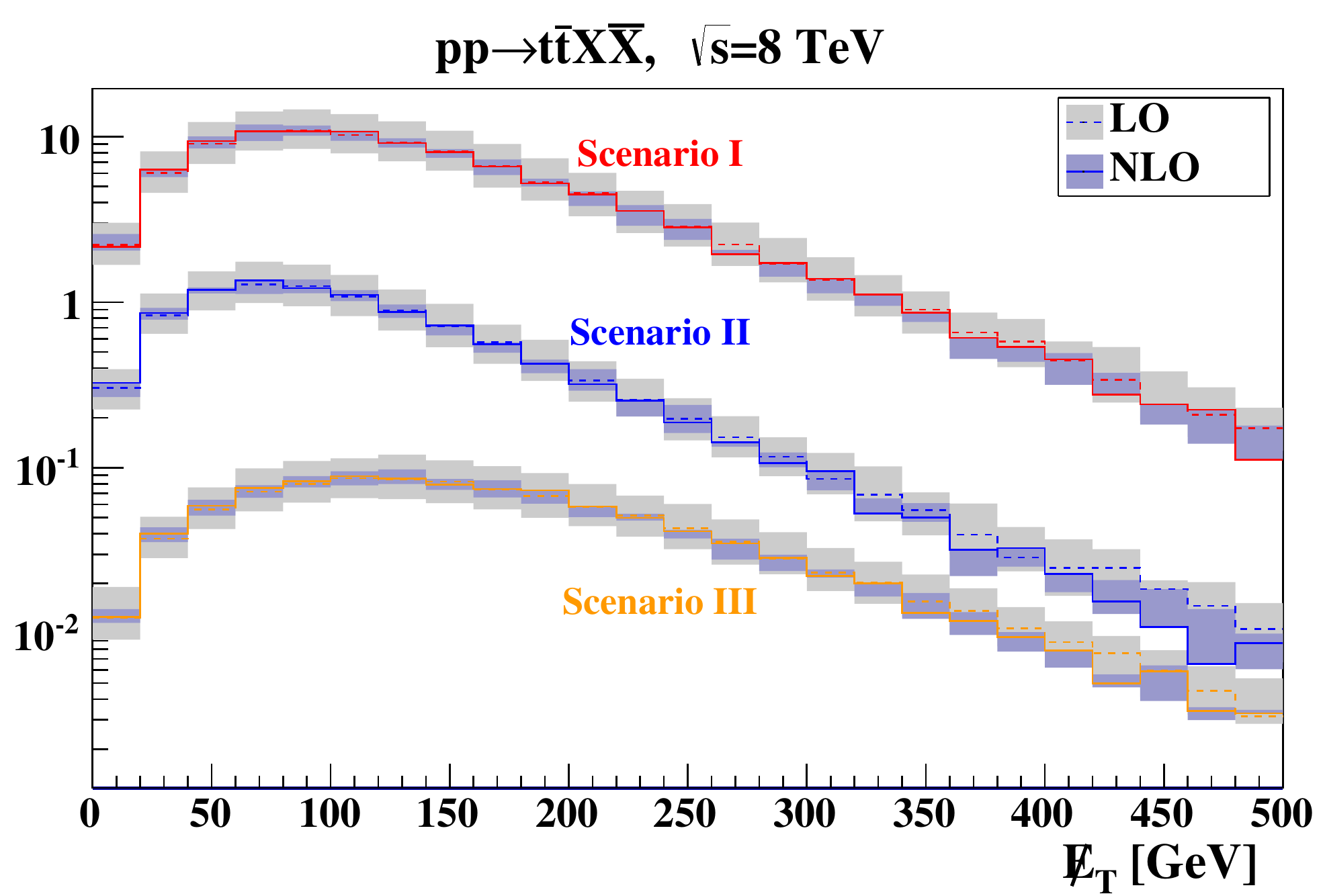}
\includegraphics[width=0.63\textwidth]{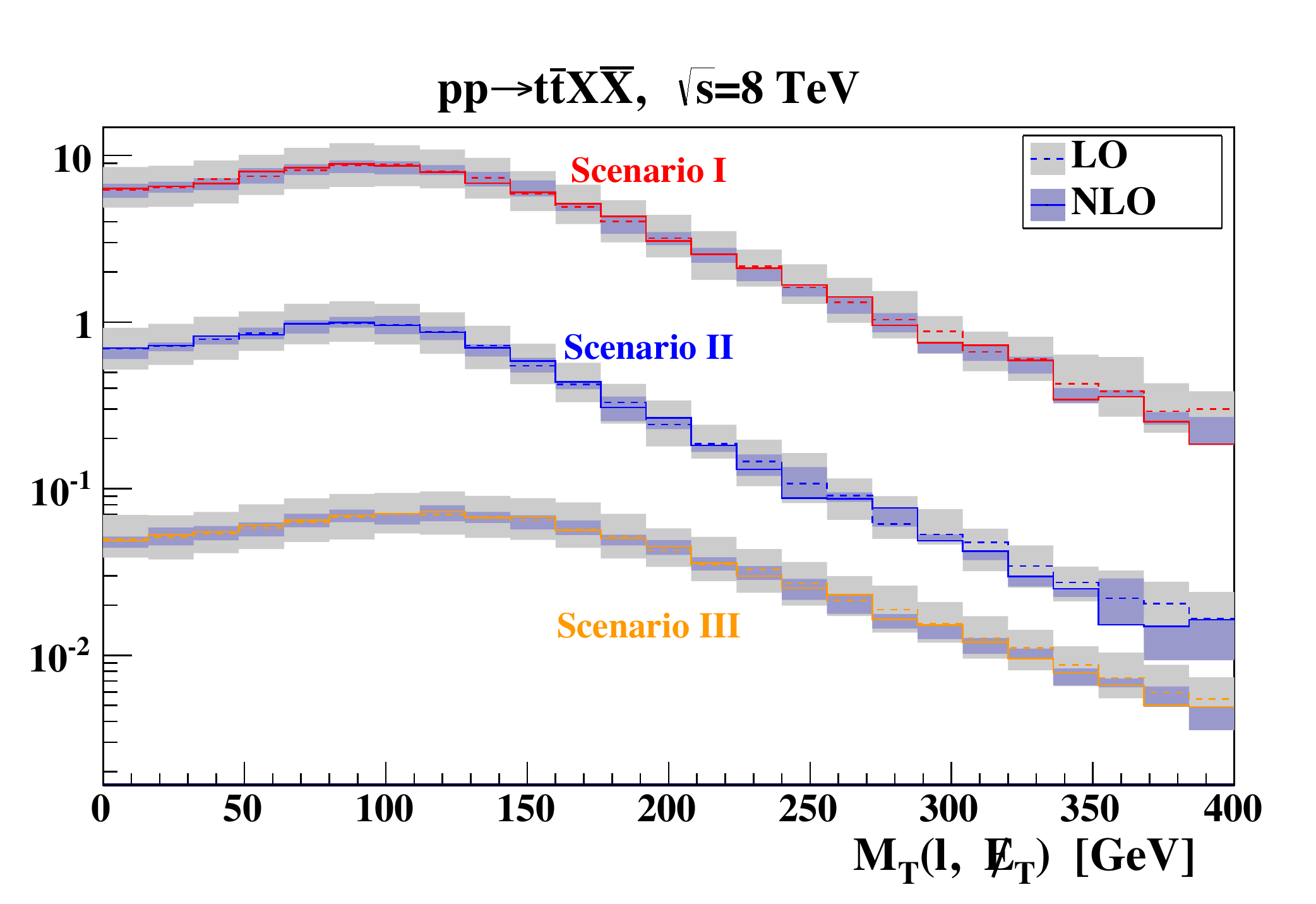}
\includegraphics[width=0.6\textwidth]{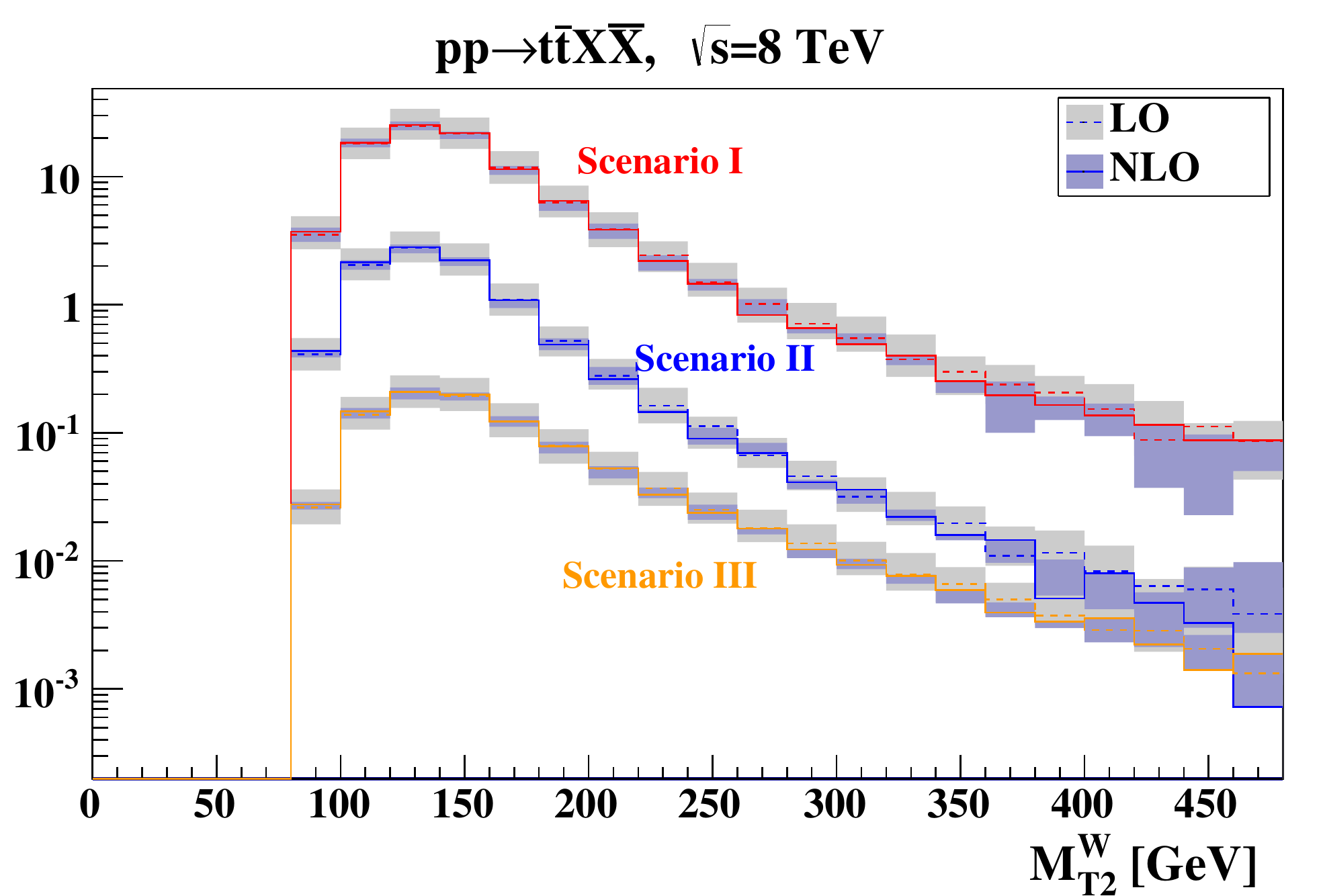} 
 \caption{Differential distributions for $\MET$, $M_T(\ell,\MET)$ and $M_{T2}^{W}$ for the three scenarios of table~\ref{tab:ttmetbench} at LO and NLO. The distributions are normalised to 100, 10 and 1 for Scenarios I, II and III respectively, and the scale uncertainty bands obtained by varying the renormalisation and factorisation scale in the range of $0.5 \mu_0 <\mu_{R,F} < 2\mu_0$ are also shown. 
 }
\label{fig:ttmet_distr}
\end{figure}

In order
to further investigate the effects of the NLO corrections, we select three
benchmark scenarios for which we perform a detailed study. These benchmarks are
defined in table~\ref{tab:ttmetbench} where they are presented along with the
corresponding LO and NLO cross-sections and the CL exclusion obtained with \MA. As discussed in appendix~\ref{sec:ttmet}, the most relevant observables for
this analysis consist of the $\MET$, $M_T(\ell,\MET)$ and $M_{T2}^{W}$ for which distributions are shown in figure~\ref{fig:ttmet_distr}. We normalise the distributions to 100, 10 and 1 for the scenarios I, II and III respectively to ensure that they are all clearly
visible in the figure. Moreover, we also indicate the scale uncertainty bands
that have been obtained from a scale variation of $0.5 \mu_0<\mu_{R,F}<2\mu_0$.
In agreement with the findings of ref.~\cite{Backovic:2015soa}, higher-order
corrections have a rather mild effect on the distribution shapes for all key
observables.
Using NLO predictions however leads to a significant reduction of the scale uncertainties compared to the LO case.
In table~\ref{tab:ttmetbench}, one can also see that the use of NLO
predictions leads to a significant reduction of the uncertainty in the cross
section which propagates down to the CLs. NLO predictions therefore allow us
to draw more reliable conclusions on whether a parameter point is excluded.

\begin{table}
\footnotesize
\begin{center} \renewcommand\arraystretch{1.2}
\begin{tabular}{clllll} 
  \hline

 & ($\my$, $\mdm$)  & $\sigma_{\textrm{LO}}$ [pb] & CL$_{\rm LO}$ [\%]
 & $\sigma_{\textrm{NLO}}$ [pb] & CL$_{\rm NLO}$ [\%] \\
  \hline\hline\
 I &   (150, 25)~GeV 
   & 0.658$^{+34.9\%}_{-24.0\%}$ & 98.7$^{+0.8\%}_{-13.0\%}$  
   & 0.773$^{+6.1\%}_{-10.1\%}$ & 95.0$^{+2.7\%}_{-0.4\%}$ \\[1mm]
 II &  (40, 30)~GeV 
   &  0.776$^{+34.2\%}_{-24.1\%}$ & 74.7$^{+19.7\%}_{-17.7\%}$  
   & 0.926$^{+5.7\%}_{-10.4\%}$ & 84.2$^{+0.4\%}_{-14.4\%}$ \\[1mm]
 III & (240, 100)~GeV 
   & 0.187$^{+37.1\%}_{-24.4\%}$ & 91.6$^{+6.4\%}_{-18.1\%}$
   & 0.216$^{+6.7\%}_{-11.4\%}$ & 86.5$^{+8.6\%}_{-5.5\%}$ \\[1mm]
\hline
\end{tabular}
\end{center}\renewcommand\arraystretch{1.0}
\caption{Benchmark scenarios used to investigate the impact of the NLO
  corrections on the $t\bar{t}+\MET$ CMS search. The LO and NLO cross sections at 8\,TeV LHC
  are shown together with the CL exclusion obtained from \MA.
  The uncertainties originating from scale variation ($0.5 \mu_0 <\mu_{R,F} < 2\mu_0$) are also shown.
  } \label{tab:ttmetbench}
\end{table}

\subsubsection{Mono-$X$ final states}\label{sec:monojres}

In addition to the constraints that can be derived by means of $t\bar t+\MET$
probes and that have been discussed in the previous section, mono-$X$ searches
can also be relevant for obtaining bounds on our top-philic dark matter model.
Monojet~\cite{Aad:2015zva,Khachatryan:2014rra,CMS:2015jdt},
mono-$Z$~\cite{Aad:2013oja,Aad:2014vka,CMS:2015jha,Khachatryan:2015bbl,%
ATLAS:monoZ} and mono-Higgs~\cite{Aad:2015dva,Aad:2015yga,monoh1,monoh2}
signals have been searched for during the first run of the LHC, and these search
results could be recast to constrain the dark matter model studied in this
work. In
contrast to tree-level dark matter production in association with a pair of top
quarks, the production of a pair of dark matter particles with a jet, a
$Z$-boson or a Higgs boson proceeds via a gluon fusion top-quark loop diagram.
Although they have been largely studied by ATLAS and CMS,
monophoton analyses cannot be
used as charge conjugation invariance forbids the existence of a monophoton
signal for the spin-0 mediator scenario.

\subsubsection*{Monojet}

We start by discussing constraints that can be imposed by the CMS 8~TeV monojet
analysis~\cite{Khachatryan:2014rra}. For this study, hard-scattering events are
generated at the LO accuracy within \MG, and the matching with parton showers is
made with \PY. The results are analysed in \MA\ that also takes care of the
detector simulation using its interface with \DEL. This recasting procedure allows us
to exclude any specific parameter space point at any desired confidence level,
our exclusion being conservatively derived on the basis of the signal region
that drives the strongest bound. This limitation is related to the lack of
public information, the statistical model used by CMS for the combination being
not available.
One can find more details for the recasting procedure in appendix~\ref{sec:monoj}.

\begin{figure}
  \center
  \includegraphics[width=0.48\textwidth]{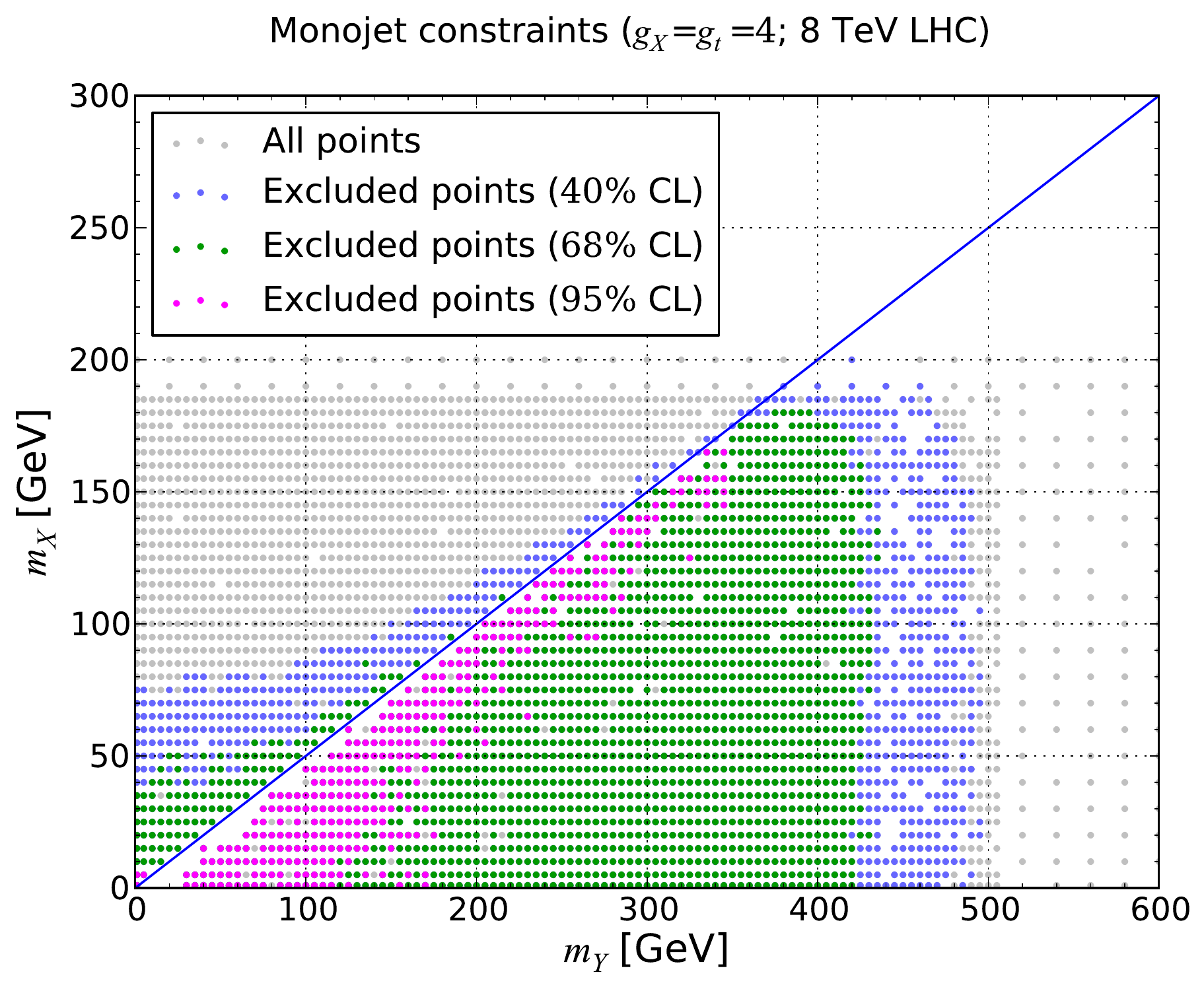}
   \caption{Constraints on the simplified top-philic dark matter model from the CMS 8~TeV monojet analysis~\cite{Khachatryan:2014rra}.
 The top and dark matter couplings to the mediator are set to 4 while the mediator and dark matter masses are allowed to vary freely.
 }
\label{fig:monojet_g4}
\end{figure}

Similar to the $t\bar t+\MET$ analysis of the previous section, we perform a
two-dimensional scan on the mediator and dark matter masses while fixing both
new physics couplings to \mbox{$\gsm=\gx=4$} (as in figure~5 in ref.~\cite{Haisch:2015ioa}).
Figure~\ref{fig:monojet_g4} shows our results,  where we represent the scenarios excluded at the
40\%, 68\% and 95\% CL. The bulk of the excluded points lie again in the
triangular low-mass region where the mediator resonantly decays into a pair of
dark matter particles. Except for the small subset of points excluded at the 40\%
and 68\% CL in the region where $m_Y < 2 \mdm$,
the extent of the exclusion region is determined by the significant reduction of
the monojet cross section below the resonant production threshold already
presented in
figure~\ref{fig:xsec}. The $pp\to Y_0 j$ cross section indeed rapidly falls with
$\my$, reaching levels beyond the sensitivity of the 8~TeV search at
$\my~\sim~500$~GeV. In addition to the decrease of the $Y_0 j$ production cross
section, the opening of the mediator decay mode into a top-antitop system when
$\my > 2 m_t$ leads to a further reduction of the monojet production rate. In
comparison with the $t\bar t+\MET$ case, the monojet search overall appears to
be more constraining, especially for higher mediator mass values thanks to
the larger monojet cross section.

\begin{figure}
\center
 \includegraphics[width=0.48\textwidth]{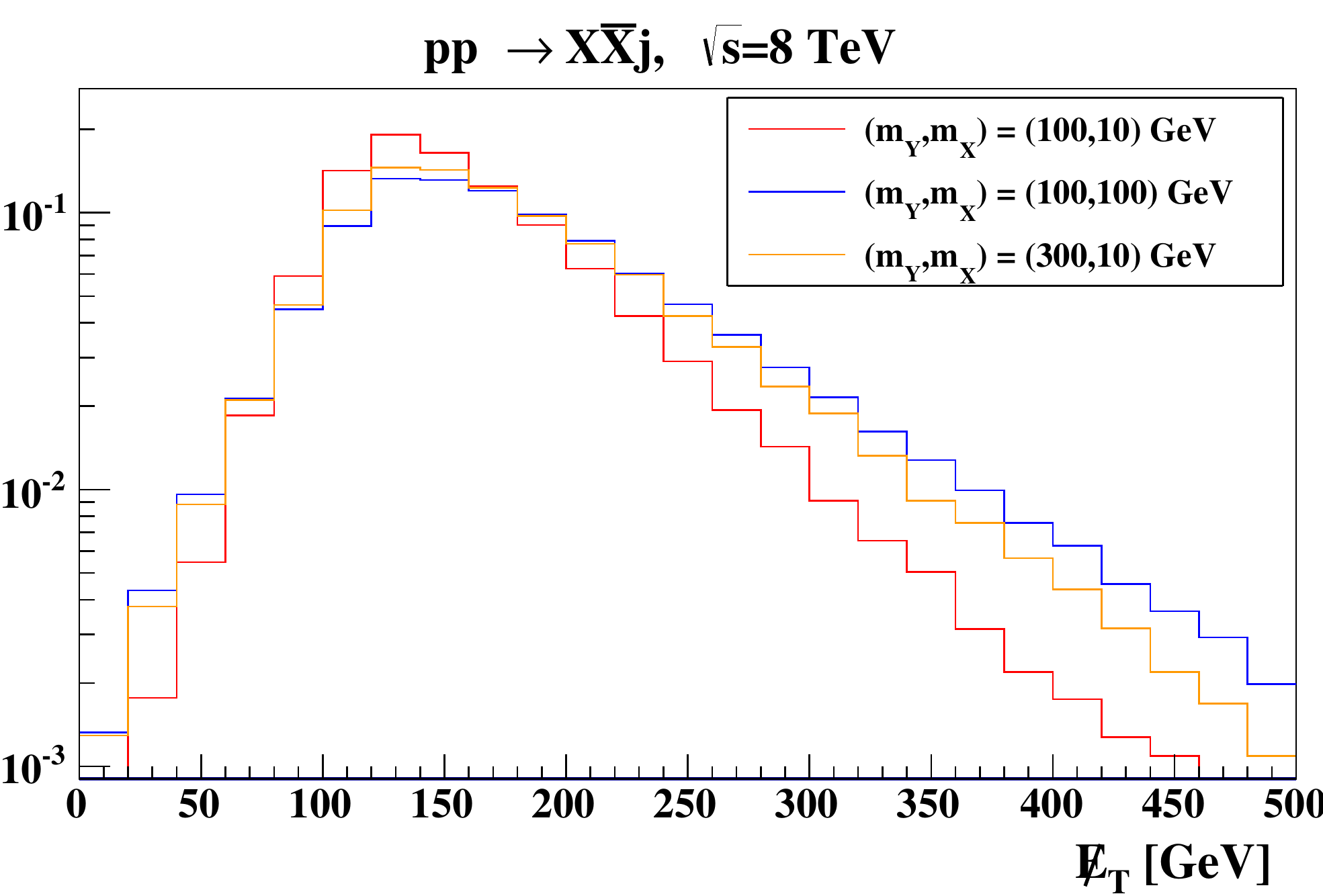}
 \includegraphics[width=0.48\textwidth]{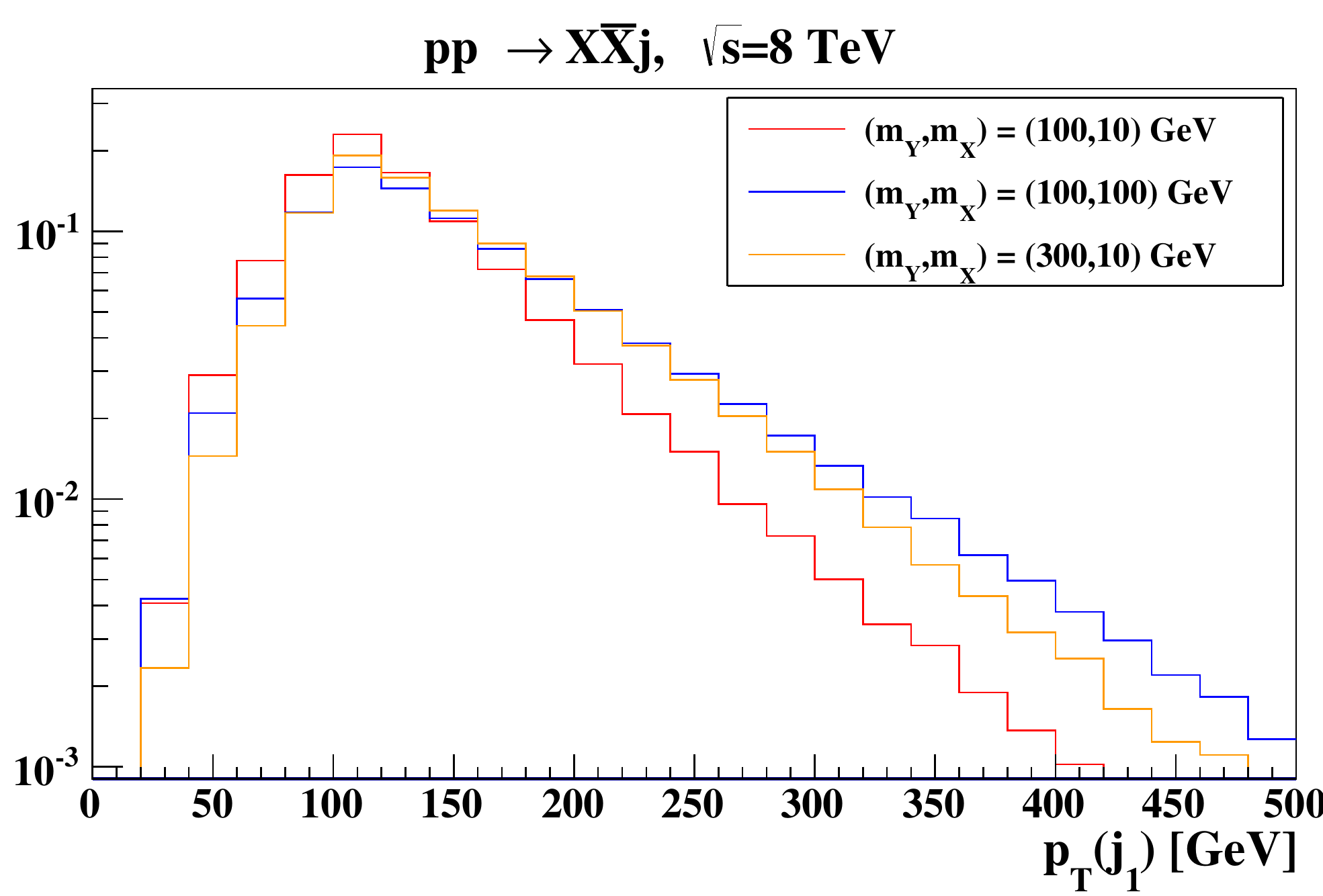}
 \caption{Differential distributions for the missing transverse energy $\MET$
   and the hardest jet transverse momentum $p_T(j_1)$ for the three scenarios
  defined in table~\ref{tab:jmetbench}. The distributions are normalised to one.
  }
\label{fig:monoj_distr}
\end{figure}

As shown in ref.~\cite{Mattelaer:2015haa}, the shape of key monojet
differential distributions differs in the resonant and in the off-shell
parameter space regions. While the total cross section falls dramatically in the
off-shell region $\my < 2 \mdm$ (as shown in figure~\ref{fig:xsec}), the
$\MET$ and jet transverse momentum distributions tend to be harder for off-shell
production. We demonstrate this feature with a detailed investigation of
three benchmark points defined in table~\ref{tab:jmetbench}. They consist of two
resonant scenarios with different mediator masses and one non-resonant scenario.
The monojet production rate is also indicated in the table, and we present
normalised distributions relevant for the monojet analysis in
figure~\ref{fig:monoj_distr}. The off-shell scenario yields harder
distributions compared to the resonant cases. This implies that a larger
fraction of events features high missing transverse energy ($\MET>$250~GeV) and
populates the different signal regions of the CMS analysis. As a result, a
better sensitivity is found than what one might expect from considering the
total cross section alone. This feature leads to the exclusion of dark matter
scenarios where $\my < 2 \mdm$, as depicted in figure~\ref{fig:monojet_g4}.

\begin{table}
\footnotesize
\begin{center} \renewcommand\arraystretch{1.2}
\begin{tabular}{lcc}
  \hline
 ($\my$, $\mdm$)  & $\sigma_{\textrm{LO}}$ [pb]  \\
  \hline\hline
(100, 10) GeV & 0.605  \\
(300, 10) GeV &  0.194 \\
(100, 100) GeV &  0.00261  \\
\hline
\end{tabular}
\end{center}\renewcommand\arraystretch{1.0}
\caption{Benchmarks used to investigate the differential distributions related
 to the CMS monojet analysis. The corresponding cross sections for a $\MET >$ 150~GeV selection are shown in the second column. } \label{tab:jmetbench}
\end{table}

In our simulation of the monojet signal, we have ignored the possible impact of
the merging of event samples featuring different final state jet multiplicities.
A reliable description of the high transverse momentum spectra of the leading
jet typically necessitates the merging of event samples including at least one and
two jets in the final state~\cite{Mattelaer:2015haa}. We have explicitly
verified that for both resonant and off-shell scenarios, employing a merged
sample does not have a big impact on the $\MET$ distribution and therefore on
the resulting exclusion contours. This originates from the analysis selection
strategy that requires one single hard jet and rather loose requirements on the
second jet, so that the configuration that dominates consists of a single hard
jet recoiling against the missing energy. Such a configuration is described
similarly by the one-jet and merged samples. We nevertheless stress that the
importance of the merging procedure has to be checked on a case-by-case basis as
this depends on the analysis, so that higher multiplicity
samples might be necessary to accurately describe the relevant distributions.

\begin{figure}
\center
 \includegraphics[width=0.48\textwidth]{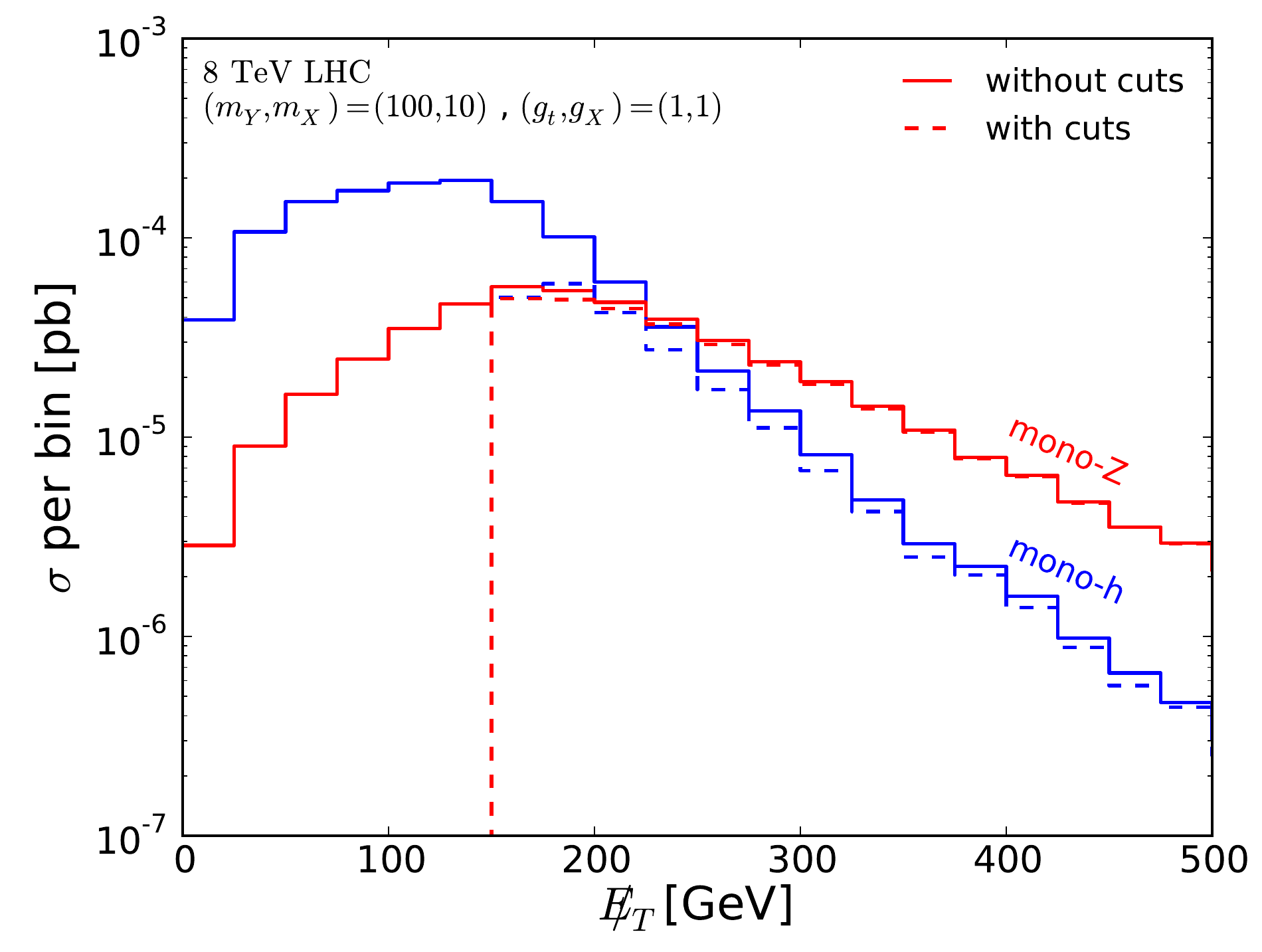}
 \includegraphics[width=0.48\textwidth]{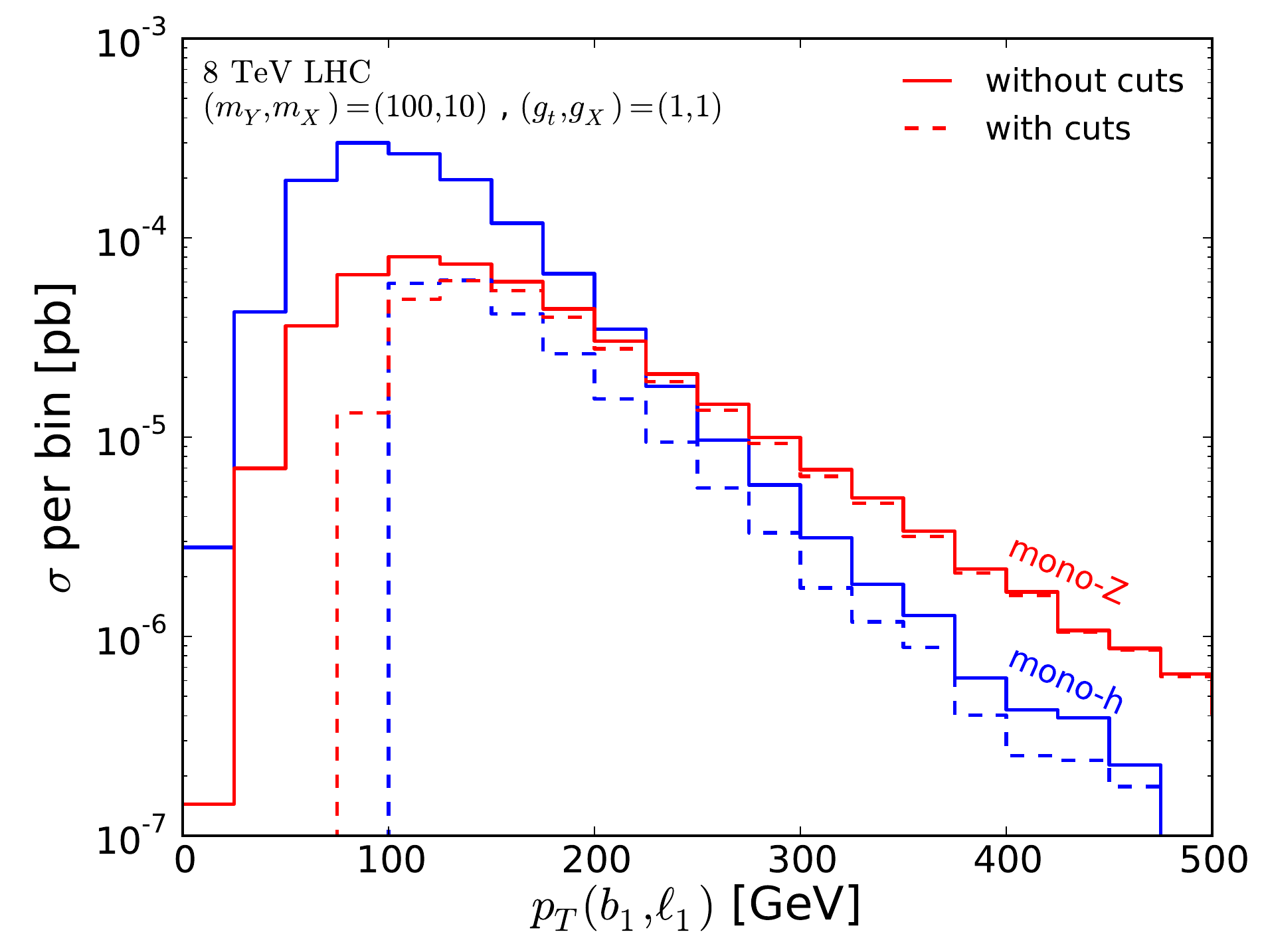}
 \caption{Distributions of missing transverse energy and of the transverse
  momentum of the leading lepton ($b$-quark) for mono-$Z$ (-Higgs) production
  at $\sqrt{s}=8$\,TeV for $(m_Y,m_X)=(100,10)$\,GeV and $(g_t,g_X)=(1,1)$, without and with including the
  analysis selections.   }
\label{fig:monozh_g4}
\end{figure}

\subsubsection*{Mono-$Z$ and mono-Higgs}

In addition to the use of monojet processes, we explore the possibility of
constraining the parameter space of our model using mono-$Z$ and mono-Higgs production. 
While the production rates are much smaller than the monojet rate as seen in figure~\ref{fig:xsec}, the backgrounds can be also small.
Therefore, these search channels can be sensitive to the top-philic simplified dark matter model,  as we will see below. 
Here, instead of employing a full recasting procedure as in the $t\bar t+\MET$ and monojet analyses, we perform parton-level analyses to provide rough estimates of the constraints on our model parameters.

We rely
on the CMS search for dark matter production in association with a $Z$-boson
that decays leptonically~\cite{Khachatryan:2015bbl}, in which a 95\% CL upper
limit on the visible cross section of 0.85~fb is obtained once a $\MET$
requirement of at least 150\,GeV and the minimal detector selection requirements for the leptons ($p_T^\ell>20$\,GeV and $|\eta^\ell|<2.5$) are considered.
We generate events for this process, and after applying the above fiducial selection requirements
we obtain a cross section of
0.30~fb for $(m_Y,m_X)=(100,10)$\,GeV and $g_t=g_X=1$.
We show in
figure~\ref{fig:monozh_g4} the $\MET$ and leading lepton transverse
momentum distributions (red lines) without and with
applying the selection strategy.
While we have not performed a detailed study, simple
estimates show good prospects for setting limits on the parameter space of the
model using the mono-$Z$ analysis results. Using the upper limit of 0.85\,fb, scenarios with couplings close to $g_t\sim2$ could be excluded
in the resonant region ($\my > 2 \mdm$) with $\my <100$\,GeV. For larger
mediator masses, the cross section starts to fall due to the reduction of the
phase space. In the off-shell region ($\my < 2 \mdm$), the mono-$Z$ cross
section suffers from the same drastic decrease seen in figure~\ref{fig:xsec} for
the $t\bar{t}+\MET$ and monojet cases.

The same procedure can be repeated to constrain the parameter space of the model
using mono-Higgs events on the basis of the results of the ATLAS search for
dark matter production in association with a Higgs boson decaying into two
bottom quarks~\cite{Aad:2015dva}. This search results in a 95\% CL upper limit
on the visible cross section of 3.6~fb for a $\MET$ threshold of 150\,GeV.
In order to estimate a limit, we generate events for $(m_Y,m_X)=(100,10)$\,GeV and $g_t=g_X=1$, and require the two $b$-quarks to have a
transverse momentum $p_T^{b_1}>100$\,GeV and $p_T^{b_2}>25$\,GeV, a
pseudorapidity $|\eta^b|<2.5$ and to be separated in the transverse plane by an
angular distance $\Delta R(b_1,b_2)<1.5$. Moreover, we only select events
exhibiting at least 150~GeV of missing transverse energy.
We show again in
figure~\ref{fig:monozh_g4} the $\MET$ and leading $b$-quark transverse
momentum distributions (blue lines) without and with
applying the above-mentioned selection requirements. 
We then include a $b$-tagging efficiency of 60\% and 
extract an upper limit on the $g_t$
coupling by comparing our results to the ATLAS limit. Coupling values of
$\gsm>2$
are found to be excluded for $\my > 2 \mdm$ with $\my <100$\,GeV. All other
parameter space regions suffer from the same limitations as the mono-$Z$
case.

From our naive parton-level analysis, we have seen that mono-$Z$ and
mono-Higgs signals show promising signs of setting constraints on the parameter
space of the model and therefore deserve dedicated studies, which will be reported elsewhere (see also ref.~\cite{Goncalves:2016bkl}).
The sensitivity to
such signals will benefit from applying more aggressive $\MET$ thresholds to
ensure the reduction of the corresponding backgrounds.
As seen in figure~\ref{fig:monozh_g4},
we obtain a rather hard $\MET$ distribution~\cite{Mattelaer:2015haa},  especially for
mono-$Z$ production. The result implies that an
increase in the $\MET$ threshold requirement in future analyses could
lead to a significant improvement of the sensitivity, especially given the
the fact that Standard Model backgrounds rapidly fall off with the increase in missing energy.

\subsection{Constraints from searches without missing transverse energy}
\label{sec:nonmet}

\subsubsection*{Dijet and diphoton resonances}
 Dijet and diphoton resonance search
results could (in principle) be used to constrain the simplified
top-philic dark matter model. Due to double-loop suppressions, mediator-induced contributions to dijet and diphoton production are only relevant in the parameter space regions where $\my < 2\mdm,2m_t$ ($i.e.$ where the mediator cannot decay into top quarks and/or dark matter particles). The partial mediator
decay rate into gluons is then always dominant (as mentioned in
section~\ref{sec:model}) since 

\be
  \frac{\Gamma(\y \rightarrow \gamma \gamma)}{\Gamma(\y \rightarrow gg)}
     \sim \frac89 \frac{\alpha_e^2}{\alpha_s^2} \approx 10^{-3}\ .\label{eq:brbr}
\ee

All LHC dijet resonance searches focus on
the dijet high invariant-mass region, leading to no useful constraints on the top-philic dark matter model. The
lowest mediator mass that is probed is $\sim500$~GeV, with a visible cross
section restricted to be smaller than 
10~pb~\cite{Khachatryan:2016ecr}.

\begin{figure}
\center
\includegraphics[width=0.6\textwidth]{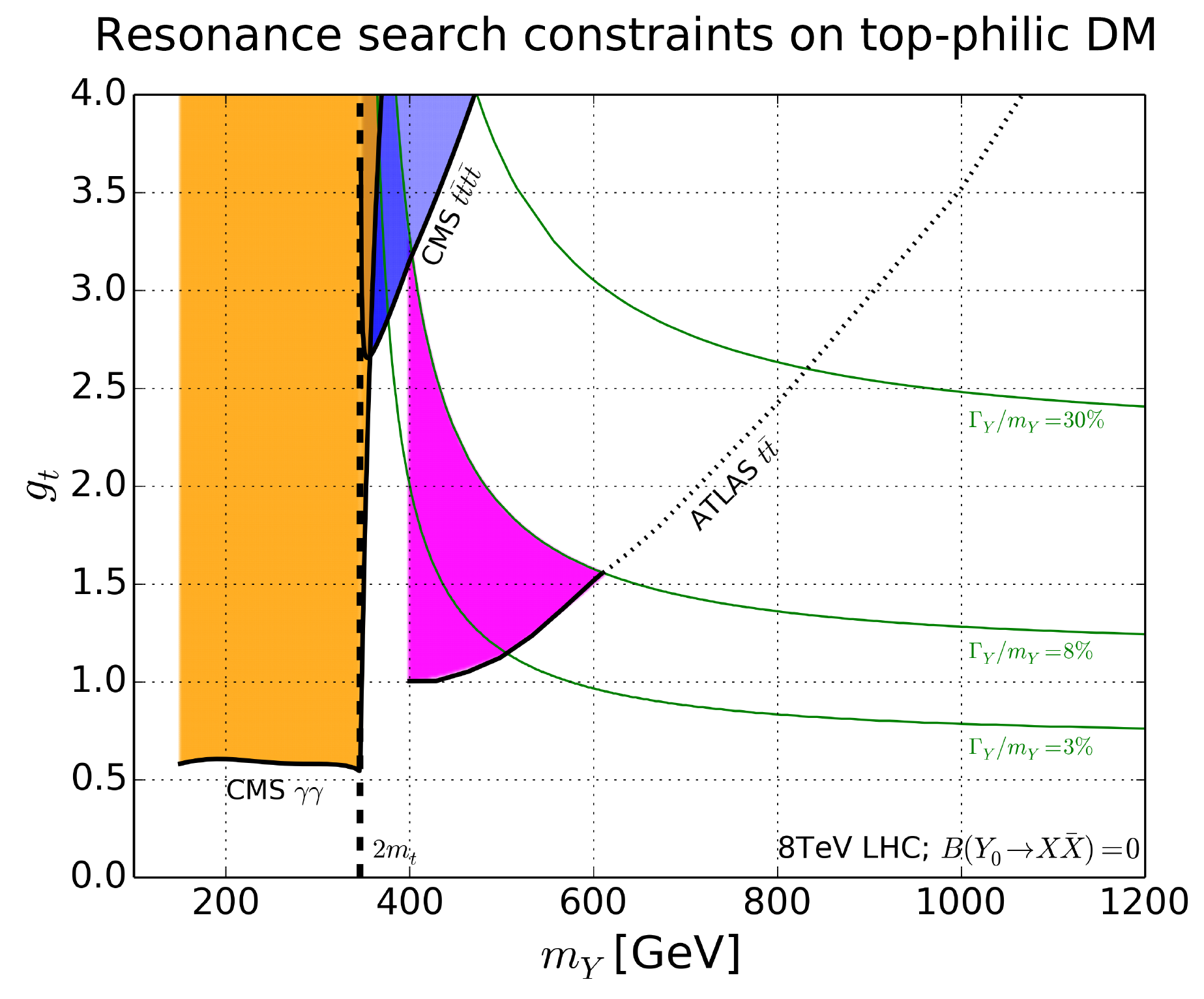}
\caption{Resonance search constraints from the LHC results at a collision
centre-of-mass energy of 8~TeV on the simplified top-philic dark matter model
presented in terms of the mediator mass $m_Y$ and the $g_t$ coupling. The
different coloured areas are excluded by the diphoton~\cite{Khachatryan:2015qba}
(orange), $t\bar t$~\cite{Aad:2015fna} (magenta) and
$t\bar tt\bar t$~\cite{Khachatryan:2014sca} (blue) searches. We include
information on the mediator width to mass ratios (green curves). We assume a negligible branching ratio to the invisible sector. }
\label{fig:nonmet}
\end{figure}

Although the branching ratio of the mediator into a photon pair is very small,
the background associated with a diphoton signal is low so that one expects to
be able to obtain stringent constraints on the model from the diphoton search
results. We focus here on the CMS
8~TeV diphoton search~\cite{Khachatryan:2015qba} that investigates resonance
masses ranging from 150~GeV to 850~GeV and derives limits on the corresponding cross
section. For instance, the 95\% CL upper bound on the mediator-induced diphoton
production cross section $\sigma(pp\to Y_0\to\gamma\gamma)$ is of 20~fb (4~fb)
for a mediator mass of 150~GeV (300~GeV). Making use of the $pp\to Y_0$ cross
section values shown in figure~\ref{fig:xsec} and the $Y_0\to\gamma\gamma$
branching ratio computed from the formulas shown in section~\ref{sec:model}, we
present diphoton constraints on the model in the $(m_Y, g_t)$ plane in
figure~\ref{fig:nonmet}. These results assume that the dark matter particle is
much heavier than the mediator that can thus not resonantly decay invisibly. The
constraints are found to be stringent below the $2m_t$ threshold, where the
$g_t$ coupling cannot be larger than $0.6$.

\subsubsection*{Top-antitop resonances}

For scenarios with mediator masses above the top-antitop threshold ($\my>2m_t$),
$t\bar{t}$ resonance searches~\cite{Aad:2015fna,Khachatryan:2015sma} can be used
as probes of the model. In our setup, loop-induced resonant mediator
contributions can indeed enhance the $t\bar t$ signal, in particular when there
is a large coupling hierarchy ($\gsm\gg g_X$) or mass hierarchy
($2m_t<\my<2\mdm$). We derive constraints on our model from the ATLAS 8~TeV
$t\bar t$ resonance search~\cite{Aad:2015fna} that relies on
the reconstruction of the invariant mass of the top-quark pair to derive a
95\% CL exclusion on the existence of a new scalar particle coupling to top
quarks. The associated cross section limits range from 3.0~pb for a mass of
400~GeV to 0.03~pb for $\my=2.5$~TeV, assuming that the narrow width
approximation is valid with a mediator width being of at most $3\%$ of its mass
and that there is no interference between the new physics and Standard Model
contributions to the $t\bar t$ signal.

Constraints are computed using the NNLO mediator production cross section
(see figure~\ref{fig:xsec}) and the relevant top-antitop mediator branching
ratio derived from the formulas presented in section~\ref{sec:model}. The latter
is in fact very close to one in the relevant region, the mediator decays into
dark matter particle pairs being kinematically forbidden and those into gluons
and photons loop-suppressed. The results are presented in the $(\my, \gsm)$
plane in figure~\ref{fig:nonmet}. This shows that scalar mediators with masses
ranging from 400~GeV to 600~GeV could be excluded for $\gsm$ couplings in the
$[1, 4]$ range, the exact details depending on $\my$ and on the fact that the
narrow-width approximation must be valid. This demonstrates the ability of
the $t\bar{t}$ channel to probe a significant portion of the $\my > 2m_t$ region
of the model parameter space. In the region where $2m_t, \,2\mdm<\my$, the partial
decay $\y \to X\bar X$ reduces the $t \bar t$ signal and therefore limits the
sensitivity of the search.

\subsubsection*{Four-top signals}
Scenarios featuring
a mediator mass above twice the top-quark mass can be probed via a four-top signal, since the mediator can be produced in association with a pair
of top quarks and further decay into a top-antitop system. Theoretically, the
Standard Model four-top cross section has been calculated with high
precision~\cite{Bevilacqua:2012em}, but the sensitivity of the 8 TeV LHC run was too low to measure the cross section. Instead, an upper limit on the cross section at a centre-of-mass energy of 8~TeV has been derived~\cite{Chatrchyan:2013fea,%
Khachatryan:2014sca}. The four-top production rate is constrained to be
below 32~fb~\cite{Khachatryan:2014sca}, a value that has to be compared to the
Standard Model prediction of about 1.3~fb. Only models with new physics
contributions well above the background (see \eg~ref.~\cite{Beck:2015cga}) can
therefore be constrained by the four-top experimental results.

In our top-philic dark matter model, the new physics contributions to the
four-top cross section can be approximated by the $t\bar{t}Y_0$ cross section,
the branching ratio $B(Y_0\to t\bar t)\sim 1$. Using the NLO cross
section (see figure~\ref{fig:xsec}), we derive limits that we represent in the
$(m_Y, g_t)$ plane in figure~\ref{fig:nonmet}. A small region of the parameter
space with $g_t >2.5$ and in which the mediator mass lies in the $[2m_t,
\sim450~{\rm GeV}]$ mass window turns out to be excluded. The weakness of the
limit is related to the steeply decreasing cross section for $pp\to Y_0t\bar t$ with the increase in $\my$.

\subsubsection*{The mediator width}
In all the above studies where the final state does not contain any missing
energy, the mediator width has been assumed narrow. Concerning the diphoton
channel, this assumption holds within the entire excluded region
as only loop-suppressed gluon and photon mediator decays are allowed. In the
region where $\my>2m_t$, the width of the mediator rises quickly with its mass,
and the width over mass ratio rapidly exceeds the 3\% value that has been
imposed in the ATLAS $t\bar t$ resonance search~\cite{Aad:2015fna} as can be
seen in figure~\ref{fig:nonmet}. The reinterpretation of the ATLAS results to a
generic $t\bar{t}$ resonance model should therefore be made carefully, as the
limit cannot be necessarily applied to scenarios featuring significantly larger
mediator widths. This is shown in figure~\ref{fig:nonmet} by a dotted line, and
we can also observe that most of the points that would have been excluded by the
ATLAS search do not fulfil the requirement of a width below 3\% of the mediator
mass. In our excluded region of the parameter space, we allow the mediator width
to reach 8\% of its mass, by the virtue of the experimental resolution on the
invariant mass of the $t \bar{t}$ system. This leads to the exclusion of
scenarios with mediator masses up to 600~GeV.

The ATLAS resonance $t \bar t$ study claims that varying the width of the
resonance from 10\% to 40\% for the massive gluon model results in a loss in
sensitivity by a factor 2 for a 1~TeV resonance. An extension of the
reinterpretation of the ATLAS
limits on our simplified top-philic dark matter model to the case of larger
resonance widths could then be performed by rescaling the limits by the
appropriate correction factor. We have nonetheless found that no additional
points are excluded even without rescaling the sensitivity of the search as
the ATLAS analysis rapidly loses sensitivity for resonance masses above 600~GeV.
Considering model points with a mediator width to mass ratio of at most about
$8\%$ therefore provides a realistic exclusion over the entire model parameter
space.

\subsubsection*{Concluding remarks on direct mediator searches}
Mediator resonance searches at 8~TeV show good prospects of constraining our
simplified top-philic dark matter model, especially in the mediator mass range
of 150--345~GeV and 400--600\,GeV by means of the diphoton and top-pair searches
respectively. So far, the $t\bar{t}$ resonance searches are strictly applicable
to a limited parameter space region of the simplified model, and considering
larger widths in the interpretation of the future results would allow for a more
straightforward reinterpretation of the limits to a wider range of parameters.
Concerning the four-top analysis, it can presently only exclude a restricted
part of the parameter space, but future measurements are expected to lead to
more competitive bounds.

Finally, the $pp \to t \overline{t}+j$ channel could also be used to probe dark
matter models coupling preferably to top quarks. This has been for instance
shown in ref.~\cite{Greiner:2014qna} where a loop-induced production of
$t\bar{t}j$  can in some cases lead to interesting constraints on
top-philic models of new physics. In our case, they are nonetheless not
expected to give more stringent constraints than the $t\bar{t}$ resonance
searches. One could also consider the
$pp \to t\bar{t} tj$ and $pp \to t\bar{t} Wt$ processes~\cite{Greiner:2014qna}.
Because of the magnitude of the electroweak couplings, these processes are
characterised by smaller cross sections than when four top quarks are involved,
and are hence not likely to set more stringent constraints on the class of
models under consideration.

%
%
%
\section{Combined constraints}\label{sec:combined}
\begin{figure}
\begin{center}
\includegraphics[scale=0.5]{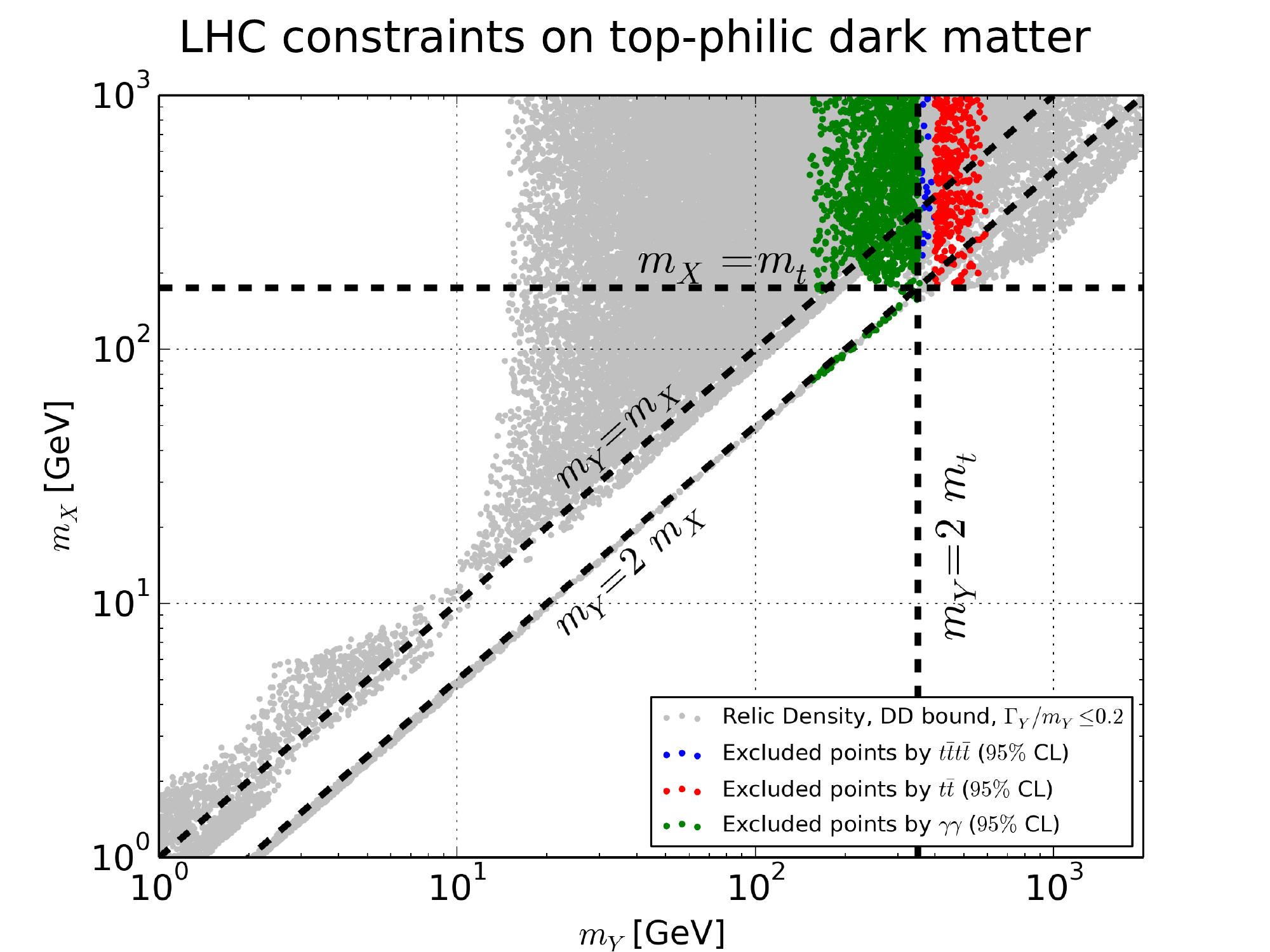}
\caption{Results of our four-dimensional parameter scan projected onto the
$(\my, \mdm)$ plane once constraints set from the LHC results are imposed. The
points excluded by the diphoton, the $t\bar{t}$ and the four-top considered
searches all satisfy the relic density, narrow width and direct detection
constraints.}
\label{fig:ttmaddm}
\end{center}
\end{figure}

The final segment of our comprehensive study of top-philic dark matter
simplified models
is a combined study of astrophysical and collider constraints. We find that in
the region where $\gx, \gsm \leq \pi$, the 8~TeV collider results that
provide relevant bounds (once the relic density and direct detection constraints
are imposed) originate from direct mediator production searches when the
mediator
further decays into a pair of Standard Model particles. Figure~\ref{fig:ttmaddm}
illustrates our results and shows the scenarios that are excluded by resonant
diphoton and top-pair searches as well as by the four-top analysis.
All points in the plot accommodate the dark matter relic density and direct detection
constraints, while the colours indicate points excluded by individual complementary collider bounds. 
The vast majority of excluded points lie in the region where $2\mdm>\my$ with
$\my \in [150, 600]$~GeV. This is the region where the mediator decay into a
pair of dark matter particles is kinematically forbidden, ensuring large
branching fractions for decays into Standard Model particles. The diphoton
resonance search excludes points below the $2m_t$ threshold, while 
$t\bar{t}$ results constrain the $400<\my<600$ GeV region. The four-top probe
is able to exclude a narrow parameter space region close to  $\my \sim 2m_t$, in
agreement with the findings shown in figure~\ref{fig:nonmet}.

\begin{figure}[h!]
\center
\includegraphics[width=0.45\textwidth]{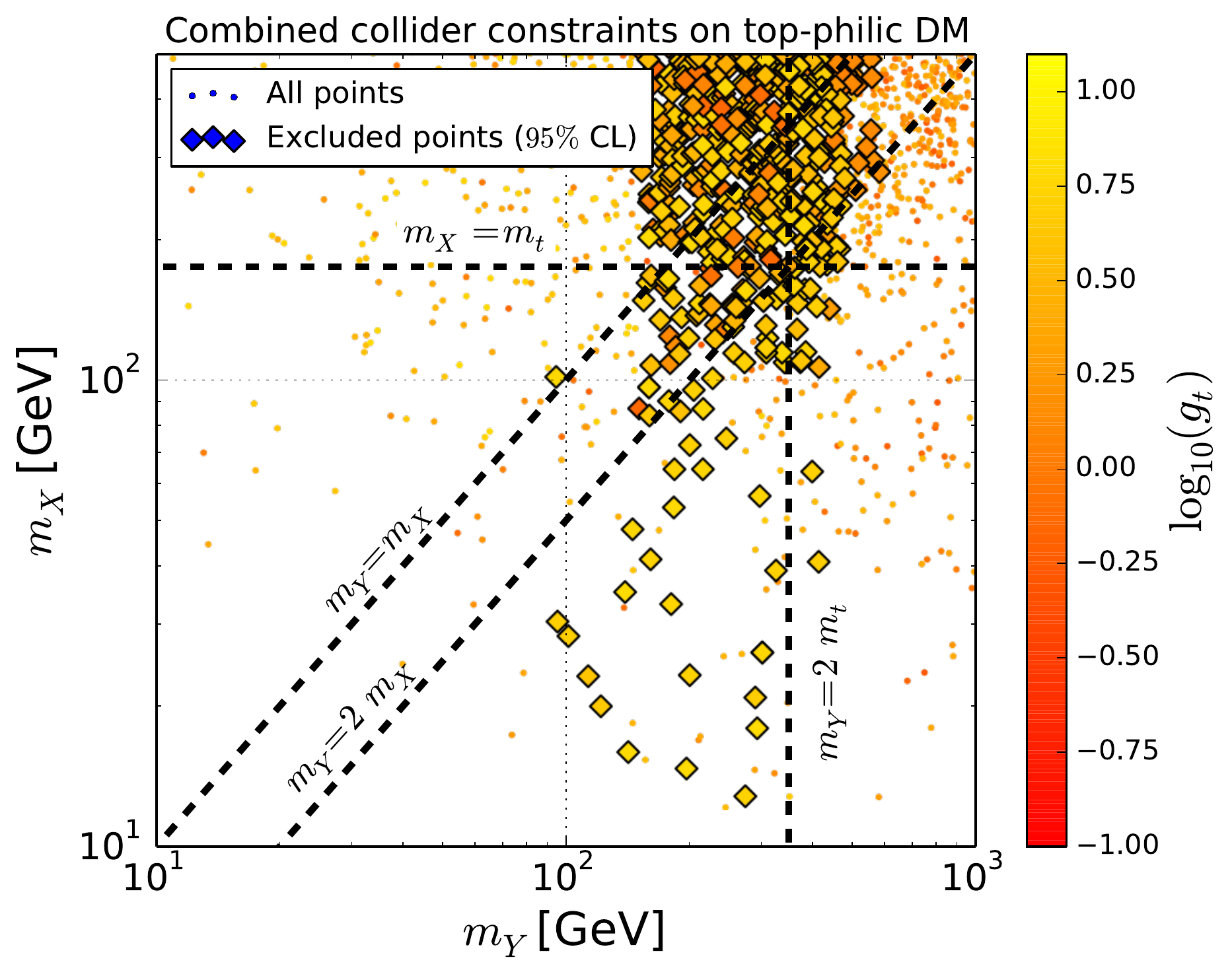}
  \includegraphics[width=0.45\textwidth]{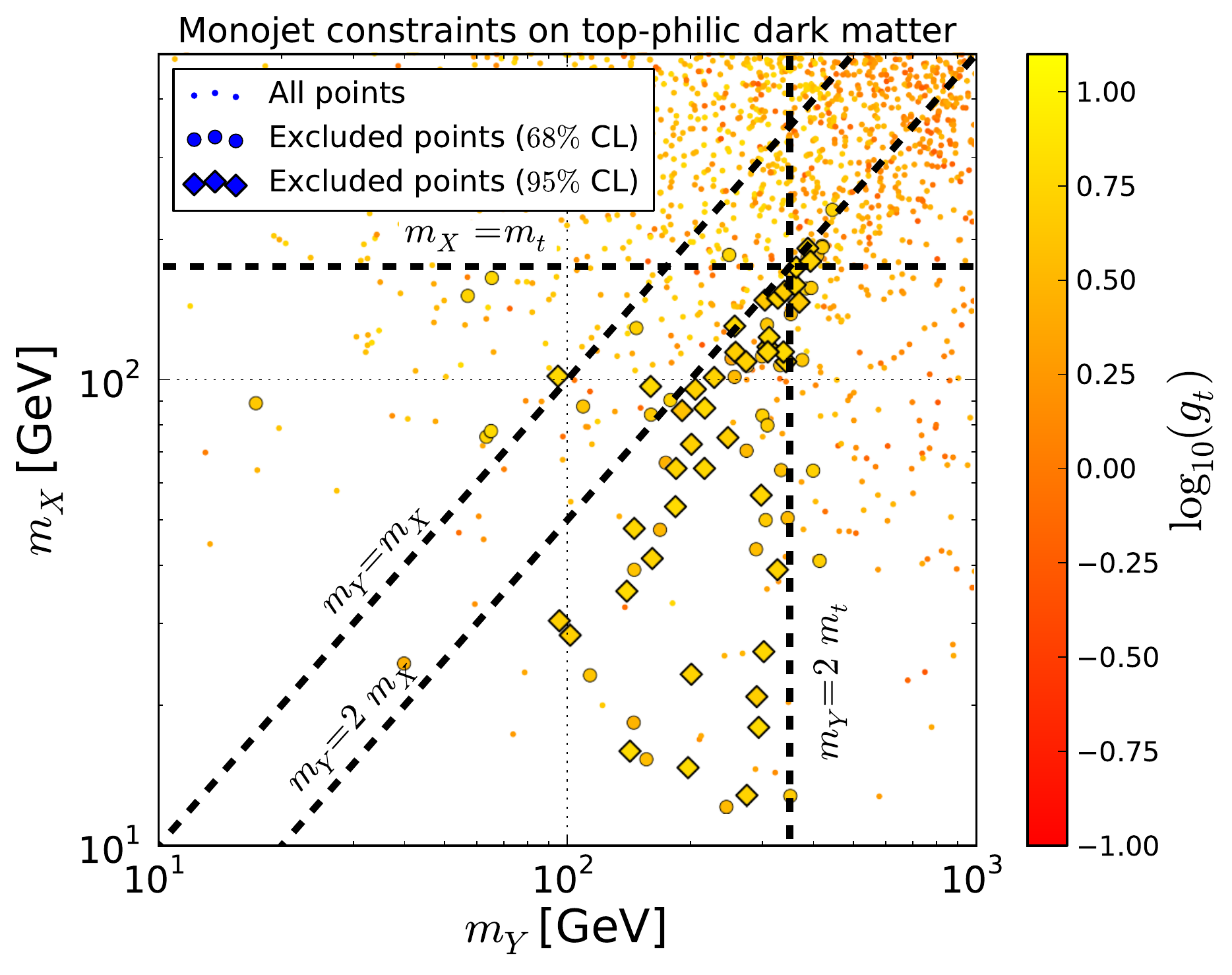}
 \includegraphics[width=0.45\textwidth]{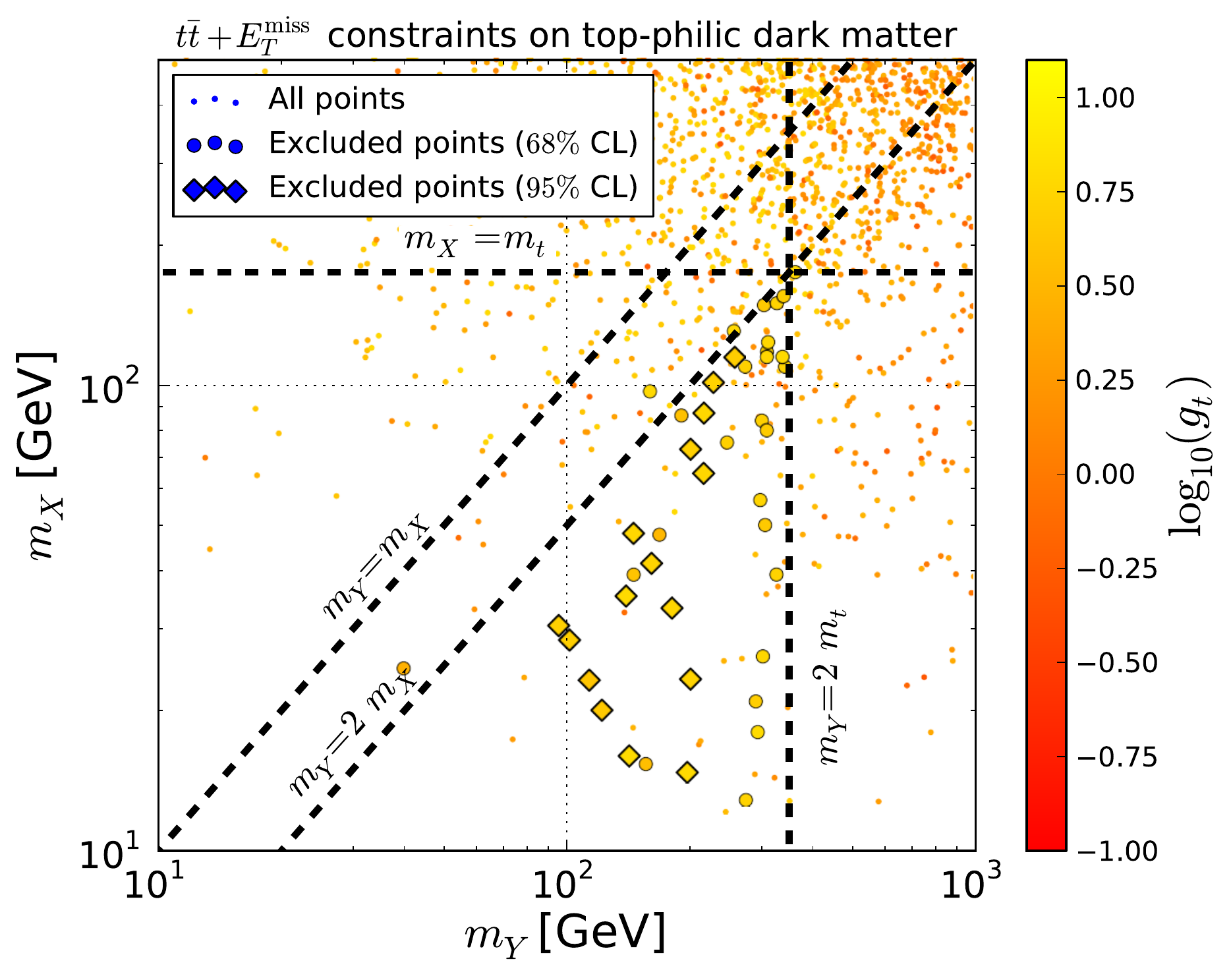}
 \includegraphics[width=0.45\textwidth]{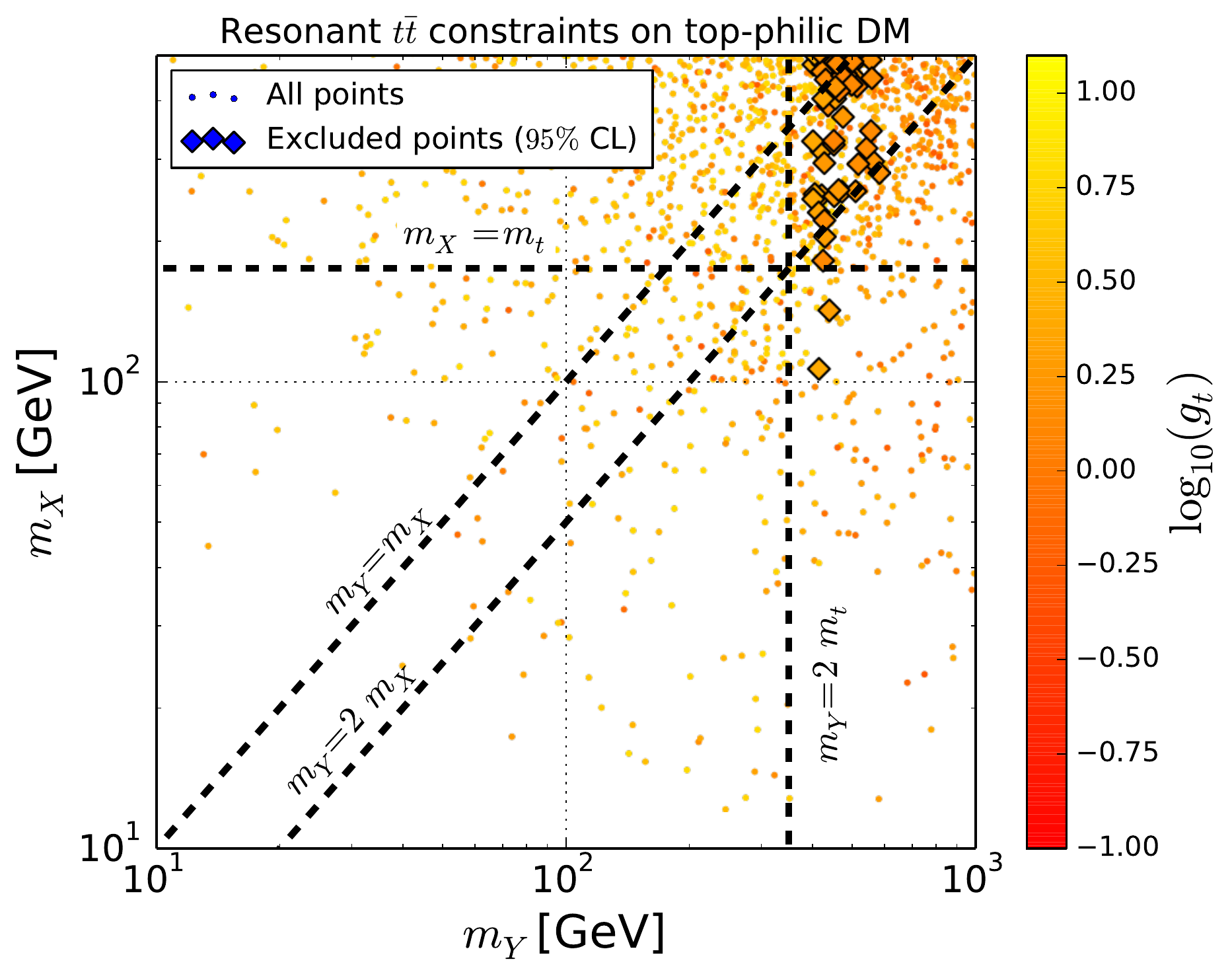}
  \includegraphics[width=0.45\textwidth]{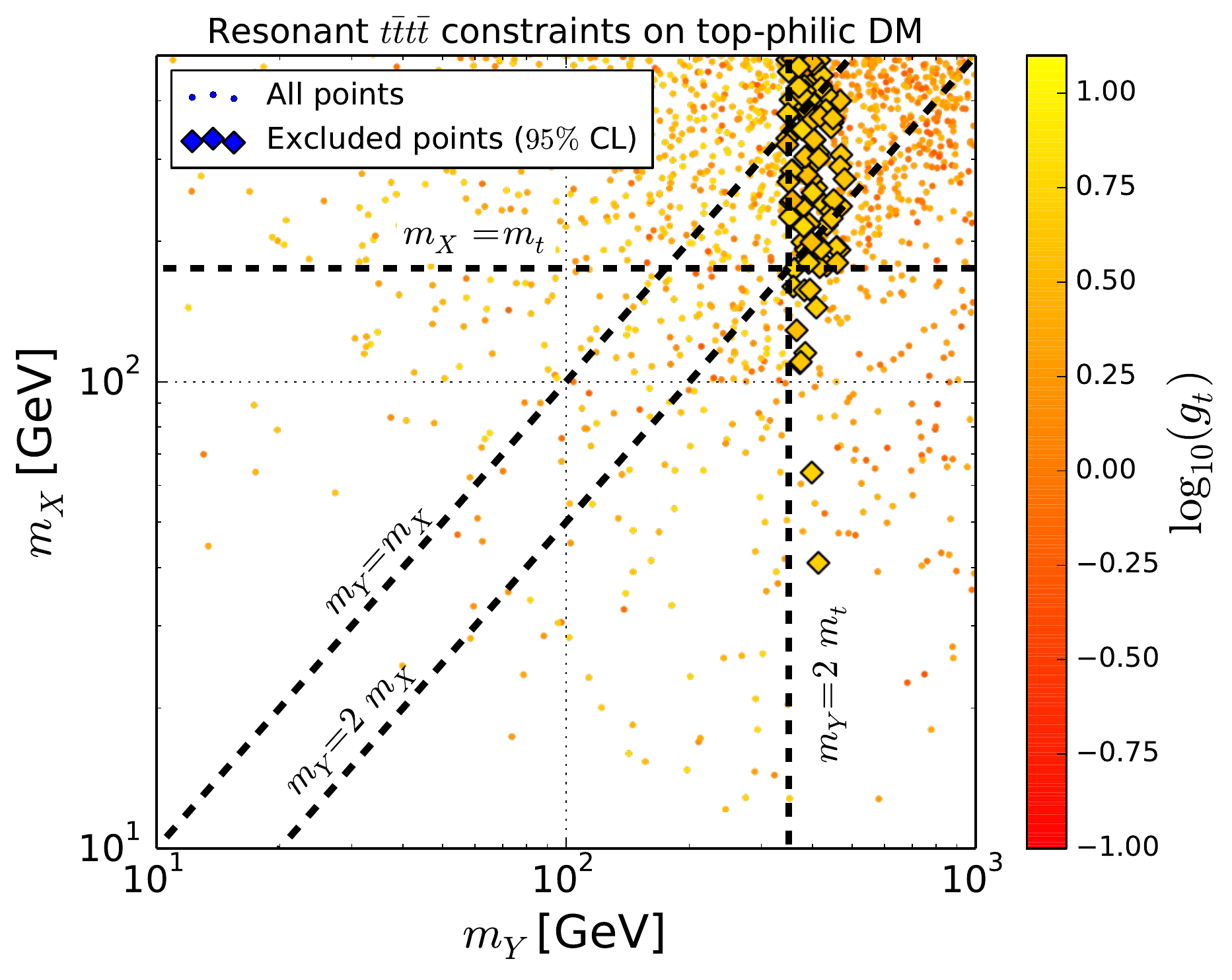}
 \includegraphics[width=0.47\textwidth]{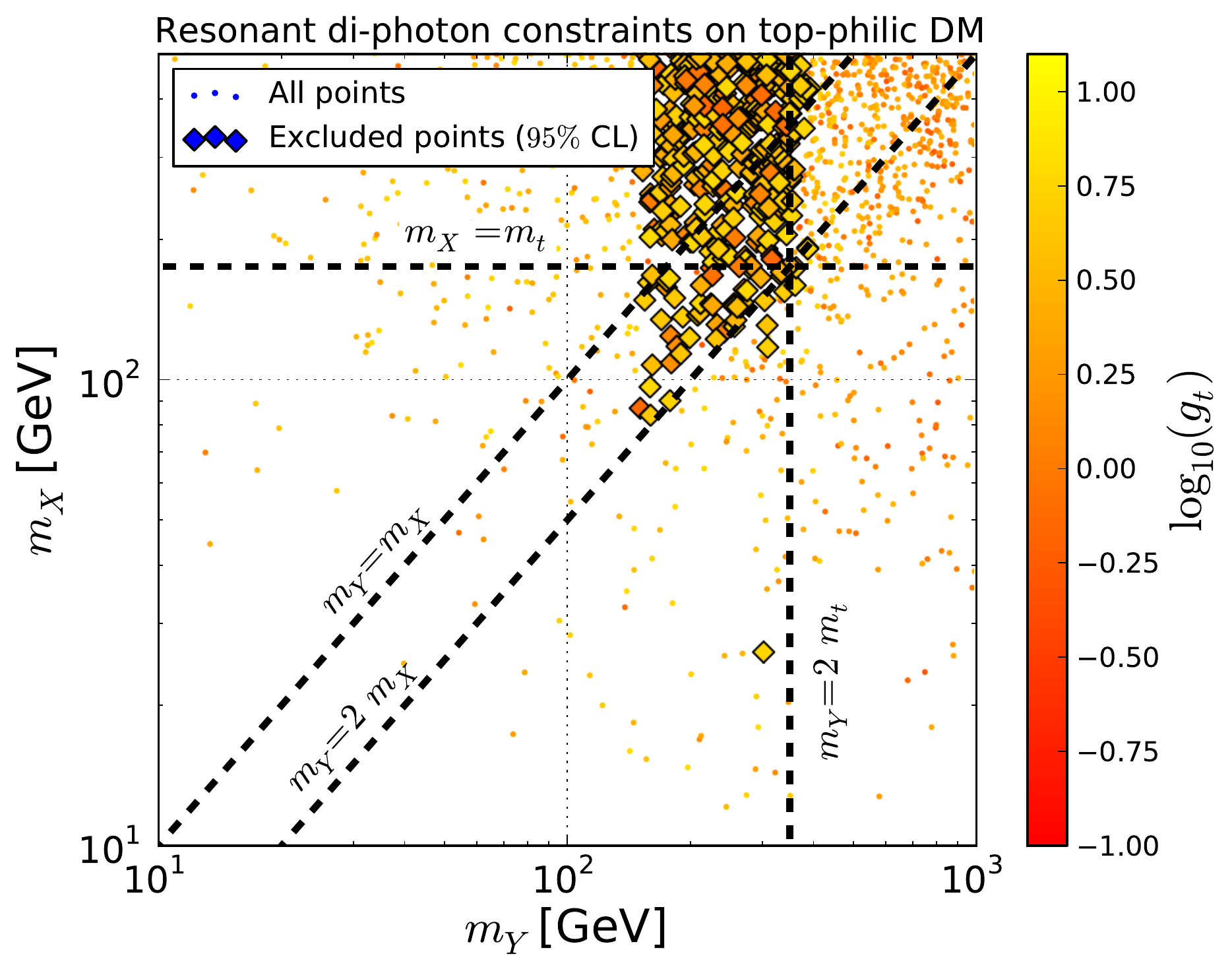}
\caption{Constraints derived from the LHC Run I results on the simplified top-philic dark
  matter model. The panels show results of a four-dimensional parameter scan, uniform on the linear scale. The upper left panel shows a combination of all relevant collider
  constraints. The upper 
  right panel shows the points excluded by monojet constraints, while the third
  panel shows the points excluded by  $t\bar{t}+\MET$
  constraints.  The resonant $t\bar{t}$ searches constraints are shown in the fourth panel, while the last two panels show the four top and the diphoton constraints.   The
  results assume couplings smaller than $2\pi$ and $\wy/\my < 0.5$, with no
  constraints from astrophysics or cosmology being imposed. In case of resonant
  $t\bar{t}$, four top, and the combined constraints, we only show the 95\% CL
  exclusion as the $t\bar{t}$ and four top results have not been obtained using a recast
  LHC analysis.}
\label{fig:colliderscan}
\end{figure}

Relaxing the requirements on the relic density, the direct detection and
the upper bound
on the coupling strengths allows for another meaningful study of combined
collider constraints. For this purpose we have performed a joint analysis of
collider bounds on the top-philic simplified dark matter model in the scope of a
four-dimensional parameter scan with a flat likelihood function over all
dimensions. We have performed the scan by restricting the couplings to be smaller
than $2 \pi$, as well as by allowing the mediator widths to reach 50\% of the
mediator mass. Figure~\ref{fig:colliderscan} shows our results, where the upper
left panel shows the model points excluded by the combination of all collider
results, and the rest of the panels show the points excluded by individual
LHC Run~I collider results. We find that the 8 TeV monojet searches
exclude model points which lie mainly in and around the triangle bounded by the
$\my = 2\mdm$ and $\my=2m_t$ lines, where the characteristic  $\gsm$ which is
excluded by the 8 TeV results is of ${\cal O}(10)$. The region in which the
excluded points are located is reasonable, as we expect any significant monojet
signal in the region where $\my > 2\mdm$. Furthermore, we expect the branching
ratio to missing energy to be lower in the region where $\my > 2m_t$ due to the
kinematically allowed decays into a pair of top quarks. This in turn leads to a
lower signal cross section in all channels with missing energy and hence a lower number of points which can be excluded by monojet searches in the $\my > 2m_t$ region.

The points excluded by the 8 TeV $t\bar{t}+\MET$ measurements lie in roughly the
same region as the points excluded by the monojet search, but with a more defined
edge of \mbox{$\my = 2 m_t$}.
Conversely, the 8 TeV  $t\bar{t}$ resonance search provides constraints in the
region of $\my \in [400, 600]$~GeV and $\mdm \gtrsim 100$~GeV, and is able to
rule out $\gsm$ couplings of ${\cal O}(1)$. The four top searches constrain roughly the same region of the $(\my, \mdm)$ parameter space as the $t\bar{t}$ searches. However, the characteristic size of the couplings four top searches are able to constrain is significantly larger than the case of $t\bar{t}$.

Finally the diphoton resonant search excludes $\my \in [150, 2m_t]$~GeV with $2\mdm>\my$, ruling out $\gsm$ couplings larger than 0.6. In the $(\my, \mdm)$ plane, we can
observe that the constraints arising from all mediator resonance searches,
\ie~the diphoton and $t\bar{t}$ analyses, are
largely complementary to those issued from searches in channels with large missing energy.

\section{750\,GeV diphoton excess}
\label{sec:diphoton}
In the light of the possible new physics signal observed in the 13 TeV
ATLAS~\cite{atlasDigamma} and CMS~\cite{CMS:2015dxe} diphoton data, we
investigate whether the simplified top-philic dark matter  model considered in
this work can accommodate the features of the observed excess. 
The excess can be interpreted as a possible signal of a new particle with
\begin{equation}
  \my \approx 750 \GeV\,,\qquad
  \wy/\my \lesssim 6 \% \qquad\text{and}\qquad
   \sigma_{\gamma \gamma} (13 \TeV) \sim 1 -10 \fb\, \label{eq:diphoton_reqs},
\end{equation}
where $\sigma_{\gamma\gamma} \equiv \sigma(pp \to \y \to \gamma \gamma)$. Near
the resonance, the diphoton cross section is analytically approximated
by~\cite{Franceschini:2015kwy}
\begin{equation}
	\sigma_{\gamma \gamma}(13~{\rm TeV}) = \frac{1}{\my\, \wy \,s} C_{gg}(13~{\rm TeV})\, \Gamma(\y \rightarrow gg)\times\Gamma(\y \rightarrow \gamma \gamma)\, , \label{eq:diphotonCS}
\end{equation}
where $C_{gg}(13 \TeV) = 2137$ is the gluonic parton luminosity factor and
$\Gamma(\y \to gg)$ and
$\Gamma(\y \to \gamma \gamma)$ are the partial decay widths given in
eqs.~\eqref{eq:y0widthgg} and \eqref{eq:y0widthph}. Eq.~\eqref{eq:diphotonCS}
should in principle also contain a contribution from the $\gamma \gamma$ initial
state. Here we neglect the photon fusion production mechanism due to fact that in the top-philic dark matter model,  the branching ratio to photons is always suppressed compared to the branching ratio to gluons (see eq.~\eqref{eq:brbr}).
The relationship between the strength of the mediator couplings to gluons and photons is also one of the
main differences between our  simplified  top-philic dark matter model and other
dark matter models that have been proposed to explain the 750 GeV diphoton
resonance excess. In the latter, the mediator couplings to gluons and photons are
typically treated as independent parameters~\cite{Backovic:2015fnp, %
Mambrini:2015wyu,Bi:2015uqd,D'Eramo:2016mgv}.

\begin{figure}
\begin{center}
\includegraphics[width=4in]{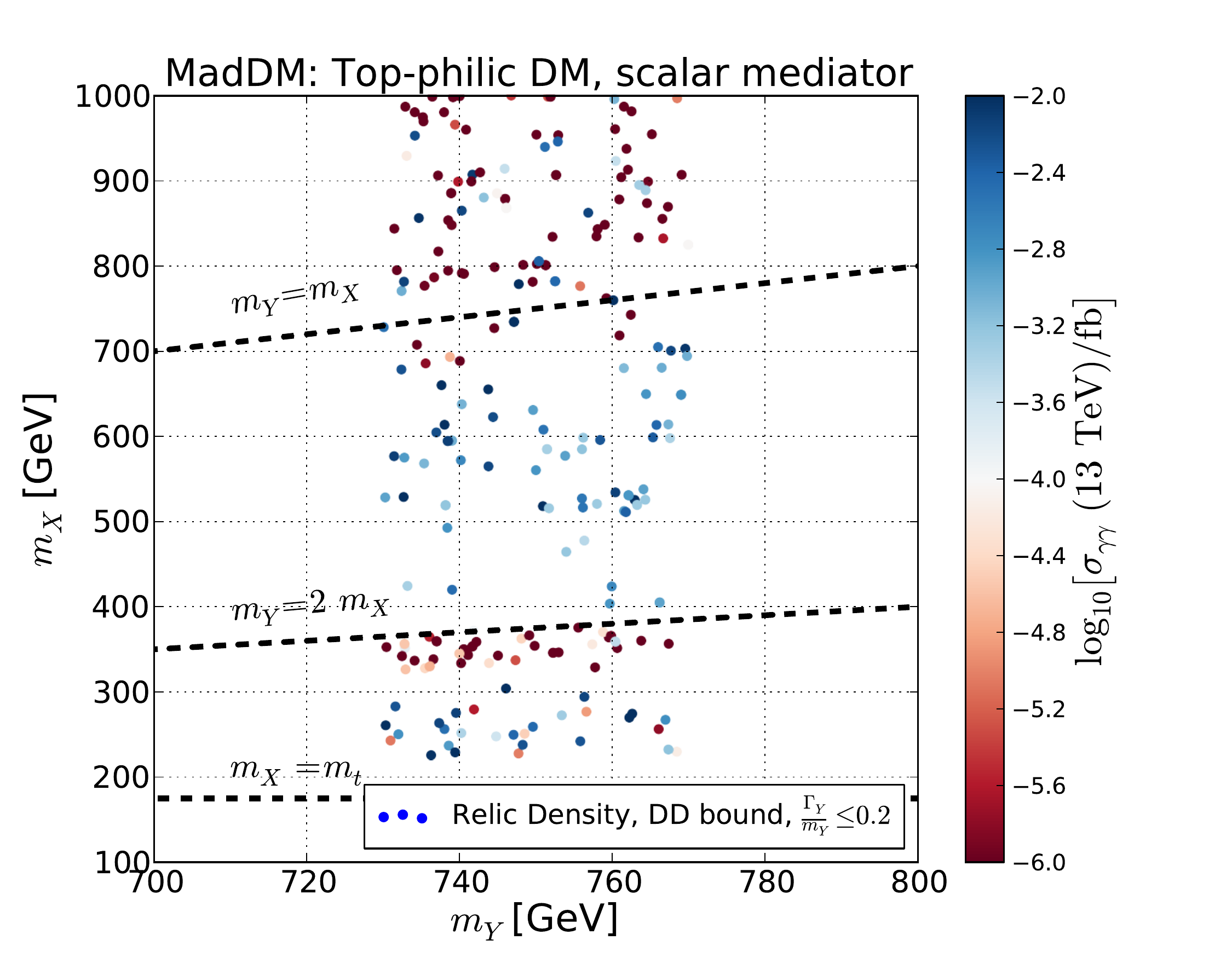}
\caption{The $pp \rightarrow \y \rightarrow \gamma \gamma$ cross section in the  simplified top-philic dark matter model. We represent scenarios allowed by relic
 density consideration, dark matter detection and a width smaller than 20 \% of the mediator mass.  The colour map shows the value of the
 diphoton cross section. We only show parameter points which satisfy
 $730 \GeV < \my < 770 \GeV$. } \label{fig:diphoton}
\end{center}
\end{figure}

The different contributions to the mediator width when $\my=750$~GeV always
include the $t\bar{t},\,  gg$ and $\gamma \gamma$ final states, while the
partial decay into a pair of dark matter particles is subject to the
value of $\mdm$. In order to determine whether our top-philic dark matter
model can explain the diphoton excess, we hence only have to address two
distinct regions of the parameter space.
\begin{itemize}
\item  $\my < 2\mdm$:  In this region, $\wy$ is obtained by summing the
contributions of the decays into $gg$, $t\bar{t}$ and $\gamma \gamma$ final
states. As the top decay channel is kinematically open, it will always dominate
over the loop-suppressed $gg$ and $\gamma \gamma$ modes, leading to
$\wy \approx \Gamma(\y \to t \bar{t})$. The mediator-induced diphoton rate at a
centre-of-mass energy of 13~TeV is then a function of the single parameter
$g_t$,
 \begin{equation}
  \sigma_{\gamma \gamma}( \my < 2\mdm) = 2.9\times 10^{-3} \, \gsm^2\, \fb\,,
  \label{eq:sigmamax}
\end{equation}
where we have fixed $\my = 750 \GeV$. In order to reproduce the
$\sigma_{\gamma \gamma} \gtrsim 1 \fb$ value that is necessary to explain the
excess, one would hence naively need $\gsm \sim 19$, which is way above the
unitarity bound. Even without considering the mediator width, cosmology or astrophysics, we find that the top-philic dark matter model is not  able to explain the diphoton excess when $\my < 2 \mdm$.

\item $\my \geq 2\mdm$: The total mediator width in this region is well
approximated by summing over the contributions originating from the decays into
$t\bar{t}$ and $\dm \dmbar$ pairs. The main difference compared to the previous
case consists of the decay channel into a $\dm \dmbar$ pair that is now
kinematically allowed, which implies possibly suppressed branching ratios to the
other final states. As the contributions of $\y$ invisible decays to a pair of
dark matter particles only appear in the denominator of
eq.~\eqref{eq:diphotonCS}, the maximum possible $\sigma_{\gamma \gamma}$ cross
section value is reached when $\Gamma(\y \to \dm \dmbar) \sim 0$, \ie~when
$\gx \approx 0$ or $\mdm \approx \my /2$. The resulting maximal diphoton rate
then turns out to be identical to the one of eq.~\eqref{eq:sigmamax} so that the
observed excess cannot be accomodated in our model.
\end{itemize}

The top-philic  simplified dark matter model that we consider cannot accommodate
the diphoton excess in any region of the model parameter space. 
Finally, we show the actual values of $\sigma_{\gamma \gamma}(13~\TeV)$ for the
scenarios that feature relic density and direct detection cross section in agreement with data in
figure~\ref{fig:diphoton}, after restricting our selection to points featuring
$730 \GeV \leq \my \leq 770 \GeV$. The largest cross section values that are
found are at least two orders of magnitude too low in order to be able to
accomodate the diphoton signal.

\section{Conclusions}\label{sec:concl}
We presented a comprehensive analysis of  simplified  top-philic dark matter models, in the scope of collider physics, astrophysics and cosmology. Our study considered the full four dimensional model parameter space, where we treated the experimental constraints on the model space both separately and in conjunction with each other. The requirement of predicting the measured relic density $\Omega_{\rm DM} h^2$~%
 gives the most stringent constraint on the viable regions
of the parameter space. Most of the region where $\my > \mdm$ cannot accommodate
the observed relic density, except near the resonance \mbox{$\my\sim 2\mdm$} and
for $\mdm > m_t$. Direct detection data complementary excludes large portions of
the parameter space in the $\my < \mdm$ region once experimental results from
 LUX and CDMSLite are accounted for.
In the context of dark matter indirect detection, we studied prospects for further model constraints  from gamma-ray flux measurements originating from
dwarf spheroidal galaxies and the gamma-ray lines issued from the inner galactic
region. In the specific
model we consider, the dark matter annihilation cross section is $p$-wave suppressed, leading to indirect detection bounds which are too weak to provide additional
constraints on the parameter space.

Collider searches from LHC Run 1 at $\sqrt{s} = 8 \TeV$ can constrain the parameter space beyond the limits obtained from the relic density and direct detection, but apply mostly in the limit of coupling values $\gtrsim 1$. We found that for couplings of $\lesssim \pi$, the resonant $t\bar{t}$ and diphoton searches are able to exclude a fraction of model points in the regions of $\my \sim 400 - 600 \GeV$ and $\my \sim 150 - 350 \GeV$  respectively, even upon assuming astrophysical and relic density constraints. 

In addition to studying collider signatures of the top-philic dark matter simplified model as a complementary way of dark matter detection, we performed a study of collider constraints without assuming relic density and direct detection (as well as extended the parameter range to include coupling values of $< 2\pi$ and $\wy \le 0.5 \my$). 
Our results for a four dimensional parameter scan show that (in the scenario where astrophysical and cosmological constraints are not relevant),$\MET+j$ and $\MET+t\bar{t}$ 8 TeV results provide meaningful bounds on the model parameter space in the $ 2 \mdm< \my < 2m_t$ region, but only for $\gsm, \gx \gtrsim \pi$. In the $\mdm > m_t$ region, the resonant $t\bar{t}$ searches are again able to exclude some model points in the $\my\sim 400-500 \GeV$ region, while $\gamma \gamma$ measurements provide constraints in the $\my < 2 m_t$ region. We have also explored the prospects of using rarer processes such as four-top production as well as mono-$Z$ and mono-Higgs production to constrain our model. While we have not performed a detailed analysis we have found that these processes show promising signs of further constraining the parameter space of our model and deserve dedicated studies. 


For the purposes of our study we have recast the CMS monojet and $\MET+t\bar{t}$ searches in the framework of \MA, which allows us to reliably extract constraints on our model, and can benefit future collider studies which go beyond our simplified model and even beyond dark matter searches. Another important aspect of our work, is the use of NLO QCD predictions for the $\MET+t\bar{t}$ process to constrain our model.  While we find that $K$-factors for this process are close to one, the importance of taking higher order effects into account lies in the reduced theoretical uncertainties of the NLO results. We have shown that the uncertainties in the CL estimates significantly reduce with the inclusion of higher order QCD terms which clearly illustrates the importance of higher order corrections on the interpretation of dark matter searches at colliders. 

In the context of the recently observed excess in the ATLAS and CMS measurements of the diphoton invariant mass spectrum, we consider the possibility that $\y$ decays to photons explain the excess of events around $m_{\gamma \gamma} = 750$ GeV. We find that due to  top-loop suppressed couplings to gluons and photons, only non-perturbative values of $g_t$  suffice to fit the features of the excess.

The work presented in this paper also represents a proof-of-concept for a unified numerical framework for dark matter studies at the interface of collider physics, astrophysics and cosmology in a generic model.


\section*{Acknowledgments}

We would like to thank the LHC dark matter forum for motivation to pursue this project and for useful discussions during the April 2016 Amsterdam workshop. We are also grateful to Lana Beck for providing her four top analysis code for cross checks of our results.
This work has been supported by the research unit New physics at the LHC of the German research foundation DFG,  the Helmholtz Alliance for Astroparticle Physics and the German Federal Ministry of Education and Research BMBF.
CA is supported by the ATTRACT - Brains back to Brussels grant from Innoviris.
MB is supported by the MOVE-IN Louvain Cofund grant. 
EC, BF and KM are supported by the Theory-LHC-France initiative of the CNRS (INP/IN2P3). EV is supported by the European Union as part of the FP7 Marie Curie
Initial Training Network MCnetITN (PITN-GA-2012-315877).
BH is supported in part by the National Fund for Scientific Research (F.R.S.-FNRSBelgium) under a FRIA grant.
The work of FM and AM is supported by the IISN ``MadGraph'' convention
4.4511.10 and the IISN ``Fundamental interactions'' convention 4.4517.08.

\appendix

\section{Mediator width}\label{sec:ywidth}

\begin{figure}[h!]
 \center 
 \includegraphics[width=0.32\textwidth]{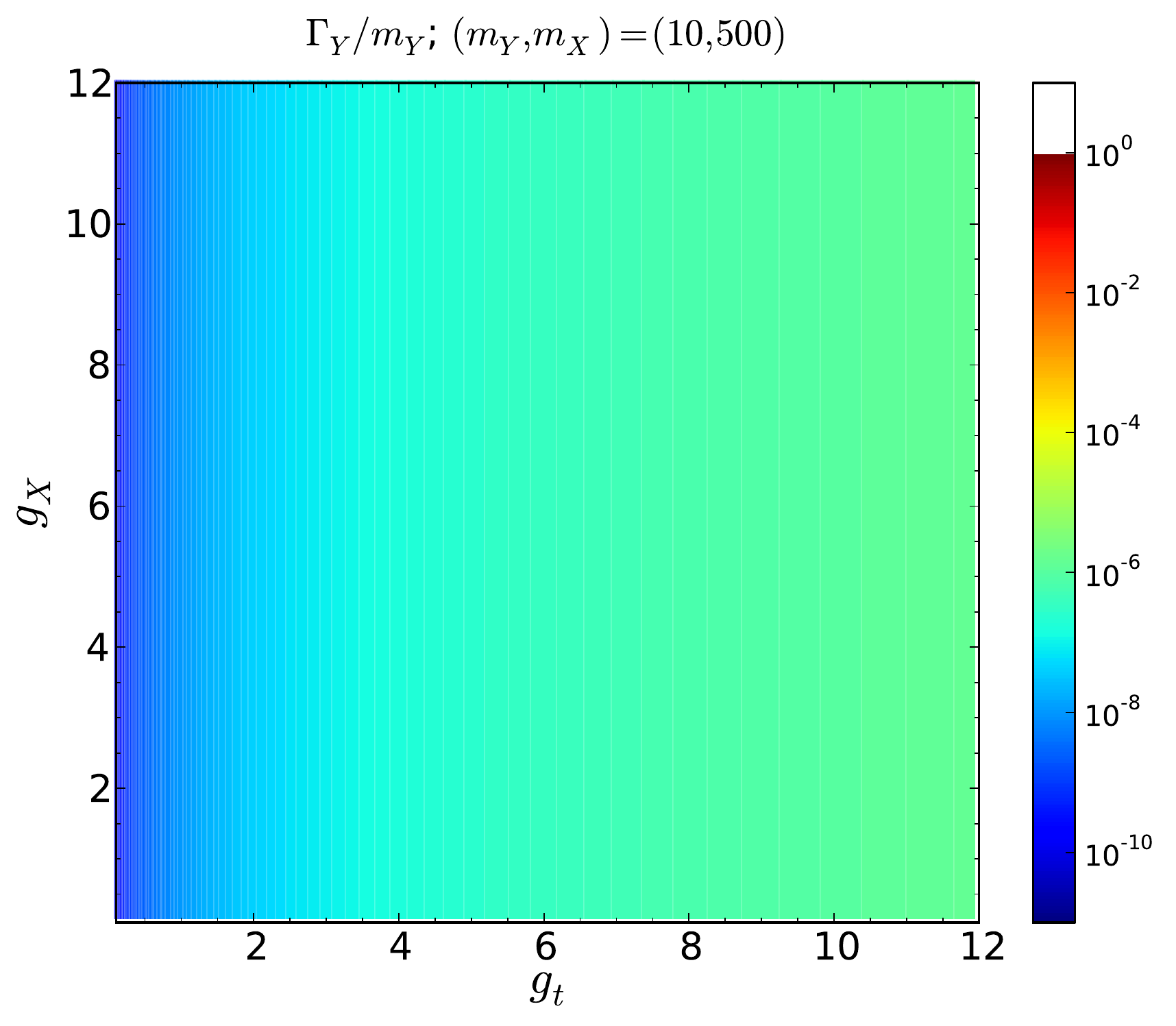}
 \includegraphics[width=0.32\textwidth]{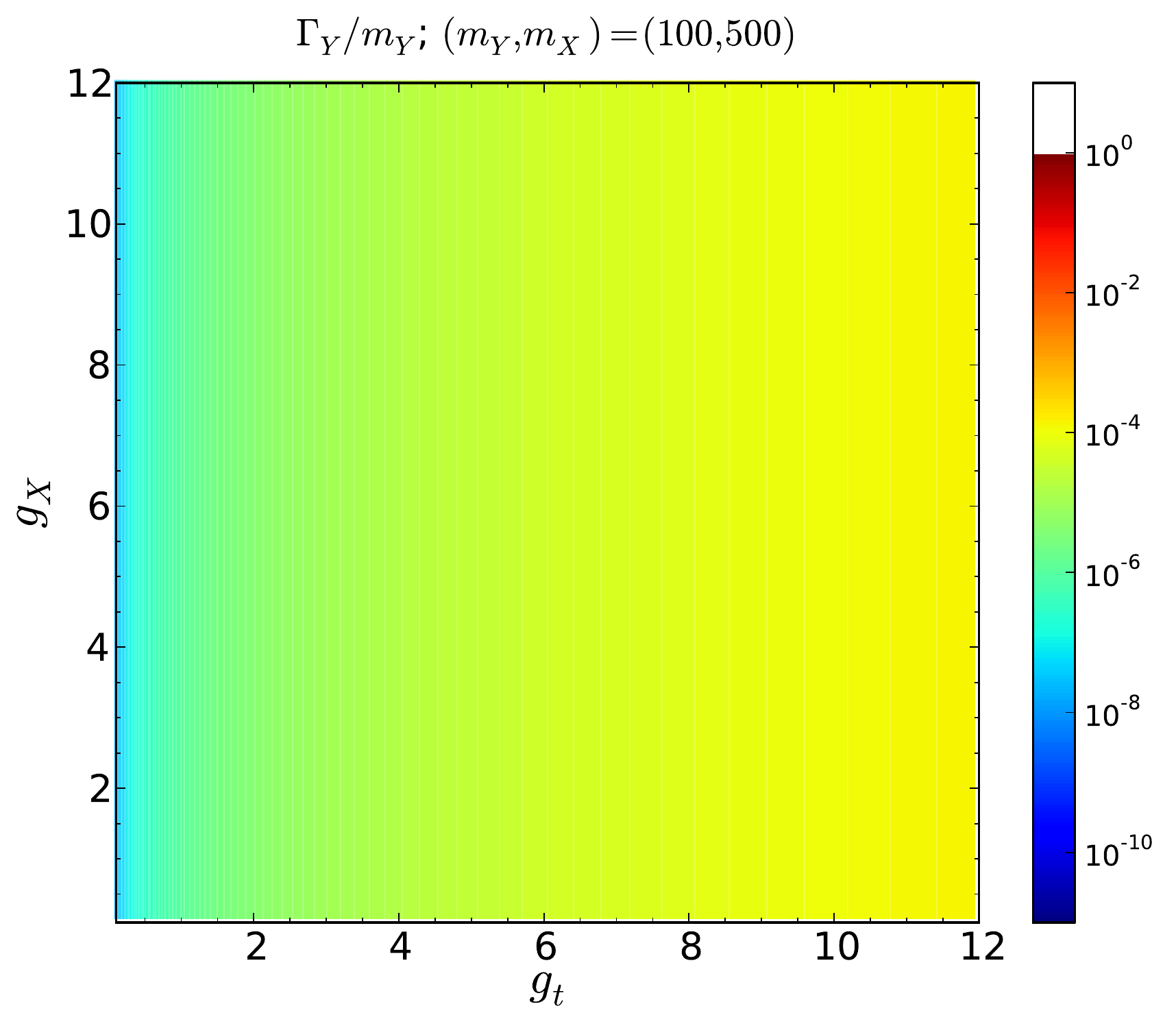}
 \includegraphics[width=0.32\textwidth]{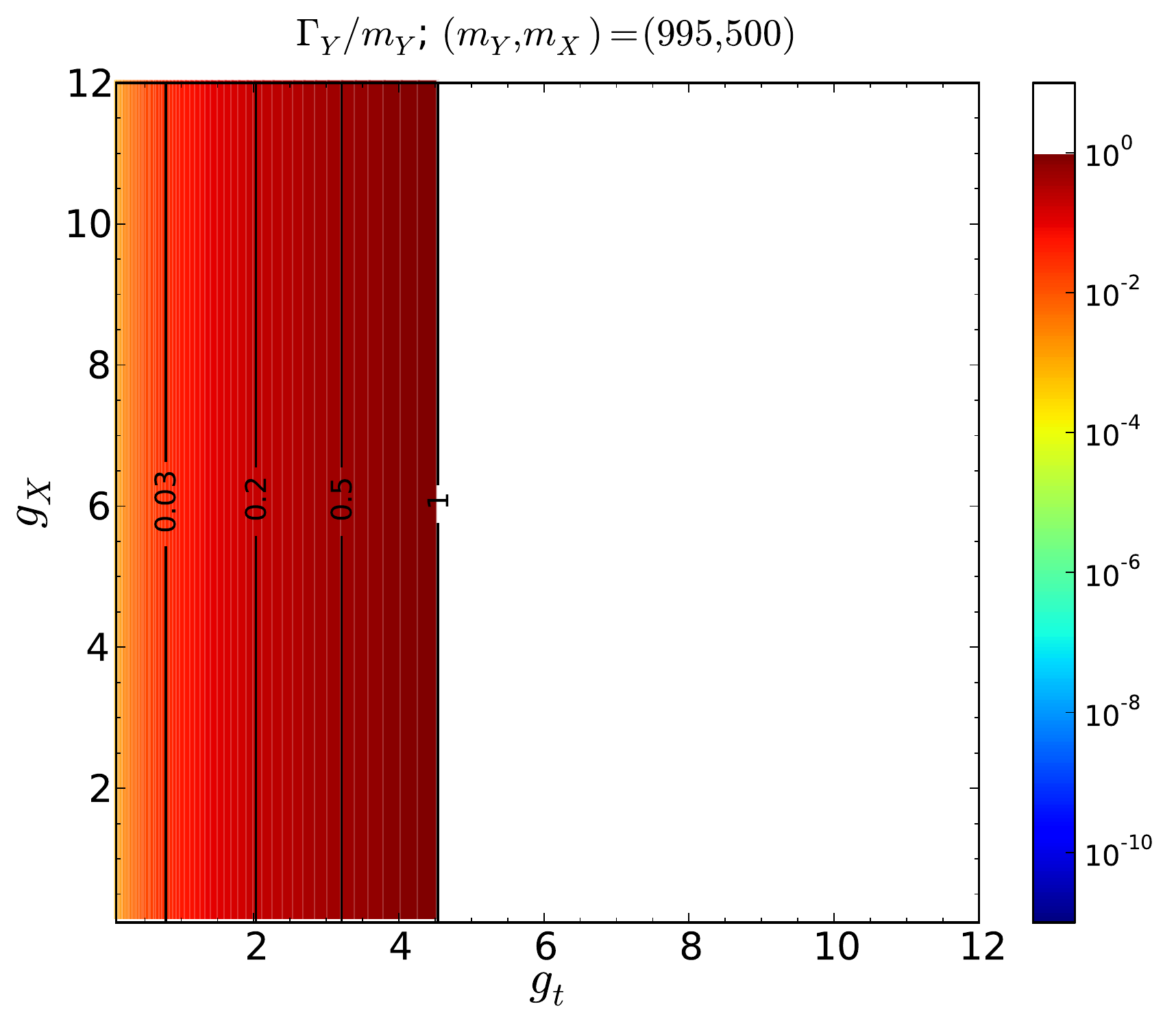}
 \includegraphics[width=0.32\textwidth]{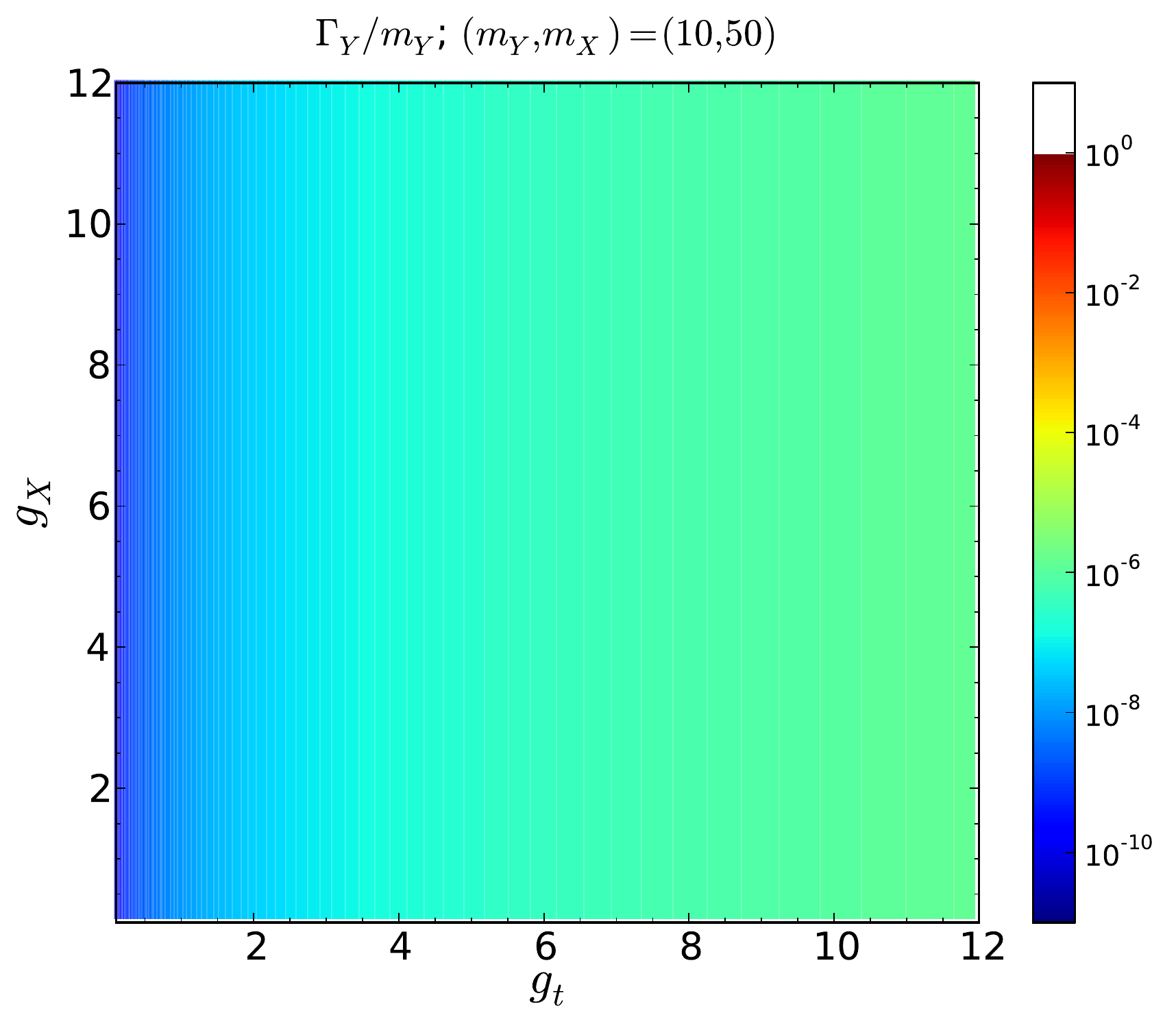}
 \includegraphics[width=0.32\textwidth]{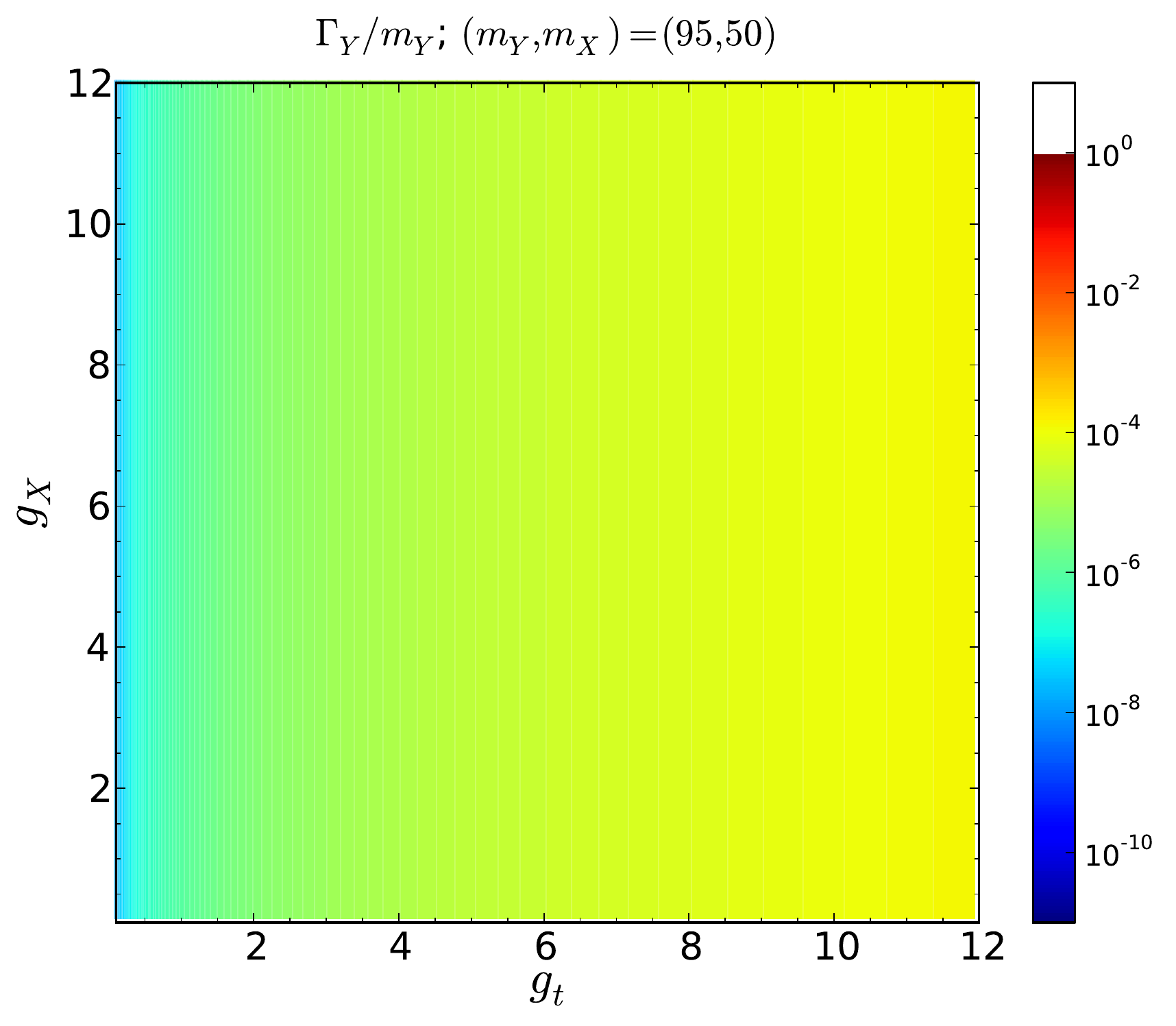}
 \includegraphics[width=0.32\textwidth]{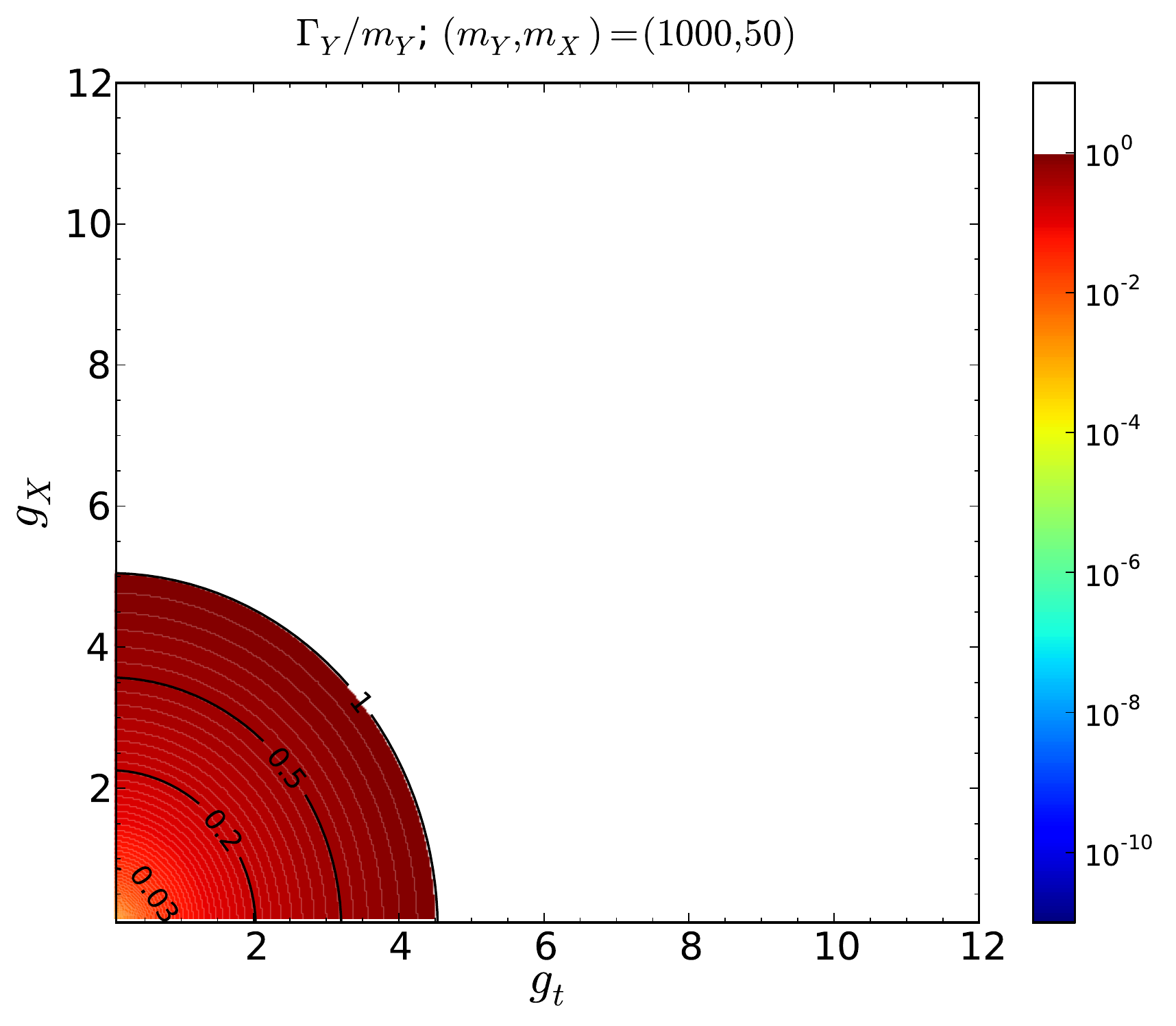}
 \includegraphics[width=0.32\textwidth]{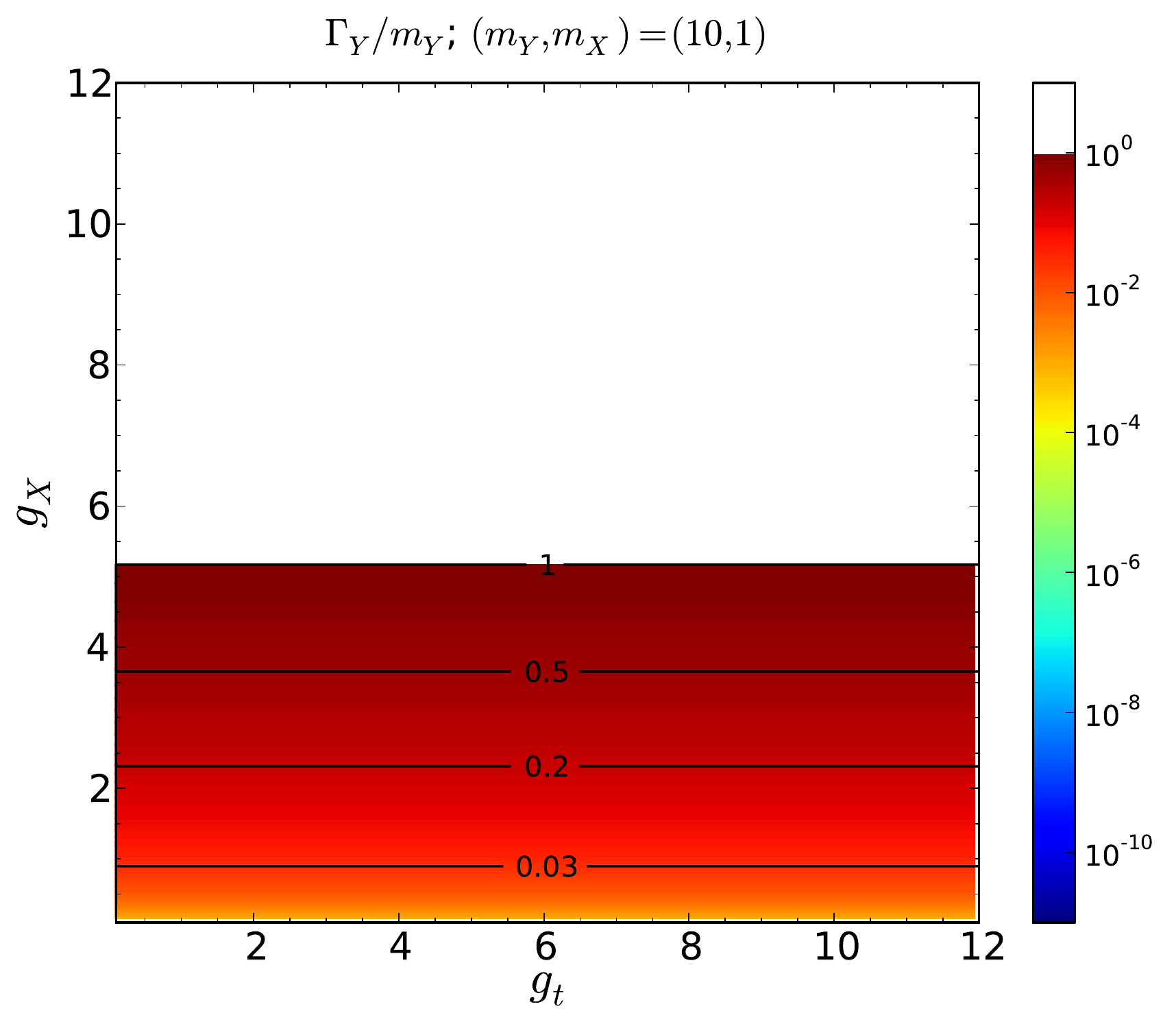}
 \includegraphics[width=0.32\textwidth]{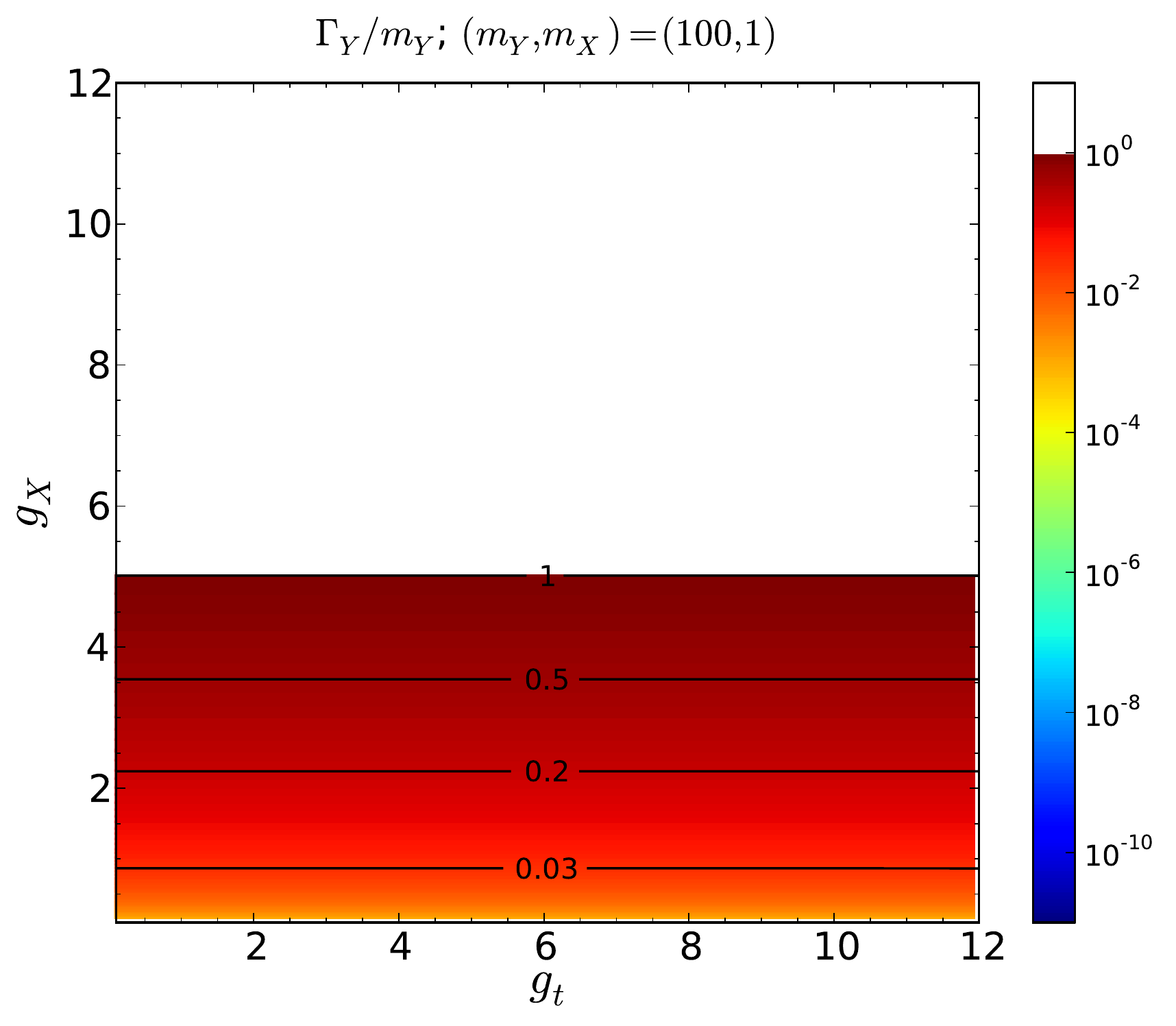}
 \includegraphics[width=0.32\textwidth]{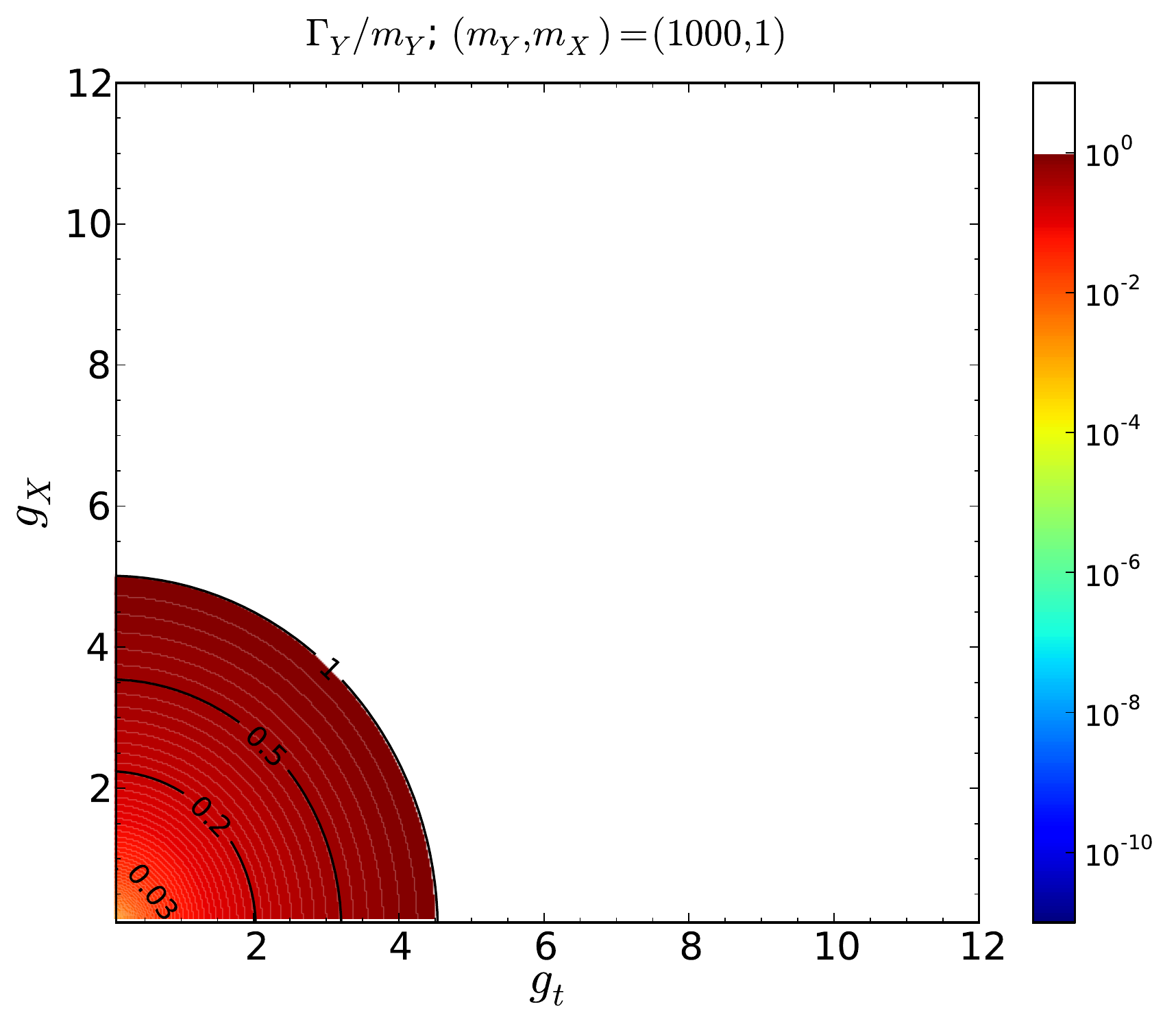}
\caption{$\Gamma_Y/m_Y$ in the ($g_{t}$, $g_X$) plane for different mass choices (expressed in $\GeV$). The colour bar shows the numerical value of the width to mass ratio.}
\label{fig:gymy_gtgx}
\end{figure}

\begin{figure}[t!]
 \center 
 \includegraphics[width=0.32\textwidth]{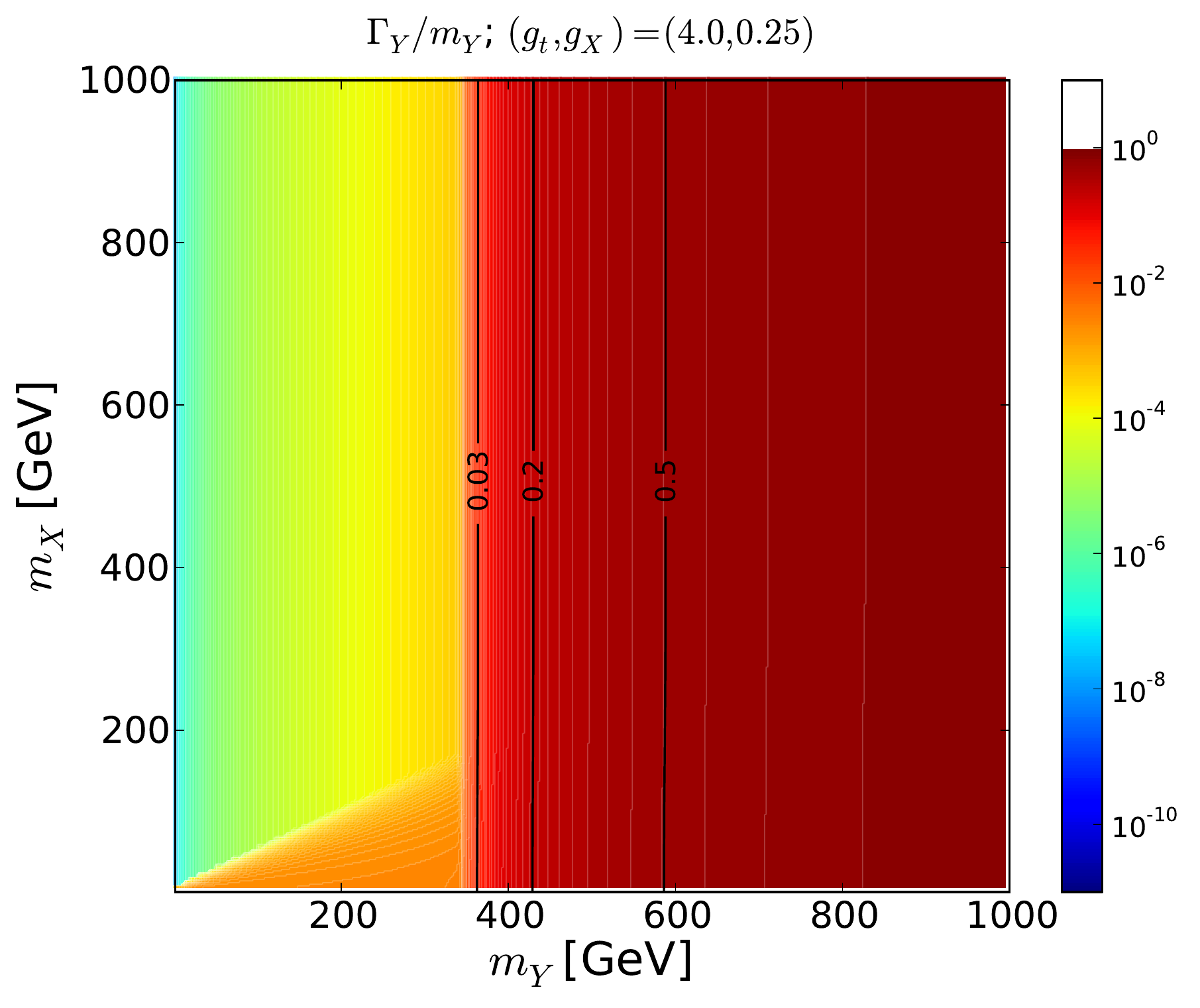}
 \includegraphics[width=0.32\textwidth]{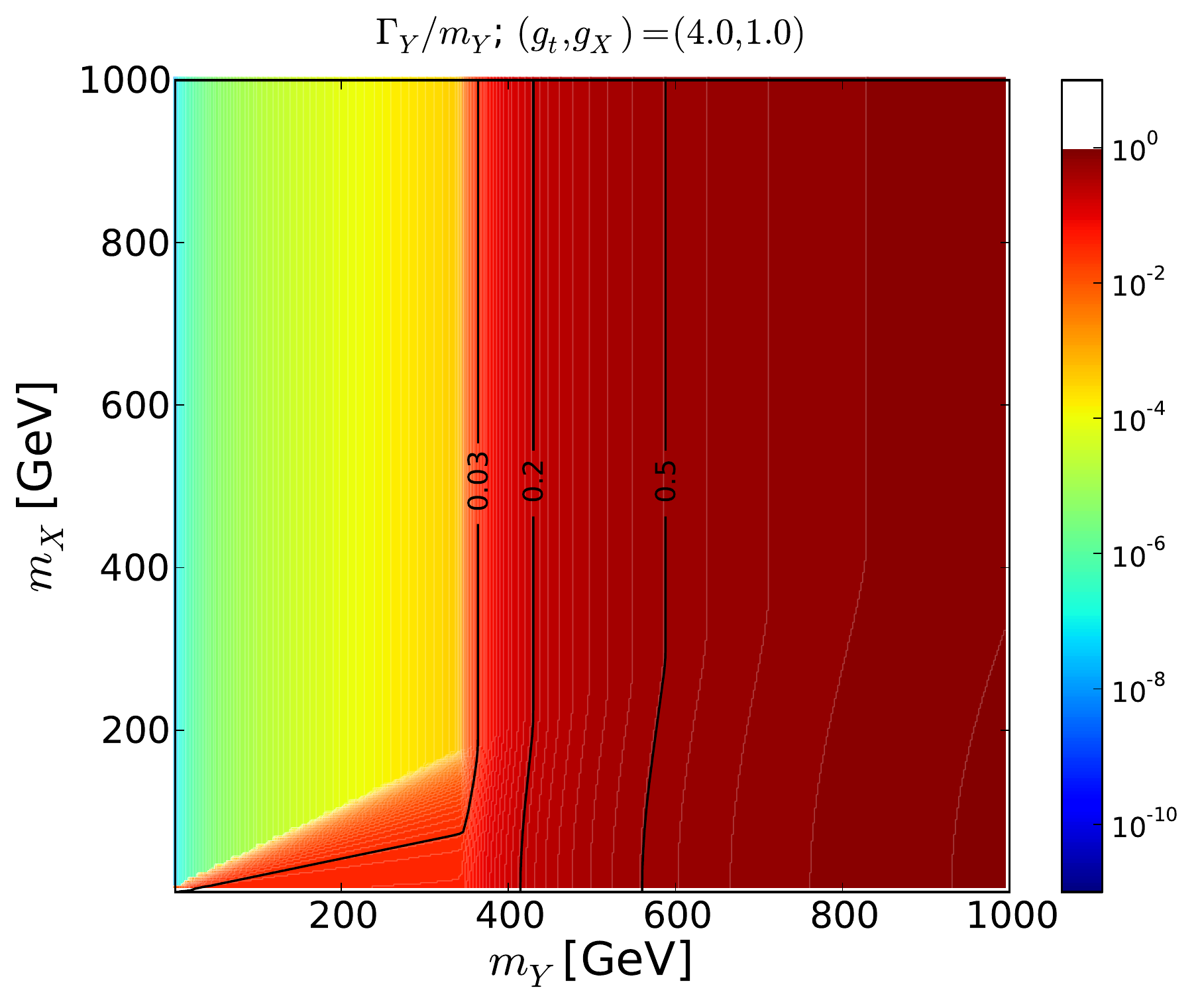}
 \includegraphics[width=0.32\textwidth]{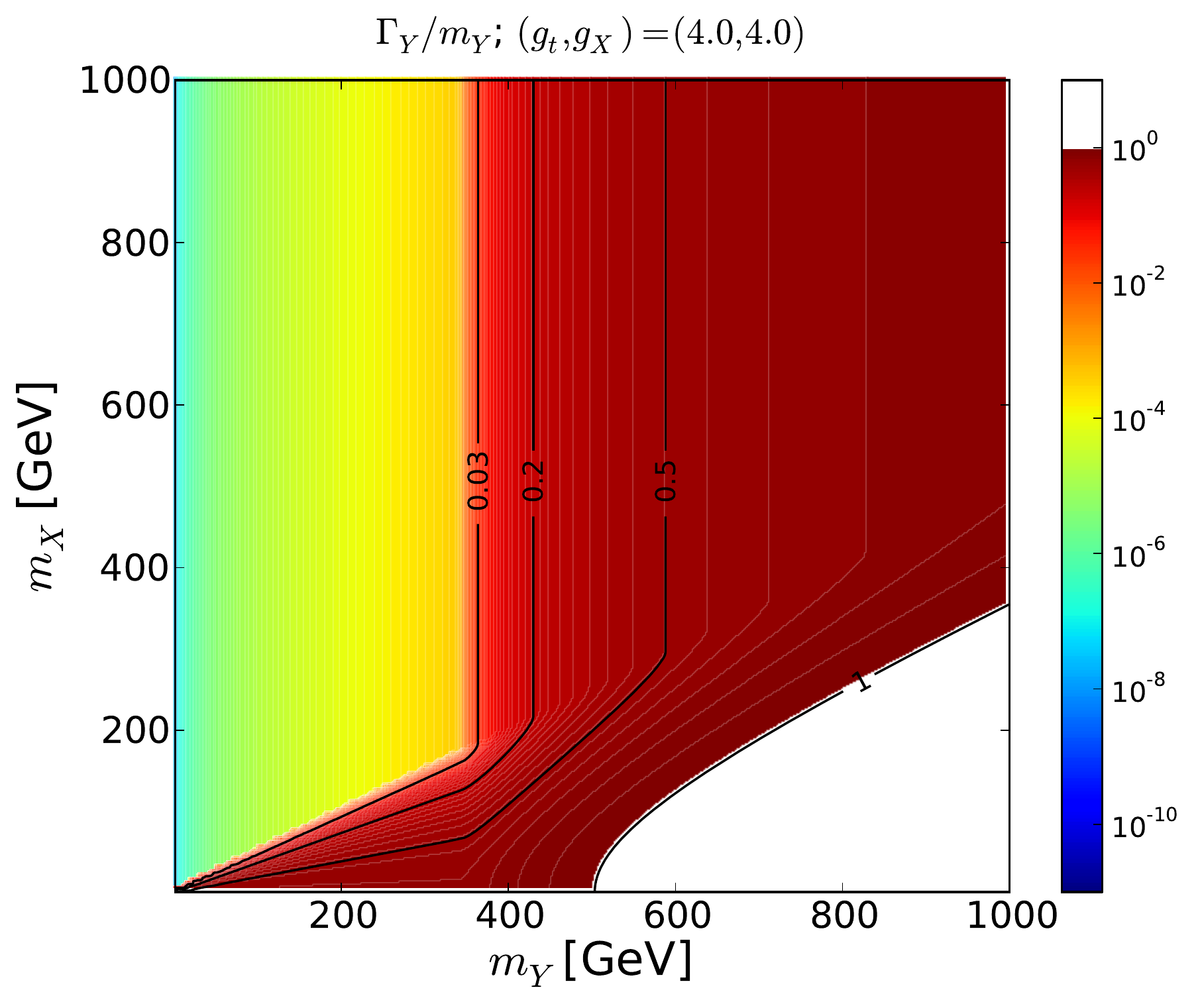}
 \includegraphics[width=0.32\textwidth]{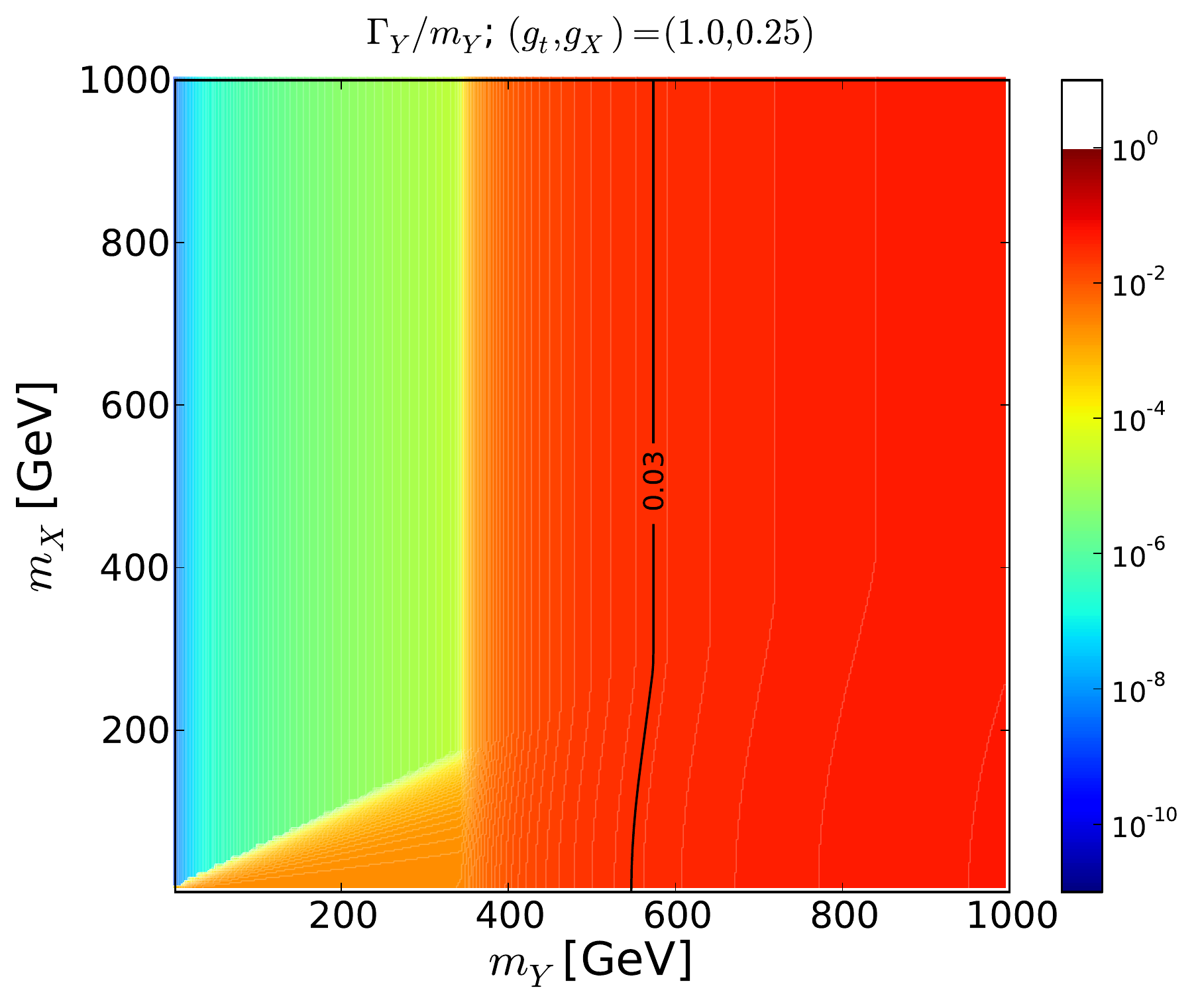}
 \includegraphics[width=0.32\textwidth]{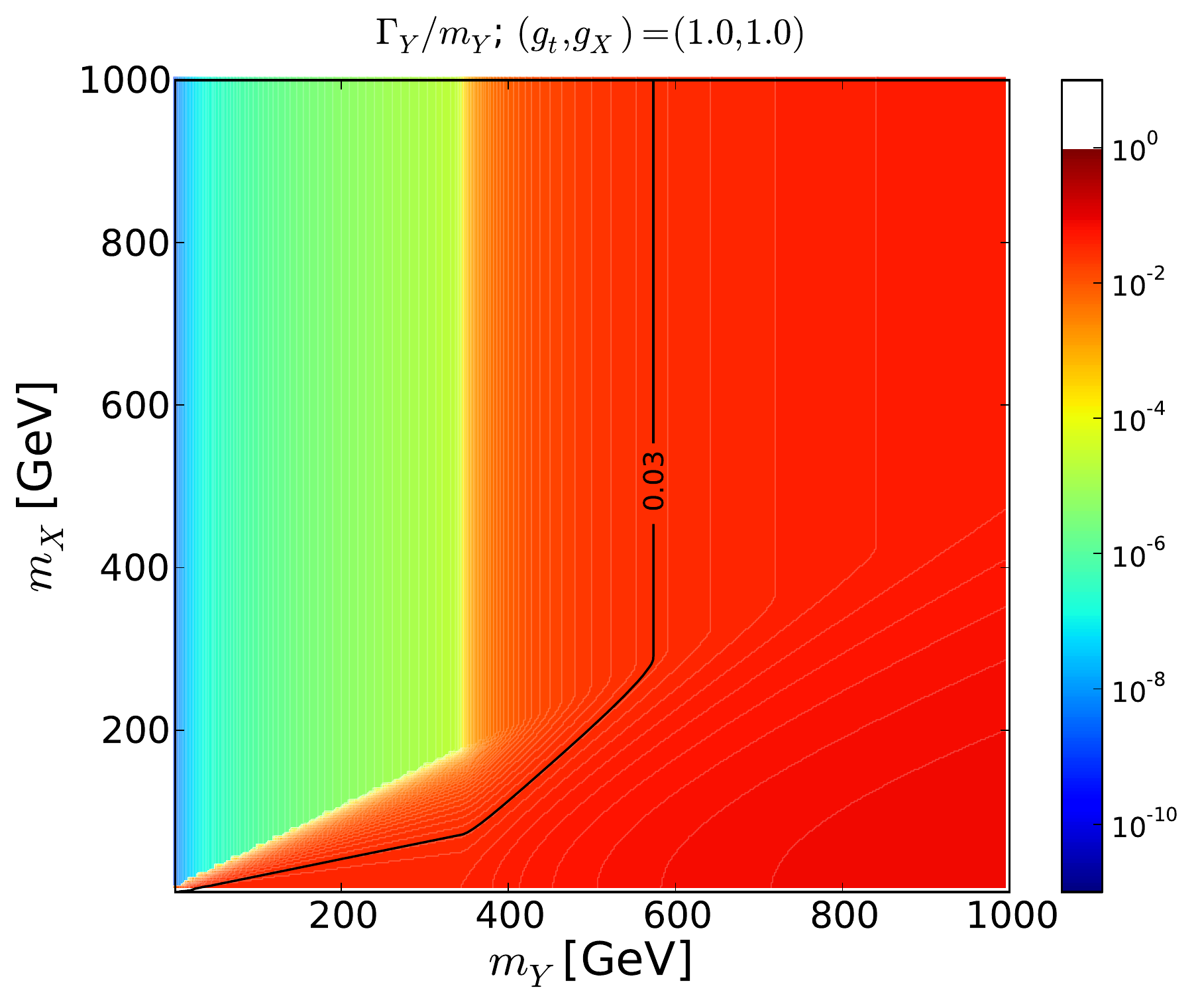}
 \includegraphics[width=0.32\textwidth]{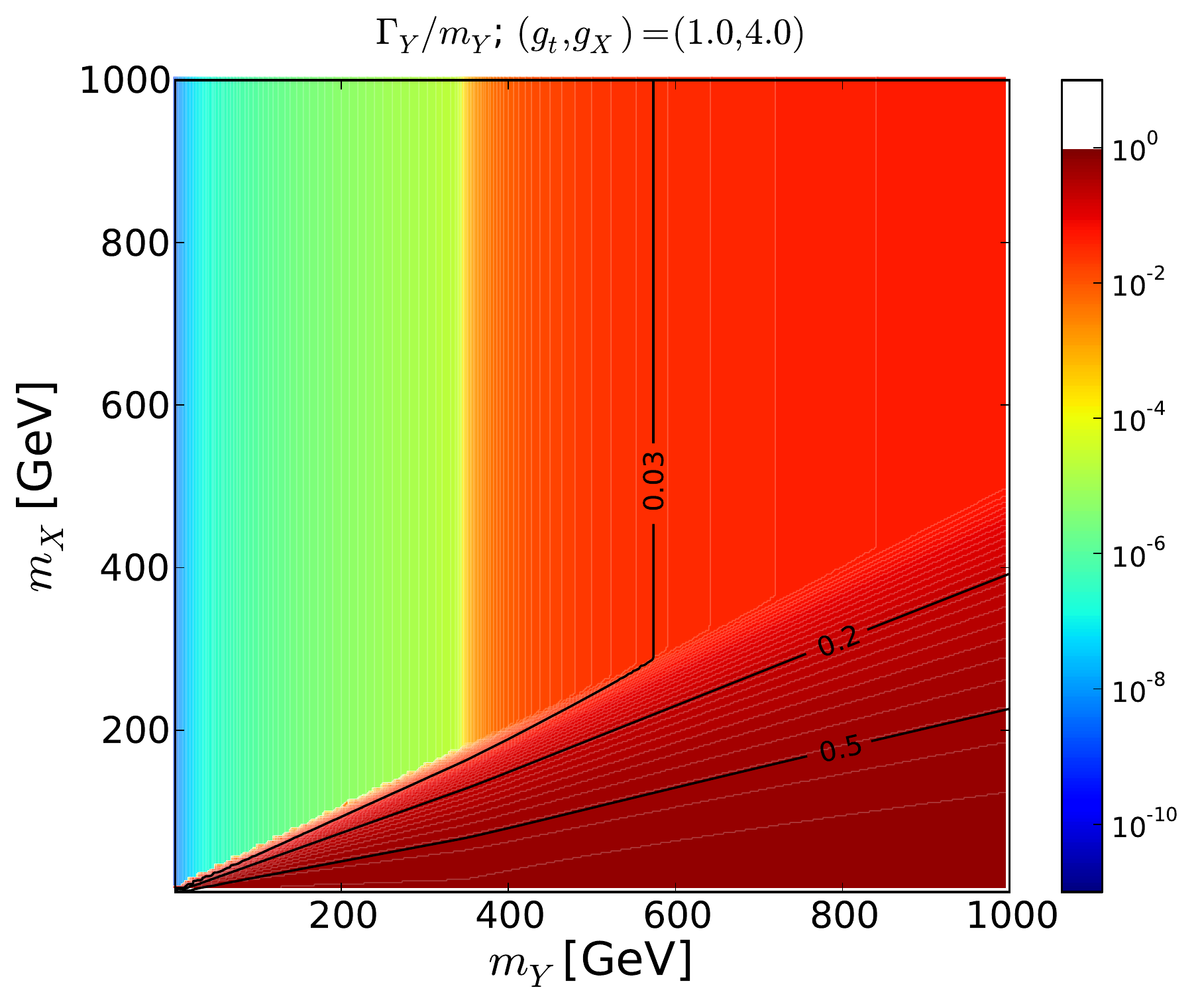}
 \includegraphics[width=0.32\textwidth]{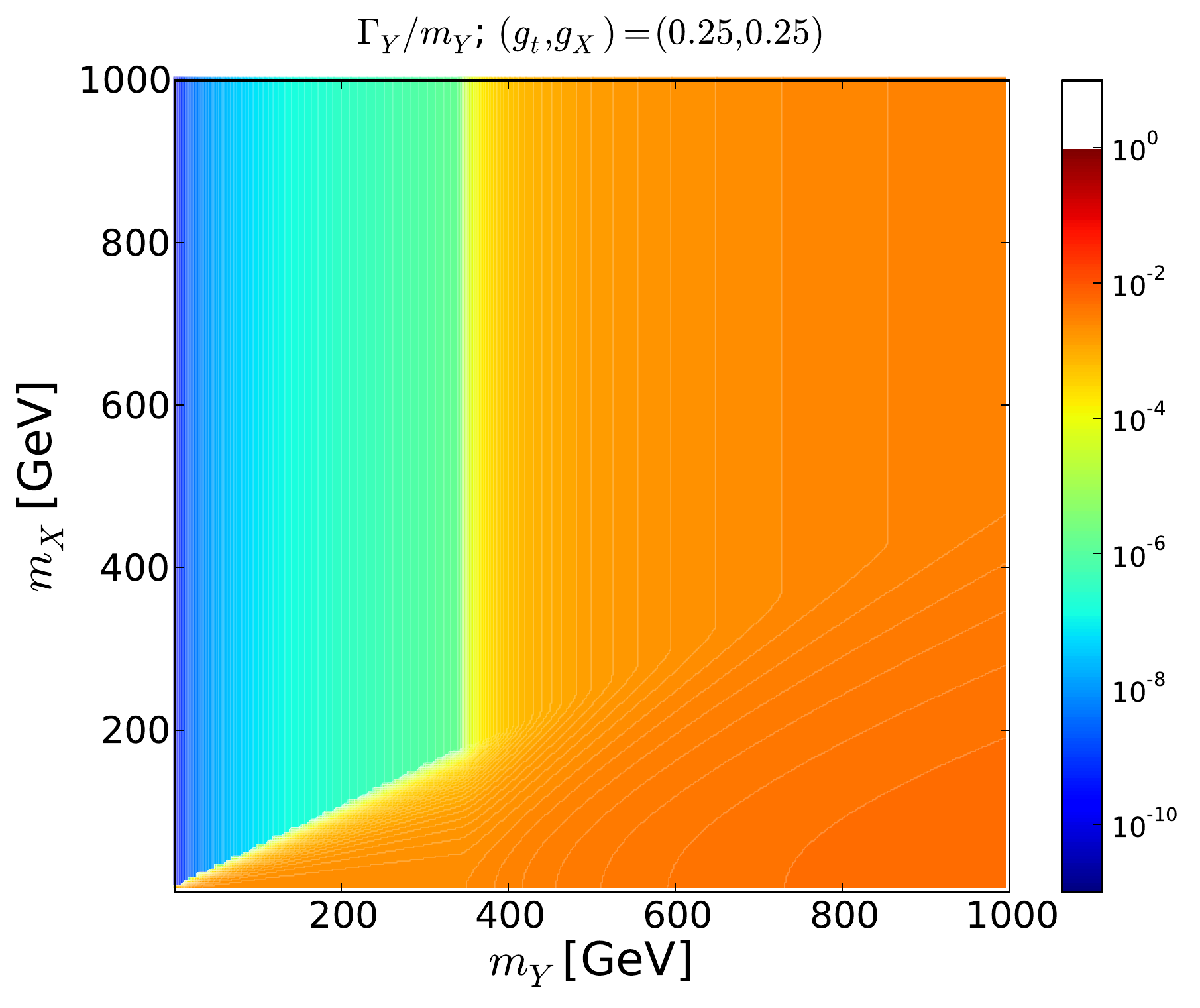}
 \includegraphics[width=0.32\textwidth]{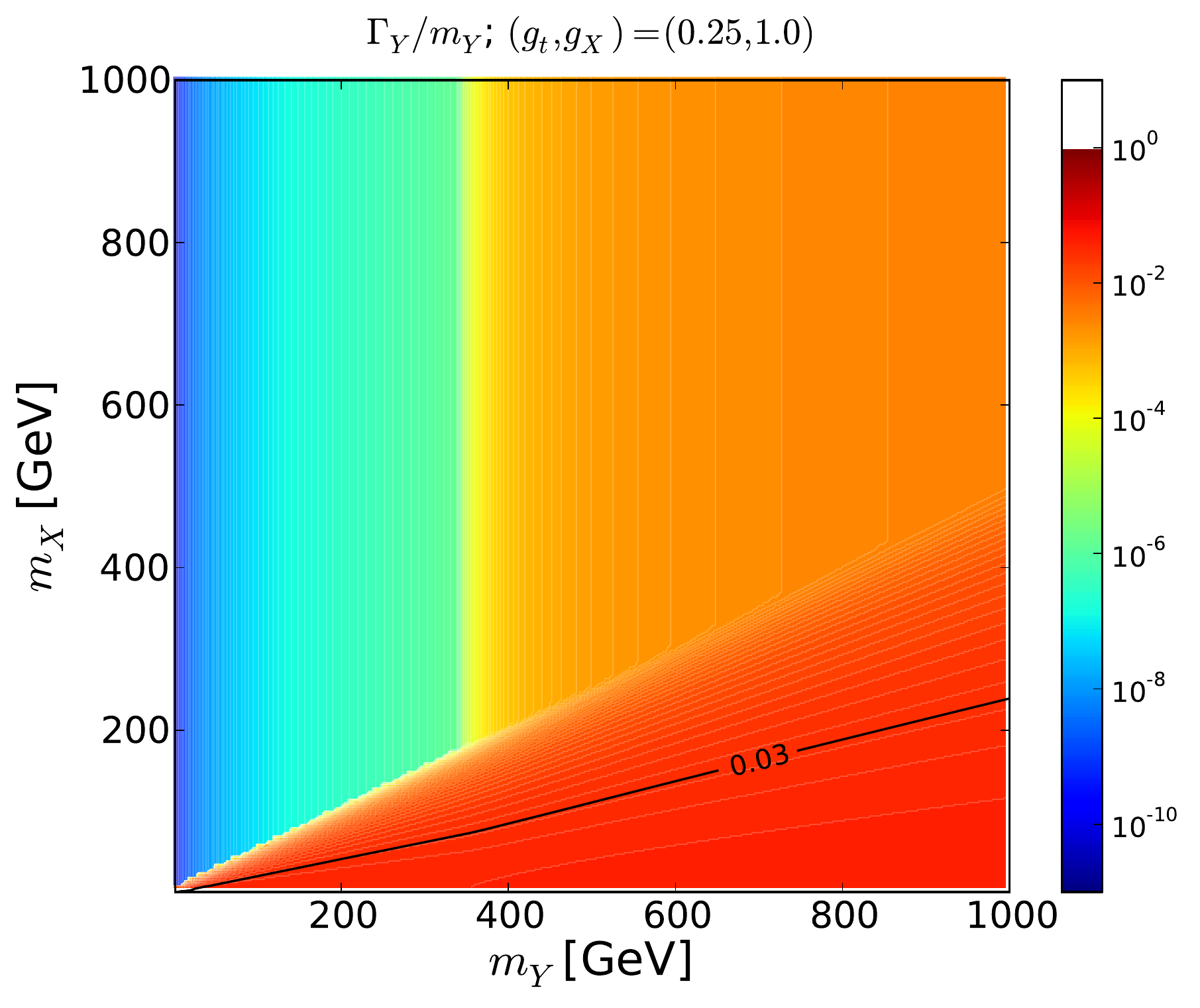}
 \includegraphics[width=0.32\textwidth]{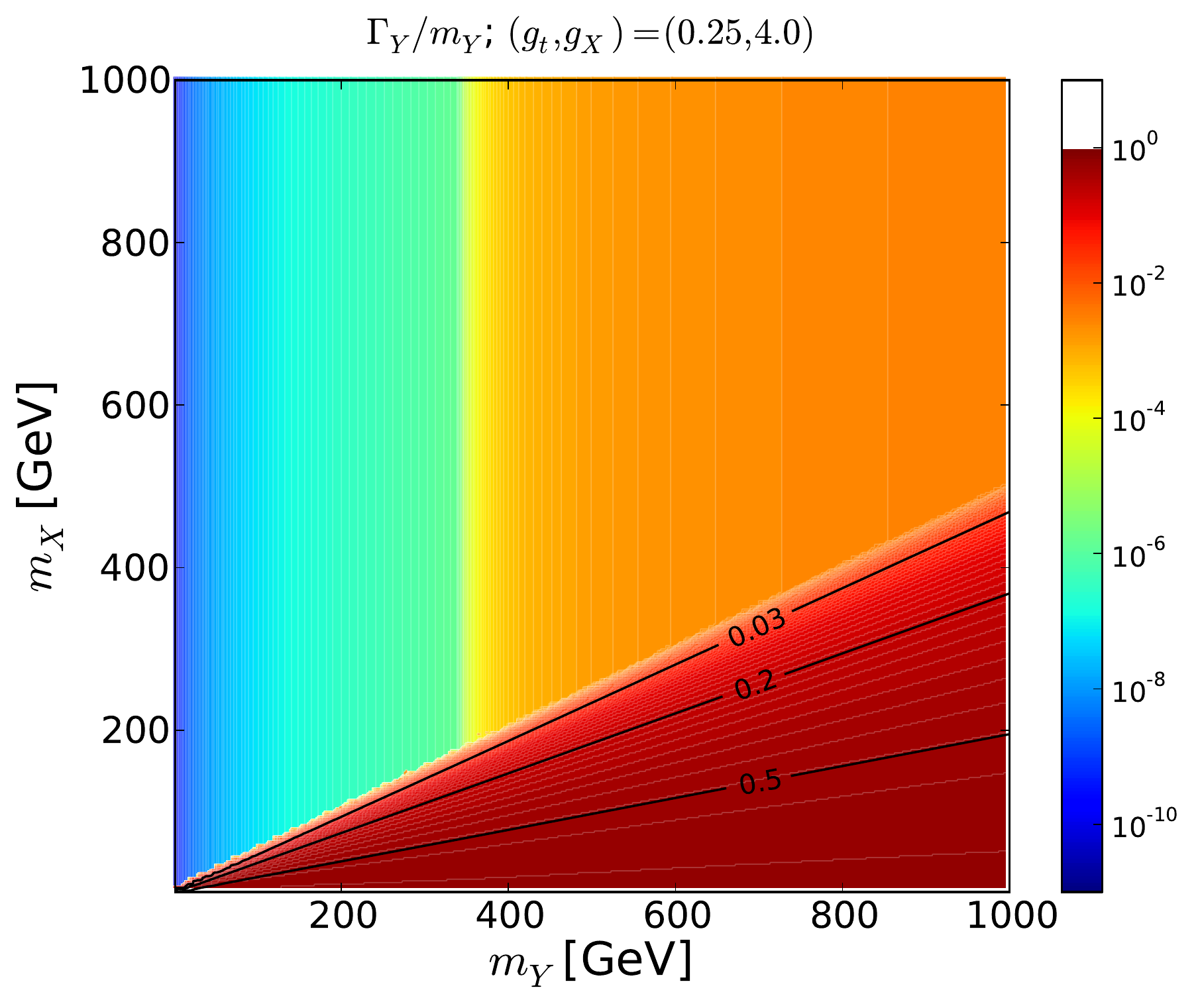}
\caption{$\Gamma_Y/m_Y$ in the ($m_Y$, $m_X$) plane for different coupling choices. The colour bar shows the numerical value of the width to mass ratio.}
\label{fig:gymy_mymx}
\end{figure}

As supplementary material, figure \ref{fig:gymy_gtgx} shows
the relative mediator width $\wy/\my$ in the ($\gsm$, $\gx$) plane for different mass choices. The magnitude of the mediator width depends on the hierarchy among the different decay processes \ie~$t\bar{t}$, $\dm \dmbar$, $gg$ and $\gamma \gamma$. Diphoton channel is negligible compared to the others and will not be discussed. Figure \ref{fig:gymy_gtgx} shows that the $\y$ resonance can be considered as narrow (\ie~$\wy/\my<0.03$) when only the $gg$ decay mode is involved. As soon as $t \bar{t}$ and $\dm \dmbar$ decay channels are opened, the ratio $\wy/\my$ grows quickly, reaching $20\%$ for $\gsm, \gx \sim 2$. The narrow width approximation is valid below couplings of $\cal{O}$(1). Figure \ref{fig:gymy_mymx} shows the relative mediator width $\wy/\my$ in the ($\my, \mdm$) plane for different coupling choices. When $\gsm, \gx \le 1$, $\wy/\my$ never exceeds $10\%$ and the narrow width approximation is reliable for a wide region of the parameter space. In the kinematic regions where $\y$ decays to $\dm$ and/or top quarks is allowed, increase in either $\gsm$ or $\gx$ quickly leads to $\wy/\my$ ratio above $20\%$.

\newpage

\section{Details of the top-philic dark matter analysis}\label{sec:dmapp}
\subsection{Consistency checks of astrophysical and cosmological dark matter signatures}\label{sec:cc}
\begin{figure}[h!]
\begin{center}
\includegraphics[width=2.5in]{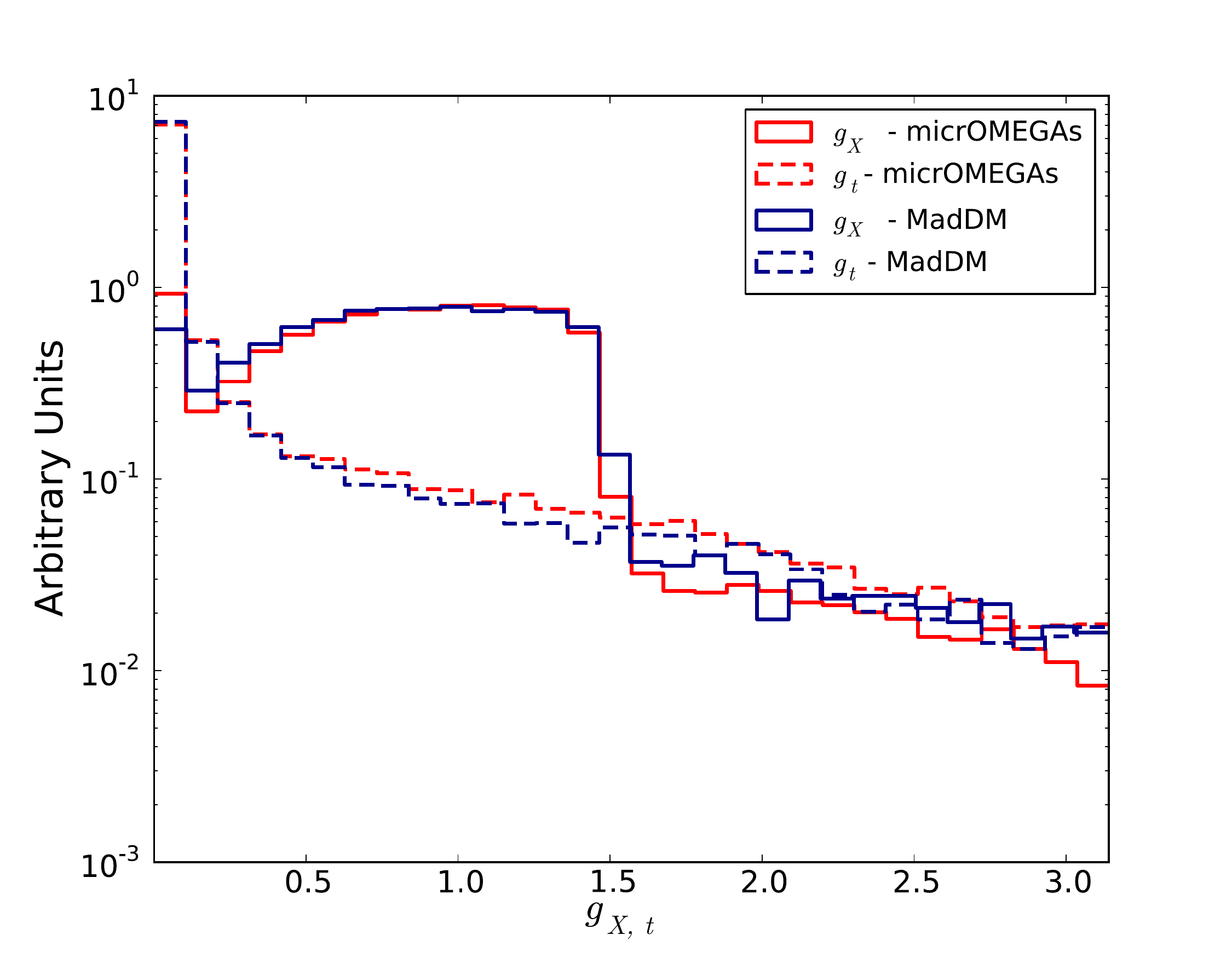}
\includegraphics[width=2.5in]{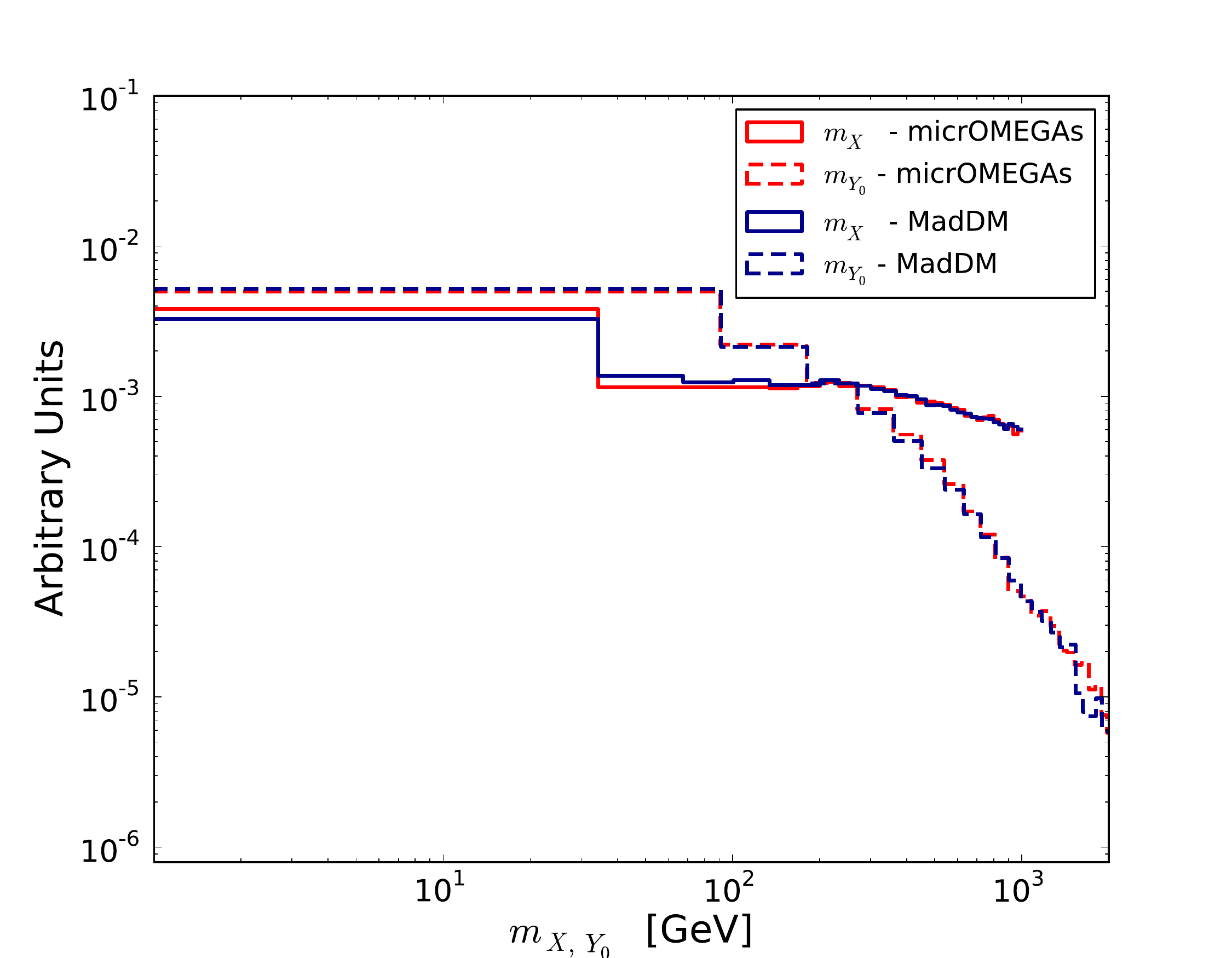}\\
\includegraphics[width=2.5in]{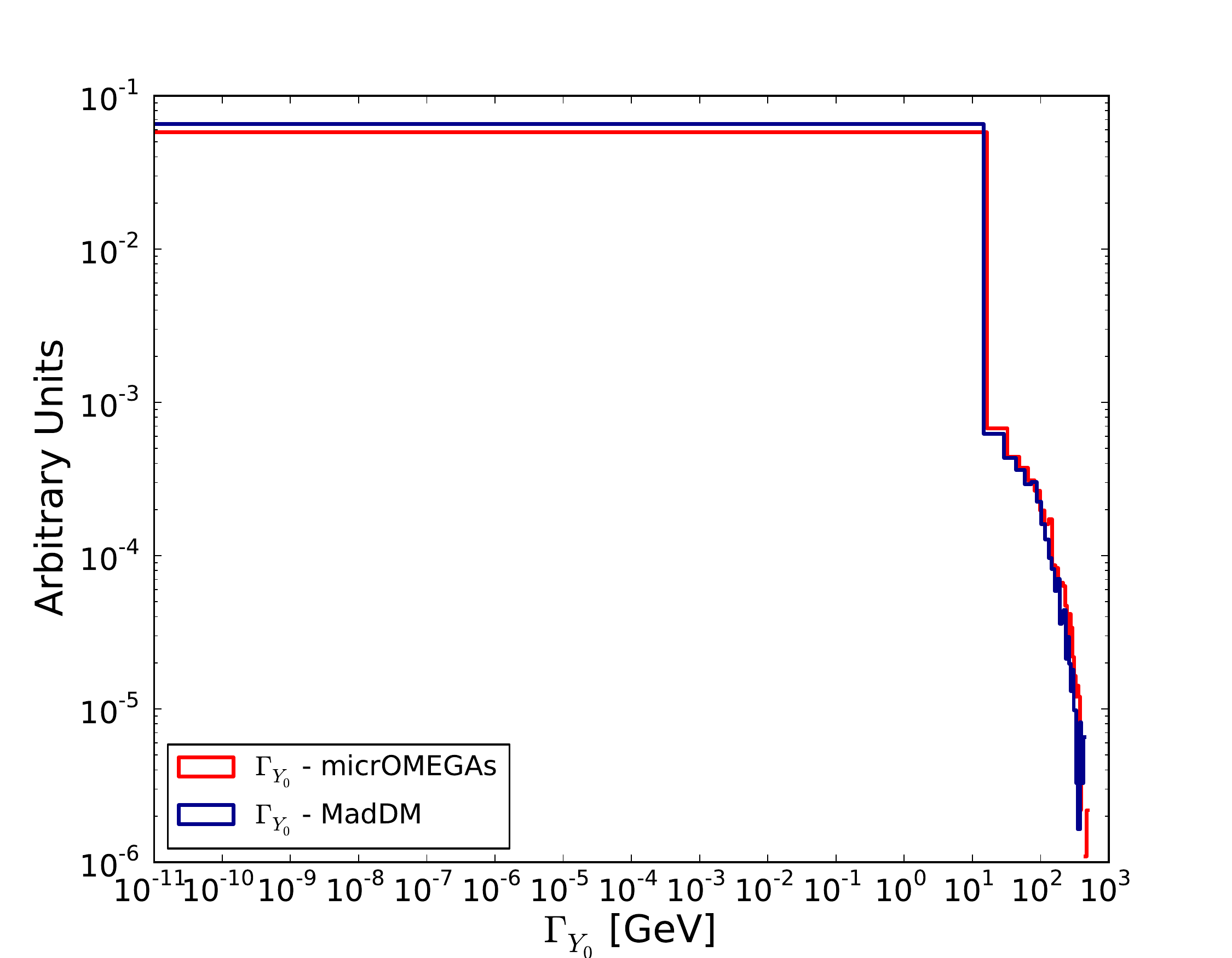}
\includegraphics[width=2.5in]{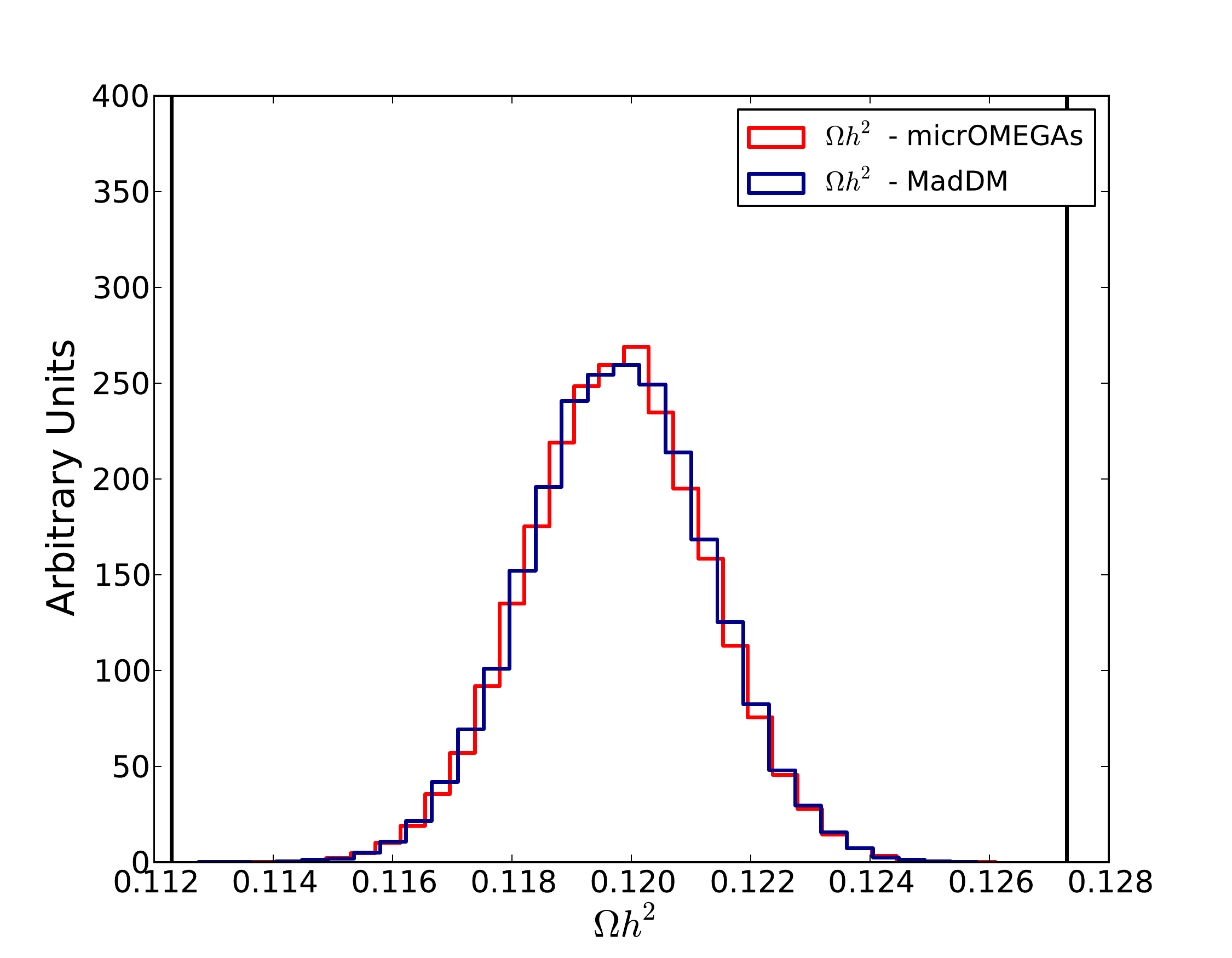}
\end{center}
\caption{Consistency check for the parameter scan.  The panels show the distribution of couplings, masses and $Y_0$ widths and relic densities resulting from a {\sc MultiNest} parameter scan. The vertical lines in the panel showing the distribution of relic density represent the $5 \sigma$ Planck bound.} \label{fig:consistency}
\end{figure}

As a part of consistency checks, we have ensured that the scan covers similar regions of the parameter space both in case of {\sc MadDM} and {\sc micrOMEGAs}. Fig.~\ref{fig:consistency} shows the results for distributions of masses and couplings in the scans, where the blue/red lines refer to {\sc MadDM}/{\sc micrOMEGAs} respectively. Similarities in the distributions of fig.~\ref{fig:consistency} indicate that parameter scanning was performed consistently between the two codes.

\begin{figure}[h!]
\begin{center}
\includegraphics[width=2.5in]{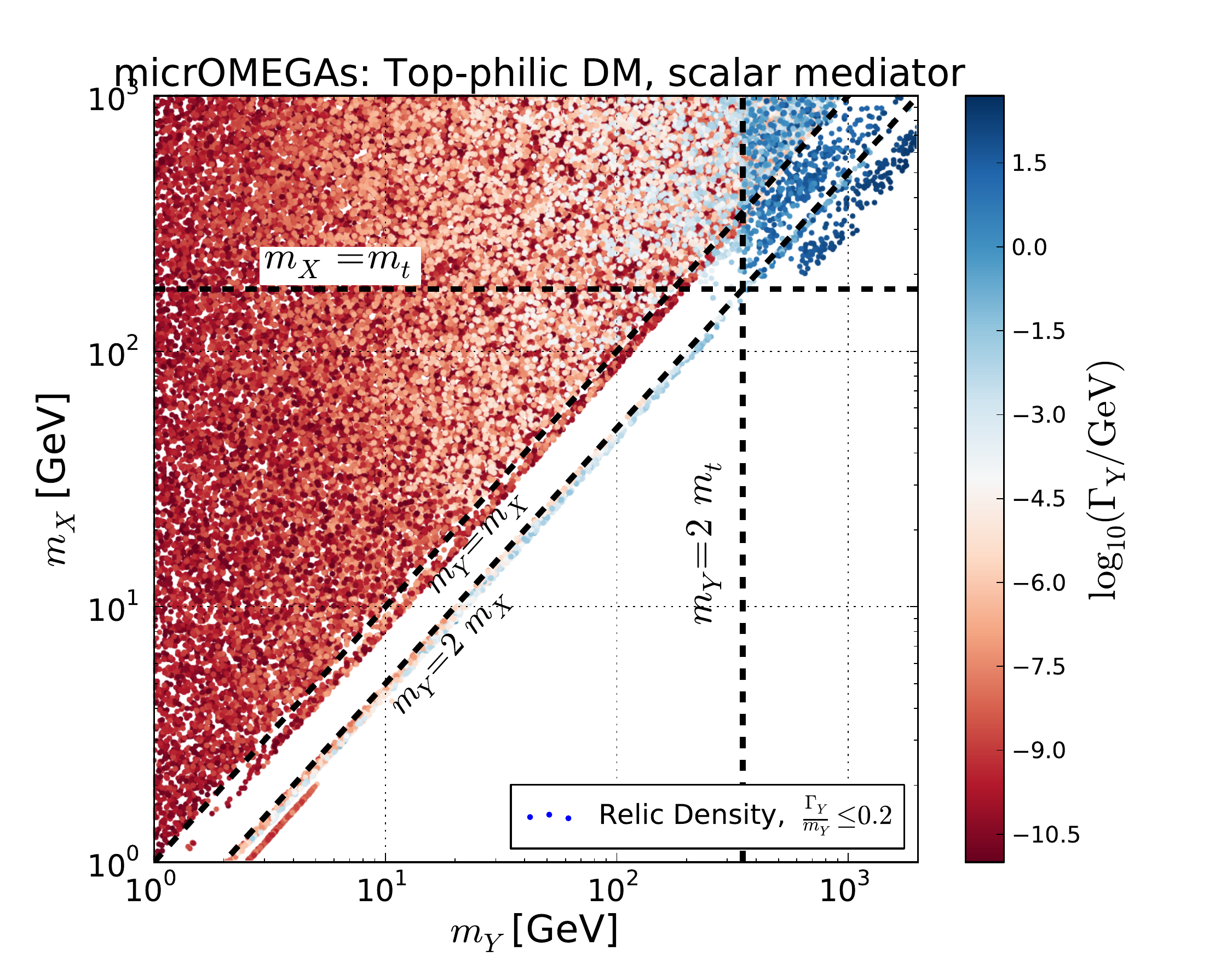}
\includegraphics[width=2.5in]{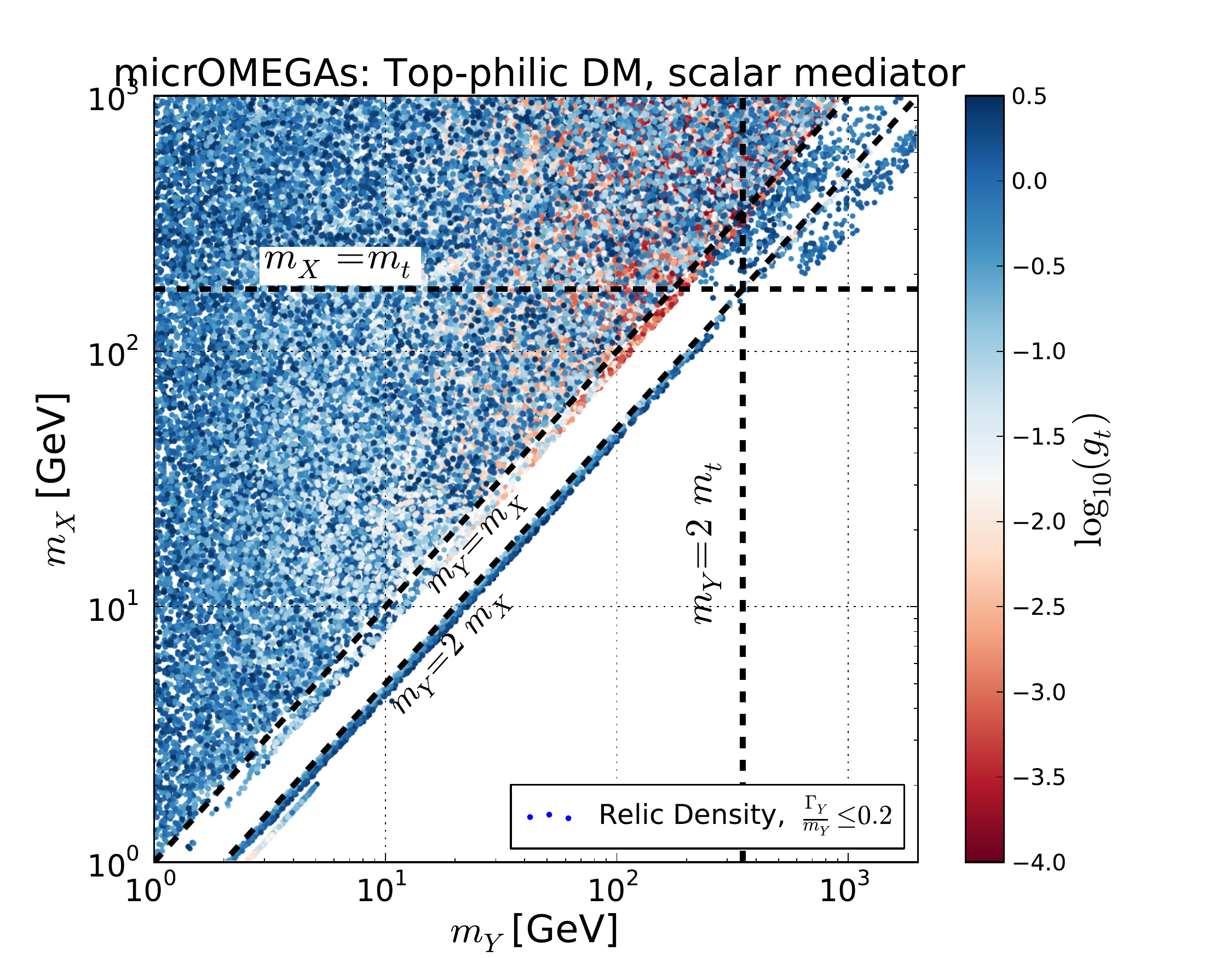}
\includegraphics[width=2.5in]{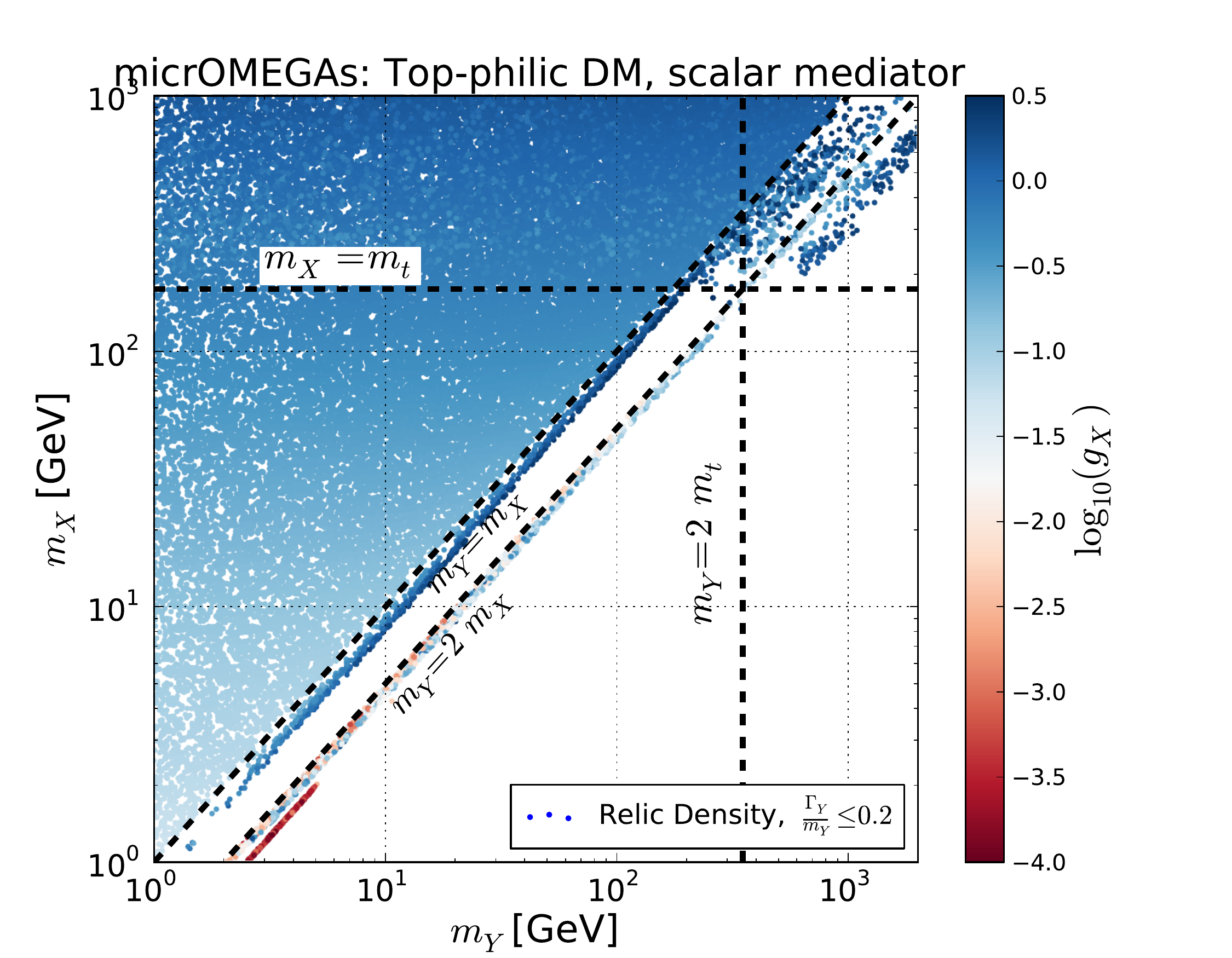}
\includegraphics[width=2.5in]{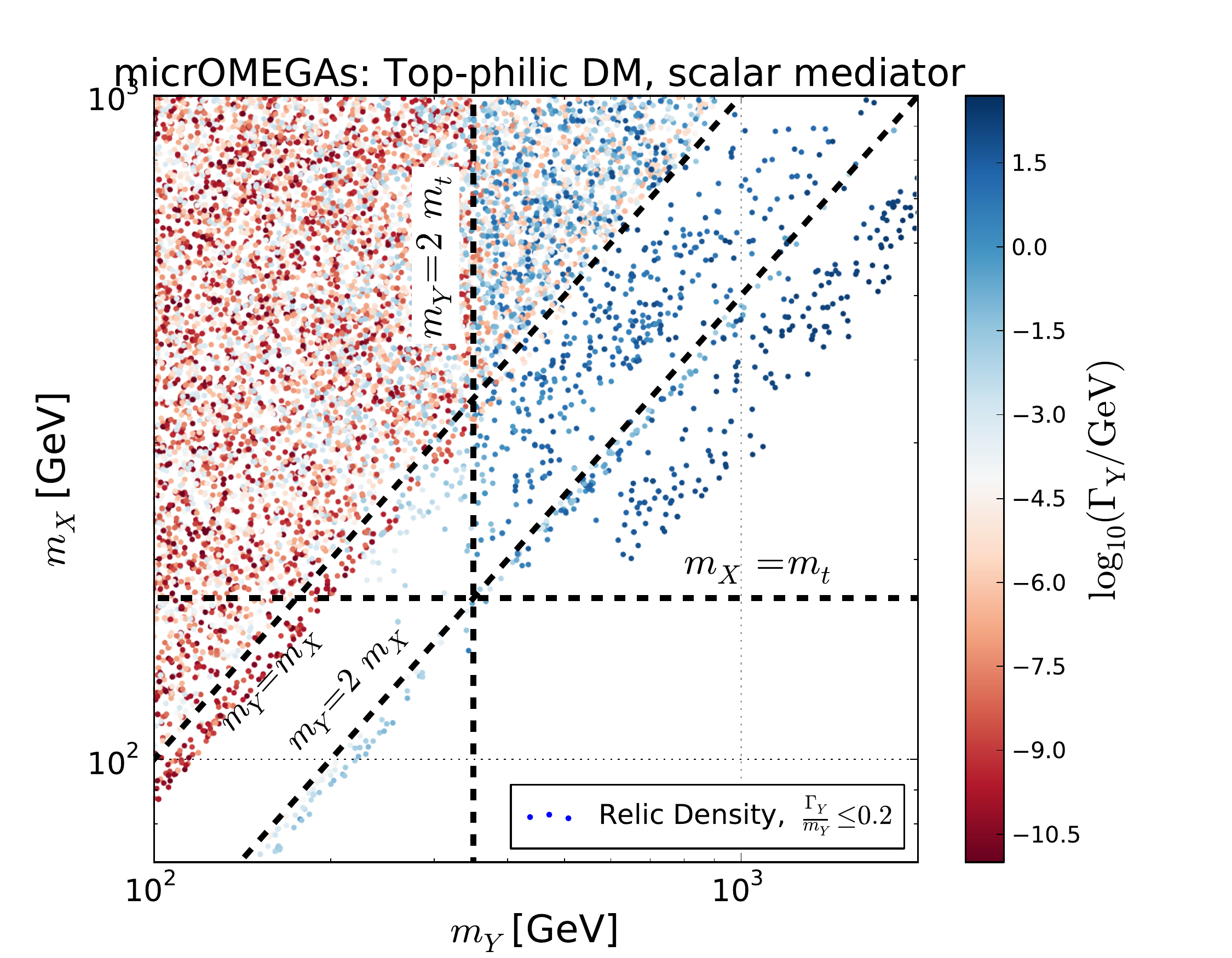}
\end{center}

We also made an explicit comparison between the projections of the four dimensional parameter scans obtained with {\sc MadDM} and {\sc micrOMEGAs} respectively. Figures~ \ref{fig:scan1momegas}  and \ref{fig:consistency2} show several examples. We don't find significant deviations between the results obtained in the two codes except in the region of $\my \sim 2 \mdm$ and $\mdm \sim  {\cal O}(1) \GeV$. In these regions we expect some discrepancies due to the possible numerical instabilities in integration of the thermally averaged cross section for amplitudes which feature resonances of extremely small widths.

\caption{Four dimensional parameter scan using {\sc micrOMEGAs}, projected onto the $\mdm, \, \my$ plane. The first three panels show the projections with the colourmap representing the values of $\wy$, $\gx$ and $\gsm$ respectively. The right-most panel shows the zoomed-in upper right region of the left-most panel. The scan assumes only relic density and narrow width constrains.} \label{fig:scan1momegas}
\end{figure}

\begin{figure}[h!]
\begin{center}
\includegraphics[width=2.5in]{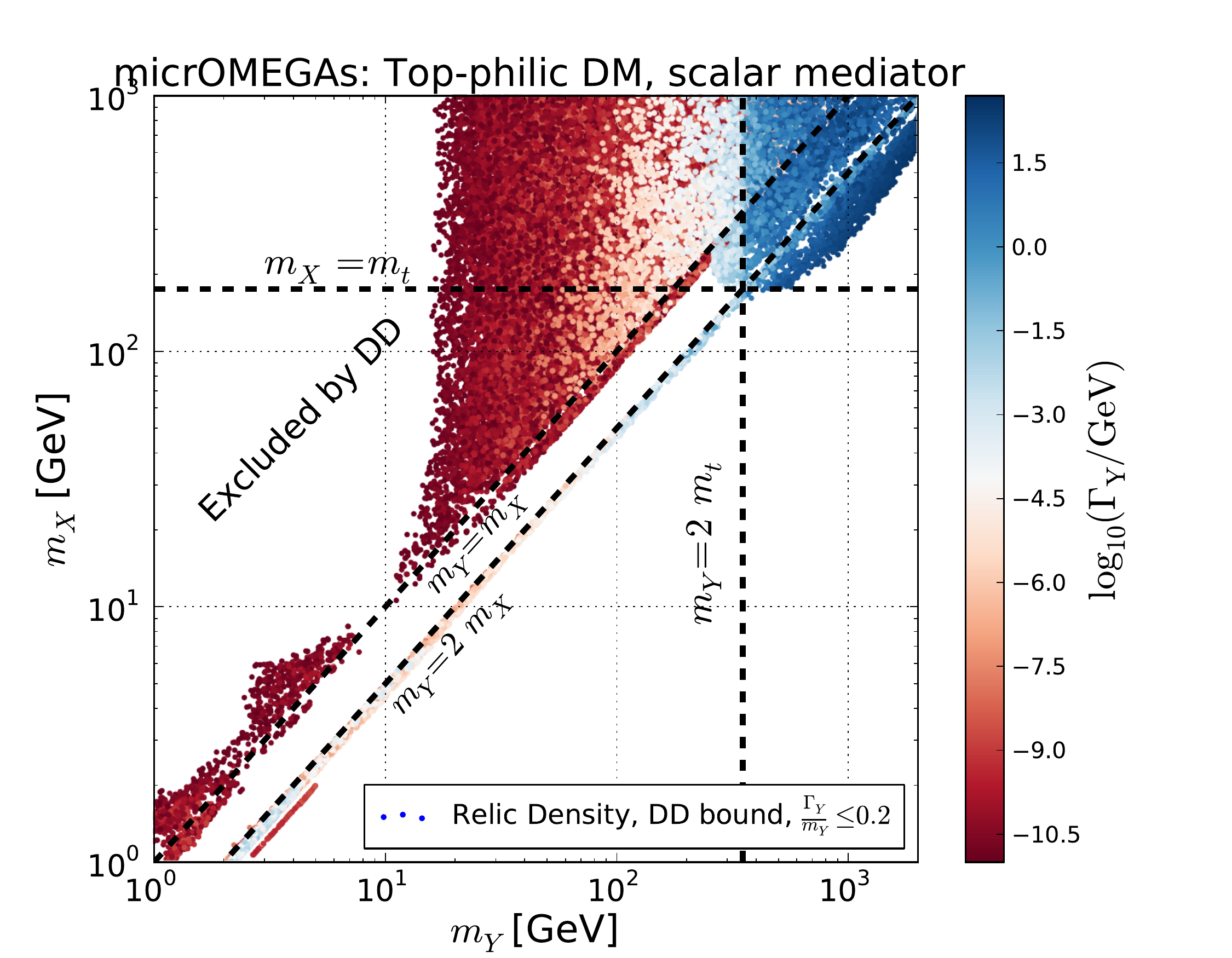}
\includegraphics[width=2.5in]{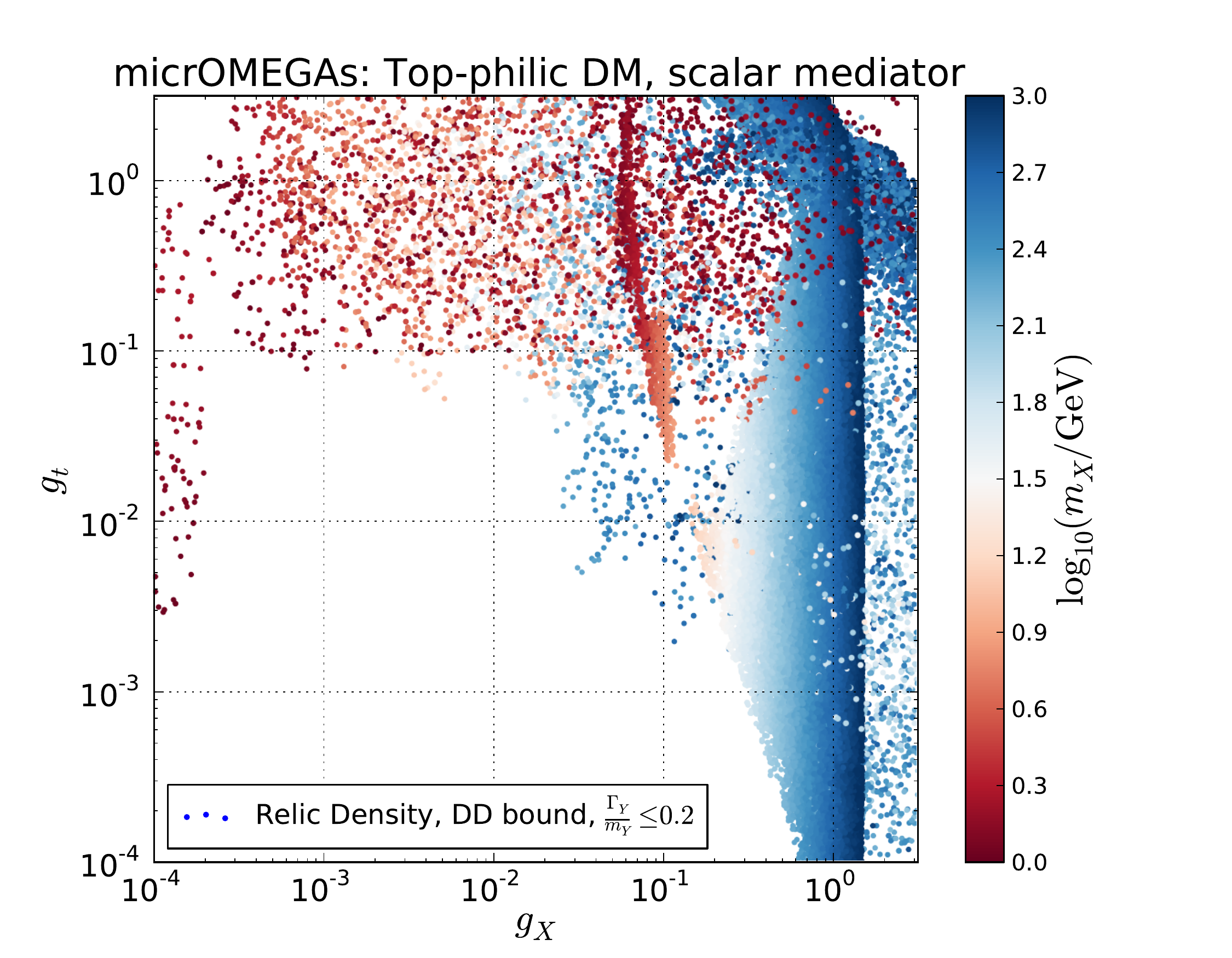}
\includegraphics[width=2.5in]{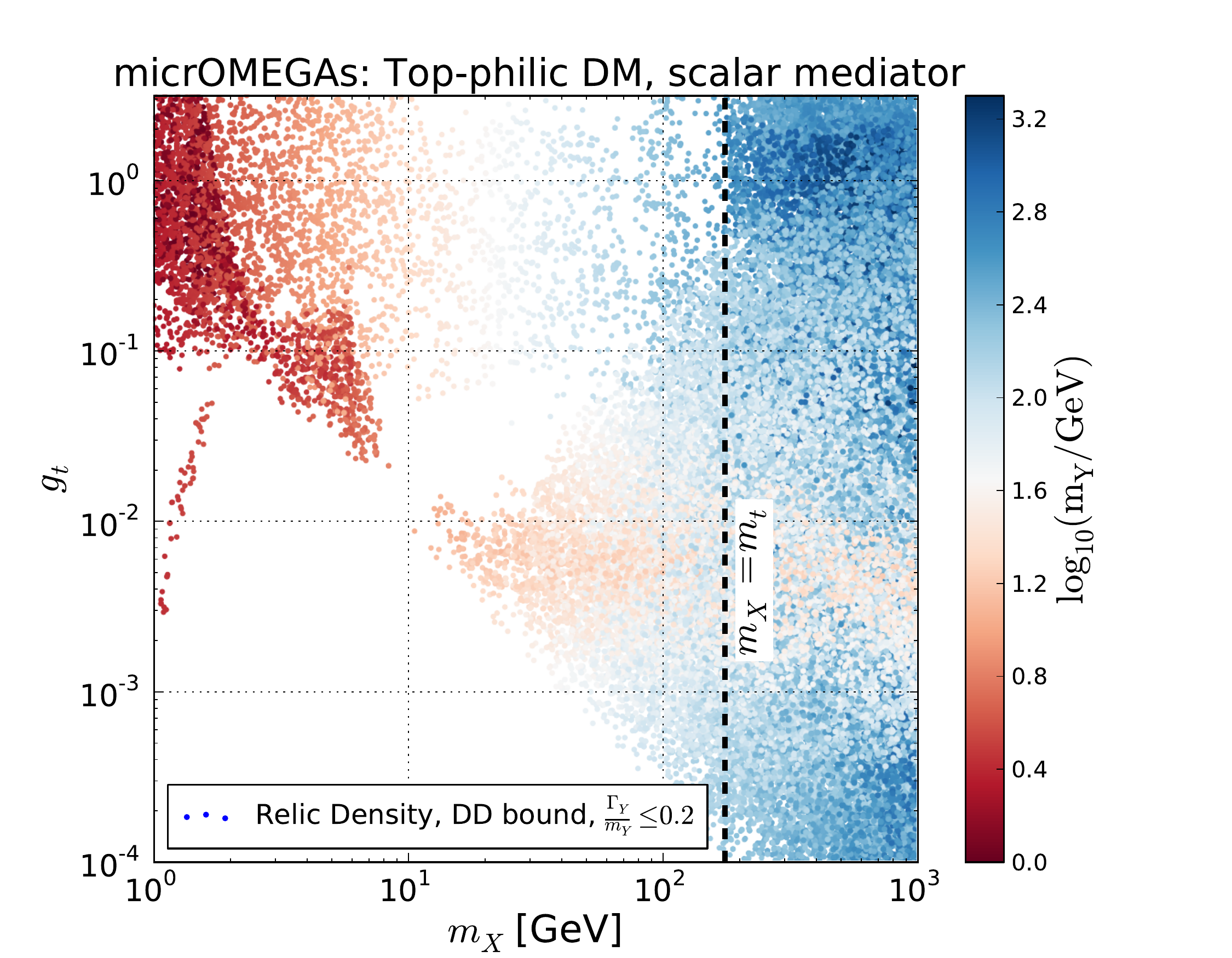}
\caption{Consistency check for the parameter scan. The panels show projections of the four dimensional parameter scan using {\sc micrOMEGAs} with the same assumptions as in fig.~\ref{fig:scan3}.\label{fig:consistency2}}
\end{center}
\end{figure}

\subsection{Details on the dark matter annihilation cross sections}\label{sec:dmann}

In this Appendix we give the detailed analytic expression of the three annihilation processes described in Sec.~\ref{sec:oh2}. 

The $s$-channel annihilation cross section $X \bar{X} \to t \bar{t}$ (process (I)) is given by:
\begin{equation}\label{eq:dmtt}
\sigma(X \bar{X} \to t \bar{t}) = \frac{3 \gx^2 \gsm^2 y_t^2}{32 \pi s} \frac{(s - 4 m_t^2)^{3/2} \sqrt{s-4 \mdm^2}}{(\my^2-s)^2+\my^2\wy^2} \,.
\end{equation}

Process (II) denotes the annihilation of dark matter into a pair of gluons via $s$-channel and is given by:
\begin{equation}\label{eq:dmgg}
\sigma(X \bar{X} \to  gg) = \frac{\gglu^2 \gx^2}{16 \pi  v^2}\frac{s^{3/2} \sqrt{s-4 \mdm^2}}{ (\my^2-s)^2+\my^2\wy^2}
\end{equation}

Finally the process (III), namely $X \bar{X} \to \y \y$ via $t$-channel is given by:
\begin{equation}\label{eq:dmyy}
\sigma(X \bar{X} \to \y \y) = \frac{\gx^4}{64 \pi} \frac{h(t_0) - h(t_1)}{s (s - 4 \mdm^2)}\,,
\end{equation}
where $t_{0,1}$ are the integration extrema:
\begin{equation}
t_{0,1} = - \frac{1}{4} \Big(\sqrt{s - 4 \mdm^2} \mp \sqrt{ s- \my^2} \Big)^2\,,
\end{equation}
and the undefined integral $h(t)$ has the form:
\begin{eqnarray}
h(t) & \equiv &  \frac{(\my^2 - 4 \mdm^2)^2}{\mdm^2 -u}  -  \frac{(\my^2 - 4 \mdm^2)^2}{\mdm^2 -t} - 4 t \nonumber\\
& + & \frac{\Big(6 \my^2 - 4\my^2 (4 \mdm^2 + s) -32 \mdm^4 + 16 \mdm^2 s + s^2 \Big)}{2 \my^2 -s} \log\left(\frac{t-\mdm^2}{\mdm^2 -u}\right)\,,
\label{eq:hoft}
\end{eqnarray}
with $t$ and $u$ Mandelstam variables such that $u= 2 \mdm^2 + 2 \my^2 -s -t$.

In general the thermally averaged cross section can be approximated in the non relativistic regime by expanding  the
cross section in powers of the dark matter relative velocity $v_\text{rel}$, with $s \simeq \mdm^2 (4 -v_\text{rel}^2)$, weighting with the appropriate Maxwell-Boltzmann distribution:
\begin{equation}\label{eq:sigmav}
\langle \sigma v_\text{rel} \rangle (x)_i = \mathcal{A}_i + \frac{3}{2} \frac{\mathcal{B}_i}{x} + \mathcal{O}(x^{-2})
\end{equation}
where the index $i$ indicated the annihilation process, $x\equiv \mdm/T$ and $T$ is the temperature of the dark matter gas. In case of $s$-channel annihilation, along the resonance the thermal average is much more complex and requires the full computation of the integral $\int {\rm d}x\,  \langle \sigma v_\text{rel} \rangle$(x). The approximation given in eq.~\eqref{eq:sigmav} holds in all regions far away from the resonance and is useful to show the dependence on $v_\text{rel}$  of $\langle \sigma v_\text{rel} \rangle (x)$ for each specific process.

For all processes (I), (II) and (III) the first coefficient is always null, $\mathcal{A}_i =0$. The first non negligible term in the expansion eq.~\eqref{eq:sigmav} is then $\mathcal{B}$:
\begin{eqnarray}
\mathcal{B}_{t\bar{t}} & = & \frac{3 \gx^2 \gsm^2 y_t^2}{16 \pi} \frac{\mdm^2 (1- m_t^2 / \mdm^2)^{3/2}}{(\my^2 - 4 \mdm^2)^2}\,,\label{eq:btt}\\
\mathcal{B}_{gg} & = &\frac{2 \gglu^2 \gx^2 \mdm^4}{2 \pi v^2 \left(\my^2-4 \mdm^2\right)^2}
 \,,\label{eq:bgg}\\
\mathcal{B}_{\y\y} & = & \frac{\gx^4}{24 \pi}\frac{\mdm^2 (9\mdm^4 -8\mdm^2\my^2+2\my^4)}{(2 \mdm^2 - \my^2)^4}\sqrt{1-\frac{\my^2}{\mdm^2}}\,.\label{eq:byy}
\end{eqnarray}
This is equivalent to say that all three process are $p$-wave suppressed for dark matter annihilation at present epoch.

The case of Dirac dark matter particles communicating with the SM via a pseudoscalar mediator has been described in~\cite{Arina:2014yna}, where analytic expressions for $\langle \sigma v_\text{rel} \rangle$ can be found. Similarly to scalar mediator $\y$ the $t$-channel process is again $p$-wave suppressed, while the $s$-channel annihilation is dominated by $s$-wave.

\section{Recasting of LHC searches within the \MA~framework}\label{sec:recast}
In this appendix, we detail the implementation, within the \MA~framework~%
\cite{Conte:2012fm,Conte:2014zja,Dumont:2014tja}, of the two dark matter
searches that we have investigated in this work. More precisely, this consists
of the CMS-B2G-14-004
analysis~\cite{Khachatryan:2015nua} that probes final states comprised of a
top-antitop system produced in association with a pair of invisible dark matter
particles (see Section~\ref{sec:ttmet}) and
the CMS-EXO-12-048 analysis~\cite{Khachatryan:2014rra} related to the
production of a pair of dark matter particles together with a hard jet (see
Section~\ref{sec:monoj}). Both recasting codes have been
validated within the version 1.3 of \MA, although the monojet search
reimplementation is also compatible with the version 1.2 of the program. The
simulation of the detector response is performed with the standard
\DEL~package that we have run from the \MA~platform.
In the monojet case, we have used the standard CMS detector parameterisation
that is the shipped with \MA, while in the top-antitop plus missing energy case,
we have designed a dedicated detector card. For both setups, jets are
reconstructed on the basis of the anti-$k_T$ algorithm~\cite{Cacciari:2008gp}
with a radius parameter set to 0.5, as implemented in
\FJ~\cite{Cacciari:2011ma}.

The validation of both our reimplementations is based on material provided by
CMS. Two UFO models~\cite{Degrande:2011ua}, one for each of the recast
analyses, have been shared so that we have been allowed to generate specific
dark matter signals for which CMS has released public cutflow charts and
differential distributions. Using \MG~\cite{Alwall:2014hca} (with the leading
order set of CTEQ6 parton densities~\cite{Pumplin:2002vw}) and
\PY~\cite{Sjostrand:2006za} (with the $Z_2^*$ tune~\cite{Field:2011iq} for the
description of the underlying events) for the simulation of the hard scattering
process and of the parton showering and hadronisation,
respectively, we have generated signal events that have been analyzed with \MA.
Our results have been confronted to the CMS official numbers, which has
allowed us to assess the validity of our recasting codes. Our simulation
procedure moreover includes the generation of matrix elements containing up
to two extra jets that we have merged according to the MLM
prescription~\cite{Mangano:2006rw,Alwall:2008qv}, the merging scale being set to
40~GeV.

All \PY, \DEL~ and \MG~configuration cards can be downloaded from
the public analysis database webpage of \MA,\\
\hspace*{.5cm}%
\url{http://madanalysis.irmp.ucl.ac.be/wiki/PublicAnalysisDatabase},\\
while the recasting {\sc C++} codes associated with the CMS-EXO-12-048
and CMS-B2G-14-004 analyses can be found on {\sc InSpire}~\cite{monojet:recast,%
ttmet:recast}.
\subsection{The CMS top-antitop plus missing energy CMS-B2G-14-004 search}
\label{sec:ttmet}

In order to validate our reimplementation of the CMS-B2G-14-004 search in \MA,
we focus on a new physics model that features the production a pair of dark
matter particle $\dm$ of mass \mbox{$m_\dm = 1$~GeV} in association with a
top-antitop pair via a four-fermion interaction. The CMS event selection
strategy requires a large amount of missing transverse energy,
a single isolated lepton and multiple jets, and uses $19.7 \fb^{-1}$ of
proton-proton collision data recorded at a center-of-mass energy of
\mbox{$\sqrt{s}=8$~TeV}.

The CMS-B2G-14-004 analysis relies on single electron and muon triggers, with
lower $p_T$ thresholds of 27~GeV and 24~GeV respectively, and the
reconstructed electron (muon) candidate is imposed to be isolated in such a way
that the sum of the transverse momenta of all objects lying in a cone of radius
\mbox{$R=0.3$} centered on the lepton has to be smaller than 10\% (12\%) of the
lepton $p_T$. Event preselection finally requires that the lepton $p_T$ is
larger than 30~GeV and pseudorapidity $|\eta|$ is smaller than 2.5 (2.1
for muons). It additionally demands the presence of at
least three jets of $p_T>30$~GeV and $|\eta|<2.4$ with one of them being
$b$-tagged, as well as missing energy $\MET > 160$~GeV. The signal region is
defined by selecting events with a large amount of missing transverse
energy \mbox{$\MET>320$} for which the transverse mass $M_T$ that is constructed
from the lepton and the missing energy is larger than 160~GeV. Moreover, the
missing transverse momentum and the two leading jets are asked to be well
separated in azimuth, $\Delta\Phi\left(j_{1,2},\MET \right)>1.2$, and the
$M_{T2}^W$ variable~\cite{Bai:2012gs} is enforced to be greater than
200~GeV.

\begin{table}
 \begin{center}
  \begin{tabular}{c|c||c|c||c|c||c}\hline
  & Selection step & CMS & $\epsilon_i^{\rm CMS}$ & MA5&  $\epsilon_i^{\rm MA5}$ & $\delta_i^{\rm rel}$ \\ \hline\hline
   0&Nominal                        &224510 &      &224510   &      &  \\
   1&Preselection                   &       &      &15468.5  & 0.069&  \\
   2&$\MET>320$~GeV                 &4220.8 &      & 4579.8  & 0.296&  \\
   3&$M_T>160$~GeV                  & 3390.1& 0.803& 3648.2 & 0.797&  0.75\%\\
   4&$\Delta\Phi(j_{1,2},\MET)>1.2$ & 2963.5& 0.874& 3124.3 & 0.856& 2.06\%\\
   5&$M_{T2}^{W}> 200$ GeV          & 2267.6& 0.765& 2403  & 0.769& -0.52\%\\
   \hline
  \end{tabular}
  \caption{\label{tab:ttbar_valid}
    Comparison of results obtained with our \MA~reimplementation (MA5) and those
    provided by the CMS collaboration (CMS). The efficiencies are defined in
    Eq.~\eqref{eq:effic} and the relative difference between the
    CMS and the \MA~results $\delta_i^{\rm rel}$ in Eq.~\eqref{eq:delma5cms}.}
 \end{center}
\end{table}

\begin{figure}
  \centering
 \includegraphics[width=0.42\textwidth]{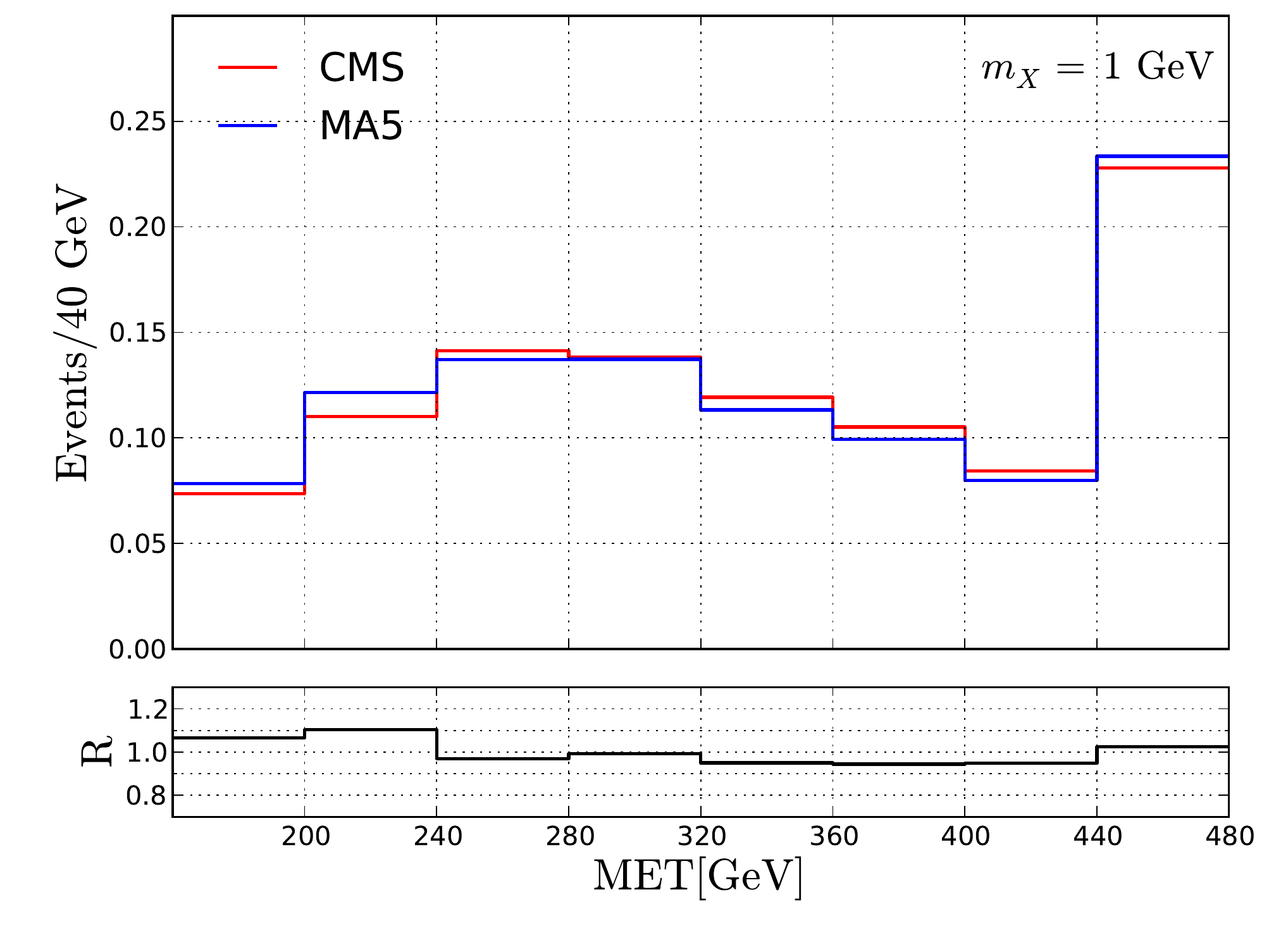}
 \includegraphics[width=0.42\textwidth]{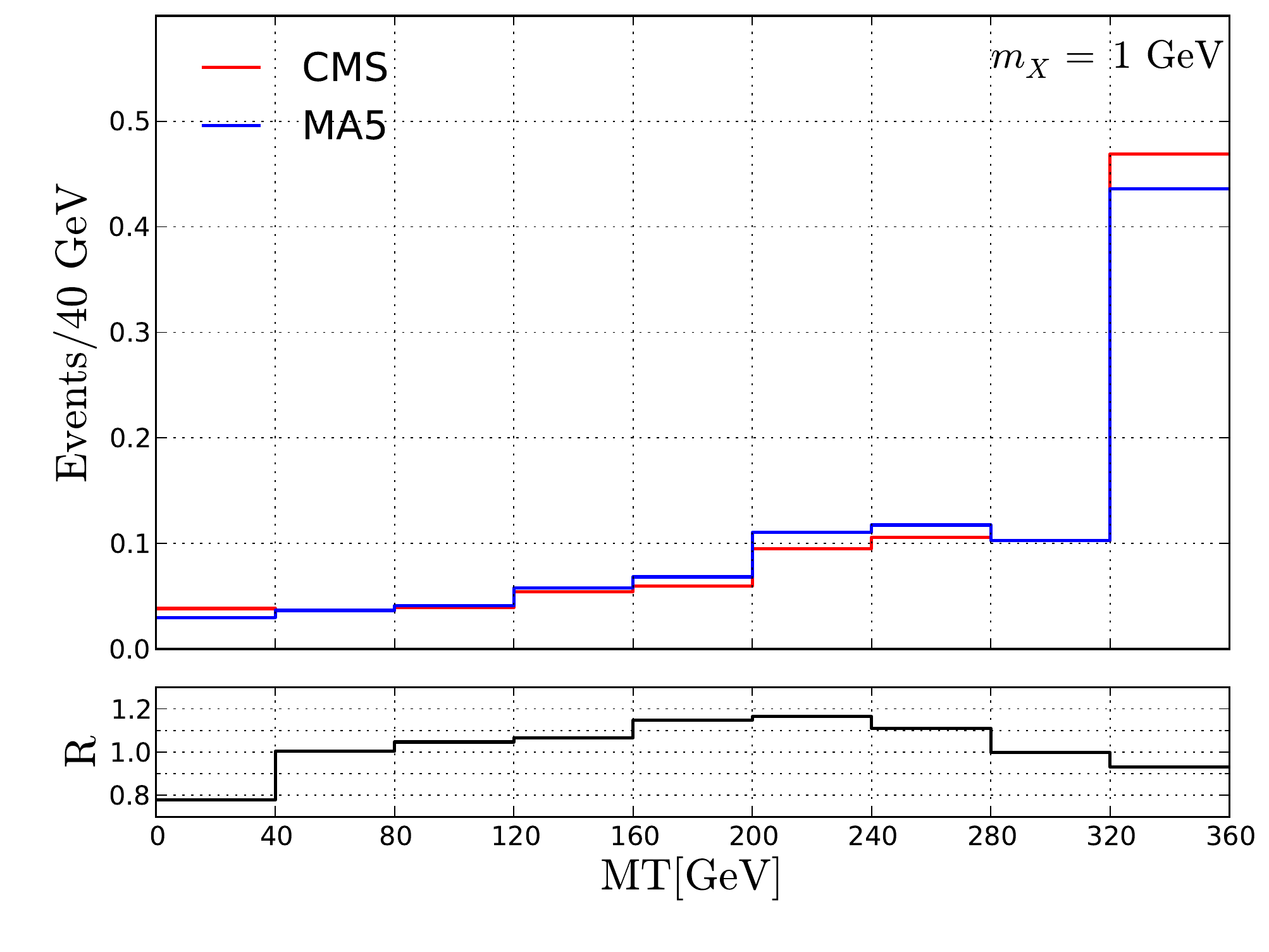}\\
 \includegraphics[width=0.42\textwidth]{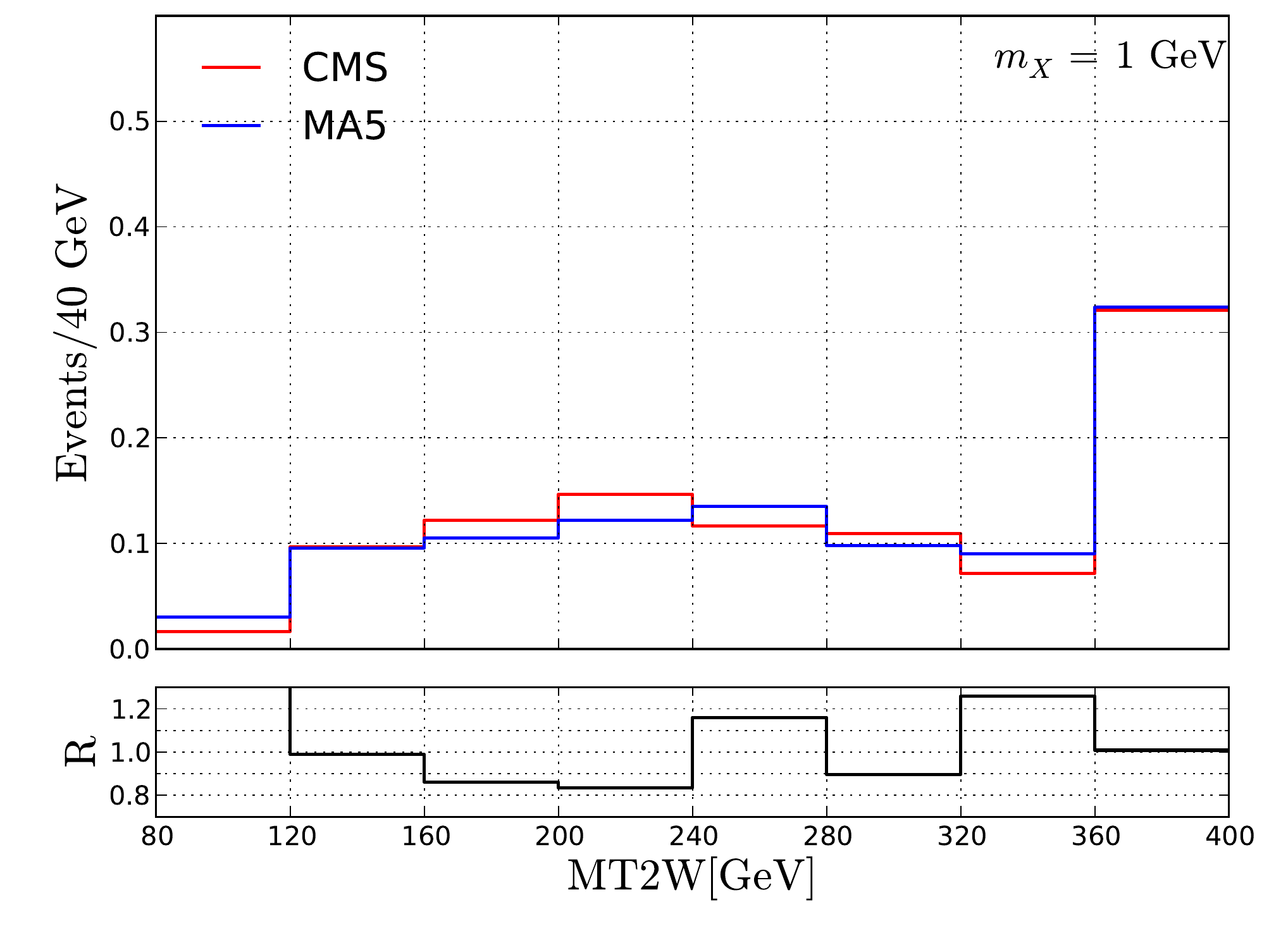}
 \includegraphics[width=0.42\textwidth]{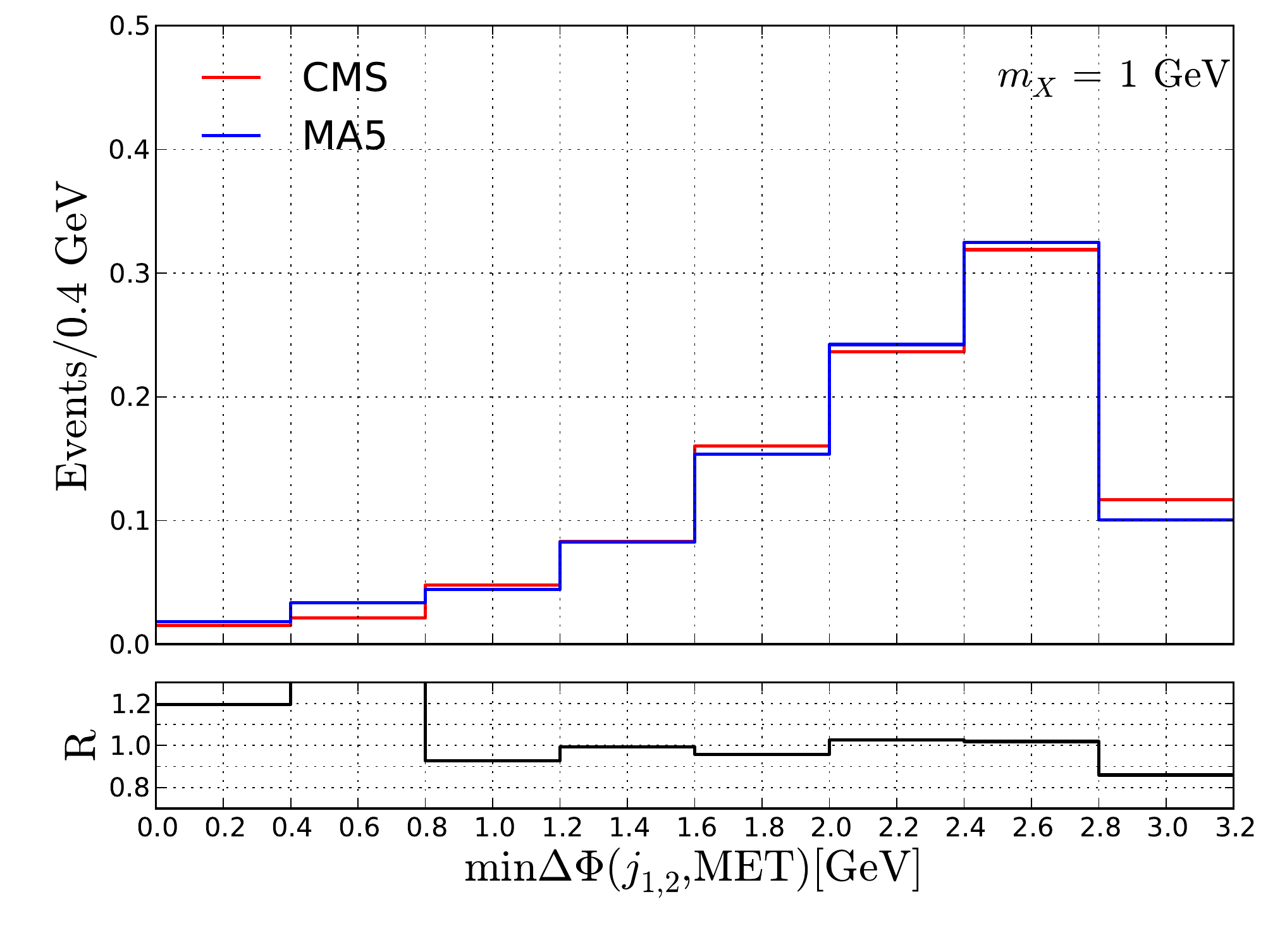}
 \caption{Missig transverse energy, $M_T$, $M_{T2}^W$ and
   $\Delta\Phi\left(j_{1,2},\MET\right)$ spectrum as obtained with \MA\ (blue)
   once all selection steps but the one related to the represented variable are
   applied, compard to the CMS official results (red).}
 \label{Fig_valid_ttbar}
\end{figure}

In Table~\ref{tab:ttbar_valid}, we confront the cutflow chart that has been
obtained with \MA\ to the official results of CMS for the benchmark scenario
under consideration.
For each step of the selection, we have calculated the related efficiency
defined as
\be
  \epsilon_i = \frac{n_i}{n_{i-1}} \ ,
\label{eq:effic}\ee
where $n_i$ and $n_{i-1}$ mean the event number after and before the considered
cut, respectively. The \textit{relative difference} information given in the
table corresponds to the difference between the \MA~and the CMS
efficiencies, normalized to the CMS result,
\be
   \delta_i^{\rm rel} = 1 - \frac{\epsilon_i^{\rm MA5}}{\epsilon_i^{\rm CMS}}
      \ .
\label{eq:delma5cms}\ee
An agreement at the percent level has been found all over the selection
procedure. Moreover, we compare several (normalized) differential
distributions as calculated with \MA\ when all selection steps but the one
related to the represented kinematic variable are included with the public CMS
results in figure~\ref{Fig_valid_ttbar}. A very good agreement can again be
observed.

\subsection{The CMS monojet CMS-EXO-12-048 search}\label{sec:monoj}

The validation of our implementation of the CMS-EXO-12-048 search in \MA~has
been achieved on the basis of a benchmark scenario that is inspired by
Refs.~\cite{Beltran:2010ww,Goodman:2010ku,Goodman:2010yf,Bai:2010hh}. In this
context, monojet events arise from the associated production of a pair of
invisible Dirac fermions of mass of 1~GeV with at least one hard jet. The
interactions of the dark particle with the Standard Model are mediated by a new
gauge boson $Z'$ of mass and width of 40~TeV and 10~GeV
respectively, and all new physics interactions have been assumed to have a
vector coupling structure and a strength set equal to 1. Concerning our signal
simulation setup, we have imposed that all parton-level jets have a
transverse momentum $p_T$ larger than 20~GeV and that the leading jet
has a $p_T > 80$~GeV. 

The CMS monojet search relies on an integrated luminosity of 19.7~fb$^{-1}$ of
proton-proton collisions at a center-of-mass energy of $\sqrt{s} = 8$ TeV. It
focuses on a signal containing a very hard jet with a transverse momentum
satisfying $p_T>110$~GeV and a pseudorapidity smaller than 4.5 in absolute
value. A second jet is moreover allowed, provided that its transverse momentum
is larger than 30~GeV, its pseudorapidity satisfies $|\eta|<4.5$ and if it is
well separated from the first jet by 2.5 radians in azimuth. Events featuring
more than two jets (with $p_T>30$~GeV and $|\eta|<4.5$), isolated electrons or
muons with a transverse momentum $p_T > 10$~GeV or hadronically decaying tau
leptons with a transverse momentum $p_T > 20$~GeV and a pseudorapidity
satisfying $|\eta|<2.3$ are discarded. The analysis then contains seven
inclusive signal regions in which the missing energy $\MET$ is required
to be above specific thresholds of 250, 300, 350, 400, 450, 500 and 550~GeV
respectively.

\begin{table}
 \begin{center}
  \begin{tabular}{c|c||c|c||c|c||c}\hline
  & Selection step & CMS & $\epsilon_i^{\rm CMS}$ & MA5&  $\epsilon_i^{\rm MA5}$ & $\delta_i^{\rm rel}$ \\ \hline\hline
   0&Nominal  &84653.7 & &84653.7& &\\
   1&One hard jet& 50817.2 & 0.6 &53431.28 & 0.631 & 5.2\% \\
   2&At most two jets&36061 & 0.7096& 38547.75 & 0.721 & 1.61\%\\
   3&Requirements if two jets& 31878.1 & 0.884& 34436.35 & 0.893 & 1.02\%\\
   4&Muon veto & 31878.1 & 1 & 34436.35 & 1.000 & 0   \\
   5&Electron veto& 31865.1 & 1 & 34436.35 & 1.000 & 0   \\
   6&Tau veto& 31695.1 & 0.995 & 34397.54 & 0.998 & 0.3\%  \\\hline
   &$\MET > 250$ GeV & 8687.22 & 0.274& 7563.04 & 0.219 & 20.00\%\\
   &$\MET > 300$ GeV & 5400.51 & 0.621& 4477.67 & 0.592 & 4.66\% \\
   &$\MET > 350$ GeV & 3394.09 & 0.628& 2813.70 & 0.628 & 0.00\%\\
   &$\MET > 400$ GeV & 2224.15 &0.6553& 1753.71 & 0.623 &4.93\%\\
   &$\MET > 450$ GeV & 1456.02 &0.654& 1110.92 & 0.633&3.21\% \\
   &$\MET > 500$ GeV & 989.806 &0.679& 722.83 & 0.650 &  4.27\%\\
   &$\MET > 550$ GeV & 671.442 &0.678& 487.54 & 0.674 &0.59\% \\
   \hline
  \end{tabular}
  \caption{\label{tab:monojet_valid}
    Comparison of results obtained with our \MA~reimplementation (MA5) and those
    provided by the CMS collaboration (CMS). The efficiencies are defined in
    Eq.~\eqref{eq:effic} and the relative difference between the
    CMS and the \MA~results $\delta_i^{\rm rel}$ in Eq.~\eqref{eq:delma5cms}.}
 \end{center}
\end{table}

The selection strategy of the CMS monojet analysis thus consists of six
preselection cuts followed by one region-dependent cut, when we ignore the first
two requirements of the analysis related to the cleaning of the events from the
detector noise that cannot be handled with \DEL. For the benchmark
scenario under consideration, we compare the results that have been derived with
our \MA~reimplementation with those provided by the CMS
collaboration in Table~\ref{tab:monojet_valid}.

\begin{figure}
 \begin{center}
 \includegraphics[trim={0 14.5cm 0 0},clip,width=.75\textwidth]{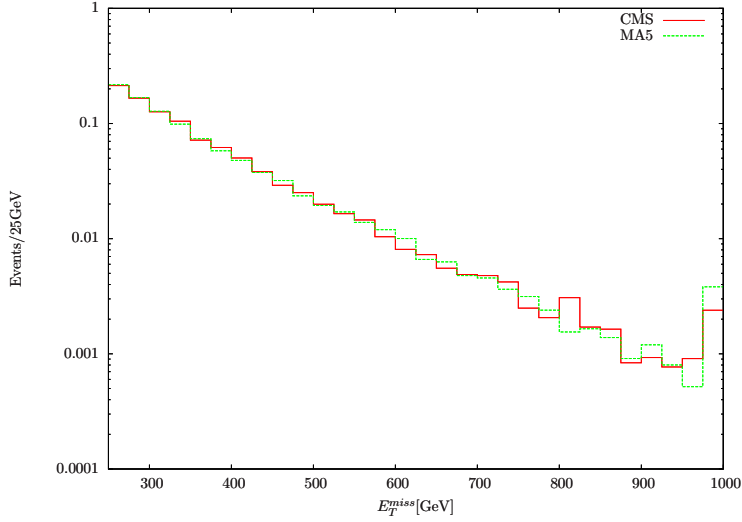}
 \end{center}\caption{Missing energy spectrum as obtained with \MA~(green dashed
  line) after the CMS-EXO-12-048 monojet preselection, compared to the CMS
  official results (red solid line). The last bin is the overflow bin.
\label{fig:histo_monojet}}
\end{figure}
We have found that all selection steps are properly described by our
implementation, with the exception the missing energy selection
$\MET > 250$~GeV for which a disagreement of about 20\% has been
observed. It is however not uncommon that low missing energy is difficult to
simulate with a fast-simulation of the detector based on \DEL. We have
verified that for missing energy values of interest, the description of the
missing energy agree relatively well with CMS,
as illustrated in figure~\ref{fig:histo_monojet} where we compare, for
a benchmark scenario where the $Z'$ mass has been set to 900 GeV, the missing
energy distribution as obtained by CMS to the one derived with
\MA.

\bibliography{Main_draft}

\providecommand{\href}[2]{#2}\begingroup\raggedright\begin{thebibliography}{10}

\bibitem{Bertone:2004pz}
G.~Bertone, D.~Hooper and J.~Silk, \emph{{Particle dark matter: Evidence,
  candidates and constraints}},
  \href{http://dx.doi.org/10.1016/j.physrep.2004.08.031}{\emph{Phys. Rept.}
  {\bf 405} (2005) 279--390}, [\href{http://arxiv.org/abs/hep-ph/0404175}{{\tt
  hep-ph/0404175}}].

\bibitem{Bertone:1235368}
G.~Bertone, \emph{{Particle dark matter: observations, models and searches}}.
\newblock Cambridge Univ. Press, Cambridge, 2010.

\bibitem{Drees:2012ji}
M.~Drees and G.~Gerbier, \emph{{Mini-Review of Dark Matter: 2012}},
  \href{http://arxiv.org/abs/1204.2373}{{\tt 1204.2373}}.

\bibitem{Clowe:2006eq}
D.~Clowe, M.~Bradac, A.~H. Gonzalez, M.~Markevitch, S.~W. Randall, C.~Jones
  et~al., \emph{{A direct empirical proof of the existence of dark matter}},
  \href{http://dx.doi.org/10.1086/508162}{\emph{Astrophys. J.} {\bf 648} (2006)
  L109--L113}, [\href{http://arxiv.org/abs/astro-ph/0608407}{{\tt
  astro-ph/0608407}}].

\bibitem{Abercrombie:2015wmb}
D.~Abercrombie et~al., \emph{{Dark Matter Benchmark Models for Early LHC Run-2
  Searches: Report of the ATLAS/CMS Dark Matter Forum}},
  \href{http://arxiv.org/abs/1507.00966}{{\tt 1507.00966}}.

\bibitem{Buckley:2014fba}
M.~R. Buckley, D.~Feld and D.~Goncalves, \emph{{Scalar Simplified Models for
  Dark Matter}},
  \href{http://dx.doi.org/10.1103/PhysRevD.91.015017}{\emph{Phys. Rev.} {\bf
  D91} (2015) 015017}, [\href{http://arxiv.org/abs/1410.6497}{{\tt
  1410.6497}}].

\bibitem{Haisch:2015ioa}
U.~Haisch and E.~Re, \emph{{Simplified dark matter top-quark interactions at
  the LHC}}, \href{http://dx.doi.org/10.1007/JHEP06(2015)078}{\emph{JHEP} {\bf
  06} (2015) 078}, [\href{http://arxiv.org/abs/1503.00691}{{\tt 1503.00691}}].

\bibitem{Heisig:2015ira}
J.~Heisig, M.~Kr{\"a}mer, M.~Pellen and C.~Wiebusch, \emph{{Constraints on
  Majorana Dark Matter from the LHC and IceCube}},
  \href{http://dx.doi.org/10.1103/PhysRevD.93.055029}{\emph{Phys. Rev.} {\bf
  D93} (2016) 055029}, [\href{http://arxiv.org/abs/1509.07867}{{\tt
  1509.07867}}].

\bibitem{Bell:2015rdw}
N.~F. Bell, Y.~Cai and R.~K. Leane, \emph{{Mono-W Dark Matter Signals at the
  LHC: Simplified Model Analysis}},
  \href{http://dx.doi.org/10.1088/1475-7516/2016/01/051}{\emph{JCAP} {\bf 1601}
  (2016) 051}, [\href{http://arxiv.org/abs/1512.00476}{{\tt 1512.00476}}].

\bibitem{Brennan:2016xjh}
A.~J. Brennan, M.~F. McDonald, J.~Gramling and T.~D. Jacques, \emph{{Collide
  and Conquer: Constraints on Simplified Dark Matter Models using Mono-X
  Collider Searches}},  \href{http://arxiv.org/abs/1603.01366}{{\tt
  1603.01366}}.

\bibitem{Pree:2016hwc}
T.~du~Pree, K.~Hahn, P.~Harris and C.~Roskas, \emph{{Cosmological constraints
  on Dark Matter models for collider searches}},
  \href{http://arxiv.org/abs/1603.08525}{{\tt 1603.08525}}.

\bibitem{Jackson:2013pjq}
C.~B. Jackson, G.~Servant, G.~Shaughnessy, T.~M.~P. Tait and M.~Taoso,
  \emph{{Gamma-ray lines and One-Loop Continuum from s-channel Dark Matter
  Annihilations}},
  \href{http://dx.doi.org/10.1088/1475-7516/2013/07/021}{\emph{JCAP} {\bf 1307}
  (2013) 021}, [\href{http://arxiv.org/abs/1302.1802}{{\tt 1302.1802}}].

\bibitem{Abdallah:2015ter}
J.~Abdallah et~al., \emph{{Simplified Models for Dark Matter Searches at the
  LHC}}, \href{http://dx.doi.org/10.1016/j.dark.2015.08.001}{\emph{Phys. Dark
  Univ.} {\bf 9-10} (2015) 8--23}, [\href{http://arxiv.org/abs/1506.03116}{{\tt
  1506.03116}}].

\bibitem{Godbole:2015gma}
R.~M. Godbole, G.~Mendiratta and T.~M.~P. Tait, \emph{{A Simplified Model for
  Dark Matter Interacting Primarily with Gluons}},
  \href{http://dx.doi.org/10.1007/JHEP08(2015)064}{\emph{JHEP} {\bf 08} (2015)
  064}, [\href{http://arxiv.org/abs/1506.01408}{{\tt 1506.01408}}].

\bibitem{Xiang:2015lfa}
Q.-F. Xiang, X.-J. Bi, P.-F. Yin and Z.-H. Yu, \emph{{Searches for dark matter
  signals in simplified models at future hadron colliders}},
  \href{http://dx.doi.org/10.1103/PhysRevD.91.095020}{\emph{Phys. Rev.} {\bf
  D91} (2015) 095020}, [\href{http://arxiv.org/abs/1503.02931}{{\tt
  1503.02931}}].

\bibitem{DiFranzo:2013vra}
A.~DiFranzo, K.~I. Nagao, A.~Rajaraman and T.~M.~P. Tait, \emph{{Simplified
  Models for Dark Matter Interacting with Quarks}},
  \href{http://dx.doi.org/10.1007/JHEP11(2013)014,
  10.1007/JHEP01(2014)162}{\emph{JHEP} {\bf 11} (2013) 014},
  [\href{http://arxiv.org/abs/1308.2679}{{\tt 1308.2679}}].

\bibitem{Alwall:2014hca}
J.~Alwall, R.~Frederix, S.~Frixione, V.~Hirschi, F.~Maltoni, O.~Mattelaer
  et~al., \emph{{The automated computation of tree-level and next-to-leading
  order differential cross sections, and their matching to parton shower
  simulations}}, \href{http://dx.doi.org/10.1007/JHEP07(2014)079}{\emph{JHEP}
  {\bf 07} (2014) 079}, [\href{http://arxiv.org/abs/1405.0301}{{\tt
  1405.0301}}].

\bibitem{Alloul:2013bka}
A.~Alloul, N.~D. Christensen, C.~Degrande, C.~Duhr and B.~Fuks,
  \emph{{FeynRules 2.0 - A complete toolbox for tree-level phenomenology}},
  \href{http://dx.doi.org/10.1016/j.cpc.2014.04.012}{\emph{Comput. Phys.
  Commun.} {\bf 185} (2014) 2250--2300},
  [\href{http://arxiv.org/abs/1310.1921}{{\tt 1310.1921}}].

\bibitem{Degrande:2014vpa}
C.~Degrande, \emph{{Automatic evaluation of UV and R2 terms for beyond the
  Standard Model Lagrangians: a proof-of-principle}},
  \href{http://dx.doi.org/10.1016/j.cpc.2015.08.015}{\emph{Comput. Phys.
  Commun.} {\bf 197} (2015) 239--262},
  [\href{http://arxiv.org/abs/1406.3030}{{\tt 1406.3030}}].

\bibitem{Conte:2012fm}
E.~Conte, B.~Fuks and G.~Serret, \emph{{MadAnalysis 5, A User-Friendly
  Framework for Collider Phenomenology}},
  \href{http://dx.doi.org/10.1016/j.cpc.2012.09.009}{\emph{Comput. Phys.
  Commun.} {\bf 184} (2013) 222--256},
  [\href{http://arxiv.org/abs/1206.1599}{{\tt 1206.1599}}].

\bibitem{Conte:2014zja}
E.~Conte, B.~Dumont, B.~Fuks and C.~Wymant, \emph{{Designing and recasting LHC
  analyses with MadAnalysis 5}},
  \href{http://dx.doi.org/10.1140/epjc/s10052-014-3103-0}{\emph{Eur. Phys. J.}
  {\bf C74} (2014) 3103}, [\href{http://arxiv.org/abs/1405.3982}{{\tt
  1405.3982}}].

\bibitem{Dumont:2014tja}
B.~Dumont, B.~Fuks, S.~Kraml, S.~Bein, G.~Chalons, E.~Conte et~al.,
  \emph{{Toward a public analysis database for LHC new physics searches using
  MADANALYSIS 5}},
  \href{http://dx.doi.org/10.1140/epjc/s10052-014-3242-3}{\emph{Eur. Phys. J.}
  {\bf C75} (2015) 56}, [\href{http://arxiv.org/abs/1407.3278}{{\tt
  1407.3278}}].

\bibitem{deFavereau:2013fsa}
{\scshape DELPHES 3} collaboration, J.~de~Favereau, C.~Delaere, P.~Demin,
  A.~Giammanco, V.~Lema\^itre, A.~Mertens et~al., \emph{{DELPHES 3, A modular
  framework for fast simulation of a generic collider experiment}},
  \href{http://dx.doi.org/10.1007/JHEP02(2014)057}{\emph{JHEP} {\bf 02} (2014)
  057}, [\href{http://arxiv.org/abs/1307.6346}{{\tt 1307.6346}}].

\bibitem{Backovic:2013dpa}
M.~Backovic, K.~Kong and M.~McCaskey, \emph{{MadDM v.1.0: Computation of Dark
  Matter Relic Abundance Using MadGraph5}},
  \href{http://dx.doi.org/10.1016/j.dark.2014.04.001}{\emph{Physics of the Dark
  Universe} {\bf 5-6} (2014) 18--28},
  [\href{http://arxiv.org/abs/1308.4955}{{\tt 1308.4955}}].

\bibitem{Backovic:2015cra}
M.~Backovic, A.~Martini, O.~Mattelaer, K.~Kong and G.~Mohlabeng, \emph{{Direct
  Detection of Dark Matter with MadDM v.2.0}},
  \href{http://dx.doi.org/10.1016/j.dark.2015.09.001}{\emph{Phys. Dark Univ.}
  {\bf 9-10} (2015) 37--50}, [\href{http://arxiv.org/abs/1505.04190}{{\tt
  1505.04190}}].

\bibitem{Feroz:2007kg}
F.~Feroz and M.~Hobson, \emph{{Multimodal nested sampling: an efficient and
  robust alternative to MCMC methods for astronomical data analysis}},
  \href{http://dx.doi.org/10.1111/j.1365-2966.2007.12353.x}{\emph{Mon.Not.Roy.Astron.Soc.}
  {\bf 384} (2008) 449}, [\href{http://arxiv.org/abs/0704.3704}{{\tt
  0704.3704}}].

\bibitem{Feroz:2008xx}
F.~Feroz, M.~Hobson and M.~Bridges, \emph{{MultiNest: an efficient and robust
  Bayesian inference tool for cosmology and particle physics}},
  \href{http://dx.doi.org/DOI:
  10.1111/j.1365-2966.2009.14548.x}{\emph{Mon.Not.Roy.Astron.Soc.} {\bf 398}
  (2009) 1601--1614}, [\href{http://arxiv.org/abs/0809.3437}{{\tt 0809.3437}}].

\bibitem{atlasDigamma}
{\scshape ATLAS} collaboration, \emph{{Search for resonances decaying to photon
  pairs in 3.2 fb$^{-1}$ of $pp$ collisions at $\sqrt{s}$ = 13 TeV with the
  ATLAS detector}},  ATLAS-CONF-2015-081.

\bibitem{CMS:2015dxe}
{\scshape CMS} collaboration, \emph{{Search for new physics in high mass
  diphoton events in proton-proton collisions at 13TeV}},  CMS-PAS-EXO-15-004.

\bibitem{Planck:2015xua}
{\scshape Planck} collaboration, P.~Ade et~al., \emph{{Planck 2015 results.
  XIII. Cosmological parameters}},  \href{http://arxiv.org/abs/1502.01589}{{\tt
  1502.01589}}.

\bibitem{Akerib:2013tjd}
{\scshape LUX} collaboration, D.~Akerib et~al., \emph{{First results from the
  LUX dark matter experiment at the Sanford Underground Research Facility}},
  \href{http://dx.doi.org/10.1103/PhysRevLett.112.091303}{\emph{Phys.Rev.Lett.}
  {\bf 112} (2014) 091303}, [\href{http://arxiv.org/abs/1310.8214}{{\tt
  1310.8214}}].

\bibitem{Agnese:2015nto}
{\scshape SuperCDMS} collaboration, R.~Agnese et~al., \emph{{WIMP-Search
  Results from the Second CDMSlite Run}}, {\emph{Submitted to: Phys. Rev.
  Lett.} (2015) }, [\href{http://arxiv.org/abs/1509.02448}{{\tt 1509.02448}}].

\bibitem{Ackermann:2015zua}
{\scshape Fermi-LAT} collaboration, M.~Ackermann et~al., \emph{{Searching for
  Dark Matter Annihilation from Milky Way Dwarf Spheroidal Galaxies with Six
  Years of Fermi Large Area Telescope Data}},
  \href{http://dx.doi.org/10.1103/PhysRevLett.115.231301}{\emph{Phys. Rev.
  Lett.} {\bf 115} (2015) 231301}, [\href{http://arxiv.org/abs/1503.02641}{{\tt
  1503.02641}}].

\bibitem{Ackermann:2015lka}
{\scshape Fermi-LAT} collaboration, M.~Ackermann et~al., \emph{{Updated search
  for spectral lines from Galactic dark matter interactions with pass 8 data
  from the Fermi Large Area Telescope}},
  \href{http://dx.doi.org/10.1103/PhysRevD.91.122002}{\emph{Phys. Rev.} {\bf
  D91} (2015) 122002}, [\href{http://arxiv.org/abs/1506.00013}{{\tt
  1506.00013}}].

\bibitem{Belanger:2014vza}
G.~B\'elanger, F.~Boudjema, A.~Pukhov and A.~Semenov, \emph{{micrOMEGAs4.1: two
  dark matter candidates}},
  \href{http://dx.doi.org/10.1016/j.cpc.2015.03.003}{\emph{Comput.Phys.Commun.}
  {\bf 192} (2015) 322--329}, [\href{http://arxiv.org/abs/1407.6129}{{\tt
  1407.6129}}].

\bibitem{Backovic:2015soa}
M.~Backovic, M.~Kramer, F.~Maltoni, A.~Martini, K.~Mawatari and M.~Pellen,
  \emph{{Higher-order QCD predictions for dark matter production at the LHC in
  simplified models with s-channel mediators}},
  \href{http://dx.doi.org/10.1140/epjc/s10052-015-3700-6}{\emph{Eur. Phys. J.}
  {\bf C75} (2015) 482}, [\href{http://arxiv.org/abs/1508.05327}{{\tt
  1508.05327}}].

\bibitem{Mattelaer:2015haa}
O.~Mattelaer and E.~Vryonidou, \emph{{Dark matter production through
  loop-induced processes at the LHC: the s-channel mediator case}},
  \href{http://dx.doi.org/10.1140/epjc/s10052-015-3665-5}{\emph{Eur. Phys. J.}
  {\bf C75} (2015) 436}, [\href{http://arxiv.org/abs/1508.00564}{{\tt
  1508.00564}}].

\bibitem{Neubert:2015fka}
M.~Neubert, J.~Wang and C.~Zhang, \emph{{Higher-Order QCD Predictions for Dark
  Matter Production in Mono-$Z$ Searches at the LHC}},
  \href{http://dx.doi.org/10.1007/JHEP02(2016)082}{\emph{JHEP} {\bf 02} (2016)
  082}, [\href{http://arxiv.org/abs/1509.05785}{{\tt 1509.05785}}].

\bibitem{FR-DMsimp:Online}
\url{http://feynrules.irmp.ucl.ac.be/wiki/DMsimp}.

\bibitem{Beringer:1900zz}
{\scshape Particle Data Group} collaboration, J.~Beringer et~al., \emph{{Review
  of Particle Physics (RPP)}},
  \href{http://dx.doi.org/10.1103/PhysRevD.86.010001}{\emph{Phys.Rev.} {\bf
  D86} (2012) 010001}.

\bibitem{D'Eramo:2014aba}
F.~D'Eramo and M.~Procura, \emph{{Connecting Dark Matter UV Complete Models to
  Direct Detection Rates via Effective Field Theory}},
  \href{http://dx.doi.org/10.1007/JHEP04(2015)054}{\emph{JHEP} {\bf 04} (2015)
  054}, [\href{http://arxiv.org/abs/1411.3342}{{\tt 1411.3342}}].

\bibitem{Vecchi:2013iza}
L.~Vecchi, \emph{{WIMPs and Un-Naturalness}},
  \href{http://arxiv.org/abs/1312.5695}{{\tt 1312.5695}}.

\bibitem{Mao:2012hf}
Y.-Y. Mao, L.~E. Strigari, R.~H. Wechsler, H.-Y. Wu and O.~Hahn,
  \emph{{Halo-to-Halo Similarity and Scatter in the Velocity Distribution of
  Dark Matter}},
  \href{http://dx.doi.org/10.1088/0004-637X/764/1/35}{\emph{Astrophys. J.} {\bf
  764} (2013) 35}, [\href{http://arxiv.org/abs/1210.2721}{{\tt 1210.2721}}].

\bibitem{Lisanti:2010qx}
M.~Lisanti, L.~E. Strigari, J.~G. Wacker and R.~H. Wechsler, \emph{{The Dark
  Matter at the End of the Galaxy}},
  \href{http://dx.doi.org/10.1103/PhysRevD.83.023519}{\emph{Phys. Rev.} {\bf
  D83} (2011) 023519}, [\href{http://arxiv.org/abs/1010.4300}{{\tt
  1010.4300}}].

\bibitem{Walker:2009zp}
M.~G. Walker, M.~Mateo, E.~W. Olszewski, J.~Penarrubia, N.~W. Evans and
  G.~Gilmore, \emph{{A Universal Mass Profile for Dwarf Spheroidal Galaxies}},
  \href{http://dx.doi.org/10.1088/0004-637X/704/2/1274,
  10.1088/0004-637X/710/1/886}{\emph{Astrophys. J.} {\bf 704} (2009)
  1274--1287}, [\href{http://arxiv.org/abs/0906.0341}{{\tt 0906.0341}}].

\bibitem{Heinemeyer:2013tqa}
{\scshape LHC Higgs Cross Section Working Group} collaboration, J.~R. Andersen
  et~al., \emph{{Handbook of LHC Higgs Cross Sections: 3. Higgs Properties}},
  \href{http://arxiv.org/abs/1307.1347}{{\tt 1307.1347}}.

\bibitem{Ball:2012cx}
R.~D. Ball et~al., \emph{{Parton distributions with LHC data}},
  \href{http://dx.doi.org/10.1016/j.nuclphysb.2012.10.003}{\emph{Nucl. Phys.}
  {\bf B867} (2013) 244--289}, [\href{http://arxiv.org/abs/1207.1303}{{\tt
  1207.1303}}].

\bibitem{Whalley:2005nh}
M.~R. Whalley, D.~Bourilkov and R.~C. Group, \emph{{The Les Houches accord PDFs
  (LHAPDF) and LHAGLUE}},  in \emph{{HERA and the LHC: A Workshop on the
  implications of HERA for LHC physics. Proceedings, Part B}}, 2005.
\newblock \href{http://arxiv.org/abs/hep-ph/0508110}{{\tt hep-ph/0508110}}.

\bibitem{Buckley:2014ana}
A.~Buckley, J.~Ferrando, S.~Lloyd, K.~Nordström, B.~Page, M.~Rüfenacht
  et~al., \emph{{LHAPDF6: parton density access in the LHC precision era}},
  \href{http://dx.doi.org/10.1140/epjc/s10052-015-3318-8}{\emph{Eur. Phys. J.}
  {\bf C75} (2015) 132}, [\href{http://arxiv.org/abs/1412.7420}{{\tt
  1412.7420}}].

\bibitem{CMS:2014pvf}
{\scshape CMS} collaboration, C.~Collaboration, \emph{{Search for the
  Production of Dark Matter in Association with Top Quark Pairs in the
  Single-lepton Final State in pp collisions at $\sqrt{s} = 8$ TeV}}, .

\bibitem{Khachatryan:2014rra}
{\scshape CMS} collaboration, V.~Khachatryan et~al., \emph{{Search for dark
  matter, extra dimensions, and unparticles in monojet events in proton?proton
  collisions at $\sqrt{s} = 8$ TeV}},
  \href{http://dx.doi.org/10.1140/epjc/s10052-015-3451-4}{\emph{Eur. Phys. J.}
  {\bf C75} (2015) 235}, [\href{http://arxiv.org/abs/1408.3583}{{\tt
  1408.3583}}].

\bibitem{Khachatryan:2015bbl}
{\scshape CMS} collaboration, V.~Khachatryan et~al., \emph{{Search for dark
  matter and unparticles produced in association with a Z boson in
  proton-proton collisions at $\sqrt s=$ 8  TeV}},
  \href{http://dx.doi.org/10.1103/PhysRevD.93.052011}{\emph{Phys. Rev.} {\bf
  D93} (2016) 052011}, [\href{http://arxiv.org/abs/1511.09375}{{\tt
  1511.09375}}].

\bibitem{Aad:2015dva}
{\scshape ATLAS} collaboration, G.~Aad et~al., \emph{{Search for dark matter
  produced in association with a Higgs boson decaying to two bottom quarks in
  $pp$ collisions at $\sqrt{s} = 8$ TeV with the ATLAS detector}},
  \href{http://dx.doi.org/10.1103/PhysRevD.93.072007}{\emph{Phys. Rev.} {\bf
  D93} (2016) 072007}, [\href{http://arxiv.org/abs/1510.06218}{{\tt
  1510.06218}}].

\bibitem{CMS:2015neg}
{\scshape CMS} collaboration, \emph{{Search for Resonances Decaying to Dijet
  Final States at $\sqrt{s} = 8$ TeV with Scouting Data}},  CMS-PAS-EXO-14-005.

\bibitem{Khachatryan:2015qba}
{\scshape CMS} collaboration, V.~Khachatryan et~al., \emph{{Search for diphoton
  resonances in the mass range from 150 to 850 GeV in pp collisions at
  $\sqrt{s} =$ 8 TeV}},
  \href{http://dx.doi.org/10.1016/j.physletb.2015.09.062}{\emph{Phys. Lett.}
  {\bf B750} (2015) 494--519}, [\href{http://arxiv.org/abs/1506.02301}{{\tt
  1506.02301}}].

\bibitem{Aad:2015fna}
{\scshape ATLAS} collaboration, G.~Aad et~al., \emph{{A search for $
  t\overline{t} $ resonances using lepton-plus-jets events in proton-proton
  collisions at $ \sqrt{s}=8 $ TeV with the ATLAS detector}},
  \href{http://dx.doi.org/10.1007/JHEP08(2015)148}{\emph{JHEP} {\bf 08} (2015)
  148}, [\href{http://arxiv.org/abs/1505.07018}{{\tt 1505.07018}}].

\bibitem{Khachatryan:2014sca}
{\scshape CMS} collaboration, V.~Khachatryan et~al., \emph{{Search for Standard
  Model Production of Four Top Quarks in the Lepton + Jets Channel in pp
  Collisions at $\sqrt{s}$ = 8 TeV}},
  \href{http://dx.doi.org/10.1007/JHEP11(2014)154}{\emph{JHEP} {\bf 11} (2014)
  154}, [\href{http://arxiv.org/abs/1409.7339}{{\tt 1409.7339}}].

\bibitem{Aad:2014vea}
{\scshape ATLAS} collaboration, G.~Aad et~al., \emph{{Search for dark matter in
  events with heavy quarks and missing transverse momentum in $pp$ collisions
  with the ATLAS detector}},
  \href{http://dx.doi.org/10.1140/epjc/s10052-015-3306-z}{\emph{Eur. Phys. J.}
  {\bf C75} (2015) 92}, [\href{http://arxiv.org/abs/1410.4031}{{\tt
  1410.4031}}].

\bibitem{Khachatryan:2015nua}
{\scshape CMS} collaboration, V.~Khachatryan et~al., \emph{{Search for the
  production of dark matter in association with top-quark pairs in the
  single-lepton final state in proton-proton collisions at $\sqrt{s}$ = 8
  TeV}}, \href{http://dx.doi.org/10.1007/JHEP06(2015)121}{\emph{JHEP} {\bf 06}
  (2015) 121}, [\href{http://arxiv.org/abs/1504.03198}{{\tt 1504.03198}}].

\bibitem{Cheung:2010zf}
K.~Cheung, K.~Mawatari, E.~Senaha, P.-Y. Tseng and T.-C. Yuan, \emph{{The Top
  Window for dark matter}},
  \href{http://dx.doi.org/10.1007/JHEP10(2010)081}{\emph{JHEP} {\bf 10} (2010)
  081}, [\href{http://arxiv.org/abs/1009.0618}{{\tt 1009.0618}}].

\bibitem{Lin:2013sca}
T.~Lin, E.~W. Kolb and L.-T. Wang, \emph{{Probing dark matter couplings to top
  and bottom quarks at the LHC}},
  \href{http://dx.doi.org/10.1103/PhysRevD.88.063510}{\emph{Phys. Rev.} {\bf
  D88} (2013) 063510}, [\href{http://arxiv.org/abs/1303.6638}{{\tt
  1303.6638}}].

\bibitem{Aad:2015zva}
{\scshape ATLAS} collaboration, G.~Aad et~al., \emph{{Search for new phenomena
  in final states with an energetic jet and large missing transverse momentum
  in pp collisions at $\sqrt{s}=$8 TeV with the ATLAS detector}},
  \href{http://dx.doi.org/10.1140/epjc/s10052-015-3517-3,
  10.1140/epjc/s10052-015-3639-7}{\emph{Eur. Phys. J.} {\bf C75} (2015) 299},
  [\href{http://arxiv.org/abs/1502.01518}{{\tt 1502.01518}}].

\bibitem{CMS:2015jdt}
{\scshape CMS} collaboration, \emph{{Search for dark matter with jets and
  missing transverse energy at 13 TeV}},  CMS-PAS-EXO-15-003.

\bibitem{Aad:2013oja}
{\scshape ATLAS} collaboration, G.~Aad et~al., \emph{{Search for dark matter in
  events with a hadronically decaying W or Z boson and missing transverse
  momentum in $pp$ collisions at $\sqrt{s} =$ 8 TeV with the ATLAS detector}},
  \href{http://dx.doi.org/10.1103/PhysRevLett.112.041802}{\emph{Phys. Rev.
  Lett.} {\bf 112} (2014) 041802}, [\href{http://arxiv.org/abs/1309.4017}{{\tt
  1309.4017}}].

\bibitem{Aad:2014vka}
{\scshape ATLAS} collaboration, G.~Aad et~al., \emph{{Search for dark matter in
  events with a Z boson and missing transverse momentum in pp collisions at
  $\sqrt{s}$=8 TeV with the ATLAS detector}},
  \href{http://dx.doi.org/10.1103/PhysRevD.90.012004}{\emph{Phys. Rev.} {\bf
  D90} (2014) 012004}, [\href{http://arxiv.org/abs/1404.0051}{{\tt
  1404.0051}}].

\bibitem{CMS:2015jha}
{\scshape CMS} collaboration, \emph{{Search for New Physics in the V-jet + MET
  final state}},  CMS-PAS-EXO-12-055.

\bibitem{ATLAS:monoZ}
{\scshape ATLAS} collaboration, \emph{{Search for dark matter produced in
  association with a hadronically decaying vector boson in $pp$ collisions at
  $\sqrt{s} = 13$ TeV with the ATLAS detector at the LHC}},
  ATLAS-CONF-2015-080.

\bibitem{Aad:2015yga}
{\scshape ATLAS} collaboration, G.~Aad et~al., \emph{{Search for Dark Matter in
  Events with Missing Transverse Momentum and a Higgs Boson Decaying to Two
  Photons in $pp$ Collisions at $\sqrt{s}=8$ TeV with the ATLAS Detector}},
  \href{http://dx.doi.org/10.1103/PhysRevLett.115.131801}{\emph{Phys. Rev.
  Lett.} {\bf 115} (2015) 131801}, [\href{http://arxiv.org/abs/1506.01081}{{\tt
  1506.01081}}].

\bibitem{monoh1}
{\scshape ATLAS} collaboration, \emph{{Search for new phenomena in events with
  missing transverse momentum and a Higgs boson decaying to two photons in p p
  collisions at $\sqrt{s} = 13$ TeV with the ATLAS detector}},
  ATLAS-CONF-2016-011.

\bibitem{monoh2}
{\scshape ATLAS} collaboration, \emph{{Search for Dark Matter in association
  with a Higgs boson decaying to $b$-quarks in $pp$ collisions at $\sqrt{s} =
  13$ TeV with the ATLAS detector}},  ATLAS-CONF-2016-019.

\bibitem{Goncalves:2016bkl}
D.~Goncalves, F.~Krauss, S.~Kuttimalai and P.~Maierhöfer, \emph{{Boosting
  invisible searches via $\boldsymbol{ZH}$: From the Higgs Boson to Dark Matter
  Simplified Models}},  \href{http://arxiv.org/abs/1605.08039}{{\tt
  1605.08039}}.

\bibitem{Khachatryan:2016ecr}
{\scshape CMS} collaboration, V.~Khachatryan et~al., \emph{{Search for narrow
  resonances in dijet final states at sqrt(s)=8 TeV with the novel CMS
  technique of data scouting}},  \href{http://arxiv.org/abs/1604.08907}{{\tt
  1604.08907}}.

\bibitem{Khachatryan:2015sma}
{\scshape CMS} collaboration, V.~Khachatryan et~al., \emph{{Search for resonant
  $t \bar t$ production in proton-proton collisions at $\sqrt{s}=8$ TeV}},
  \href{http://dx.doi.org/10.1103/PhysRevD.93.012001}{\emph{Phys. Rev.} {\bf
  D93} (2016) 012001}, [\href{http://arxiv.org/abs/1506.03062}{{\tt
  1506.03062}}].

\bibitem{Bevilacqua:2012em}
G.~Bevilacqua and M.~Worek, \emph{{Constraining BSM Physics at the LHC: Four
  top final states with NLO accuracy in perturbative QCD}},
  \href{http://dx.doi.org/10.1007/JHEP07(2012)111}{\emph{JHEP} {\bf 07} (2012)
  111}, [\href{http://arxiv.org/abs/1206.3064}{{\tt 1206.3064}}].

\bibitem{Chatrchyan:2013fea}
{\scshape CMS} collaboration, S.~Chatrchyan et~al., \emph{{Search for new
  physics in events with same-sign dileptons and jets in pp collisions at
  $\sqrt{s}$ = 8 TeV}}, \href{http://dx.doi.org/10.1007/JHEP01(2015)014,
  10.1007/JHEP01(2014)163}{\emph{JHEP} {\bf 01} (2014) 163},
  [\href{http://arxiv.org/abs/1311.6736}{{\tt 1311.6736}}].

\bibitem{Beck:2015cga}
L.~Beck, F.~Blekman, D.~Dobur, B.~Fuks, J.~Keaveney and K.~Mawatari,
  \emph{{Probing top-philic sgluons with LHC Run I data}},
  \href{http://dx.doi.org/10.1016/j.physletb.2015.04.043}{\emph{Phys. Lett.}
  {\bf B746} (2015) 48--52}, [\href{http://arxiv.org/abs/1501.07580}{{\tt
  1501.07580}}].

\bibitem{Greiner:2014qna}
N.~Greiner, K.~Kong, J.-C. Park, S.~C. Park and J.-C. Winter,
  \emph{{Model-Independent Production of a Top-Philic Resonance at the LHC}},
  \href{http://dx.doi.org/10.1007/JHEP04(2015)029}{\emph{JHEP} {\bf 04} (2015)
  029}, [\href{http://arxiv.org/abs/1410.6099}{{\tt 1410.6099}}].

\bibitem{Franceschini:2015kwy}
R.~Franceschini, G.~F. Giudice, J.~F. Kamenik, M.~McCullough, A.~Pomarol,
  R.~Rattazzi et~al., \emph{{What is the gamma gamma resonance at 750 GeV?}},
  \href{http://arxiv.org/abs/1512.04933}{{\tt 1512.04933}}.

\bibitem{Backovic:2015fnp}
M.~Backovic, A.~Mariotti and D.~Redigolo, \emph{{Di-photon excess illuminates
  Dark Matter}},  \href{http://arxiv.org/abs/1512.04917}{{\tt 1512.04917}}.

\bibitem{Mambrini:2015wyu}
Y.~Mambrini, G.~Arcadi and A.~Djouadi, \emph{{The LHC diphoton resonance and
  dark matter}},
  \href{http://dx.doi.org/10.1016/j.physletb.2016.02.049}{\emph{Phys. Lett.}
  {\bf B755} (2016) 426--432}, [\href{http://arxiv.org/abs/1512.04913}{{\tt
  1512.04913}}].

\bibitem{Bi:2015uqd}
X.-J. Bi, Q.-F. Xiang, P.-F. Yin and Z.-H. Yu, \emph{{The 750 GeV diphoton
  excess at the LHC and dark matter constraints}},
  \href{http://arxiv.org/abs/1512.06787}{{\tt 1512.06787}}.

\bibitem{D'Eramo:2016mgv}
F.~D'Eramo, J.~de~Vries and P.~Panci, \emph{{A 750 GeV Portal: LHC
  Phenomenology and Dark Matter Candidates}},
  \href{http://arxiv.org/abs/1601.01571}{{\tt 1601.01571}}.

\bibitem{Arina:2014yna}
C.~Arina, E.~Del~Nobile and P.~Panci, \emph{{Dark Matter with
  Pseudoscalar-Mediated Interactions Explains the DAMA Signal and the Galactic
  Center Excess}},
  \href{http://dx.doi.org/10.1103/PhysRevLett.114.011301}{\emph{Phys. Rev.
  Lett.} {\bf 114} (2015) 011301}, [\href{http://arxiv.org/abs/1406.5542}{{\tt
  1406.5542}}].

\bibitem{Cacciari:2008gp}
M.~Cacciari, G.~P. Salam and G.~Soyez, \emph{{The Anti-k(t) jet clustering
  algorithm}},
  \href{http://dx.doi.org/10.1088/1126-6708/2008/04/063}{\emph{JHEP} {\bf 0804}
  (2008) 063}, [\href{http://arxiv.org/abs/0802.1189}{{\tt 0802.1189}}].

\bibitem{Cacciari:2011ma}
M.~Cacciari, G.~P. Salam and G.~Soyez, \emph{{FastJet User Manual}},
  \href{http://dx.doi.org/10.1140/epjc/s10052-012-1896-2}{\emph{Eur. Phys. J.}
  {\bf C72} (2012) 1896}, [\href{http://arxiv.org/abs/1111.6097}{{\tt
  1111.6097}}].

\bibitem{Degrande:2011ua}
C.~Degrande, C.~Duhr, B.~Fuks, D.~Grellscheid, O.~Mattelaer and T.~Reiter,
  \emph{{UFO - The Universal FeynRules Output}},
  \href{http://dx.doi.org/10.1016/j.cpc.2012.01.022}{\emph{Comput. Phys.
  Commun.} {\bf 183} (2012) 1201--1214},
  [\href{http://arxiv.org/abs/1108.2040}{{\tt 1108.2040}}].

\bibitem{Pumplin:2002vw}
J.~Pumplin, D.~R. Stump, J.~Huston, H.~L. Lai, P.~M. Nadolsky and W.~K. Tung,
  \emph{{New generation of parton distributions with uncertainties from global
  QCD analysis}},
  \href{http://dx.doi.org/10.1088/1126-6708/2002/07/012}{\emph{JHEP} {\bf 07}
  (2002) 012}, [\href{http://arxiv.org/abs/hep-ph/0201195}{{\tt
  hep-ph/0201195}}].

\bibitem{Sjostrand:2006za}
T.~Sjostrand, S.~Mrenna and P.~Z. Skands, \emph{{PYTHIA 6.4 Physics and
  Manual}}, \href{http://dx.doi.org/10.1088/1126-6708/2006/05/026}{\emph{JHEP}
  {\bf 05} (2006) 026}, [\href{http://arxiv.org/abs/hep-ph/0603175}{{\tt
  hep-ph/0603175}}].

\bibitem{Field:2011iq}
R.~Field, \emph{{Min-Bias and the Underlying Event at the LHC}},
  \href{http://dx.doi.org/10.5506/APhysPolB.42.2631}{\emph{Acta Phys. Polon.}
  {\bf B42} (2011) 2631--2656}, [\href{http://arxiv.org/abs/1110.5530}{{\tt
  1110.5530}}].

\bibitem{Mangano:2006rw}
M.~L. Mangano, M.~Moretti, F.~Piccinini and M.~Treccani, \emph{{Matching matrix
  elements and shower evolution for top-quark production in hadronic
  collisions}},
  \href{http://dx.doi.org/10.1088/1126-6708/2007/01/013}{\emph{JHEP} {\bf 01}
  (2007) 013}, [\href{http://arxiv.org/abs/hep-ph/0611129}{{\tt
  hep-ph/0611129}}].

\bibitem{Alwall:2008qv}
J.~Alwall, S.~de~Visscher and F.~Maltoni, \emph{{QCD radiation in the
  production of heavy colored particles at the LHC}},
  \href{http://dx.doi.org/10.1088/1126-6708/2009/02/017}{\emph{JHEP} {\bf 02}
  (2009) 017}, [\href{http://arxiv.org/abs/0810.5350}{{\tt 0810.5350}}].

\bibitem{monojet:recast}
J.~Guo, E.~Conte and B.~Fuks, \emph{{MadAnalysis5 implementation of the CMS
  monojet search (EXO-12-048)}},  10.7484/INSPIREHEP.DATA.JAN2.UNDA.

\bibitem{ttmet:recast}
B.~Fuks and A.~Martini, \emph{{MadAnalysis5 implementation of the CMS search
  for dark matter production with top quark pairs in the single lepton channel
  (CMS-B2G-14-004)}},  10.7484/INSPIREHEP.DATA.MIHA.JR4G.

\bibitem{Bai:2012gs}
Y.~Bai, H.-C. Cheng, J.~Gallicchio and J.~Gu, \emph{{Stop the Top Background of
  the Stop Search}},
  \href{http://dx.doi.org/10.1007/JHEP07(2012)110}{\emph{JHEP} {\bf 07} (2012)
  110}, [\href{http://arxiv.org/abs/1203.4813}{{\tt 1203.4813}}].

\bibitem{Beltran:2010ww}
M.~Beltran, D.~Hooper, E.~W. Kolb, Z.~A.~C. Krusberg and T.~M.~P. Tait,
  \emph{{Maverick dark matter at colliders}},
  \href{http://dx.doi.org/10.1007/JHEP09(2010)037}{\emph{JHEP} {\bf 09} (2010)
  037}, [\href{http://arxiv.org/abs/1002.4137}{{\tt 1002.4137}}].

\bibitem{Goodman:2010ku}
J.~Goodman, M.~Ibe, A.~Rajaraman, W.~Shepherd, T.~M.~P. Tait and H.-B. Yu,
  \emph{{Constraints on Dark Matter from Colliders}},
  \href{http://dx.doi.org/10.1103/PhysRevD.82.116010}{\emph{Phys. Rev.} {\bf
  D82} (2010) 116010}, [\href{http://arxiv.org/abs/1008.1783}{{\tt
  1008.1783}}].

\bibitem{Goodman:2010yf}
J.~Goodman, M.~Ibe, A.~Rajaraman, W.~Shepherd, T.~M.~P. Tait and H.-B. Yu,
  \emph{{Constraints on Light Majorana dark Matter from Colliders}},
  \href{http://dx.doi.org/10.1016/j.physletb.2010.11.009}{\emph{Phys. Lett.}
  {\bf B695} (2011) 185--188}, [\href{http://arxiv.org/abs/1005.1286}{{\tt
  1005.1286}}].

\bibitem{Bai:2010hh}
Y.~Bai, P.~J. Fox and R.~Harnik, \emph{{The Tevatron at the Frontier of Dark
  Matter Direct Detection}},
  \href{http://dx.doi.org/10.1007/JHEP12(2010)048}{\emph{JHEP} {\bf 12} (2010)
  048}, [\href{http://arxiv.org/abs/1005.3797}{{\tt 1005.3797}}].

\end{thebibliography}\endgroup
\bibliographystyle{JHEP}

\end{document}